\documentclass[a4paper,onecolumn,showpacs,notitlepage,floatfix,nofootinbib,superscriptaddress,prd]{revtex4-1}
\usepackage{graphicx}
\usepackage{amsfonts}
\usepackage{amsmath}
\usepackage{hyperref}
\usepackage{float}
\usepackage{amssymb,amsbsy}
\usepackage{epstopdf}
\usepackage{color}
\newcommand*\Bell{\ensuremath{\boldsymbol\ell}}
\begin{document}

\title{Derivation of anisotropic dissipative fluid dynamics from the Boltzmann equation}
\author{Etele Moln\'ar}
\affiliation{Institut f\"ur Theoretische Physik, Johann Wolfgang Goethe--Universit\"at,
Max-von-Laue-Str.\ 1, D--60438 Frankfurt am Main, Germany}

\author{Harri Niemi}
\affiliation{Institut f\"ur Theoretische Physik, Johann Wolfgang Goethe--Universit\"at,
Max-von-Laue-Str.\ 1, D--60438 Frankfurt am Main, Germany}

\author{Dirk H.\ Rischke}
\affiliation{Institut f\"ur Theoretische Physik, Johann Wolfgang Goethe--Universit\"at,
Max-von-Laue-Str.\ 1, D--60438 Frankfurt am Main, Germany}


\begin{abstract}
Fluid-dynamical equations of motion can be derived from the Boltzmann equation in terms of an expansion around
a single-particle distribution function which is in local thermodynamical equilibrium, i.e., isotropic
in momentum space in the rest frame of a fluid element. However, in situations where the single-particle
distribution function is highly anisotropic in momentum space, such as the initial stage of heavy-ion collisions 
at relativistic energies, such an expansion is bound to break down. Nevertheless, one can still derive a fluid-dynamical theory, called anisotropic dissipative fluid dynamics,
in terms of an expansion around a single-particle distribution function, $\hat{f}_{0\bf k}$, which incorporates 
(at least parts of) the momentum anisotropy via a suitable parametrization. We construct such an
expansion in terms of polynomials in energy and momentum in the direction of the anisotropy
and of irreducible tensors in the two-dimensional momentum subspace orthogonal to both the fluid velocity and the
direction of the anisotropy. From the Boltzmann equation we then derive the set of equations of motion 
for the irreducible moments of the deviation of the single-particle distribution function from $\hat{f}_{0\bf k}$.
Truncating this set via the 14-moment approximation, 
we obtain the equations of motion of anisotropic dissipative fluid dynamics.
\end{abstract}

\maketitle

\section{Introduction}
\label{Introduction}

Fluid dynamics is the effective theory for the long-wavelength, small-frequency dynamics 
of a given system. As such, it finds widespread application in many different areas of physics.
Its basic equations represent nothing but the conservation of particle number\footnote{At relativistic 
energy scales, pair-creation processes require that particle-number conservation is replaced
by net-charge conservation. Bearing this in mind, for the sake of notational simplicity, we nevertheless 
keep the former in the present work.} and energy-momentum, 
i.e., in special-relativistic notation:
\begin{equation} \label{eom}
\partial_\mu N^\mu =0\;,\;\;\;  \partial_\mu T^{\mu \nu} = 0 \;,
\end{equation}
where $N^\mu$ is the particle four-current and
$T^{\mu \nu}$ is the energy-momentum tensor.
These five equations are not closed, since they involve in general 14 unknowns, the four components of
the particle four-current and the ten components of the (symmetric) energy-momentum tensor. Closure, and thus a
unique solution, is possible under certain simplifying assumptions.

The most drastic assumption is to reduce $N^\mu$ and $T^{\mu \nu}$ to the so-called \emph{ideal-fluid form\/},
\begin{equation} \label{idealfluidform}
N_0^\mu = n_0\, u^\mu\;,\;\;\; T_0^{\mu \nu} = e_0\, u^\mu u^\nu - P_0 \, \Delta^{\mu \nu}\;, 
\end{equation}
where $n_0,e_0,$ and $P_0$ are the particle number density, the energy density, and the thermodynamic pressure, 
respectively, $u^\mu=\gamma (1,\mathbf{v})$ is the fluid four-velocity [${\bf v}$ is the fluid three-velocity and 
$\gamma =(1-\mathbf{v}^{2})^{-1/2}$, such that $u^{\mu }u_{\mu }=1$, in units where $c=1$ and where
the metric tensor is $g^{\mu \nu} ={\rm diag}(+,-,-,-)$], 
and $\Delta^{\mu \nu} = g^{\mu \nu} - u^\mu u^\nu$
is the projector onto the three-dimensional subspace orthogonal to $u^\mu$, 
$\Delta^{\mu \nu}u_\nu = u_{\mu} \Delta^{\mu \nu} =0$. Now the equations of motion (\ref{eom}) contain
only six unknowns, and providing a thermodynamic equation of state (EoS) of the form
$P_0(e_0,n_0)$ closes them.

The microscopic picture underlying ideal fluid dynamics is that the fluid is in \emph{local thermodynamical equilibrium}, 
e.g.\ for dilute gases at each space-time point $x^\mu$ the single-particle distribution function assumes the form
\begin{equation}
f_{0\mathbf{k}}\left( \alpha _{0},\beta _{0}E_{\mathbf{k}u}\right) =\left[
\exp (-\alpha _{0}+\beta _{0}E_{\mathbf{k}u})+a\right] ^{-1}\;.
\label{Equilibrium_dist}
\end{equation}
Here, $\alpha _{0}= \mu \beta _{0}$, where 
$\mu$ is the chemical potential associated with the particle density $n_0$,  $\beta _{0}=1/T$ is the inverse
temperature (in units where $k_B=1$) and $E_{\mathbf{k}u}\equiv k^{\mu }u_{\mu }$, where
$k^\mu = (k_0, {\bf k})$ is the particle four-momentum. In the local rest (LR) frame of the fluid, 
$u_{LR}^{\mu }=\left(1,0,0,0\right)$, $E_{\mathbf{k}u}= \sqrt{\mathbf{k}^{2}+m_{0}^{2}}$ 
is the (relativistic on-shell) energy of a particle with mass $m_{0}$.  
In local thermodynamical equilibrium the quantities 
$\alpha_{0},\,\beta _{0}$, and $u^{\mu }$ are functions of $x^\mu$.  
In global thermodynamical equilibrium, they assume constant values at all $x^\mu$. 
The quantity $a=1$ for Fermi--Dirac and $a=-1$ for
Bose--Einstein statistics, respectively. Boltzmann statistics is obtained in
the limit $a\rightarrow 0$. Equation (\ref{Equilibrium_dist}) represents the
well-known J\"{u}ttner distribution function \cite{Juttner,Juttner_quantum},
which for $a=0$ is identical to the relativistic Maxwell-Boltzmann
distribution function. 

The particle four-current and the energy-momentum tensor are simply the
first and second moments of $f_{0\mathbf{k}}$,
\begin{equation} \label{moments_in_equilibrium}
N_0^\mu = \left\langle k^\mu \right\rangle_0 \;, \;\;\; T_0^{\mu \nu} = \left\langle k^\mu k^\nu \right\rangle_0\;,
\end{equation}
where (in units where $\hbar = 1$)
\begin{equation} \label{average_in_equilibrium}
\langle \cdots \rangle_0 = \int dK\, (\cdots)\; f_{0\mathbf{k}}\;.
\end{equation}
Here, $dK = g d^3\mathbf{k}/[(2\pi)^3k_0]$, where $g$ counts the number of internal degrees of freedom. 
Inserting Eq.\ (\ref{Equilibrium_dist}) into Eq.\ (\ref{moments_in_equilibrium}) it can be shown that $N_0^\mu$
and $T_0^{\mu \nu}$ assume the ideal-fluid form (\ref{idealfluidform}). The thermodynamic variables
$e_0,P_0,n_0$ are functions of $\alpha_0, \beta_0$, such that, together with the three independent components 
of $u^\mu$, there are in total five independent quantities, which (for given initial conditions) 
are uniquely determined by the five equations of motion (\ref{eom}).

In general, the equations of motion (\ref{eom}) can also be closed by providing nine additional equations involving
no further unknowns than the 14 components of $N^\mu$ and $T^{\mu \nu}$.
However, the choice of these additional equations needs further assumptions. 
One possibility is to use the second law of thermodynamics, which requires that any
closed system approaches global thermodynamical equilibrium. If the microscopic processes driving the system
towards equilibrium happen on space-time scales that are much smaller than the ones on which the
fluid-dynamical fields $N^\mu$, $T^{\mu \nu}$ vary, it is reasonable to assume that the system will rapidly
approach a state which is sufficiently close to local thermodynamical equilibrium. In other words,
defining
\begin{equation}
\delta N^\mu \equiv N^\mu - N_0^\mu\;, \;\;\; \delta T^{\mu \nu} \equiv T^{\mu \nu} - T_0^{\mu \nu}\;,
\end{equation}
the deviations of particle four-current and energy-momentum tensor from the ideal-fluid form become small, 
$|\delta N^\mu| \ll |N_0^\mu|$, $|\delta T^{\mu \nu}|\ll |T_0^{\mu \nu}|$. 
As explained above, $N_0^\mu$ and $T_0^{\mu \nu}$ are completely determined
by five independent quantities (for a given equation of state $P_0(e_0,n_0)$). Therefore,
$\delta N^\mu$ and $\delta T^{\mu \nu}$ involve in total nine independent quantities. 
Thus, one has to specify nine additional equations in order to close the equations of motion (\ref{eom}).
These $14$ equations then define a theory of \emph{dissipative\/}  (or  \emph{viscous}) \emph{fluid dynamics}. 

Unlike ideal fluid dynamics, in dissipative fluid dynamics 
the fluid four-velocity, and thus the choice of the LR frame of the fluid, is not uniquely defined. 
The two most popular choices are the Eckart frame \cite{Eckart:1940te}, where 
\begin{equation}
u^\mu = \frac{N^\mu}{\sqrt{N^\alpha N_\alpha}}
\label{LR_Eckart_1_2}
\end{equation}
is proportional to the flow of particles, and the Landau frame \cite{Landau_book},
where 
\begin{equation}
u^\mu = \frac{T^{\mu \nu}u_\nu}{\sqrt{u_\alpha T^{\alpha \beta} T_{\beta \gamma} u^\gamma}}
\label{LR_Landau_1_2}
\end{equation} 
is proportional to the flow of energy. In the Eckart frame, the nine independent variables, or
\emph{dissipative currents},
which enter besides $n_0,e_0$, and $u^\mu$, are
the \emph{bulk viscous pressure\/} $\Pi$ (which is the difference between the isotropic pressure 
$P=-\frac{1}{3} \Delta_{\mu \nu} T^{\mu \nu}$ and the thermodynamic pressure 
$P_0=-\frac{1}{3} \Delta_{\mu \nu} T_0^{\mu \nu} $, $\Pi \equiv P- P_0$, or, equivalently, up to a factor
$-1/3$ is equal to the trace of $\delta T_{\mu \nu}$, $\Pi = - \frac{1}{3} \Delta_{\mu \nu} \delta T^{\mu \nu}$), the
\emph{shear-stress tensor\/} $\pi^{\mu \nu}$ [which is the trace-free part of $\delta T^{\mu \nu}$, 
$\pi^{\mu \nu} = \Delta^{\mu \nu}_{\alpha \beta} \delta T^{\alpha \beta}$, where
$\Delta^{\mu \nu}_{\alpha \beta} = \frac{1}{2} (\Delta^\mu_\alpha \Delta^\nu_\beta + \Delta^\mu_\beta \Delta^\nu_\alpha)
- \frac{1}{3} \Delta^{\mu \nu} \Delta_{\alpha \beta}$], and the \emph{energy diffusion current\/} $W^{\mu}$ 
(which is the flow of energy relative to the flow of particles, 
$W^{\mu} = \Delta^{\mu \alpha} \delta T_{\alpha \beta} u^\beta$). 
In the Landau frame, the latter is replaced by the
\emph{particle diffusion current} $V^\mu$ (which is the flow of particles relative to the flow of energy, 
$V^\mu = \Delta^{\mu \alpha} \delta N_\alpha$).

Providing equations for the dissipative currents closes the equations of motion (\ref{eom}) and defines
a theory of dissipative fluid dynamics. However, there is 
considerable freedom in choosing these additional equations.
Using the second law of thermodynamics 
one can show \cite{Israel} that, to leading order, the dissipative currents must be proportional to gradients of
$\alpha_0$, $\beta_0$, and $u^\mu$, e.g.\ in the Landau frame,
\begin{eqnarray}
\Pi & = & - \zeta\, \partial_\mu u^\mu\;, \notag \\
V^\mu & = & \kappa_n\, \Delta^{\mu \nu} \partial_\nu \alpha_0\;, \notag \\
\pi^{\mu \nu} & = & 2 \eta \,\Delta^{\mu \nu}_{\alpha \beta} \partial^\alpha u^\beta\;, \label{NS}
\end{eqnarray}
where $\zeta$ is the bulk-viscosity, $\eta$ the shear-viscosity, and $\kappa_n$ the particle-diffusion coefficient, 
respectively.
Equations (\ref{NS}) are the relativistic analogues of the {\em Navier-Stokes equations\/} 
\cite{Eckart:1940te,Landau_book}. Since the dissipative
currents are of first order in gradients, one also speaks of a {\em first-order theory\/} of dissipative fluid dynamics. 
Note that the dissipative currents are expressed solely in terms of the quantities $\alpha_0, \beta_0$,
and $u^\mu$, which are already present in ideal fluid dynamics. Inserting Eqs.\ (\ref{NS}) into the equations 
of motion (\ref{eom}) thus closes the latter.

Unfortunately, directly using Eqs.\ (\ref{NS}) in the equations of motion (\ref{eom}) renders them parabolic,
leading to an unstable and acausal theory \cite{his}, at least in the relativistic case. 
The reason is that, in Eqs.\ (\ref{NS}), the dissipative currents (the left-hand sides of these equations) 
react {\em instantaneously\/} to the dissipative forces (the right-hand sides). 
This unphysical behavior can be cured \cite{Pu:2009fj} by assuming a non-vanishing
relaxation time for the dissipative currents, as suggested in Refs.\ 
\cite{Israel,Muller_67,Stewart:1972hg,Stewart:1977,Israel:1979wp}. In its most simple form, these equations read
\begin{eqnarray}
\tau_\Pi D \Pi + \Pi & = & - \zeta\, \partial_\mu u^\mu\;, \notag \\
\tau_n \Delta^{\mu \nu} D V_\nu + V^\mu & = & \kappa_n\, \Delta^{\mu \nu} \partial_\nu \alpha_0\;, \notag \\
\tau_\pi \Delta^{\mu \nu}_{\alpha \beta} D \pi^{\alpha \beta} + \pi^{\mu \nu} 
& = & 2 \eta \,\Delta^{\mu \nu}_{\alpha \beta} \partial^\alpha u^\beta\;, \label{IS}
\end{eqnarray}
where $D \equiv u^\mu \partial_\mu$ is the comoving derivative and $\tau_\Pi, \tau_n$, and $\tau_\pi$ are
the relaxation times associated with the individual dissipative currents. Note that these equations promote the
dissipative currents to dynamical variables, which relax towards their Navier-Stokes values (\ref{NS}) on
timescales given by $\tau_\Pi, \tau_n,$ and $\tau_\pi$, respectively. On account of the Navier-Stokes equations
(\ref{NS}), the dissipative currents themselves are already of first order in gradients.
Since the equations (\ref{IS}) contain derivatives of the dissipative currents, these equations are formally
of second order in gradients. One refers to formulations of dissipative fluid dynamics which contain
terms of second order in gradients as {\em second-order\/} theories of dissipative fluid dynamics.

The microscopic picture behind dissipative fluid dynamics is that the single-particle distribution function
$f_{\bf k}$ is, albeit not identical to the local-equilibrium form (\ref{Equilibrium_dist}), still sufficiently close
to it, i.e., 
\begin{equation} \label{f0+df}
f_{\bf k} = f_{0{\bf k}} + \delta f_{\bf k}\;,
\end{equation}
with the deviation
$\delta f_{\bf k}$ from local thermodynamical equilibrium being small, $|\delta f_{\bf k}| \ll 1$. The 
equations of motion of dissipative fluid dynamics can be derived from this microscopic picture using 
the Boltzmann equation \cite{deGroot,Cercignani_book}, 
\begin{equation}
k^{\mu }\partial _{\mu }f_{\mathbf{k}}=C\left[ f\right] \;,
\label{Boltzmann_eq_2}
\end{equation}%
where $C[f]$ is the collision integral. Employing the microscopic definitions of the particle four-current and
the energy-momentum tensor,
\begin{equation} \label{moments_in_nonequilibrium}
N^\mu = \left\langle k^\mu \right\rangle \;, \;\;\; T^{\mu \nu} = \left\langle k^\mu k^\nu \right\rangle\;,
\end{equation}
where, analogous to Eq.\ (\ref{average_in_equilibrium}), 
\begin{equation} \label{average_in_nonequilibrium}
\langle \cdots \rangle = \int dK\; (\cdots)\, f_{\mathbf{k}}\;,
\end{equation}
the equations of motion (\ref{eom}) are nothing but the
zeroth and first moments of the Boltzmann equation (assuming that the collision integral respects 
particle-number and energy-momentum conservation). Closure of the equations of motion (\ref{eom}) can
be achieved by considering higher moments of the Boltzmann equation. This strategy has been pioneered by
the authors of Refs.\ \cite{Stewart:1972hg,Stewart:1977,Israel:1979wp}. Honoring their work, the resulting
equations [the most simple form of which reads as given in Eqs.\ (\ref{IS})] are commonly referred to as
Israel-Stewart (IS) equations. Similar approaches have been
pursued in Refs.\ \cite{Geroch:1990bw,Grmela:1997zz,Betz:2008me}. 

Recently, following the original idea of Grad \cite{Grad}, the deviation $\delta f_{\bf k}$ has
been expanded in a set of orthogonal polynomials in energy, $E_{{\bf k}u} = k^\mu u_\mu$, and 
irreducible tensors in momentum, $1, k^{\langle \mu \rangle}, k^{\langle \mu} k^{\nu \rangle}, \ldots$,
where the angular brackets denote the symmetrized and, for more than one Lorentz index, tracefree
projection orthogonal to $u^\mu$, for details see Refs.\ \cite{deGroot,Denicol:2012cn}.  
Subsequently, the equations of motion of dissipative fluid dynamics have 
been derived by a systematic truncation (based on power counting in Knudsen and
inverse Reynolds numbers) of the set of moments
of the Boltzmann equation \cite{Denicol:2012cn,Denicol:2012es,Molnar:2013lta}. The lowest-order truncation gives the
IS equations (\ref{IS}), including all other terms that are of second order in gradients (equivalent to terms
of second order in either Knudsen or inverse Reynolds number, or of first order in the product of these).
The advantage of this approach is that it can be systematically improved and applied to situations where
the Knudsen or inverse Reynolds numbers are not small \cite{Denicol:2012vq}.

As a power series in gradients (or in Knudsen and inverse Reynolds number), the range of applicability of dissipative 
fluid dynamics is (at least formally) restricted to situations where the deviation $\delta f_{\bf k}$ of the single-particle
distribution function $f_{\bf k}$ from the one in local thermodynamical equilibrium, $f_{0{\bf k}}$, is small.
There are, however, situations where gradients, or $\delta f_{\bf k}$, respectively, become so large that
a power-series expansion is expected to break down. In this case, one should modify the above described 
approach to (dissipative) fluid dynamics, explicitly taking into account deviations from local thermodynamical
equilibrium to all orders. One of these situations is the initial stage of ultrarelativistic heavy-ion collisions. In this
case, the gradient of the fluid velocity in beam ($z-$)direction is of the order of the inverse lifetime of the system,
$\partial_z v_z \sim 1/t$ \cite{Bjorken:1982qr}, which can in principle become arbitrarily large as one approaches 
the moment of impact of the colliding nuclei (at $t=0$). This large gradient is reflected in a 
single-particle distribution function which is highly anisotropic in the $z-$direction in momentum space.

Fluid dynamics for anisotropic single-particle distribution functions have been studied a long time ago \cite{Barz:1987pq}.
Incidentally, the physics motivation was very similar to the case described above, 
namely to account for  momentum-space
anisotropies in the initial stage of heavy-ion collisions, although at that time the available beam energies were orders
of magnitude smaller. Recently, Florkowski, Martinez, Ryblewski, and Strickland
\cite{Martinez:2009ry,Martinez:2010sc,Martinez:2010sd,Florkowski:2010cf,Ryblewski:2010bs,
Ryblewski:2011aq,Ryblewski:2012rr,Martinez:2012tu} proposed
a formulation of anisotropic fluid dynamics based on a 
single-particle distribution function, which is anisotropic in momentum space and
whose specific form is motivated by the 
gluon fields created in the initial stages of a heavy-ion collision \cite{Romatschke:2003ms}. 
In this formulation, $f_{\bf k}$ is assumed to have a spheroidal shape in the LR frame,
which is deformed in the $z-$direction with respect to a spherically symmetric $f_{0{\bf k}}$. 
The degree of anisotropy is quantified by a single parameter. On the other hand,
having in mind applying their formalism to the evolution of the fluid in the
mid-rapidity region of a heavy-ion collision, where there is no conserved
net-charge, the authors of Refs.\ 
\cite{Martinez:2009ry,Martinez:2010sc,Martinez:2010sd,Florkowski:2010cf,Ryblewski:2010bs,
Ryblewski:2011aq,Ryblewski:2012rr,Martinez:2012tu}
assumed that there is no parameter like $\alpha_0$ which controls the particle-number (or more precisely, 
net-charge) density.
Thus, besides energy-momentum conservation [the second equation (\ref{eom})], a single
additional equation is necessary to close the system of equations of motion. The simplest possibility is
the zeroth moment of the Boltzmann equation, with the collision integral taken in relaxation-time
approximation \cite{Martinez:2010sc}. Another possibility is the entropy equation (with a non-vanishing
source term, as entropy must grow in non-equilibrium situations) \cite{Florkowski:2010cf}. But in principle,
also higher moments of the Boltzmann equation could serve to determine the anisotropy 
parameter \cite{Nopoush:2014pfa,Alqahtani:2015qja,Florkowski:2015cba}.

In this formulation of anisotropic fluid dynamics, while the anisotropy parameter is a function of space-time, 
the form of the single-particle distribution function as a function of the
anisotropy parameter always remains the same. Even though one may generalize
this idea by introducing additional parameters to capture the anisotropy to an even better degree \cite{tinti},
this does not change the fact that one is always restricted by the chosen form 
of the anisotropic distribution function.
In this sense, this approach is rather a generalization of
{\em ideal\/} fluid dynamics, where it is assumed that the single-particle distribution always has
the form (\ref{Equilibrium_dist}), than of {\em dissipative\/} fluid dynamics. 
Even so, when compared to ideal fluid dynamics, this approach includes dissipative 
effects due to the anisotropic single-particle distribution function.

Generalizing dissipative fluid dynamics to a theory of anisotropic dissipative fluid dynamics, 
one should take an anisotropic single-particle distribution function, called $\hat{f}_{0{\bf k}}$ 
in the following, as the starting point for an expansion of the general single-particle distribution function, i.e., 
\begin{equation}
f_{\mathbf{k}} \equiv \hat{f}_{0\mathbf{k}}+\delta \hat{f}_{\mathbf{k}}\;.  \label{kinetic:f=f0+df_2}
\end{equation}
While this looks similar to Eq.\ (\ref{f0+df}), the rationale behind an expansion around 
$\hat{f}_{0\mathbf{k}}$ instead of
around $f_{0\mathbf{k}}$ as in Eq.\ (\ref{f0+df}) is the following: in the case of a pronounced anisotropy, 
$\delta f_{\bf k}$ in Eq.\ (\ref{f0+df}) may be of similar magnitude (or even larger) than $f_{0{\bf k}}$,
i.e., an expansion around the local equilibrium distribution (\ref{Equilibrium_dist}) converges badly.
However, taking a suitably chosen $\hat{f}_{0\mathbf{k}}$, we ensure that
$|\delta \hat{f}_{\mathbf{k}}|\ll |\delta f_{\mathbf{k}}|$, so that the convergence properties of 
the series expansion are vastly improved. 

This strategy has been applied in Ref.\ \cite{Bazow:2013ifa} to derive equations of motion for 
anisotropic dissipative fluid dynamics (or, as called there, ``viscous anisotropic hydrodynamics''), 
based on the method presented in Refs.\ \cite{Grad, Denicol:2010xn}. 
However, the form of $\delta \hat{f}_{\bf k}$ was a simple linear function of the tensors $1, k^\mu, k^\mu k^\nu$ 
with 14 unknown coefficients. These were then expressed in terms
of the 14 fluid-dynamical variables $N^\mu, T^{\mu \nu}$ (or an equivalent set of variables) via a linear
mapping procedure (which employs the so-called Landau-matching conditions and the choice of LR frame).
This strategy is analogous to that of Ref.\ \cite{Israel:1979wp} for the derivation
of ``ordinary'' dissipative fluid dynamics. 

The disadvantages of this approach were explained in the
introduction of Ref.\ \cite{Denicol:2012cn}, the main one being that it is not systematically improvable. 
In order to provide an improvable framework in the case of ``ordinary'' dissipative fluid dynamics, 
it was essential to use a set of orthogonal polynomials in energy, $E_{{\bf k}u} = k^\mu u_\mu$, 
and irreducible tensors in momentum, $1, k^{\langle \mu \rangle}, 
k^{\langle \mu} k^{\nu \rangle}, \ldots$, in the expansion of $\delta f_{\bf k}$. 
This set was used for an expansion
of $\delta \hat{f}_{\bf k}$ in deriving equations of motion 
for anisotropic dissipative fluid dynamics in Ref.\ \cite{,Bazow:2015zca}. However, in the case
of anisotropic dissipative fluid dynamics, besides $u^\mu$ there is an additional space-like four-vector, 
$l^\mu$, which defines the direction of the anisotropy (as explained above, usually taken to be the $z-$direction) 
and can be chosen to be orthogonal to $u^\mu$, $l^\mu u_\mu =0$. 
In place of $\delta f_{\bf k}$ in Eq.\ (\ref{f0+df}), one
now needs to expand $\delta \hat{f}_{\bf k}$ in Eq.\ (\ref{kinetic:f=f0+df_2}). This expansion involves
orthogonal polynomials in the {\em two\/} variables 
$E_{{\bf k}u}$ and the particle momentum in the direction of the anisotropy, 
$E_{{\bf k}l} = - k^\mu l_\mu$, as well as irreducible tensors
which are orthogonal to {\em both\/} $u^\mu$ {\em and\/} $l^\mu$. 
A derivation of anisotropic dissipative fluid dynamics along these lines
is the main goal of the present paper. In this way, we provide a systematically improvable 
framework for anisotropic dissipative fluid dynamics.

This paper is organized as follows. 
In Sec.\ \ref {sect:general_fluid_variables} we introduce the tensor decomposition of
fluid-dynamical variables with respect to the time-like fluid four-velocity $u^{\mu }$ 
and the space-like four-vector $l^{\mu }$ ($l^{\mu}l_{\mu }=-1$), which 
is usually chosen to point into the direction of the spatial anisotropy. In
Secs.\ \ref{sect:isotropic_state} and \ref{sect:anisotropic_state} we study
two limiting cases of this tensor decomposition. The first is the well-known
ideal-fluid limit where only tensor structures proportional to $u^{\mu }$
and $\Delta ^{\mu \nu}$ appear in the moments of the
single-particle distribution function \cite{deGroot,Cercignani_book}. The
second is the anisotropic case, where the single-particle distribution
function also depends on $l^{\mu }$ besides $u^{\mu }$ and where tensor
structures proportional to $u^{\mu }$, $l^{\mu }$, their direct product, and
the two-space projector orthogonal to both $u^{\mu }$ and $l^{\mu }$ 
\cite{Gedalin_1991,Gedalin_1995,Huang:2009ue,Huang:2011dc}, 
\begin{equation}
\Xi ^{\mu\nu }\equiv \Delta^{\mu \nu} + l^\mu l^\nu = g^{\mu \nu }-u^{\mu }u^{\nu }+l^{\mu }l^{\nu }\;,
\end{equation} 
appear in the moments of the single-particle distribution function.
In Sec.\ \ref{sect:expansion} we present the expansion of the
single-particle distribution function $f_{\bf k}$ around the anisotropic state $\hat{f}_{0{\bf k}}$. In
analogy to Refs.\ \cite{deGroot,Denicol:2012cn}, this is done in terms of an
orthogonal basis of irreducible tensors in momentum space. However, in
contrast to previous work, these tensors are not only orthogonal to $u^{\mu} $ 
but also to $l^{\mu }$. Then, in Sec.\ \ref{sect:Eqs_of_motion}, taking
moments of the Boltzmann equation we derive the equations of motion for the
irreducible moments of the single-particle distribution function up to
tensor-rank two. These equations are not yet closed and need to be truncated
in order to derive the fluid-dynamical equations of motion in terms of
conserved quantities, i.e., the particle four-current $N^{\mu }$ and the
energy-momentum tensor $T^{\mu \nu }$. In Sec.\ \ref{sect:Collision integral}, 
we study the explicit form of the collision integral. 
Finally, in Sec.\ \ref{fluideom} we give the derivation of the fluid-dynamical 
equations of motion in the 14--moment approximation. 
Section \ref{conclusions} concludes this work with a summary and an outlook.
Details of our calculations are delegated to various appendices.

We adopt natural units, $\hbar = c = k_{B}=1$, throughout this work. 
The symmetrization of tensor indices is denoted by 
$\left( \ \right) $ around Greek indices, e.g., $A^{\left( \mu \nu \right)}
=\left( A^{\mu \nu }+A^{\nu \mu }\right) /2$ while $\left[ \ \right] $
means the antisymmetrization of indices, i.e., $A^{\left[ \mu \nu \right]}
=\left( A^{\mu \nu }-A^{\nu \mu }\right) /2$. The projection of an
arbitrary four-vector $A^{\mu }$ onto the directions orthogonal to $u^{\mu }$
will be denoted by $\left\langle \ \right\rangle $ around Greek indices, 
$A^{\left\langle \mu \right\rangle }=\Delta ^{\mu \nu }A_{\nu }$. The
projection of an arbitrary four-vector $A^{\mu }$ onto the directions
orthogonal to both $u^{\mu }$ and $l^{\mu }$ will be denoted by $\left\{ \
\right\} $ around indices, $A^{\left\{ \mu \right\} }=\Xi ^{\mu \nu }A_{\nu} $. 
Projections of higher-rank tensors will be denoted in a similar manner.
In case of an arbitrary anisotropy the four-momentum of particles is 
$k^{\mu}=E_{\mathbf{k}u}u^{\mu }+E_{\mathbf{k}l}l^{\mu }+k^{\left\{ \mu \right\} }$, 
where $k^{\left\{ \mu\right\} }=\Xi ^{\mu \nu }k_{\nu }$ 
are the components of the momentum orthogonal to $u^{\mu }$ and $l^{\mu }$.

\section{Fluid-dynamical variables}
\label{sect:general_fluid_variables}

In this section we introduce the tensor decomposition of the
fluid-dynamical variables with respect to the time-like fluid four-velocity $u^{\mu }$, as well as
the decomposition with respect to both $u^\mu$ and the space-like four-vector $l^{\mu }$. 
Details of the calculation can be found in Apps.\ \ref{Quantities} and \ref{appB}.

Equation (\ref{moments_in_nonequilibrium}) gives
the microscopic definition of the particle 4-current and the energy-momentum tensor in terms of the 
first and second moment of the single-particle distribution function $f_{\mathbf{k}}$.
Using Eqs.\ (\ref{4_vector_decomposition_u}), (\ref{isotropic_decomposition}), 
and the fact that the energy-momentum tensor is
symmetric, the tensor decomposition of $N^\mu$ and $T^{\mu \nu}$ with 
respect to $u^{\mu }$ and $\Delta^{\mu \nu }$ reads 
\begin{eqnarray}
N^{\mu } & \equiv & \langle k^\mu \rangle = n \,u^{\mu }+V^{\mu }\;,
\label{kinetic:N_mu} \\
T^{\mu \nu } & \equiv & \langle k^\mu k^\nu \rangle = e\, u^{\mu}u^{\nu }-P\, \Delta ^{\mu \nu }
+2\,W^{\left( \mu \right. }u^{\left. \nu \right)}+\pi^{\mu \nu }\; , \label{kinetic:T_munu}
\end{eqnarray}
where the angular brackets denote the momentum-space average defined in 
Eq.\ (\ref{average_in_nonequilibrium}). 
On the one hand, the various quantities appearing in the tensor decomposition on the right-hand side 
can be expressed in terms of different projections of the particle four-current and the energy-momentum tensor.
On the other hand,  employing Eqs.\ (\ref{k_mu_iso}), 
(\ref{B7}), these quantities can be identified as 
moments of the single-particle distribution function $f_{\bf k}$. Following this strategy,
the scalar coefficients in Eqs.\ (\ref{kinetic:N_mu}), 
(\ref{kinetic:T_munu}), i.e., the particle density $n$, the energy density $e$, and the isotropic pressure $P$, read
\begin{eqnarray}
n &\equiv &\left\langle E_{\mathbf{k}u}\right\rangle =N^{\mu }u_{\mu }\;,
\label{kinetic:n} \\
e &\equiv &\left\langle E_{\mathbf{k}u}^{2}\right\rangle =T^{\mu \nu }u_{\mu}u_{\nu }\;,  \label{kinetic:e} \\
P &\equiv &-\frac{1}{3}\left\langle \Delta ^{\mu \nu }k_{\mu }k_{\nu}\right\rangle 
=-\frac{1}{3}T^{\mu \nu }\Delta _{\mu \nu }\;,
\label{kinetic:P}
\end{eqnarray}
while the particle and energy-momentum diffusion currents $V^{\mu }$ and $W^{\mu }$, 
respectively, are 
\begin{eqnarray}
V^{\mu } &\equiv &\left\langle k^{\left\langle \mu \right\rangle}\right\rangle 
=\Delta _{\nu }^{\mu }N^{\nu }\;,  \label{kinetic:V_mu} \\
W^{\mu } &\equiv &\left\langle E_{\mathbf{k}u}k^{\left\langle \mu\right\rangle }\right\rangle 
=\Delta _{\alpha }^{\mu }T^{\alpha \beta}u_{\beta }\;.  \label{kinetic:W_mu}
\end{eqnarray}
Both are orthogonal to the flow velocity, $V^{\mu }u_{\mu }=W^{\mu}u_{\mu}=0 $.
Finally, the shear-stress tensor 
\begin{equation}
\pi ^{\mu \nu }\equiv \left\langle k^{\left\langle \mu \right. }k^{\left. \nu \right\rangle }\right\rangle 
=\Delta_{\alpha \beta }^{\mu \nu}T^{\alpha \beta } \; , \label{kinetic:pi_munu}
\end{equation}
is the part of the energy-momentum tensor that is symmetric, $\pi ^{\mu \nu}=\pi ^{\nu \mu }$, 
traceless, $\pi ^{\mu \nu }g_{\mu \nu }=0$, and
orthogonal to the flow velocity $\pi ^{\mu \nu }u_{\mu }=0$. Here 
$\Delta _{\alpha \beta }^{\mu \nu}$ is the projection tensor
defined in Eq.\ (\ref{Delta_traceless}), such that 
$k^{\langle\mu} k^{\nu \rangle } =\Delta _{\alpha \beta }^{\mu \nu} k^{\alpha} k^{\beta}$.

The particle four-current and the energy-momentum tensor can also be
decomposed with respect to $u^{\mu }$, the space-like four-vector 
$l^{\mu }$, and the projection tensor $\Xi ^{\mu \nu }$. Using Eqs.\  (\ref{4_vector_decomposition_u_l}),
(\ref{anisotropic_decomposition}) and the fact that the energy-momentum tensor is symmetric,
we obtain 
\begin{eqnarray}
N^{\mu } &=&n\, u^{\mu }+n_{l}\, l^{\mu }+V_{\perp }^{\mu }\;,
\label{kinetic:N_mu_u_l} \\
T^{\mu \nu } &=&e\, u^{\mu }u^{\nu }+2\, M\, u^{\left( \mu \right. }l^{\left. \nu\right) }
+P_{l}\, l^{\mu }l^{\nu }-P_{\perp }\, \Xi ^{\mu \nu }+2\, W_{\perp u}^{\left( \mu \right. }
u^{\left. \nu \right) }+2\, W_{\perp l}^{\left( \mu\right. }l^{\left. \nu \right) }
+\pi _{\perp }^{\mu \nu }\;.
\label{kinetic:T_munu_u_l}
\end{eqnarray}
Here, in addition to the previously defined quantities, we have denoted the part
of the particle diffusion current in the $l^{\mu }$ direction by $n_{l}$, while the particle
diffusion current orthogonal to both four-vectors is denoted by $V_{\perp}^{\mu }$. 
The pressure in the transverse direction is denoted by $P_{\perp} $, 
while the pressure in the longitudinal direction is 
$P_{l}$. The projection of the energy-momentum tensor in both $u^{\mu }$ and 
$l^{\nu }$ direction is denoted by $M$. The projection in either
the $u^{\mu }$ or $l^{\mu }$ direction and orthogonal to both
directions is denoted by $W_{\perp u}^{\mu }$ or $W_{\perp l}^{\mu }$, respectively. The
only rank-two tensor in the subspace orthogonal to both distinguished
four-vectors is given by the "transverse" shear-stress tensor $\pi _{\perp }^{\mu \nu }$.
Note that the various subscripts \emph{u} (for projection onto the direction
of $u^{\mu }$), \emph{l} (for projection onto the direction of $l^{\mu }$)
and $\perp $ (for projection onto the direction "perpendicular" to both 
$u^{\mu }$ and $l^{\mu }$) serve as reminders of the directions that the
various quantities are projected onto.

These newly defined quantities can either be expressed in terms of different
projections of the conserved particle four-current and energy-momentum
tensor or, with the help of Eqs.\ (\ref{k_mu_aniso}), (\ref{B8}), identified as moments
of $f_{\bf k}$, i.e.,
\begin{eqnarray}
n_{l} &\equiv &\left\langle E_{\mathbf{k}l}\right\rangle =-N^{\mu }l_{\mu}\;,  \label{kinetic:n_l} \\
M &\equiv &\left\langle E_{\mathbf{k}u}E_{\mathbf{k}l}\right\rangle =-T^{\mu\nu }u_{\mu }l_{\nu }\;,  \label{kinetic:M} \\
P_{l} &\equiv &\left\langle E_{\mathbf{k}l}^{2}\right\rangle =T^{\mu \nu}l_{\mu }l_{\nu }\;,  \label{kinetic:P_l} \\
P_{\perp } &\equiv &-\frac{1}{2} \left\langle \Xi ^{\mu \nu }k_{\mu }k_{\nu}\right\rangle 
=-\frac{1}{2}\, T^{\mu \nu }\Xi _{\mu \nu }\;, \label{kinetic:P_t}
\end{eqnarray}
and 
\begin{eqnarray}
V_{\perp }^{\mu } &\equiv &\left\langle k^{\left\{ \mu \right\}}\right\rangle 
=\Xi _{\nu }^{\mu }N^{\nu }\;,  \label{kinetic:Vt_mu} \\
W_{\perp u}^{\mu } &\equiv &\left\langle E_{\mathbf{k}u}k^{\left\{ \mu\right\} }\right\rangle 
=\Xi _{\alpha }^{\mu }T^{\alpha \beta }u_{\beta }\;, \label{kinetic:Wu_mu} \\
W_{\perp l}^{\mu } &\equiv &\left\langle E_{\mathbf{k}l}k^{\left\{ \mu\right\} }\right\rangle 
=-\Xi _{\alpha }^{\mu }T^{\alpha \beta }l_{\beta }\;,\label{kinetic:Wl_mu} \\
\pi _{\perp }^{\mu \nu } &\equiv &\left\langle k^{\left\{ \mu \right.}k^{\left. \nu \right\} }\right\rangle 
=\Xi _{\alpha \beta }^{\mu \nu}T^{\alpha \beta }\;.  \label{kinetic:pit_munu}
\end{eqnarray}
From these definitions it is evident that $V_{\perp }^{\mu }u_{\mu}=V_{\perp }^{\mu }l_{\mu }=0$ 
as well as $W_{\perp u}^{\mu }u_{\mu}=W_{\perp u}^{\mu }l_{\mu }=0$ and 
$W_{\perp l}^{\mu }u_{\mu }=W_{\perp l}^{\mu }l_{\mu }=0$. The transverse shear-stress tensor 
$\pi _{\perp }^{\mu \nu }$ is the part of the energy-momentum tensor that is symmetric, 
$\pi _{\perp}^{\mu \nu }=\pi _{\perp }^{\nu \mu }$, traceless, $\pi _{\perp }^{\mu \nu}g_{\mu \nu }=0$, 
and orthogonal to both preferred four-vectors, $\pi_{\perp }^{\mu \nu }u_{\mu }=\pi _{\perp }^{\mu \nu }l_{\mu }=0$.
Here $\Xi _{\alpha \beta }^{\mu \nu}$ is the projection tensor 
defined in Eq.\ (\ref{Xi_traceless}), such that
$k^{\{\mu} k^{ \nu \} } =\Xi _{\alpha \beta }^{\mu \nu} k^{\alpha} k^{\beta}$.

Note that the latter decomposition of the conserved quantities with respect to $u^\mu$, $l^\mu$,
and $\Xi^{\mu \nu}$ was, to our knowledge, 
already given by Barz, K\"{a}mpfer, Luk\'{a}cs, Martin\'{a}s, and Wolf as early as the late 1980's 
\cite{Barz:1987pq}, and was later on used in Refs.\ \cite{Huang:2009ue,Huang:2011dc}. 
Recently, Florkowski, Martinez, Ryblewski, and Strickland
\cite{Martinez:2009ry,Martinez:2010sc,Martinez:2010sd,Martinez:2012tu,Florkowski:2010cf,Ryblewski:2010bs,Ryblewski:2011aq,Ryblewski:2012rr} 
proposed the so-called anisotropic hydrodynamics formalism based on a specific
distribution function that has an anisotropic spheroidal shape in momentum space in the LR frame 
\cite{Romatschke:2003ms}. This particular ansatz leads to a 
less general form of the energy-momentum tensor, which only features a
pressure anisotropy, $P_l \neq P_\perp$, while other terms listed in Eqs.\ (\ref{kinetic:n_l}) -- (\ref{kinetic:pit_munu}) 
vanish equivalently. 
This was later improved in Ref.\ \cite{Bazow:2013ifa} based on Eq.\ (\ref{kinetic:f=f0+df_2}) but 
the decomposition still differs from Eqs.\ (\ref{kinetic:N_mu_u_l}), (\ref{kinetic:T_munu_u_l}). For example 
$\tilde{\pi}^{\mu \nu}$ from Eq.\ (25g) of Ref.\ \cite{Bazow:2013ifa} is not orthogonal to the four-vector specifying the
direction of the anisotropy (i.e., in our case $l^\mu$).

Comparing Eq.\ (\ref{kinetic:P}) to Eqs.\ (\ref{kinetic:P_l}) -- (\ref{kinetic:P_t}) 
the longitudinal and transverse pressure components are
related to the isotropic pressure as 
\begin{equation}
P=\frac{1}{3}\left( P_{l}+ 2P_{\perp }\right) \;.  \label{P_iso_relation}
\end{equation}
This result is independent on how far off the system is from local
thermodynamical equilibrium. In case that $P=P_{\perp }=P_{l}$ the pressure
is isotropic, but the system may not be in local thermodynamical
equilibrium, because the bulk viscous pressure 
\begin{equation} \label{Pi_iso}
\Pi_{iso} \equiv P-P_{0} \; ,
\end{equation}
may be non-zero. Here, $P_{0}$ is the pressure in local thermodynamical equilibrium.

Furthermore, not only the isotropic pressure separates into two parts but
also the diffusion currents $V^{\mu }$ and $W^{\mu }$ are split according to the
direction defined by $l^{\mu }$ and the direction perpendicular to it, 
\begin{eqnarray}
V^{\mu } &=& n_{l}\, l^{\mu }+V_{\perp }^{\mu }\;,  \label{V=V_t+n_l} \\
W^{\mu } &=& M\, l^{\mu }+W_{\perp u}^{\mu }\;.  \label{W=Wu+M}
\end{eqnarray}
Finally, using Eq.\ (\ref{Delta_Xi_relation}), we can show that 
\begin{equation}
\pi ^{\mu \nu }=\pi _{\perp }^{\mu \nu }+2\, W_{\perp l}^{\left( \mu \right.}l^{\left. \nu \right) }
+\frac{1}{3}\left( P_{l}-P_{\perp }\right) \left(2\, l^{\mu }l^{\nu }+\Xi ^{\mu \nu }\right) \;.  \label{pi=pi_t+W_l+P_tP_l}
\end{equation}
Equations (\ref{P_iso_relation}) -- (\ref{pi=pi_t+W_l+P_tP_l}) relate the
fluid-dynamical quantities decomposed with respect to $u^{\mu},\,\Delta^{\mu \nu }$ 
to those decomposed with respect to $u^{\mu },\,l^{\mu }$, and 
$\Xi^{\mu \nu }$. Note that, in terms of independent degrees of freedom,
these two decompositions are completely equivalent. In general, $N^{\mu }$
has four while $T^{\mu \nu }$ has ten independent components. The
decomposition with respect to (a given four-vector) $u^{\mu}$ and the
projector $\Delta^{\mu \nu }$ also contains 14 independent dynamical
variables. These are the three scalars $n$, $e$, and $P$, the two vectors 
$V^{\mu }$ and $W^{\mu }$, each with three independent components, while the
shear-stress tensor $\pi ^{\mu \nu }$ has five independent components. On
the other hand, the decomposition with respect to (given) $u^{\mu},\,l^{\mu} $, and $\Xi^{\mu \nu }$ 
has the six scalars $n$, $e$, $n_{l}$,  $M $, $P_{l}$, and $P_{\perp}$, the three vectors $V_{\perp}^{\mu }$, 
$W_{\perp u}^{\mu }$, and $W_{\perp l}^{\mu }$, with two independent
components each, whereas the shear-stress tensor in the transverse
direction, $\pi_{\perp}^{\mu \nu}$, possesses only two independent components.

So far, the four-vectors $u^{\mu }$ and $l^{\mu}$ were arbitrary quantities.
However, commonly the (time-like) four-vector $u^\mu$ is supposed to have a
physical meaning, namely the fluid four-velocity. In this case, it becomes a
dynamical quantity with three independent degrees of freedom. The choice of
the fluid four-velocity is not unique. The two most popular choices to
fix the LR frame of the fluid have already been discussed in the Introduction,
the Eckart frame \cite{Eckart:1940te}, Eq.\ (\ref{LR_Eckart_1_2}), which follows the flow
of particles, and the Landau frame \cite{Landau_book}, Eq.\ (\ref{LR_Landau_1_2}), 
which follows the flow of energy. 
Consequently, in the Eckart frame, there is no diffusion of particles (or charges) relative 
to $u^\mu$, so that
\begin{equation}\label{LR_Eckart_2}
V^{\mu }=0\;,\;\;\; V_{\perp }^{\mu }=0\;,\;\;\; n_{l}=0\;,
\end{equation}
where we used Eq.\ (\ref{V=V_t+n_l}). In the Landau frame, the energy diffusion current vanishes, 
\begin{equation} \label{LR_Landau_2}
W^{\mu }=0\;,\;\;\; W_{\perp u}^{\mu }=0\;,\;\;\; M=0\;,
\end{equation}
where we used Eq.\ (\ref{W=Wu+M}). 
In both cases, three of the 14
independent fluid-dynamical variables are replaced by the three independent
components of $u^\mu$, so the total number of independent variables is still 14. 

It is also possible to assign a physical meaning to the (so far fixed)
four-vector $l^{\mu }$. This vector would then become a dynamical variable
with two independent components. A clever choice of frame could then be used
to eliminate two of the 14 independent fluid-dynamical variables and replace them
with the two independent components of $l^{\mu }$. This is, however, not
what is commonly done. In the initial stage of a heavy-ion collision, the
single-particle distribution function is highly anisotropic in the
beam ($z$--) direction. Therefore, $l^{\mu }$ is usually taken to be
\begin{equation}
l^{\mu }=\gamma _{z}(v^{z},0,0,1)\;,
\end{equation}
see Refs.\ \cite{Florkowski:2010cf,Ryblewski:2011aq,Ryblewski:2012rr} and
Eq.\ (\ref{l_mu}).
Here, $v^{z}$ is the $z$--component of the fluid three-velocity $\mathbf{v}$, 
and $\gamma _{z}=(1-v_{z}^{2})^{-1/2}$. One can easily convince oneself
that $l^{\mu }l_{\mu }=-1$ and $l^{\mu }u_{\mu}=0$. Since $v^{z}$ is
uniquely determined by $u^{\mu}$, this choice of $l^{\mu}$ does not
represent a new dynamical quantity, and $l^{\mu}$ is completely fixed once 
$u^{\mu}$ is known.

\section{Local thermodynamical equilibrium}
\label{sect:isotropic_state}

In this section, we discuss the moments of the single-particle distribution function in local thermodynamical
equilibrium. The equilibrium moments of tensor-rank $n$ are defined as
\begin{equation}
\mathcal{I}_{i}^{\mu _{1}\cdots \mu _{n}}=\left\langle E_{\mathbf{k}u}^{i}\,
k^{\mu _{1}}\cdots k^{\mu _{n}}\right\rangle _{0}\;,  \label{I_n_tens}
\end{equation}
where the angular brackets denote the average over momentum space 
defined in Eq.\ (\ref{average_in_equilibrium}). The subscript $i$ on this quantity
reflects the power of $E_{\mathbf{k}u} $ in the definition of the moment. Due to the fact that the equilibrium
distribution function depends only on the quantities $\alpha _{0},\,\beta_{0}$, and the flow velocity $u^{\mu }$, 
the equilibrium moments can be expanded in terms of $u^{\mu }$ and the projector $\Delta ^{\mu \nu }$ as
\begin{equation}
\mathcal{I}_{i}^{\mu _{1}\cdots \mu _{n}}=\sum_{q=0}^{\left[ n/2\right]}\left( -1\right)^{q}\, b_{nq}\, I_{i+n,q}\, 
\Delta ^{\left( \mu _{1}\mu _{2}\right.}\cdots \Delta ^{\mu _{2q-1}\mu _{2q}}u^{\mu _{2q+1}}\cdots 
u^{\left. \mu_{n}\right) }\;,  \label{I_nq_moment}
\end{equation}
where $n$, $q$ are natural numbers while the sum runs over $0\leq q\leq \left[ n/2\right] $. 
Here, $[n/2]$ denotes the largest integer which is
less than or equal to $n/2$. The symmetrized tensors 
$\Delta ^{\left( \mu_{1}\mu _{2}\right. }\cdots \Delta ^{\mu _{2q-1}\mu _{2q}}u^{\mu_{2q+1}}\cdots 
u^{\left. \mu _{n}\right) }$ are discussed in App.\ \ref{appendix_thermo_integrals}.
The symmetrization yields
\begin{equation}
b_{nq} = \frac{n!}{2^{q}q!\left( n-2q\right) !}  \label{b_nq}
\end{equation}
distinct terms, see App.\ \ref{appendix_thermo_integrals}. Finally,
$I_{nq}$ are the so-called relativistic thermodynamic integrals that only 
depend on the equilibrium variables $\alpha _{0}$ and $\beta _{0}$,
\begin{equation}
I_{nq}\left( \alpha _{0},\beta _{0}\right) =\frac{\left( -1\right) ^{q}}{\left( 2q+1\right) !!}
\left\langle E_{\mathbf{k}u}^{n-2q}\left( \Delta^{\alpha \beta }k_{\alpha }k_{\beta }\right)^{q}\right\rangle _{0}\;,
\label{I_nq}
\end{equation}
where the double factorial is defined as $\left( 2q+1\right) !!=\left(2q+1\right) !/\left( 2^{q}q!\right) $.
Note that the thermodynamic integrals with index $q=0$ are directly given by the moments 
(\ref{I_n_tens}) of rank zero, 
\begin{equation}
I_{i0}\equiv \mathcal{I}_{i}\;.  \label{I_i0=I_i}
\end{equation}
Analogously, the thermodynamic integrals with index $q=1$ can be obtained
from a projection of the moments (\ref{I_n_tens}) of rank two, 
\begin{equation}
I_{i+2,1}\equiv -\frac{1}{3}\,\mathcal{I}_{i}^{\mu \nu }\Delta _{\mu \nu }=-
\frac{1}{3}\left( m_{0}^{2}\, \mathcal{I}_{i}-\mathcal{I}_{i+2}\right) \;,
\label{I_i+2,1=I_i_Delta}
\end{equation}
where we made use of the explicit form of $\Delta _{\mu \nu }$, of the
on-shell condition $k^{\mu }k_{\mu }=m_{0}^{2}$, and again employed Eq.\ (\ref{I_n_tens}). 

The so-called auxiliary thermodynamic integrals are defined as \cite{Israel:1979wp} 
\begin{equation}
J_{nq}\equiv \left( \frac{\partial I_{nq}}{\partial \alpha _{0}}\right)_{\beta _{0}}
=\frac{\left( -1\right) ^{q}}{\left( 2q+1\right) !!}\int dK\, E_{\mathbf{k}u}^{n-2q}
\left( \Delta ^{\alpha \beta }k_{\alpha }k_{\beta}\right)^{q}f_{0\mathbf{k}}
\left( 1-af_{0\mathbf{k}}\right) \;.
\label{J_nq}
\end{equation}
Note that
\begin{equation} \label{J_n+1q}
\left( \frac{\partial I_{nq}}{\partial \beta_{0}}\right)_{\alpha_{0}} 
= - \frac{\left( -1\right) ^{q}}{\left( 2q+1\right) !!}\int dK\, E_{\mathbf{k}u}^{n+1-2q}
\left( \Delta ^{\alpha \beta }k_{\alpha }k_{\beta}\right)^{q}f_{0\mathbf{k}}
\left( 1-af_{0\mathbf{k}}\right) \equiv - J_{n+1,q}\;.
\end{equation}

Using the definition (\ref{moments_in_equilibrium}) of the conserved quantities and
Eqs.\ (\ref{I_n_tens}) -- (\ref{b_nq}),
\begin{eqnarray}
N_{0}^{\mu } &\equiv &\mathcal{I}_{0}^{\mu }=I_{10}\, u^{\mu }\,,
\label{N_0_mu} \\
T_{0}^{\mu \nu } &\equiv &\mathcal{I}_{0}^{\mu \nu }=I_{20}\, u^{\mu }u^{\nu} 
- I_{21}\, \Delta ^{\mu \nu }\;.  \label{T_0_munu}
\end{eqnarray}
Tensor-projecting these quantities, we obtain 
\begin{eqnarray}
I_{10} &\equiv &N_{0}^{\mu }u_{\mu }=n_{0}=\left\langle E_{\mathbf{k}u}\right\rangle_{0}
\equiv \mathcal{I}_{1}\;,  \label{n_0_equilibrium} \\
I_{20} &\equiv &T_{0}^{\mu \nu }u_{\mu }u_{\nu }=e_{0}=\left\langle E_{\mathbf{k}u}^{2}\right\rangle_{0}
\equiv \mathcal{I}_{2}\;, \label{e_0_equilibrium} \\
I_{21} &\equiv &-\frac{1}{3}\,T_{0}^{\mu \nu }\Delta _{\mu \nu }=P_{0}
=-\frac{1}{3}\left\langle \Delta ^{\mu \nu }k_{\mu }k_{\nu }\right\rangle_{0}
\equiv -\frac{1}{3}\left( m_{0}^{2}\, \mathcal{I}_{0}-\mathcal{I}_{2}\right) \;.  \label{P_0_equilibrium}
\end{eqnarray}
Here we used Eqs.\ (\ref{kinetic:n}) -- (\ref{kinetic:P}), the explicit form of 
$\Delta^{\mu \nu }$, as well as Eq.\ (\ref{I_n_tens}). We note that the
left- and right-hand sides of these equations are consistent with the
relations (\ref{I_i0=I_i}) and (\ref{I_i+2,1=I_i_Delta}). Using Eqs.\ (\ref{n_0_equilibrium}) -- (\ref{P_0_equilibrium}), 
we see that the tensor decompositions (\ref{N_0_mu}) and (\ref{T_0_munu}) correspond to the usual
ideal-fluid form (\ref{idealfluidform}) of the conserved quantities.

\section{Anisotropic state}
\label{sect:anisotropic_state}

In this section, we discuss the case where the single-particle distribution function
has an anisotropic shape in momentum space.
Let us assume that this function differs from
the local equilibrium distribution function (\ref{Equilibrium_dist}) such
that it is a function of the scalars $\hat{\alpha}$, $\hat{\beta}_{u}$, and 
$\hat{\beta}_{l}$ as well as of two distinct four-vectors, the flow velocity 
$u^{\mu }$ and the vector $l^{\mu }$ parametrizing the direction of the
anisotropy. All these quantities are functions of $x^\mu$. 
Such distribution functions are common in plasma physics where the 
presence of magnetic fields introduces a momentum anisotropy and so the particle 
momenta parallel and perpendicular to the magnetic field are different; as in the case of 
the bi-Maxwellian, the drifting Maxwellian, or the loss-cone distribution functions \cite{Plasma_Physics_book}.
Analogously, $\hat{\beta}_{l}$ can be thought of as an additional parameter characterizing
the temperature difference between the directions parallel and perpendicular to
the $z$--axis.

As discussed in the Introduction, we will denote the anisotropic single-particle distribution function as 
$\hat{f}_{0\mathbf{k}}=\hat{f}_{0\mathbf{k}}\left(\hat{\alpha},\hat{\beta}_{u} 
E_{\mathbf{k}u},\hat{\beta}_{l}E_{\mathbf{k}l}\right)$. 
At this point, the functional dependence on $\hat{\beta}_l E_{\mathbf{k}l}$ 
does not need to be specified. All we need to know is that this combination of 
variables parametrizes the momentum anisotropy. 

We will also assume that 
\begin{equation}
\lim_{\hat{\beta}_l \rightarrow 0} \hat{f}_{0\mathbf{k}}\left( \hat{\alpha},
\hat{\beta}_{u}E_{\mathbf{k}u},\hat{\beta}_{l}E_{\mathbf{k}l}\right) 
= f_{0\mathbf{k}}\left( \hat{\alpha},\hat{\beta}_{u}E_{\mathbf{k}u}\right)\;,
\label{hat_f->f_0}
\end{equation}
i.e., that in the limit of vanishing anisotropy the single-particle
distribution function assumes the local-equilibrium form (\ref{Equilibrium_dist}). 
The assumption (\ref{hat_f->f_0}) has no impact 
on our formulation of anisotropic dissipative fluid dynamics, it is merely
physically natural. We furthermore demand that 
\begin{equation}
\left( \frac{ \partial \hat{f}_{0\mathbf{k}}}{\partial \hat{\alpha}}
\right)_{\hat{\beta}_{u},\hat{\beta}_{l}}=\hat{f}_{0\mathbf{k}}\left( 1-a \hat{f}_{0\mathbf{k}}\right) \;.
\label{hat_f_alpha}
\end{equation}
This further constraint on the form of $\hat{f}_{0\mathbf{k}}$ is naturally
respected in the limit of vanishing anisotropy, cf.\ Eq.\ (\ref{Equilibrium_dist}). 
There is no real physical reason that we should require
it also for $\hat{\beta}_l \neq 0$, but it simplifies the following calculations. 
For instance, the spheroidal distribution function proposed in Ref.\ \cite{Romatschke:2003ms} and used in 
Refs.\ \cite{Martinez:2009ry,Martinez:2010sc,Martinez:2010sd,Martinez:2012tu,Florkowski:2010cf,Ryblewski:2010bs,Ryblewski:2011aq,Ryblewski:2012rr} 
satisfies the above constraints and can be used for explicit calculations.

In analogy to Eq.\ (\ref{I_n_tens}) we now introduce a set of generalized moments of $\hat{f}_{0\mathbf{k}}$ 
of tensor-rank $n$, 
\begin{equation}
\hat{\mathcal{I}}_{ij}^{\mu _{1}\cdots \mu _{n}}\equiv \left\langle E_{\mathbf{k}u}^{i}\, E_{\mathbf{k}l}^{j}\,
k^{\mu _{1}}\cdots k^{\mu_{n}}\right\rangle _{\hat{0}}\;,  \label{I_ij_tens}
\end{equation}
where, similarly to Eqs.\ (\ref{average_in_equilibrium}) and (\ref{average_in_nonequilibrium}),
\begin{equation}
\left\langle \cdots \right\rangle _{\hat{0}}=\int dK\left( \cdots \right) \hat{f}_{0\mathbf{k}}\;,
\end{equation}
and the subscripts $i$ and $j$ denote the
powers of $E_{\mathbf{k}u}$ and $E_{\mathbf{k}l}$, respectively.
These generalized moments can be expanded in terms of the two four-vectors $u^\mu$, $l^\mu$, and the
tensor $\Xi^{\mu\nu}$,
\begin{equation}
\hat{\mathcal{I}}_{ij}^{\mu _{1}\cdots \mu _{n}}=\sum_{q=0}^{\left[ n/2\right] }
\sum_{r=0}^{n-2q}\left( -1\right)^{q}\, b_{nrq}\, \hat{I}_{i+j+n,j+r,q}\,
\Xi^{\left( \mu _{1}\mu _{2}\right. }\cdots \Xi ^{\mu_{2q-1}\mu _{2q}}l^{\mu _{2q+1}}\cdots 
l^{\mu _{2q+r}}u^{\mu_{2q+r+1}}\cdots u^{\left. \mu _{n}\right) }\;,  \label{I_nrq_moment}
\end{equation}
where $n$, $r$, and $q$ are natural numbers; $r$ counts the number of four-vectors $l^{\mu }$ and
$q$ the number of $\Xi$ projectors in the expansion. The
symmetrized tensor products $\Xi^{\left( \mu _{1}\mu _{2}\right. }\cdots \Xi ^{\mu_{2q-1}\mu _{2q}}l^{\mu _{2q+1}}\cdots 
l^{\mu _{2q+r}}u^{\mu_{2q+r+1}}\cdots u^{\left. \mu _{n}\right) }$ are discussed in App.\ \ref{appendix_thermo_integrals}.
The symmetrization yields
\begin{equation}
b_{nrq}\equiv \frac{n!}{2^{q}q!\,r!\left( n-r-2q\right) !}   \label{b_nrq}
\end{equation}
distinct terms, see App.\ \ref{appendix_thermo_integrals}. 
Finally, the generalized thermodynamic integrals $\hat{I}_{nrq}$ are defined as 
\begin{equation}
\hat{I}_{nrq}\left( \hat{\alpha},\hat{\beta}_{u},\hat{\beta}_{l}\right) =
\frac{\left( -1\right) ^{q}}{\left( 2q\right) !!}\left\langle E_{\mathbf{k}u}^{n-r-2q}\,
E_{\mathbf{k}l}^{r}\left( \Xi ^{\mu \nu }k_{\mu }k_{\nu}\right)^{q}\right\rangle_{\hat{0}}\;,  \label{I_nrq}
\end{equation}
where the double factorial of an even number is $\left( 2q\right) !!\equiv 2^{q} q!$. 
The corresponding generalized auxiliary thermodynamic integrals
are defined with the help of Eq.\ (\ref{hat_f_alpha}) and similarly to Eq.\ (\ref{J_nq}), 
\begin{equation}
\hat{J}_{nrq}\equiv \left( \frac{\partial \hat{I}_{nrq}}{\partial \hat{\alpha}}\right)_{\hat{\beta}_{u},\hat{\beta}_{l}}
=\frac{\left( -1\right)^{q}} {\left( 2q\right) !!}\int dK\,E_{\mathbf{k}u}^{n-r-2q}\,E_{\mathbf{k}l}^{r}
\left( \Xi ^{\mu \nu }k_{\mu }k_{\nu }\right) ^{q}\hat{f}_{0\mathbf{k}}\left( 1-a\hat{f}_{0\mathbf{k}}\right) \;.  \label{J_nrq}
\end{equation}

As in Eqs.\ (\ref{I_i0=I_i}) and (\ref{I_i+2,1=I_i_Delta}) one can easily
show that, in the case $q=0$,  comparison of Eqs.\ (\ref{I_ij_tens}) and (\ref{I_nrq}) leads to 
\begin{equation}
\hat{I}_{i+j,j,0}\equiv \hat{\mathcal{I}}_{ij}\;,  \label{I_i+j,j,0=I_ij}
\end{equation}
while for $q=1$, using the explicit form of $\Xi ^{\mu \nu }$ and the
on-shell condition $k^{\mu }k_{\mu }=m_{0}^{2}$, comparison of 
Eqs.\ (\ref{I_ij_tens}) and (\ref{I_nrq}) yields 
\begin{equation}
\hat{I}_{i+j+2,j,1}\equiv -\frac{1}{2}\, \hat{\mathcal{I}}_{ij}^{\mu \nu }\Xi_{\mu \nu }
=-\frac{1}{2}\left( m_{0}^{2}\,\hat{\mathcal{I}}_{ij}-\hat{\mathcal{I}}_{i+2,j}+\hat{\mathcal{I}}_{i,j+2}\right) \;.
\label{I_i+2,j,1=I_ij_Delta}
\end{equation}

Note that these quantities can also be used in (local) thermodynamic
equilibrium, see for example Eq.\ (\ref{I_ij_moment_equilibrium}). 
The conventional relativistic thermodynamic integrals (\ref{I_nq}), (\ref{J_nq}) 
are then recovered as linear combinations of those defined in 
Eqs.\ (\ref{I_nrq}), (\ref{J_nrq}), as shown in App.\ \ref{appendix_thermo_integrals_equilibrium}. 
Furthermore, also note that the moment defined in the first line of Eq.\ (66) of 
Ref.\ \cite{Bazow:2013ifa} is (because of the $\Delta$ instead of the $\Xi$ projector in the integrand) 
not equivalent to the one defined in Eq.\ (\ref{I_ij_tens}).

It is instructive to explicitly write down the tensor decomposition of the
generalized moments (\ref{I_nrq_moment}) of $\hat{f}_{0\mathbf{k}}$. The
conserved quantities read 
\begin{eqnarray}
\hat{N}^{\mu } &\equiv &\hat{\mathcal{I}}_{00}^{\mu }=\hat{I}_{100}\, u^{\mu }+
\hat{I}_{110}\, l^{\mu }\;,  \label{N_0_mu_u_l} \\
\hat{T}^{\mu \nu } &\equiv &\hat{\mathcal{I}}_{00}^{\mu \nu}
=\hat{I}_{200}\, u^{\mu }u^{\nu }+2\, \hat{I}_{210}\, u^{\left( \mu \right. }
l^{\left. \nu\right) }+\hat{I}_{220}\, l^{\mu }l^{\nu }-\hat{I}_{201}\, \Xi ^{\mu \nu }\;.
\label{T_0_munu_u_l}
\end{eqnarray}
The coefficients can be obtained by appropriate tensor projections of these
quantities. According to Eqs.\ (\ref{kinetic:n}), (\ref{kinetic:e}) and Eqs.\ (\ref{kinetic:n_l}) -- (\ref{kinetic:P_t}) we obtain 
\begin{eqnarray}
\hat{I}_{100} &\equiv &\hat{N}^{\mu }u_{\mu }=\hat{n}
=\left\langle E_{\mathbf{k}u}\right\rangle _{\hat{0}}\equiv \hat{\mathcal{I}}_{10}\;,
\label{n_aniso} \\
\hat{I}_{110} &\equiv &-\hat{N}^{\mu }l_{\mu }=\hat{n}_{l}
=\left\langle E_{\mathbf{k}l}\right\rangle _{\hat{0}}\equiv \hat{\mathcal{I}}_{01}\;,
\label{n_l_aniso} \\
\hat{I}_{200} &\equiv &\hat{T}^{\mu \nu }u_{\mu }u_{\nu }
=\hat{e}=\left\langle E_{\mathbf{k}u}^{2}\right\rangle _{\hat{0}}
\equiv \hat{\mathcal{I}}_{20}\;,  \label{e_aniso} \\
\hat{I}_{210} &\equiv &-\hat{T}^{\mu \nu }u_{\mu }l_{\nu }
=\hat{M}=\left\langle E_{\mathbf{k}u}E_{\mathbf{k}l}\right\rangle _{\hat{0}}\equiv 
\hat{\mathcal{I}}_{11}\;,  \label{M_aniso} \\
\hat{I}_{220} &\equiv &\hat{T}^{\mu \nu }l_{\mu }l_{\nu }
=\hat{P}_{l}=\left\langle E_{\mathbf{k}l}^{2}\right\rangle _{\hat{0}}
\equiv \hat{\mathcal{I}}_{02}\;,  \label{P_l_aniso} \\
\hat{I}_{201} &\equiv &-\frac{1}{2}\hat{T}^{\mu \nu }\Xi _{\mu \nu }
=\hat{P}_{\perp }=-\frac{1}{2}\left\langle \Xi ^{\alpha \beta }k_{\alpha }k_{\beta
}\right\rangle _{\hat{0}}\equiv -\frac{1}{2}\left( m_{0}^{2}\,
\hat{\mathcal{I}}_{00}-\hat{\mathcal{I}}_{20}+\hat{\mathcal{I}}_{02}\right) \;.
\label{P_t_aniso}
\end{eqnarray}
Note that $\hat{e}$, $\hat{P}_{l}$, and $\hat{P}_\perp$ are
related to each other, see Eq.\ (\ref{P_t_aniso}), hence they are not
independent variables.

In general, the conserved quantities (\ref{N_0_mu_u_l}), (\ref{T_0_munu_u_l})
contain eleven unknowns: the scalar quantities $\hat{n},\,\hat{e},\,\hat{n}_{l},\,\hat{M},\,\hat{P}_{l}$, 
and $\hat{P}_{\perp }$, and the vectors $u^{\mu }$ (three independent components) and 
$l^{\mu }$ (two independent components). At the end of Sec.\ \ref{sect:general_fluid_variables}, we had
already discussed our choice of $l^{\mu }$ which is completely determined by 
$u^{\mu }$, so there remain nine unknowns. The choice of a LR frame (Eckart
or Landau) eliminates either $\hat{n}_{l}$ or $\hat{M}$, hence leaving eight unknowns. However, once 
$\hat{f}_{0\mathbf{k}}\left( \hat{\alpha},\hat{\beta}_{u}E_{\mathbf{k}u},\hat{\beta}_{l}E_{\mathbf{k}l}\right)$ is
specified, the remaining five scalar unknowns ($\hat{n},\,\hat{e}\,,\hat{P}_{l},\,\hat{P}_{\perp }$, 
and -- depending on the choice of the LR frame --
either $\hat{n}_{l}$ or $\hat{M}$) are not independent variables anymore;
they are functions of the three independent variables $\hat{\alpha},\,\hat{\beta}_{u},\,\hat{\beta}_{l}$.
This reduces the number of independent variables to six. 
Five constraints are provided by the five equations of
motion $\partial _{\mu }\hat{N}^{\mu }=0$ and $\partial _{\mu }\hat{T}^{\mu\nu }=0$. 
In the ideal-fluid limit, $\hat{\beta}_{l}\rightarrow 0$, and the system of equations of motion is closed. 
For arbitrary $\hat{\beta}_{l}$, however, we need an additional equation of
motion to close the system of equations. 
This will effectively describe the decay of the momentum 
anisotropy of the distribution function and the approach of the system to 
local thermal equilibrium.
This auxiliary equation can be provided, for example, from the higher moments of the Boltzmann 
equation as is usually done in kinetic theory.

It is instructive to repeat this discussion from a slightly different
perspective. For very large times, any closed system described by the Boltzmann
equation will reach global thermodynamical equilibrium. If, in this process,
the system first reaches local thermodynamical equilibrium, the evolution
towards global equilibrium is governed by ideal fluid dynamics. In this
case, it is advantageous to explicitly exhibit the equilibrium EoS $P_0(e_0,n_0)$ 
in the fluid-dynamical equations of motion.
Usually, this is done via the Landau matching conditions. These conditions
require that particle density $n$ and energy density $e$ in a general
non-equilibrium state are equal to those of a fictitious (local)
thermodynamical equilibrium state, $n= n_0(\alpha_0, \beta_0)$, 
$e=e_0(\alpha_0, \beta_0)$. These equations implicitly determine the intensive
parameters $\alpha_0,\, \beta_0$ in the distribution function 
(\ref{Equilibrium_dist}) pertaining to the fictitious (local) equilibrium state.

Analogously, for the anisotropic state the Landau matching conditions read 
$\hat{n}\left( \hat{\alpha},\hat{\beta}_{u},\hat{\beta}_{l}\right)
=n_{0}\left( \alpha _{0},\beta _{0}\right) $ and 
$\hat{e}\left( \hat{\alpha},\hat{\beta}_{u},\hat{\beta}_{l}\right) =e_{0}\left( \alpha _{0},\beta_{0}\right)$, which
is equivalent to
\begin{eqnarray}\label{Landau_matching_n0}
\left(\hat{N}^{\mu} - N^{\mu}_0 \right)u_{\mu} =0 \; , \\ 
\left(\hat{T}^{\mu \nu} - T^{\mu \nu}_0 \right)u_{\mu} u_{\nu} =0\;,  \label{Landau_matching_e0}
\end{eqnarray}
where $N^{\mu}_0$ and $T^{\mu \nu}_0$ were defined in Eq.\ (\ref{idealfluidform}).
These dynamical matching conditions will determine the two parameters of a fictitious 
equilibrium state, $\alpha_0=\alpha_0(\hat{\alpha},\hat{\beta}_{u},\hat{\beta}_{l})$
and $\beta_0=\beta_0(\hat{\alpha},\hat{\beta}_{u},\hat{\beta}_{l})$, as function of
the three scalar parameters $\hat{\alpha},\hat{\beta}_{u},\hat{\beta}_{l}$ pertaining to the anisotropic state.

In principle, there are infinitely many possibilities to extend an 
equilibrium distribution function by an additional free parameter, $\hat{\beta}_{l} $, 
each one resulting in a different set of
thermodynamical relations, i.e., the latter are not universal. 
For example, choosing $\hat{\beta}_{l} $ as a free intensive variable that is related to some new 
conjugate extensive quantity, as in Ref.\ \cite{Barz:1987pq}, 
we recover the conventional laws of
thermodynamics only in the equilibrium limit, 
$\hat{\beta}_l \rightarrow 0$. 
Therefore, without specifying the exact form of the anisotropic
distribution function we cannot derive any thermodynamic relation, and in
particular the EoS, from kinetic theory.

Using the Landau matching conditions (\ref{Landau_matching_n0}), 
(\ref{Landau_matching_e0}) the tensor decomposition of the
conserved quantities reads 
\begin{eqnarray} \label{N_mu_aniso_2}
\hat{N}^{\mu } &=&n_{0}\, u^{\mu }+\hat{n}_{l}\,l^{\mu }\;, \\ 
\hat{T}^{\mu \nu } &=&e_{0}\,u^{\mu }u^{\nu }+2\,\hat{M}\,u^{\left( \mu \right.}l^{\left. \nu \right) }
+\hat{P}_{l}\, l^{\mu }l^{\nu }-\hat{P}_{\perp }\, \Xi^{\mu \nu }\;.  \label{T_munu_aniso_2}
\end{eqnarray}
The equilibrium EoS is now introduced by writing the isotropic pressure 
(\ref{P_iso_relation}) in the form 
\begin{equation}
\hat{P} \equiv \frac{1}{3} \left( \hat{P}_l + 2 \hat{P}_\perp \right) \equiv
P_0\left(\alpha_0, \beta_0\right)  + \hat{\Pi}\left(\hat{\alpha}, \hat{\beta}_u,\hat{\beta}_l \right) \;,  \label{bulk_aniso}
\end{equation}
which at the same time defines the bulk viscous pressure $\hat{\Pi}$ with respect to the
pressure in local equilibrium and hence can be used to
eliminate either $\hat{P}_l$ or $\hat{P}_\perp$ in Eq.\ (\ref{T_munu_aniso_2}). 

Making the connection to the equilibrium EoS is advantageous when the
system is close to the ideal-fluid limit. However, for the anisotropic state
it does not solve the problem that the five conservation equations do not
determine all independent variables. 
We are thus left with six independent variables, say $n_0, \,e_0,$ and $\hat{P}_l$ 
(or $\hat{\Pi}$) and the three components of $u^\mu$, which means that we must supply 
one additional equation of motion.

As mentioned above, the tensor decompositions (\ref{N_mu_aniso_2}), (\ref{T_munu_aniso_2}) 
were obtained previously in 
Refs.\ \cite{Martinez:2009ry,Martinez:2010sc,Martinez:2010sd,Martinez:2012tu,Florkowski:2010cf,Ryblewski:2010bs,Ryblewski:2011aq,Ryblewski:2012rr}
based on the distribution function given in Ref.\ \cite{Romatschke:2003ms}. 
This spheroidal single-particle distribution function leads to 
$\hat{n}_{l}=\hat{M}=0$, hence exclusively to a pressure anisotropy.
Furthermore, in these refs.\ the case of vanishing $\hat{\alpha}$
was considered, which for a massless ideal gas EoS also leads to a vanishing bulk viscous pressure. 
This also leaves the freedom to use the zeroth moment of the
Boltzmann equation as additional input to close the equations of motion.

\section{Expansion around the anisotropic distribution function}
\label{sect:expansion}

In principle, $f_{\bf k}$ is a solution of the Boltzmann equation. 
However, if we are only interested in the low-frequency,
large-wavenumber limit of the latter, we may consider the (much simpler) fluid-dynamical 
equations of motion. In order to derive them from the Boltzmann equation, the
method of moments is particularly well suited \cite{Denicol:2012cn,Grad,Israel:1979wp},
where $f_{\mathbf{k}}$ is expanded around the distribution function 
$f_{0 \mathbf{k}}$ of a fictitious (local) equilibrium state. 
The corrections are written in terms of the irreducible
moments of $\delta f_{\mathbf{k}} \equiv f_{\mathbf{k}} - f_{0 \mathbf{k}}$.
Then, the infinite set of moments of the Boltzmann equation provide an
infinite set of equations of motion for these irreducible moments.
Conventional dissipative fluid dynamics then emerges by truncating this set
and expressing the irreducible moments in terms of the fluid-dynamical variables, for more
details see Ref.\ \cite{Denicol:2012cn}.

Here we will follow the same strategy, except expanding $f_{\mathbf{k}}$
around $\hat{f}_{0\mathbf{k}}$ instead of $f_{0\mathbf{k}}$, see also Refs.\ 
\cite{Bazow:2013ifa,Heinz:2014zha}. Nevertheless, at each step it is 
instructive to also recall the conventional expansion of $f_{\mathbf{k}}$
around $f_{0\mathbf{k}}$. Hence, 
\begin{equation}
f_{\mathbf{k}}=f_{0\mathbf{k}}+\delta f_{\mathbf{k}} \equiv \hat{f}_{0\mathbf{k}}
+\delta \hat{f}_{\mathbf{k}}\;,  \label{kinetic:f=f0+df}
\end{equation}
where it is implicitly assumed that the corrections $\delta f_{\mathbf{k}},\,\delta \hat{f}_{\mathbf{k}}$ 
fulfill $|\delta f_{\mathbf{k}}|\ll f_{0\mathbf{k}}$ and $|\delta \hat{f}_{\mathbf{k}}|\ll \hat{f}_{0\mathbf{k}}$.
The rationale behind the expansion around $\hat{f}_{0\mathbf{k}}$ instead of
around $f_{0\mathbf{k}}$ is that, in the case of a pronounced anisotropy, 
$|\delta \hat{f}_{\mathbf{k}}|\ll |\delta f_{\mathbf{k}}|$, so that the
convergence properties of the former series expansion are vastly improved
over those of the latter. Without loss of generality we write these
corrections as follows,
\begin{eqnarray}
\delta f_{\mathbf{k}} &=&f_{0\mathbf{k}}\left( 1-af_{0\mathbf{k}}\right)
\phi _{\mathbf{k}}\;, \\
\delta \hat{f}_{\mathbf{k}} &=&\hat{f}_{0\mathbf{k}}
\left( 1-a\hat{f}_{0\mathbf{k}}\right) \hat{\phi}_{\mathbf{k}}\;,  \label{kinetic:df}
\end{eqnarray}
where $\phi_{\mathbf{k}}$ and $\hat{\phi}_{\mathbf{k}}$ are measures of the
deviation of $f_{\mathbf{k}}$ from $f_{0\mathbf{k}}$ or $\hat{f}_{0\mathbf{k}}$, respectively.

We recall \cite{Denicol:2012cn,Denicol:2012es} that $\phi _{\mathbf{k}}$ can
be expanded in terms of a complete and orthogonal set of irreducible tensors 
$1,\,k^{\left\langle \mu \right\rangle },\,k^{\left\langle \mu \right.}
k^{\left. \nu \right\rangle },\,k^{\left\langle \mu \right. }k^{\nu}k^{\left. \lambda \right\rangle },\cdots $, 
where 
\begin{equation}
k^{\left\langle \mu _{1}\right. }\cdots k^{\left. \mu _{\ell }\right\rangle}
=\Delta _{\nu _{1}\cdots \nu _{\ell }}^{\mu _{1}\cdots \mu _{\ell }}k^{\nu_{1}}\cdots k^{\nu _{\ell }}\;.  
\label{iso_irreducible_momenta}
\end{equation}
The symmetric and traceless projection tensor 
$\Delta _{\nu _{1}\cdots \nu_{\ell }}^{\mu _{1}\cdots \mu _{\ell }}$ is defined in Eq.\ (\ref{Deltaproj}). 
By definition, this tensor and thus $k^{\left\langle \mu _{1}\right.}\cdots k^{\left. \mu _{\ell }\right\rangle }$ 
are orthogonal to $u^{\mu }$.
For an arbitrary function $\mathrm{F}(E_{\mathbf{k}u})$ the irreducible
tensors satisfy the following orthogonality condition 
\begin{equation}
\int dK\text{ }\mathrm{F}(E_{\mathbf{k}u})\ k^{\left\langle \mu _{1}\right.}\cdots 
k^{\left. \mu _{\ell }\right\rangle }k^{\left\langle \nu _{1}\right.}\cdots 
k^{\left. \nu _{n}\right\rangle }=\frac{\ell !\ \delta _{\ell n}}{\left( 2\ell +1\right) !!}
\Delta ^{\mu _{1}\cdots \mu _{\ell }\nu _{1}\cdots
\nu _{\ell }}\int dK\ \mathrm{F}(E_{\mathbf{k}u})\ \left( \Delta ^{\alpha\beta }k_{\alpha }k_{\beta }\right) ^{\ell }\;,
\label{normalization_isotropic}
\end{equation}
for the proof see App.\ \ref{appendix_orhogonality}. 
The expansion of $\phi_{\mathbf{k}}$ in terms of a complete and orthogonal set of irreducible tensors 
(\ref{iso_irreducible_momenta}) now reads 
\begin{equation}
\phi _{\mathbf{k}}=\sum_{\ell =0}^{\infty }\sum_{n=0}^{N_{\ell}}c_{n}^{\left\langle \mu _{1}\cdots 
\mu _{\ell }\right\rangle}k_{\left\langle \mu _{1}\right. }\cdots k_{\left. \mu _{\ell }\right\rangle}
P_{\mathbf{k}n}^{\left( \ell \right) }\;. \label{phi_k_iso}
\end{equation}
Here, the sum over $n$ runs in principle from zero to infinity, but in
practice we have to truncate it at some natural number $N_{\ell }$.
Furthermore, $c_{n}^{\left\langle \mu _{1}\cdots \mu _{\ell }\right\rangle }$
are coefficients given in Eq.\ (26) of Ref.\ \cite{Denicol:2012cn}, and the polynomials 
in $E_{\mathbf{k}u}$ are defined as 
\begin{equation}
P_{\mathbf{k}n}^{\left( \ell \right) }=\sum_{i=0}^{n}a_{ni}^{(\ell )}
E_{\mathbf{k}u}^{i}\;.  \label{P_n_polynomials}
\end{equation}
Finally, we can write the distribution function as 
\begin{equation}
f_{\mathbf{k}}=f_{0\mathbf{k}}+f_{0\mathbf{k}}\left( 1-af_{0\mathbf{k}}\right) 
\sum_{\ell =0}^{\infty }\sum_{n=0}^{N_{\ell }}\rho _{n}^{\mu_{1}\cdots \mu _{\ell }}
k_{\left\langle \mu _{1}\right. }\cdots k_{\left.\mu _{\ell }\right\rangle }\mathcal{H}_{\mathbf{k}n}^{(\ell )}\;,
\label{f_iso_expansion}
\end{equation}
see Eq.\ (30) of Ref.\ \cite{Denicol:2012cn}. Here, 
\begin{equation}
\mathcal{H}_{\mathbf{k}n}^{(\ell )}=\frac{W^{(\ell )}}{\ell !}
\sum_{i=n}^{N_{\ell }}a_{in}^{(\ell )}P_{\mathbf{k}i}^{(\ell )}\;,
\label{H_coeff}
\end{equation}
with $W^{(\ell )}=(-1)^{\ell }/J_{2\ell ,\ell }$, while the irreducible
moments of $\delta f_{\mathbf{k}}$ are defined as 
\begin{equation}
\rho _{i}^{\mu _{1}\cdots \mu _{\ell }}\equiv \left\langle E_{\mathbf{k}u}^{i}
k^{\left\langle \mu _{1}\right. }\cdots k^{\left. \mu _{\ell}\right\rangle }\right\rangle _{\delta }\;,  
\label{irr_mom_iso}
\end{equation}
where 
\begin{equation}
\left\langle \cdots \right\rangle_{\delta }\equiv \left\langle
\cdots \right\rangle -\left\langle \cdots \right\rangle _{0}
=\int dK\left(\cdots \right) \delta f_{\mathbf{k}}\;.
\end{equation} 
One can easily show substituting 
Eq.\ (\ref{f_iso_expansion}) into Eq.\ (\ref{irr_mom_iso}) and
using the orthogonality condition (\ref{normalization_isotropic}) and
the definition of the auxiliary thermodynamic integrals (\ref{J_nq}) that all irreducible
moments are linearly related to each other [for more details see, Eq.\ (72)
of Ref.\ \cite{Molnar:2013lta}], 
\begin{equation}
\rho _{i}^{\mu _{1}\cdots \mu _{\ell }}\equiv \left( -1\right) ^{\ell }\ell!
\sum_{n=0}^{N_{\ell }}\rho _{n}^{\mu _{1}\cdots \mu _{\ell }}\gamma_{in}^{\left( \ell \right) }\;,  
\label{irr_mom_iso_relations}
\end{equation}
where 
\begin{equation}
\gamma _{in}^{\left( \ell \right) }=\frac{W^{(\ell )}}{\ell !}
\sum_{n^{\prime }=n}^{N_{\ell }}\sum_{i^{\prime }=0}^{n^\prime}
a_{n^{\prime}n}^{(\ell )}a_{n^{\prime }i^{\prime }}^{(\ell )}
J_{i+i^{\prime }+2\ell,\ell }\;.
\end{equation}
Note that these relations are also valid for moments with $\textit{negative}$ $i$, 
hence it is possible to express the irreducible moments with negative powers of 
$E_{\mathbf{k}u}$ in terms 
of the ones with positive $i$, for details see, Eq.\ (65) of Ref.\ \cite{Denicol:2012cn}. 

Similarly, the correction $\hat{\phi}_{\mathbf{k}}$ to the anisotropic state
is expanded in terms of \textit{another\/} complete, orthogonal set of
irreducible tensors, $1,\ k^{\left\{ \mu \right\} },\ k^{\left\{ \mu \right.}k^{\left. \nu \right\} },\ 
k^{\left\{ \mu \right. }k^{\nu }k^{\left. \lambda \right\} },\ \cdots $, where 
\begin{equation}
k^{\left\{ \mu _{1}\right. }\cdots k^{\left. \mu _{\ell }\right\} }
=\Xi_{\nu _{1}\cdots \nu _{\ell }}^{\mu _{1}\cdots \mu _{\ell }}k^{\nu_{1}}\cdots k^{\nu _{\ell }}\;.  
\label{aniso_irreducible_momenta}
\end{equation}
Here, the symmetric and traceless projection tensor $\Xi _{\nu _{1}\cdots
\nu _{\ell }}^{\mu _{1}\cdots \mu _{\ell }}$ is defined in Eq.\ (\ref{Xiproj}). 
By definition, this tensor and thus $k^{\left\{ \mu _{1}\right. }\cdots
k^{\left. \mu _{\ell }\right\} }$ are orthogonal to \textit{both\/} $u^{\mu} $ and $l^{\mu }$. 
For an arbitrary function of \textit{both\/} $E_{\mathbf{k}u}$ and $E_{\mathbf{k}l}$, say
$\mathrm{\hat{F}}(E_{\mathbf{k}u},E_{\mathbf{k}l}) $, the irreducible tensors satisfy the 
following orthogonality condition, 
\begin{equation}
\int dK\text{ }\mathrm{\hat{F}}(E_{\mathbf{k}u},E_{\mathbf{k}l})\ k^{\left\{\mu _{1}\right. }\cdots 
k^{\left. \mu _{\ell }\right\} }k^{\left\{ \nu_{1}\right. }\cdots k^{\left. \nu _{n}\right\} }
=\frac{\delta _{\ell n}}{2^{\ell }}\, \Xi ^{\mu _{1}\cdots \mu _{\ell }\nu _{1}\cdots \nu _{\ell }}
\int dK\ \mathrm{\hat{F}}(E_{\mathbf{k}u},E_{\mathbf{k}l})\ \left( \Xi ^{\alpha \beta }
k_{\alpha }k_{\beta }\right) ^{\ell }\;,
\label{normalization_anisotropic}
\end{equation}
for the proof see App.\ \ref{appendix_orhogonality}. The expansion of 
$\hat{\phi}_{\mathbf{k}}$ now reads 
\begin{equation}
\hat{\phi}_{\mathbf{k}}=\sum_{\ell =0}^{\infty }\sum_{n=0}^{N_{\ell}}
\sum_{m=0}^{N_{\ell }-n}c_{nm}^{\left\{ \mu _{1}\cdots \mu _{\ell}\right\} }
k_{\left\{ \mu _{1}\right. }\cdots k_{\left. \mu _{\ell }\right\}}
P_{\mathbf{k}nm}^{\left( \ell \right) }\;.  \label{kinetic:phi}
\end{equation}
Here, similarly to Eq.\ (\ref{phi_k_iso}), $N_{\ell }$ truncates the (in principle) 
infinite sums over $n$ and $m$ at some natural number. 
Furthermore, $c_{nm}^{\left\{ \mu _{1}\cdots \mu _{\ell }\right\} }$
are some coefficients that will be determined later, while the 
$P_{{\bf k}nm}^{\left( \ell \right) }$ form an orthogonal set of polynomials in 
\textit{both\/} $E_{\mathbf{k}u}$ and $E_{\mathbf{k}l}$, 
\begin{equation}
P_{\mathbf{k}nm}^{\left( \ell \right)}=\sum_{i=0}^{n}\sum_{j=0}^{m}a_{nimj}^{(\ell )}
E_{\mathbf{k}u}^{i}E_{\mathbf{k}l}^{j}\;,  \label{kinetic:P_nm}
\end{equation}
where $a_{nimj}^{(\ell )}$ are coefficients that are independent of 
$E_{\mathbf{k}u}$ and $E_{\mathbf{k}l}$. These multivariate polynomials in
$E_{\mathbf{k}u}$ and $E_{\mathbf{k}l}$ are constructed to satisfy the following orthonormality
relation,
\begin{equation}
\int dK\,\hat{\omega}^{\left( \ell \right) }P_{\mathbf{k}nm}^{\left( \ell \right) }
P_{\mathbf{k}n^{\prime }m^{\prime }}^{\left( \ell \right) }=\delta_{nn^{\prime }}\delta _{mm^{\prime }}\;,  
\label{kinetic:orthonormality}
\end{equation}
where the weight is defined as 
\begin{equation}
\hat{\omega}^{\left( \ell \right) }=\frac{\hat{W}^{\left( \ell \right) }}{\left( 2\ell \right) !!}
\left( \Xi ^{\alpha \beta }k_{\alpha }k_{\beta}\right) ^{\ell }\hat{f}_{0\mathbf{k}}
\left( 1-a\hat{f}_{0\mathbf{k}}\right)\;.  \label{kinetic:weight}
\end{equation}
The normalization constant $\hat{W}^{\left( \ell \right) }$ and the
coefficients $a_{nimj}^{(\ell )}$ can be found via the Gram-Schmidt
orthogonalization procedure, see App.\ \ref{polynomial_coefficients}. Note
that for $m=0$, the multivariate polynomials $P_{\mathbf{k}nm}^{\left( \ell \right) }$ 
defined in Eq.\ (\ref{kinetic:P_nm}) naturally reduce to the
polynomials $P_{\mathbf{k}n}^{\left( \ell \right) }$, for more details see
App.\ \ref{polynomial_coefficients}.

Finally, in complete analogy to the expansion (\ref{f_iso_expansion}), the
distribution function can be written as 
\begin{equation}
f_{\mathbf{k}}=\hat{f}_{0\mathbf{k}}+\hat{f}_{0\mathbf{k}}\left( 1-a\hat{f}_{0\mathbf{k}}\right) 
\sum_{\ell =0}^{\infty }\sum_{n=0}^{N_{\ell}}\sum_{m=0}^{N_{\ell }-n}
\hat{\rho}_{nm}^{\mu _{1}\cdots \mu _{\ell}}k_{\left\{ \mu _{1}\right. }\cdots k_{\left. \mu _{\ell }\right\} }
\hat{\mathcal{H}}_{\mathbf{k}nm}^{(\ell )}\;.  \label{kinetic:f_full_anisotropic}
\end{equation}
Here we introduced the irreducible moments of $\delta \hat{f}_{\mathbf{k}}$,
\begin{equation}
\hat{\rho}_{ij}^{\mu _{1}\cdots \mu _{\ell }}\equiv \left\langle E_{\mathbf{k}u}^{i}\,
E_{\mathbf{k}l}^{j}\, k^{\left\{ \mu _{1}\right. }\cdots k^{\left. \mu_{\ell }\right\} }\right\rangle _{\hat{\delta}}\;,  
\label{kinetic:rho_ij}
\end{equation}
where 
\begin{equation}
\left\langle \cdots \right\rangle _{\hat{\delta}}\equiv \left\langle
\cdots \right\rangle -\left\langle \cdots \right\rangle _{\hat{0}}=\int dK\left(
\cdots \right) \delta \hat{f}_{\mathbf{k}}\;.
\end{equation} 
Note the differences to the
irreducible moments (\ref{irr_mom_iso}): apart from the weight factor $\delta \hat{f}_{\mathbf{k}}$ 
instead of $\delta f_{\mathbf{k}}$ they carry 
\textit{two} indices $i$ and $j$, to indicate the different powers of 
$E_{\mathbf{k}u}$ \textit{and} $E_{\mathbf{k}l}$ and finally, the irreducible
tensors $k^{\left\{ \mu _{1}\right. }\cdots k^{\left. \mu _{\ell }\right\} }$
appear instead of $k^{\left\langle \mu _{1}\right. }\cdots k^{\left. \mu_{\ell }\right\rangle }$. 
Furthermore, the coefficients $\hat{\mathcal{H}}_{\mathbf{k}nm}^{(\ell )}$ are defined as 
\begin{equation}
\hat{\mathcal{H}}_{\mathbf{k}nm}^{(\ell )}=\frac{\hat{W}^{(\ell )}}{\ell !}
\sum_{i=n}^{N_{\ell }-m}\sum_{j=m}^{N_{\ell }-i}a_{injm}^{(\ell )}P_{\mathbf{k}ij}^{\left( \ell \right) }\;.  
\label{kinetic:H_nm}
\end{equation}

For the proof of Eq.\ (\ref{kinetic:f_full_anisotropic}) we first determine the 
coefficients $c_{nm}^{\left\{ \mu _{1}\cdots \mu _{\ell}\right\}}$ in Eq.\ (\ref{kinetic:phi}). 
With the help of Eqs.\ (\ref{normalization_anisotropic}), (\ref{kinetic:phi}), 
and (\ref{kinetic:orthonormality}) one can prove that 
\begin{equation}
c_{nm}^{\left\{ \mu _{1}\cdots \mu _{\ell }\right\} }\equiv \frac{\hat{W}^{\left( \ell \right) }}{\ell !}
\left\langle P_{\mathbf{k}nm}^{\left( \ell \right) }k^{\left\{ \mu _{1}\right. }\cdots 
k^{\left. \mu _{\ell }\right\}}\right\rangle _{\hat{\delta}}\;.  \label{kinetic:c_nm}
\end{equation}
Inserting this into Eq.\ (\ref{kinetic:phi}) and employing 
Eq.\ (\ref{kinetic:P_nm}), we obtain 
\begin{equation}
\hat{\phi}_{\mathbf{k}}=\sum_{\ell =0}^{\infty }\frac{\hat{W}^{\left( \ell\right) }}{\ell !}
\sum_{n=0}^{N_{\ell }}\sum_{m=0}^{N_{\ell}-n}\sum_{i=0}^{n}\sum_{j=0}^{m}a_{nimj}^{(\ell )}
\hat{\rho}_{ij}^{\mu_{1}\cdots \mu _{\ell }}k_{\left\{ \mu _{1}\right. }\cdots 
k_{\left. \mu_{\ell }\right\} }P_{\mathbf{k}nm}^{\left( \ell \right) }\;,
\label{kinetic:phi_2}
\end{equation}
where we have used the definition (\ref{kinetic:rho_ij}) of $\hat{\rho}_{ij}^{\mu _{1}\cdots \mu _{\ell }}$. 
Now, by renaming $n\leftrightarrow i$
and $m\leftrightarrow j$ and cleverly reordering the sums, we obtain 
Eq.\ (\ref{kinetic:f_full_anisotropic}) with $\hat{\mathcal{H}}_{\mathbf{k}nm}^{(\ell )}$ 
as defined in Eq.\ (\ref{kinetic:H_nm}).

Note that, similarly to Eq.\ (\ref{irr_mom_iso_relations}), once we truncate 
the expansion at some finite $N_\ell$, any irreducible moment of 
$\delta \hat{f}_{\mathbf{k}}$ with corresponding tensor rank is linearly related 
to the moments that appear in the truncated expansion, 
\begin{equation}
\hat{\rho}_{ij}^{\mu _{1}\cdots \mu _{\ell }}\equiv \left( -1\right) ^{\ell}\ell !
\sum_{n=0}^{N_{\ell }}\sum_{m=0}^{N_{\ell }-n}\hat{\rho}_{nm}^{\mu_{1}\cdots \mu _{\ell }}
\gamma _{injm}^{\left( \ell \right) }\;,
\end{equation}
where 
\begin{equation}
\gamma _{injm}^{\left( \ell \right) }=\frac{\hat{W}^{\left( \ell \right) }}{\ell !}
\sum_{n^{\prime }=n}^{N_{\ell }-m}\sum_{m^{\prime }=m}^{N_{\ell}-n^{\prime }}
\sum_{i^{\prime }=0}^{n^{\prime }}\sum_{j^{\prime}=0}^{m^{\prime }}a_{n^{\prime }nm^{\prime }m}^{(\ell )}
a_{n^{\prime}i^{\prime }m^{\prime }j^{\prime }}^{(\ell )}\ \hat{J}_{i+i^{\prime}+j+j^{\prime }+2\ell ,j+j^{\prime },\ell }\;.  
\label{kinetic:rho_ij_rho_nm}
\end{equation}
We also remark that the irreducible moments with 
negative powers $i$ and $j$ of $E_{\mathbf{k}u}$ 
and $E_{\mathbf{k}l}$, respectively, can be expressed in terms of the ones with positive indices.

\section{Equations of motion for the irreducible moments}
\label{sect:Eqs_of_motion}

The space-time evolution of the single-particle distribution function $f_{\mathbf{k}}$ of a 
single-component, weakly interacting, dilute gas is given
by the relativistic Boltzmann equation (\ref{Boltzmann_eq_2}).
Considering only binary collisions, the collision term on the right-hand side is 
\begin{equation}
C\left[ f\right] =\frac{1}{2}\int dK^{\prime }dPdP^{\prime }
\left[ W_{\mathbf{pp}^{\prime }\rightarrow \mathbf{kk}^{\prime }}f_{\mathbf{p}}f_{\mathbf{p}^{\prime }}
\left( 1-af_{\mathbf{k}}\right) \left( 1-af_{\mathbf{k}^{\prime }}\right) 
-W_{\mathbf{kk}^{\prime }\rightarrow \mathbf{pp}^{\prime}}f_{\mathbf{k}}f_{\mathbf{k}^{\prime }}
\left( 1-af_{\mathbf{p}}\right)\left( 1-af_{\mathbf{p}^{\prime }}\right) \right] \;.  \label{COLL_INT}
\end{equation}
Here the factors $1-af_{\mathbf{k}}$ represent the corrections from quantum
statistics. The factor $1/2$ appears if the colliding particles are
indistinguishable. The invariant transition rate $W_{\mathbf{kk}^{\prime}\rightarrow \mathbf{pp}^{\prime }}$ 
satisfies detailed balance, $W_{\mathbf{kk}^{\prime }\rightarrow \mathbf{pp}^{\prime }}
=W_{\mathbf{pp}^{\prime}\rightarrow \mathbf{kk}^{\prime }}$, and is symmetric with respect to
exchange of momenta, $W_{\mathbf{kk}^{\prime }\rightarrow \mathbf{pp}^{\prime }}
=W_{\mathbf{k}^{\prime }\mathbf{k}\rightarrow \mathbf{pp}^{\prime}}
=W_{\mathbf{kk}^{\prime }\rightarrow \mathbf{p}^{\prime }\mathbf{p}}$.

In order to derive equations of motion for the irreducible moments of $\delta \hat{f}_{\mathbf{k}}$, 
we proceed along the lines of 
Ref.\ \cite{Denicol:2012cn}, except that we express the derivatives using 
Eq.\ (\ref{partial_mu_aniso}) instead of Eq.\ (\ref{partial_mu_iso}) and thus we rewrite the Boltzmann equation 
(\ref{Boltzmann_eq_2}) as an evolution equation for the correction $\delta \hat{f}_{\mathbf{k}}$ 
instead of $\delta f_{\mathbf{k}}$,
\begin{equation}
D\delta \hat{f}_{\mathbf{k}}=-D\hat{f}_{0\mathbf{k}}+E_{\mathbf{k}u}^{-1}
\left( E_{\mathbf{k}l}D_{l}\hat{f}_{0\mathbf{k}}+E_{\mathbf{k}l}D_{l}\delta \hat{f}_{\mathbf{k}}
-k^{\mu }\tilde{\nabla}_{\mu }\hat{f}_{0\mathbf{k}}-k^{\mu }\tilde{\nabla}_{\mu }
\delta \hat{f}_{\mathbf{k}}\right)+E_{\mathbf{k}u}^{-1}C\left[ \hat{f}_{0\mathbf{k}}
+\delta \hat{f}_{\mathbf{k}}\right] \;.  \label{Ddeltahatf}
\end{equation}
Here $D=u^{\mu }\partial_{\mu }$, $D_{l}=-l^{\mu }\partial_{\mu }$, and 
$\tilde{\nabla}_{\mu }=\Xi_{\mu \nu }\partial ^{\nu }$.

Now we form moments of Eq.\ (\ref{Ddeltahatf}), which leads to an infinite
set of equations of motion for the irreducible moments (\ref{kinetic:rho_ij}). Defining 
\begin{equation}
D\hat{\rho}_{ij}^{\left\{ \mu _{1}\cdots \mu _{\ell }\right\} }
\equiv \Xi_{\nu _{1}\cdots \nu _{\ell }}^{\mu _{1}\cdots \mu _{\ell }}\,
D\hat{\rho}_{ij}^{\nu _{1}\cdots \nu _{\ell }}\;,  \label{D_anisotropic}
\end{equation}
as well as 
\begin{equation}
\mathcal{C}_{ij}^{\left\{ \mu _{1}\cdots \mu _{\ell }\right\} }
=\Xi _{\nu_{1}\cdots \nu _{\ell }}^{\mu _{1}\cdots \mu _{\ell }}\int dK\, E_{\mathbf{k}u}^{i}\, 
E_{\mathbf{k}l}^{j}\, k^{\nu _{1}}\cdots k^{\nu _{\ell }}C\left[ f\right]
\;,  \label{Aniso_coll_int}
\end{equation}
we obtain, after a long, but straightforward calculation, the equation of
motion for the irreducible moments of tensor-rank zero 
\begin{align}
D\hat{\rho}_{ij}& =\mathcal{C}_{i-1,j}-D\hat{\mathcal{I}}_{ij}
+D_{l}\hat{\mathcal{I}}_{i-1,j+1}+\left( i\,\hat{\mathcal{I}}_{i-1,j+1}
+j\,\hat{\mathcal{I}}_{i+1,j-1}\right) l_{\alpha }Du^{\alpha }-\left[ \left( i-1\right) 
\hat{\mathcal{I}}_{i-2,j+2}+\left( j+1\right) \hat{\mathcal{I}}_{ij}\right]
l_{\alpha }D_{l}u^{\alpha }  \notag \\
& +\frac{1}{2}\left[ m_{0}^{2}\left( i-1\right) \hat{\mathcal{I}}_{i-2,j}
-\left( i+1\right) \hat{\mathcal{I}}_{ij}+\left( i-1\right) \hat{\mathcal{I}}_{i-2,j+2}\right] 
\tilde{\theta}-\frac{1}{2}\left[ m_{0}^{2}\, j\, \hat{\mathcal{I}}_{i-1,j-1}
-j\, \hat{\mathcal{I}}_{i+1,j-1}+\left( j+2\right) 
\hat{\mathcal{I}}_{i-1,j+1}\right] \tilde{\theta}_{l}  \notag \\
& +D_{l}\hat{\rho}_{i-1,j+1}-\tilde{\nabla}_{\mu }\hat{\rho}_{i-1,j}^{\mu }+
\left[ \left( i-1\right) \hat{\rho}_{i-2,j+1}^{\mu }+j\,\hat{\rho}_{i,j-1}^{\mu }\right] 
l_{\alpha }\tilde{\nabla}_{\mu }u^{\alpha }+\left( i \,\hat{\rho}_{i-1,j+1}
+j\,\hat{\rho}_{i+1,j-1}\right) l_{\alpha }Du^{\alpha } 
\notag \\
& -\left[ \left( i-1\right) \hat{\rho}_{i-2,j+2}+\left( j+1\right) \hat{\rho}_{ij}\right] 
l_{\alpha }D_{l}u^{\alpha }+i\, \hat{\rho}_{i-1,j}^{\mu }Du_{\mu}
-\left( i-1\right) \hat{\rho}_{i-2,j+1}^{\mu }D_{l}u_{\mu }  \notag \\
& -j\, \hat{\rho}_{i,j-1}^{\mu }Dl_{\mu }+\left( j+1\right) \hat{\rho}_{i-1,j}^{\mu }
D_{l}l_{\mu }+\frac{1}{2}\left[ m_{0}^{2}\left( i-1\right) 
\hat{\rho}_{i-2,j}-\left( i+1\right) \hat{\rho}_{ij}+\left( i-1\right) \hat{\rho}_{i-2,j+2}\right] \tilde{\theta}  \notag \\
& -\frac{1}{2}\left[ m_{0}^{2}\, j\, \hat{\rho}_{i-1,j-1}-j\, \hat{\rho}_{i+1,j-1}
+\left( j+2\right) \hat{\rho}_{i-1,j+1}\right] \tilde{\theta}_{l}
+\left( i-1\right) \hat{\rho}_{i-2,j}^{\mu \nu }\tilde{\sigma}_{\mu \nu}
-j\, \hat{\rho}_{i-1,j-1}^{\mu \nu }\tilde{\sigma}_{l,\mu \nu }\;,
\label{eq_aniso_scalar}
\end{align}
where $\tilde{\theta}=\tilde{\nabla}_{\mu }u^{\mu }$, $\tilde{\theta}_{l}=
\tilde{\nabla}_{\mu }l^{\mu }$, $\tilde{\sigma}^{\mu \nu }=\partial ^{\{\mu}u^{\nu \}}$, 
and $\tilde{\sigma}_{l}^{\mu \nu }=\partial ^{\{\mu }l^{\nu \}}$.

Similarly, the time-evolution equation for the irreducible moments of tensor-rank one is 
\begin{align}
D\hat{\rho}_{ij}^{\left\{ \mu \right\} }& =\mathcal{C}_{i-1,j}^{\left\{ \mu \right\} }
-\frac{1}{2}\tilde{\nabla}^{\mu }\left( m_{0}^{2}\, 
\hat{\mathcal{I}}_{i-1,j}-\hat{\mathcal{I}}_{i+1,j}+\hat{\mathcal{I}}_{i-1,j+2}\right)  \notag \\
& +\frac{1}{2}\left[ m_{0}^{2}\, i\, \hat{\mathcal{I}}_{i-1,j}-\left( i+2\right) 
\hat{\mathcal{I}}_{i+1,j}+i\, \hat{\mathcal{I}}_{i-1,j+2}\right] \Xi^\mu_\alpha Du^{\alpha}  
-\frac{1}{2}\left[ m_{0}^{2}\, j\, \hat{\mathcal{I}}_{i,j-1}-j\, \hat{\mathcal{I}}_{i+2,j-1}
+\left( j+2\right) \hat{\mathcal{I}}_{i,j+1}\right] \Xi^\mu_\alpha Dl^{\alpha} \notag \\
& -\frac{1}{2}\left[ m_{0}^{2}\left( i-1\right) \hat{\mathcal{I}}_{i-2,j+1}-\left( i+1\right) \hat{\mathcal{I}}_{i,j+1}+\left( i-1\right) 
\hat{\mathcal{I}}_{i-2,j+3}\right] \Xi^\mu_\alpha D_l u^{\alpha}  \notag \\
& +\frac{1}{2}\left[ m_{0}^{2}\left( j+1\right) \hat{\mathcal{I}}_{i-1,j}
-\left( j+1\right) \hat{\mathcal{I}}_{i+1,j}+\left( j+3\right) \hat{\mathcal{I}}_{i-1,j+2}\right] 
\Xi^\mu_\alpha D_l l^{\alpha}\notag \\
& +\frac{1}{2}\left[ (i-1) \left( m_{0}^{2}\, \hat{\mathcal{I}}_{i-2,j+1}-\hat{\mathcal{I}}_{i,j+1}
+\hat{\mathcal{I}}_{i-2,j+3}\right)
+ j\left( m_{0}^{2}\, \hat{\mathcal{I}}_{i,j-1}-\hat{\mathcal{I}}_{i+2,j-1}
+\hat{\mathcal{I}}_{i,j+1}\right) \right] l_{\alpha }\tilde{\nabla}^{\mu }u^{\alpha }  \notag \\
& +\frac{1}{2}\left[ m_{0}^{2}\, i\, \hat{\rho}_{i-1,j}-\left( i+2\right) \hat{\rho}_{i+1,j}
+i\, \hat{\rho}_{i-1,j+2}\right] \Xi^\mu_\alpha Du^{\alpha}
 -\frac{1}{2}\left[ m_{0}^{2}\, j\, \hat{\rho}_{i,j-1}-j\, \hat{\rho}_{i+2,j-1}
+\left( j+2\right) \hat{\rho}_{i,j+1}\right]  \Xi^\mu_\alpha D l^{\alpha}  \notag \\
& -\frac{1}{2}\left[ m_{0}^{2}\left( i-1\right) \hat{\rho}_{i-2,j+1}-\left(
i+1\right) \hat{\rho}_{i,j+1}+\left( i-1\right) \hat{\rho}_{i-2,j+3}\right]
 \Xi^\mu_\alpha D_l u^{\alpha}  \notag \\
& +\frac{1}{2}\left[ m_{0}^{2}\left( j+1\right) \hat{\rho}_{i-1,j}-\left(
j+1\right) \hat{\rho}_{i+1,j}+\left( j+3\right) \hat{\rho}_{i-1,j+2}\right]
 \Xi^\mu_\alpha D_l l^{\alpha}   \notag \\
& -\frac{1}{2}\tilde{\nabla}^{\mu }\left( m_{0}^{2}\, \hat{\rho}_{i-1,j}-\hat{\rho}_{i+1,j}
+\hat{\rho}_{i-1,j+2}\right) +\Xi _{\alpha }^{\mu }D_{l}\hat{\rho}_{i-1,j+1}^{\alpha }  \notag \\
& +\frac{1}{2}\left[ (i-1) \left( m_{0}^{2}\, \hat{\rho}_{i-2,j+1}-\hat{\rho}_{i,j+1}
+\hat{\rho}_{i-2,j+3}\right) + j
\left( m_{0}^{2}\, \hat{\rho}_{i,j-1}-\hat{\rho}_{i+2,j-1}+\hat{\rho}_{i,j+1}\right) \right]
l_{\alpha }\tilde{\nabla}^{\mu }u^{\alpha }  \notag \\
& +\left[ i\, \hat{\rho}_{i-1,j+1}^{\mu }+j\, \hat{\rho}_{i+1,j-1}^{\mu }\right]
l_{\alpha } Du^{\alpha }+\hat{\rho}_{ij,\nu }\, \tilde{\omega}^{\mu \nu }
+\hat{\rho}_{i-1,j+1,\nu }\, \tilde{\omega}_{l}^{\mu \nu }  \notag \\
& +\frac{1}{2}\left[ m_{0}^{2}\left( i-1\right) \hat{\rho}_{i-2,j}^{\mu}
-\left( i+2\right) \hat{\rho}_{ij}^{\mu }+\left( i-1\right) \hat{\rho}_{i-2,j+2}^{\mu }\right] 
\tilde{\theta}-\frac{1}{2}\left[ m_{0}^{2}\, j\, \hat{\rho}_{i-1,j-1}^{\mu }
-j\, \hat{\rho}_{i+1,j-1}^{\mu }+\left( j+3\right) \hat{\rho}_{i-1,j+1}^{\mu }\right] \tilde{\theta}_{l}  \notag \\
& +\frac{1}{2}\left[ m_{0}^{2}\left( i-1\right) \hat{\rho}_{i-2,j,\nu}
-\left( i+1\right) \hat{\rho}_{ij,\nu }+\left( i-1\right) \hat{\rho}_{i-2,j+2,\nu }\right] \tilde{\sigma}^{\mu \nu }  \notag \\
& -\frac{1}{2}\left[ m_{0}^{2}\, j\, \hat{\rho}_{i-1,j-1,\nu }-j\, \hat{\rho}_{i+1,j-1,\nu }
+\left( j+2\right) \hat{\rho}_{i-1,j+1,\nu }\right] \tilde{\sigma}_{l}^{\mu \nu }
-\left[ \left( i-1\right) \hat{\rho}_{i-2,j+2}^{\mu}
+\left( j+1\right) \hat{\rho}_{ij}^{\mu}\right] l_{\alpha }D_{l}u^{\alpha }  \notag \\
& -\Xi _{\alpha }^{\mu }\tilde{\nabla}_{\nu }\hat{\rho}_{i-1,j}^{\alpha \nu}
- \left( i-1\right) \hat{\rho}_{i-2,j+1}^{\mu \nu }D_{l}u_{\nu }+\left(
j+1\right) \hat{\rho}_{i-1,j}^{\mu \nu }D_{l}l_{\nu }+i\, \hat{\rho}_{i-1,j}^{\mu \nu }
Du_{\nu }-j\, \hat{\rho}_{i,j-1}^{\mu \nu }Dl_{\nu }  \notag \\
& +\left[ \left( i-1\right) \hat{\rho}_{i-2,j+1}^{\mu \nu }
+j\, \hat{\rho}_{i,j-1}^{\mu \nu }\right] l_{\alpha }\tilde{\nabla}_{\nu }u^{\alpha }
+\left( i-1\right) \hat{\rho}_{i-2,j}^{\mu \nu \lambda }
\tilde{\sigma}_{\nu \lambda }-j\, \hat{\rho}_{i-1,j-1}^{\mu \nu \lambda }
\tilde{\sigma}_{l,\nu \lambda }\;,  \label{eq_aniso_vector}
\end{align}
where $\tilde{\omega}^{\mu \nu }=\Xi ^{\mu \alpha }\Xi ^{\nu \beta }\partial_{\lbrack \alpha }u_{\beta ]}$ and 
$\tilde{\omega}_{l}^{\mu \nu }=\Xi ^{\mu \alpha }\Xi ^{\nu \beta }\partial _{\lbrack \alpha }l_{\beta ]}$.

Finally, the equation of motion for the irreducible moments of tensor-rank two is 
\begin{align}
\lefteqn{D\hat{\rho}_{ij}^{\left\{ \mu \nu \right\} } =\mathcal{C}_{i-1,j}^{\left\{\mu \nu \right\} } }\notag \\
& +\frac{1}{4}\left\{ m_{0}^{4}\left( i-1\right) \hat{\mathcal{I}}_{i-2,j}-2m_{0}^{2}\left[ \left( i+1\right) 
\hat{\mathcal{I}}_{ij}-\left( i-1\right) \hat{\mathcal{I}}_{i-2,j+2}\right] - 2\left( i+1\right) \hat{\mathcal{I}}_{i,j+2}
+\left(i+3\right) \hat{\mathcal{I}}_{i+2,j}+\left( i-1\right) \hat{\mathcal{I}}_{i-2,j+4}\right\} \tilde{\sigma}^{\mu \nu }  \notag \\
& -\frac{1}{4}\left\{ m_{0}^{4}\, j\, \hat{\mathcal{I}}_{i-1,j-1}-2m_{0}^{2}\left[ j\, \hat{\mathcal{I}}_{i+1,j-1}
-\left( j+2\right) \hat{\mathcal{I}}_{i-1,j+1}\right] - 2\left( j+2\right) \hat{\mathcal{I}}_{i+1,j+1}+j\, \hat{\mathcal{I}}_{i+3,j-1}
+\left( j+4\right) \hat{\mathcal{I}}_{i-1,j+3}\right\} \tilde{\sigma}_{l}^{\mu \nu }  \notag \\
& +\frac{1}{4}\left\{ m_{0}^{4}\left( i-1\right) \hat{\rho}_{i-2,j}-2m_{0}^{2}\left[ \left( i+1\right) \hat{\rho}_{ij}
-\left(i-1\right) \hat{\rho}_{i-2,j+2}\right]  - 2\left( i+1\right) \hat{\rho}_{i,j+2}+\left( i+3\right) 
\hat{\rho}_{i+2,j}+\left( i-1\right) \hat{\rho}_{i-2,j+4}\right\} \tilde{\sigma}^{\mu \nu }  \notag \\
& -\frac{1}{4}\left\{ m_{0}^{4}\, j\, \hat{\rho}_{i-1,j-1}-2m_{0}^{2}\left[ j\, \hat{\rho}_{i+1,j-1}
-\left( j+2\right) \hat{\rho}_{i-1,j+1}\right] - 2\left( j+2\right) \hat{\rho}_{i+1,j+1}+j\, \hat{\rho}_{i+3,j-1}
+\left( j+4\right) \hat{\rho}_{i-1,j+3}\right\} \tilde{\sigma}_{l}^{\mu \nu }  \notag \\
& +\frac{1}{2}\left[ m_{0}^{2}\, i\, \hat{\rho}_{i-1,j}^{\left\{ \mu \right.}
-\left( i+4\right) \hat{\rho}_{i+1,j}^{\left\{ \mu \right. }+i\, \hat{\rho}_{i-1,j+2}^{\left\{ \mu \right. }\right] 
Du^{\left. \nu \right\} }-\frac{1}{2}\left[ m_{0}^{2}\, j\, \hat{\rho}_{i,j-1}^{\left\{ \mu \right. }
-j\, \hat{\rho}_{i+2,j-1}^{\left\{ \mu \right. }+\left( j+4\right) \hat{\rho}_{i,j+1}^{\left\{ \mu \right. }\right] 
Dl^{\left. \nu \right\} }  \notag \\
& +\frac{1}{2}\Xi _{\alpha \beta }^{\mu \nu }\left[ (i-1) \left( m_{0}^{2}\, \hat{\rho}_{i-2,j+1}^{\alpha }
-\hat{\rho}_{i,j+1}^{\alpha }+\hat{\rho}_{i-2,j+3}^{\alpha }\right) 
+ j\left( m_{0}^{2}\, \hat{\rho}_{i,j-1}^{\alpha }
-\hat{\rho}_{i+2,j-1}^{\alpha }+\hat{\rho}_{i,j+1}^{\alpha}\right) \right]
l_{\gamma }\tilde{\nabla}^{\beta }u^{\gamma }  \notag \\
& -\frac{1}{2}\Xi _{\alpha \beta }^{\mu \nu }\left[ m_{0}^{2}\left(i-1\right) \hat{\rho}_{i-2,j+1}^{\alpha }
-\left( i+3\right) \hat{\rho}_{i,j+1}^{\alpha }+\left( i-1\right) \hat{\rho}_{i-2,j+3}^{\alpha }\right]
D_{l}u^{\beta }  \notag \\
& +\frac{1}{2}\Xi _{\alpha \beta }^{\mu \nu }\left[ m_{0}^{2}\left(j+1\right) \hat{\rho}_{i-1,j}^{\alpha }
-\left( j+1\right) \hat{\rho}_{i+1,j}^{\alpha }+\left( j+5\right) \hat{\rho}_{i-1,j+2}^{\alpha }\right]
D_{l}l^{\beta }  \notag \\
& -2\, \tilde{\omega}_{\lambda }^{\hspace*{0.1cm}\left\{ \mu \right. }\hat{\rho}_{ij}^{\left.
\nu \right\} \lambda }-2\, \tilde{\omega}_{l,\lambda }^{\hspace*{0.2cm}\left\{ \mu \right. } 
\hat{\rho}_{i-1,j+1}^{\left. \nu \right\} \lambda }-\frac{1}{2}\tilde{\nabla}^{\left\{ \mu \right. }
\left( m_0^{2}\, \hat{\rho}_{i-1,j}^{\left. \nu \right\} }-
\hat{\rho}_{i+1,j}^{\left. \nu \right\} }+\hat{\rho}_{i-1,j+2}^{\left. \nu \right\} }\right)   \notag \\
& +\Xi _{\alpha \beta }^{\mu \nu }D_{l}\hat{\rho}_{i-1,j+1}^{\alpha \beta }+\left[ i\, 
\hat{\rho}_{i-1,j+1}^{\mu \nu }+j\, \hat{\rho}_{i+1,j-1}^{\mu \nu }\right] l_{\alpha }
Du^{\alpha }-\left[ \left( i-1\right) \hat{\rho}_{i-2,j+2}^{\mu \nu }
+\left( j+1\right) \hat{\rho}_{ij}^{\mu \nu }\right] l_{\alpha }D_{l}u^{\alpha }  \notag \\
& +\frac{1}{2}\left[ m_{0}^{2}\left( i-1\right) \hat{\rho}_{i-2,j}^{\mu \nu}
-\left( i+3\right) \hat{\rho}_{ij}^{\mu \nu }+\left( i-1\right) \hat{\rho}_{i-2,j+2}^{\mu \nu }\right] \tilde{\theta}  
+\frac{2}{3}\left[ m_{0}^{2}\left( i-1\right) \hat{\rho}_{i-2,j}^{\kappa
\left\{ \mu \right. }-\left( i+2\right) \hat{\rho}_{ij}^{\kappa \left\{ \mu\right. }
+\left( i-1\right) \hat{\rho}_{i-2,j+2}^{\kappa \left\{ \mu \right.}\right] 
\tilde{\sigma}_{\kappa }^{\left. \nu \right\} }  \notag \\
& -\frac{1}{2}\left[ m_{0}^{2}\, j\, \hat{\rho}_{i-1,j-1}^{\mu \nu }-j\, \hat{\rho}_{i+1,j-1}^{\mu \nu }
+\left( j+4\right) \hat{\rho}_{i-1,j+1}^{\mu \nu }\right] \tilde{\theta}_{l}  
-\frac{2}{3}\left[ m_{0}^{2}\, j\, \hat{\rho}_{i-1,j-1}^{\kappa \left\{ \mu\right. }
-j\, \hat{\rho}_{i+1,j-1}^{\kappa \left\{ \mu \right. }+\left(j+3\right) 
\hat{\rho}_{i-1,j+1}^{\kappa \left\{ \mu \right. }\right] \tilde{\sigma}_{l,\kappa }^{\left. \nu \right\} }  \notag \\
& -\Xi _{\alpha \beta }^{\mu \nu }\tilde{\nabla}_{\lambda }\hat{\rho}_{i-1,j}^{\alpha \beta \lambda }
+i\, \hat{\rho}_{i-1,j}^{\mu \nu \gamma}Du_{\gamma }-j\, \hat{\rho}_{i,j-1}^{\mu \nu \gamma }
Dl_{\gamma }-\left(i-1\right) \hat{\rho}_{i-2,j+1}^{\mu \nu \lambda }D_{l}u_{\lambda }
+\left(j+1\right) \hat{\rho}_{i-1,j}^{\mu \nu \lambda }D_{l}l_{\lambda }  \notag \\
& +\left[ \left( i-1\right) \hat{\rho}_{i-2,j+1}^{\mu \nu \lambda }+j\, \hat{\rho}_{i,j-1}^{\mu \nu \lambda}
\right] l_{\alpha } \tilde{\nabla}_{\lambda }u^{\alpha }+\left( i-1\right) \hat{\rho}_{i-2,j}^{\mu \nu \lambda \kappa }
\tilde{\sigma}_{\lambda \kappa }-j\hat{\rho}_{i-1,j-1}^{\mu \nu \lambda \kappa }\tilde{\sigma}_{l,\lambda \kappa }\;.
\label{eq_aniso_tensor}
\end{align}
Since the fluid-dynamical equations of motion do not contain quantities of
tensor rank higher than two, we do not explicitly quote the equations of
motion for the irreducible moments $\hat{\rho}_{ij}^{\left\{ \mu _{1}\cdots
\mu _{\ell }\right\} }$ with $\ell \geq 3$. The equations of motion of
relativistic dissipative fluid dynamics for an anisotropic reference state
can now be obtained from these general equations for different values of $i$
and $j$. This will be further elaborated in Sec.\ \ref{fluideom}.

\section{Collision integrals}
\label{sect:Collision integral}

In order to derive the fluid-dynamical equations, we still need to consider
the collision terms (\ref{Aniso_coll_int}), which appear in 
Eqs.\ (\ref{eq_aniso_scalar}) -- (\ref{eq_aniso_tensor}). Exchanging integration variables 
$({\bf p},{\bf p}^\prime) \leftrightarrow ({\bf k}, {\bf k}^\prime)$, we can rewrite Eq.\
(\ref{Aniso_coll_int}) as 
\begin{equation} \label{collint}
\mathcal{C}_{ij}^{\left\{ \mu _{1}\cdots \mu _{\ell }\right\} }=\frac{1}{2}
\int dKdK^{\prime }dPdP^{\prime }f_{\mathbf{k}}f_{\mathbf{k}^{\prime}}
\left( 1-af_{\mathbf{p}}\right) \left( 1-af_{\mathbf{p}^{\prime }}\right)
W_{\mathbf{kk}\prime \rightarrow \mathbf{pp}\prime }
\left( E_{\mathbf{p}u}^{i}E_{\mathbf{p}l}^{j}p^{\left\{ \mu _{1}\right. }\cdots p^{\left. \mu_{\ell }\right\} }
-E_{\mathbf{k}u}^{i}E_{\mathbf{k}l}^{j}k^{\left\{ \mu_{1}\right. }\cdots k^{\left. \mu _{\ell }\right\} }\right) \;.
\end{equation}
As a consequence of the conservation of particle number as well as energy
and momentum in binary collisions, we have 
\begin{equation}  \label{vanishing_coll_int}
\mathcal{C}_{00}=\mathcal{C}_{10}=\mathcal{C}_{01}
= \mathcal{C}_{00}^{\left\{ \mu \right\} }=0
\end{equation}
for any distribution function $f_{\mathbf{k}}$.

Now inserting the distribution function from Eq.\ (\ref{kinetic:f_full_anisotropic}) 
into Eq.\ (\ref{collint}) and neglecting terms proportional to 
$\delta \hat{f}_{\mathbf{k}}\delta \hat{f}_{\mathbf{k}^{\prime }}$ we obtain 
\begin{equation}  \label{cchatl}
\mathcal{C}_{ij}^{\left\{ \mu _{1}\cdots \mu _{\ell }\right\} }
=\hat{\mathcal{C}}_{ij}^{\left\{ \mu _{1}\cdots \mu _{\ell }\right\} }
+\hat{\mathcal{L}}_{ij}^{\left\{ \mu _{1}\cdots \mu _{\ell }\right\} }\;.
\end{equation}
Here,
\begin{equation}  \label{chat_coll_int}
\hat{\mathcal{C}}_{ij}^{\left\{ \mu _{1}\cdots \mu _{\ell }\right\} }\equiv \frac{1}{2}
\int dKdK^{\prime }dPdP^{\prime }W_{\mathbf{kk}\prime \rightarrow 
\mathbf{pp}\prime }\hat{f}_{0\mathbf{k}}\hat{f}_{0\mathbf{k}^{\prime }}( 1-a 
\hat{f}_{0\mathbf{p}}) ( 1-a\hat{f}_{0\mathbf{p}^{\prime }}) 
\left( E_{\mathbf{p}u}^{i}E_{\mathbf{p}l}^{j}p^{\left\{ \mu _{1}\right. }\cdots
p^{\left. \mu _{\ell }\right\} }-E_{\mathbf{k}u}^{i}E_{\mathbf{k}l}^{j}
k^{\left\{ \mu _{1}\right. }\cdots k^{\left. \mu _{\ell }\right\} }\right) \;.
\end{equation}
In local thermodynamical equilibrium, i.e., replacing $\hat{f}_{0\mathbf{k}}$
by $f _{0\mathbf{k}}$, such a collision integral vanishes due to the symmetry
of the collision rate, $W_{\mathbf{kk}^{\prime }\rightarrow \mathbf{pp}^{\prime }}
=W_{\mathbf{pp}^{\prime}\rightarrow \mathbf{kk}^{\prime }}$, and energy conservation
in binary elastic collisions, $E_{{\bf p}u} + E_{{\bf p}^\prime u} =E_{{\bf k}u} + E_{{\bf k}^\prime u}$. 
However, for the anisotropic state characterized by $\hat{f}_{0\mathbf{k}}$ this is a priori
not the case. Thus, if we consider the fluid dynamics of such a system,
without additional corrections from the irreducible moments of $\delta \hat{f}_{\mathbf{k}}$, as was done in 
Refs.\ \cite{Florkowski:2010cf,Ryblewski:2010bs,Ryblewski:2011aq,Ryblewski:2012rr,Martinez:2009ry,Martinez:2010sc,Martinez:2010sd,Martinez:2012tu}, 
the microscopic collision dynamics contained in the term (\ref{chat_coll_int}) 
is solely responsible for the approach towards local
thermodynamic equilibrium.

Let us now turn to the second term in Eq.\ (\ref{cchatl}) and, in order to simplify the discussion, consider
the case of Boltzmann statistics, $a =0$. Then,
\begin{align}
\hat{\mathcal{L}}_{ij}^{\left\{ \mu _{1}\cdots \mu _{\ell }\right\} }& 
= \frac{1}{2}\sum_{r=0}^{\infty }\sum_{n=0}^{N_{r}}\sum_{m=0}^{N_{r}-n} 
\hat{\rho}_{nm}^{\nu _{1}\cdots \nu _{r}} \int dKdK^{\prime }dPdP^{\prime }
W_{\mathbf{kk}\prime \rightarrow \mathbf{pp}\prime }\hat{f}_{0\mathbf{k}}\hat{f}_{0\mathbf{k}^{\prime }}
\notag \\
& \times \left( k_{\left\{ \nu _{1}\right. }\cdots k_{\left. \nu_{r}\right\} }\hat{\mathcal{H}}_{\mathbf{k}nm}^{(r)}
+k_{\left\{ \nu_{1}\right. }^{\prime }\cdots k_{\left. \nu _{r}\right\} }^{\prime }
\hat{\mathcal{H}}_{\mathbf{k}^{\prime }nm}^{(r)}\right) \left( E_{\mathbf{p}u}^{i}E_{\mathbf{p}l}^{j}
p^{\left\{ \mu _{1}\right. }\cdots p^{\left. \mu_{\ell }\right\} }-E_{\mathbf{k}u}^{i}E_{\mathbf{k}l}^{j}
k^{\left\{ \mu_{1}\right. }\cdots k^{\left. \mu _{\ell }\right\} }\right) \;.
\label{L_ij_full}
\end{align}
In analogy to Eqs.\ (50) and (51) of Ref.\ \cite{Denicol:2012cn}, this
expression can be rewritten as, see App.\ \ref{appendix_orhogonality}
for details, 
\begin{equation}
\hat{\mathcal{L}}_{ij}^{\left\{ \mu _{1}\cdots \mu _{\ell }\right\} }\equiv
\sum_{n=0}^{N_{\ell }}\sum_{m=0}^{N_{\ell }-n}\hat{\rho}_{nm}^{\mu_{1}\cdots \mu _{\ell }} 
\mathcal{A}_{injm}^{\left( \ell \right) } \;,
\label{L_ij_simplified}
\end{equation}
where 
\begin{align}
\mathcal{A}_{injm}^{\left( \ell \right) } & =\frac{1}{2} \int dKdK^{\prime
}dPdP^{\prime }W_{\mathbf{kk}\prime \rightarrow \mathbf{pp}\prime }
\hat{f}_{0\mathbf{k}}\hat{f}_{0\mathbf{k}^{\prime }}
\notag \\
& \times \left( k_{\left\{ \mu _{1}\right. }\cdots k_{\left. \mu _{\ell}\right\} }
\hat{\mathcal{H}}_{\mathbf{k}nm}^{(\ell )}+k_{\left\{ \mu_{1}\right. }^{\prime }\cdots 
k_{\left. \mu _{\ell }\right\} }^{\prime }\hat{\mathcal{H}}_{\mathbf{k}^{\prime }nm}^{(\ell )}\right) 
\left( E_{\mathbf{p}u}^{i}E_{\mathbf{p}l}^{j}p^{\left\{ \mu _{1}\right. }\cdots p^{\left. \mu_{\ell }\right\} }
-E_{\mathbf{k}u}^{i}E_{\mathbf{k}l}^{j}k^{\left\{ \mu_{1}\right. }\cdots k^{\left. \mu _{\ell }\right\} }\right) \;.
\label{A_l_injm}
\end{align}
Similar to the tensor $\hat{\mathcal{C}}_{ij}^{\left\{ \mu _{1}\cdots \mu_{\ell }\right\} }$ 
in Eq.\ (\ref{chat_coll_int}), the coefficients 
$\mathcal{A}_{injm}^{\left( \ell \right) }$ contain information about the
microscopic interactions. The difference is that the tensor 
$\hat{\mathcal{L}}_{ij}^{\left\{ \mu _{1}\cdots \mu _{\ell }\right\} }$ is linearly
proportional to the irreducible moments $\hat{\rho}_{nm}^{\mu _{1}\cdots \mu_{\ell}}$. 
This means on the one hand that $\hat{\mathcal{L}}_{ij}^{\left\{ \mu_{1}\cdots \mu _{\ell }\right\} }=0$ 
if we consider a state characterized by 
$\hat{f}_{0\mathbf{k}}$ without corrections $\sim \delta \hat{f}_{\mathbf{k}}$, 
such as in Refs.\ 
\cite{Florkowski:2010cf,Ryblewski:2010bs,Ryblewski:2011aq,Ryblewski:2012rr,Martinez:2009ry,Martinez:2010sc,Martinez:2010sd,Martinez:2012tu}. 
On the other hand, the linear proportionality of 
$\hat{\mathcal{L}}_{ij}^{\left\{ \mu _{1}\cdots \mu _{\ell }\right\} }$ to 
$\hat{\rho}_{nm}^{\mu _{1}\cdots \mu _{\ell}}$ means that the coefficients 
$\mathcal{A}_{injm}^{\left( \ell \right) }$ are inversely proportional to the relaxation
time scales for the irreducible moments.

In the remainder of this section, we discuss a widely used simplification
for the collision integral, the so-called relaxation-time approximation
(RTA) \cite{Bhatnagar:1954zz}. In this case, the full collision integral is
replaced by the following relativistically invariant expression 
\cite{Anderson_Witting},
\begin{equation}
C\left[ f\right] \equiv -\frac{E_{\mathbf{k}u}}{\tau _{rel}}\left( f_{\mathbf{k}}-f_{0\mathbf{k}}\right) 
=-\frac{E_{\mathbf{k}u}}{\tau _{rel}}\left( \hat{f}_{0\mathbf{k}}+\delta \hat{f}_{\mathbf{k}}
-f_{0\mathbf{k}}\right) \;,
\end{equation}
where we used Eq.\ (\ref{kinetic:f=f0+df}) and $\tau _{rel}$ is the
so-called relaxation time, which is assumed to be independent of the particle
four-momenta and determines the time scale over which $f_{\mathbf{k}}$
relaxes towards $f_{0\mathbf{k}}$. In our case this approximation translates
with the help of Eq.\ (\ref{Aniso_coll_int}) and the definitions 
(\ref{I_ij_tens}), (\ref{kinetic:rho_ij}) into 
\begin{equation}
\mathcal{C}_{i-1,j}^{\left\{ \mu _{1}\cdots \mu _{\ell }\right\} }
=-\frac{1}{\tau _{rel}}\left( \hat{\mathcal{I}}_{ij}^{\left\{ \mu _{1}\cdots \mu _{\ell}\right\} }
+\hat{\rho}_{ij}^{\left\{ \mu _{1}\cdots \mu _{\ell }\right\} }-
\mathcal{I}_{ij}^{\left\{ \mu _{1}\cdots \mu _{\ell }\right\} }\right) \;,
\end{equation}
where we defined the generalized moments of the local equilibrium
distribution function $f_{0\mathbf{k}}$ of tensor-rank $\ell $, 
\begin{equation}
\lim_{\hat{\beta}_{l}\rightarrow 0}\hat{\mathcal{I}}_{ij}^{\mu _{1}\cdots
\mu _{\ell }}\equiv \mathcal{I}_{ij}^{\mu _{1}\cdots \mu _{\ell}}
=\left\langle E_{\mathbf{k}u}^{i}E_{\mathbf{k}l}^{j}k^{\mu _{1}}\cdots
k^{\mu _{\ell }}\right\rangle _{0}\;,  \label{I_ij_moment_equilibrium}
\end{equation}
for more details see App.\ \ref{appendix_thermo_integrals_equilibrium}. Note that for symmetry reasons
$\mathcal{I}_{ij}^{\mu _{1}\cdots \mu _{\ell}}\equiv 0$ for all odd $j$.
For the scalar collision integrals, this simplifies to 
\begin{equation} \label{RTA_1}
\mathcal{C}_{i-1,j}=-\frac{1}{\tau _{rel}}\left( \hat{\mathcal{I}}_{ij}
+\hat{\rho}_{ij}-\mathcal{I}_{ij}\right) \;.
\end{equation}
However, for any $\ell \geq 1$, $\hat{\mathcal{I}}_{ij}^{\left\{ \mu_{1}\cdots \mu _{\ell }\right\} }
=\mathcal{I}_{ij}^{\left\{ \mu _{1}\cdots \mu _{\ell }\right\} }=0$, and we have 
\begin{equation}
\mathcal{C}_{i-1,j}^{\left\{ \mu _{1}\cdots \mu _{\ell }\right\} }
=-\frac{1}{\tau _{rel}}\hat{\rho}_{ij}^{\left\{ \mu _{1}\cdots \mu _{\ell }\right\}
}\;,\;\;\;\ell \geq 1\;.
\end{equation}
This means that, in RTA, for any $\ell \geq 1$ we have 
$\hat{\mathcal{C}}_{ij}^{\left\{ \mu _{1}\cdots \mu _{\ell }\right\} }\equiv 0$
and $\mathcal{A}_{injm}^{\left( \ell \right) }= - \delta _{in}\delta _{jm}/\tau_{rel}$.

\section{Fluid-dynamical equations of motion}
\label{fluideom}

In this section, we derive the fluid-dynamical equations
of motion from the equations of motion for the irreducible moments, 
Eqs.\ (\ref{eq_aniso_scalar}) -- (\ref{eq_aniso_tensor}). We split the discussion
into two subsections. In the first, we derive the conservation equations
for particle number and energy-momentum. These equations are special, since
for them the collision integrals vanish identically. In the second subsection, we
discuss the remaining equations which are necessary to close the system of
equations of motion. As expected, these equations correspond to relaxation
equations for the dissipative quantities.

\subsection{Conservation equations}

For the conservation equations, the collision integrals vanish on account of
particle-number and energy-momentum conservation in binary collisions, cf.\
Eq.\ (\ref{vanishing_coll_int}). We shall show below that, from this
equation, we can identify

\begin{enumerate}
\item[(i)] the conservation equation for particle number as the equation of
motion (\ref{eq_aniso_scalar}) for $(i,j)=(1,0)$, because 
$\mathcal{C}_{00}=0 $,

\item[(ii)] the conservation of energy as the equation of motion 
(\ref{eq_aniso_scalar}) for $(i,j)=(2,0)$, because $\mathcal{C}_{10}=0$,

\item[(iii)] the conservation equation for the momentum in $l^\mu$ direction
as the equation of motion (\ref{eq_aniso_scalar}) for $(i,j)=(1,1)$, because 
$\mathcal{C}_{01}=0$, and

\item[(iv)] the conservation equation for the momentum in the direction
transverse to $l^\mu$ as the equation of motion (\ref{eq_aniso_vector}) for 
$(i,j)=(1,0)$, because $\mathcal{C}_{00}^{\{\mu\}} =0$.
\end{enumerate}

In order to prove this, we first have to write the respective equations for
the irreducible moments of $\delta \hat{f}_{\mathbf{k}}$ in terms of the
usual fluid-dynamical variables. Using Eq.\ (\ref{kinetic:rho_ij}), the
definition of $\delta \hat{f}_{\mathbf{k}}$, and Eqs.\ (\ref{kinetic:n}), 
(\ref{kinetic:e}), (\ref{kinetic:n_l}) -- (\ref{kinetic:P_l}) and 
(\ref{kinetic:Vt_mu}) -- (\ref{kinetic:pit_munu}), as well as 
Eqs.\ (\ref{n_aniso}) -- (\ref{P_l_aniso}) we obtain 
\begin{eqnarray}
\hat{\rho}_{10} &=&n-\hat{n}\;,  \label{hat_rho_n} \\
\hat{\rho}_{20} &=&e-\hat{e}\;,  \label{hat_rho_e} \\
\hat{\rho}_{01} &=&n_{l}-\hat{n}_{l}\;,  \label{hat_rho_n_l} \\
\hat{\rho}_{11} &=&M-\hat{M}\;,  \label{hat_rho_M} \\
\hat{\rho}_{02} &=&P_{l}-\hat{P}_{l}\;,  \label{hat_rho_P_l} \\
\hat{\rho}_{00}^{\mu } &=&V_{\perp }^{\mu },\   \label{hat_rho_V} \\
\hat{\rho}_{10}^{\mu } &=&W_{\perp u}^{\mu },\   \label{hat_rho_W_u} \\
\hat{\rho}_{01}^{\mu } &=&W_{\perp l}^{\mu }\;,  \label{hat_rho_W_l} \\
\hat{\rho}_{00}^{\mu \nu } &=&\pi _{\perp }^{\mu \nu }\;.
\label{hat_rho_pi_t}
\end{eqnarray}
Inserting these identities into Eq.\ (\ref{eq_aniso_scalar}) for 
$(i,j)=(1,0) $ leads to the conservation equation for particle number, 
\begin{equation}
\partial _{\mu }N^{\mu }\equiv Dn-D_{l}n_{l}+n\,\tilde{\theta}+n_{l}\,\tilde{\theta}_{l}
+n\,l_{\mu }D_{l}u^{\mu }-n_{l}\,l_{\mu }Du^{\mu }-V_{\perp}^{\mu }Du_{\mu }
-V_{\perp }^{\mu }D_{l}l_{\mu }+\tilde{\nabla}_{\mu}V_{\perp }^{\mu }=0\;.  \label{conservation_eq_particle}
\end{equation}
Analogously, inserting them into Eq.\ (\ref{eq_aniso_scalar}) for $(i,j)=(2,0)$ we obtain 
the energy conservation equation, 
\begin{align}
u_{\nu }\partial _{\mu }T^{\mu \nu }& \equiv De-D_{l}M+\left( e+P_{\perp}\right) \tilde{\theta}
+M\tilde{\theta}_{l}+\left( e+P_{l}\right) l_{\mu}D_{l}u^{\mu }-2M\,l_{\mu }Du^{\mu }  \notag \\
& -2W_{\perp u}^{\mu }Du_{\mu }-W_{\perp u}^{\mu }D_{l}l_{\mu }+W_{\perp l}^{\nu }D_{l}u_{\nu }
-W_{\perp l}^{\mu }\,l_{\nu }\tilde{\nabla}_{\mu}u^{\nu }+\tilde{\nabla}_{\mu }W_{\perp u}^{\mu }
-\pi _{\perp }^{\mu \nu }
\tilde{\sigma}_{\mu \nu }=0\;.  \label{conservation_eq_energy}
\end{align}
Repeating this for Eq.\ (\ref{eq_aniso_scalar}) in the case $(i,j)=(1,1)$
yields the conservation of momentum in $l^{\mu }$ direction, 
\begin{align}
l_{\nu }\partial _{\mu }T^{\mu \nu }& \equiv -DM+D_{l}P_{l}-M\tilde{\theta}
+\left( P_{\perp }-P_{l}\right) \tilde{\theta}_{l}-2M\,l_{\mu }D_{l}u^{\mu}
+\left( e+P_{l}\right) l_{\mu }Du^{\mu }  \notag \\
& +W_{\perp l}^{\mu }Du_{\mu }+2W_{\perp l}^{\mu }D_{l}l_{\mu }-W_{\perp u}^{\mu }
Dl_{\mu }+W_{\perp l}^{\mu }\,l_{\nu }\tilde{\nabla}_{\mu }u^{\nu }-
\tilde{\nabla}_{\mu }W_{\perp l}^{\mu }-\pi _{\perp }^{\mu \nu }\tilde{\sigma}_{l,\mu \nu }=0\;.  \label{conservation_eq_momenta_l}
\end{align}
Finally, using Eq.\ (\ref{eq_aniso_vector}) for $(i,j)=(1,0)$, we obtain the
conservation of momentum in the direction transverse to $l^{\mu }$, 
\begin{align}
\Xi _{\nu }^{\alpha }\partial _{\mu }T^{\mu \nu }& \equiv -\tilde{\nabla}^{\alpha }
P_{\perp }+DW_{\perp u}^{\alpha }-\Xi _{\nu }^{\alpha}D_{l}W_{\perp l}^{\nu }
+\frac{3}{2}W_{\perp u}^{\alpha }\ \tilde{\theta}+
\frac{3}{2}W_{\perp l}^{\alpha }\ \tilde{\theta}_{l}+W_{\perp u}^{\alpha }\
l_{\mu }D_{l}u^{\mu }-W_{\perp l}^{\alpha }\ l_{\mu }Du^{\mu }  \notag \\
& +\left( e+P_{\perp }\right) \left( Du^{\alpha }+l^{\alpha }l_{\nu }Du^{\nu}\right) 
+M\left( Dl^{\alpha }+u^{\alpha }l_{\nu }Du^{\nu }\right) 
-M\left(D_{l}u^{\alpha }+l^{\alpha }l_{\nu }D_{l}u^{\nu }\right)  \notag \\
& +\left( P_{\perp }-P_{l}\right) \left( D_{l}l^{\alpha }+u^{\alpha }l_{\nu
}D_{l}u^{\nu }\right) +W_{\perp u,\nu }\left( \tilde{\sigma}^{\alpha \nu }-
\tilde{\omega}^{\alpha \nu }\right) +W_{\perp l,\nu }\left( \tilde{\sigma}_{l}^{\alpha \nu }
-\tilde{\omega}_{l}^{\alpha \nu }\right)  \notag \\
& -\pi _{\perp }^{\mu \alpha }Du_{\mu }-\pi _{\perp }^{\mu \alpha}D_{l}l_{\mu }
+\Xi _{\nu }^{\alpha }\tilde{\nabla}_{\mu }\pi _{\perp }^{\mu\nu }=0\;.  \label{conservation_eq_momenta_t}
\end{align}
Of course, the conservation equations (\ref{conservation_eq_particle}) -- 
(\ref{conservation_eq_momenta_t}) also follow from taking the four-derivative
of the tensor decompositions (\ref{kinetic:N_mu_u_l}) and 
(\ref{kinetic:T_munu_u_l}) and projecting onto the directions given by 
$u^{\mu }$, $l^{\mu }$, and $\Xi ^{\mu \nu }$.
Note that irreducible moments of higher tensor rank, like $\hat{\rho}_{ij}^{\mu \nu \lambda }$ 
do not appear in Eq.\ (\ref{conservation_eq_momenta_t}), since the coefficients of these 
terms in Eq.\ (\ref{eq_aniso_vector}) vanish for $i=1$ and $j=0$. 

Now we need to find the connection between the actual state of the fluid
to a fictitious anisotropic state where the single-particle distribution function
is given by $\hat{f}_{0{\bf k}}$, i.e., we need to determine the three 
intensive quantities $\hat{\alpha}$, $\hat{\beta}_{u}$, and $\hat{\beta}_{l}$ of this
state in terms of the fluid-dynamical variables of the actual state. 
As usual, we impose the Landau matching conditions for particle-number and energy density
$n=\hat{n}(\hat{\alpha},\hat{\beta}_u,\hat{\beta}_l)$ and 
$e=\hat{e}(\hat{\alpha},\hat{\beta}_u,\hat{\beta}_l)$, which
are equivalent to
\begin{eqnarray}\label{Landau_matching_rho_10}
\left(N^{\mu} - \hat{N}^{\mu} \right)u_{\mu}
\equiv \hat{\rho}_{10}=0 \; , \\ 
\left(T^{\mu \nu} - \hat{T}^{\mu \nu} \right)u_{\mu} u_{\nu} \equiv \hat{\rho}_{20}=0\;.  \label{Landau_matching_rho_20}
\end{eqnarray}
Since these are only two constraints, but we have three intensive quantities, we require an additional 
matching condition.
In analogy to the Landau matching conditions (\ref{Landau_matching_rho_10}), (\ref{Landau_matching_rho_20}), 
we may assume that 
\begin{eqnarray}
\left(T^{\mu \nu} - \hat{T}^{\mu \nu} \right)l_{\mu} l_{\nu} \equiv \hat{\rho}_{02}=0\;,  \label{matching_condition_rho_02}
\end{eqnarray}
meaning that the pressure in the direction of the anisotropy is equal to its value in the fictitious 
anisotropic reference state, i.e., $P_l=\hat{P}_l(\hat{\alpha},\hat{\beta}_u,\hat{\beta}_l)$,
and hence any correction from $\delta \hat{f}_{\mathbf{k}}$ to this pressure component 
vanishes. However, the choice for the third matching condition is not unique 
[neither are the usual Landau matching conditions
(\ref{Landau_matching_rho_10}), (\ref{Landau_matching_rho_20}), 
see e.g.\ Refs. \cite{Tsumura:2007ji,Tsumura:2009vm}], 
because we could have demanded for example also the equivalence of the transverse pressures, 
$P_\perp=\hat{P}_\perp(\hat{\alpha},\hat{\beta}_u,\hat{\beta}_l)$, together with Eqs.\ (\ref{Landau_matching_rho_10}),
(\ref{Landau_matching_rho_20}) to infer the parameters $\hat{\alpha}$, $\hat{\beta}_{u}$, and $\hat{\beta}_{l}$. 
Another alternative could be to enforce $\hat{n}_l = n_l$, or $\hat{M} = M$, or even 
$P_\perp - P_l= \hat{P}_\perp - \hat{P}_l$ as advocated in Ref.\ \cite{Bazow:2013ifa}.

Furthermore, there is the choice of the LR frame of the fluid.
Eckart's definition (\ref{LR_Eckart_2}) leads to
\begin{equation}
\hat{\rho}_{00}^{\mu }=0\;,\;\;\;\hat{\rho}_{01} \equiv -\hat{n}_{l}\; ,
\label{LR_Eckart_rho}
\end{equation}
which is equivalent to $V^{\mu}_\perp = 0$ and $n_l = 0$.
On the other hand, Landau's definition (\ref{LR_Landau_2}) implies
\begin{equation}
\hat{\rho}_{10}^{\mu }=0\;,\;\;\;\hat{\rho}_{11}\equiv -\hat{M}\;,
\label{LR_Landau_rho}
\end{equation}
which leads to $W^{\mu}_{\perp u} = 0$ and $M=0$. Thus, for both choices of the LR frame, we obtain a
simplification of the conservation equations (\ref{conservation_eq_particle}) -- (\ref{conservation_eq_momenta_t}),
as several terms vanish identically.
Note that the functional form of the anisotropic distribution function 
might explicitly lead to $\hat{n}_l=0$ and/or $\hat{M}=0$, 
as is the case in Refs.\ \cite{Martinez:2009ry,Martinez:2010sc,Martinez:2010sd,Martinez:2012tu}.
In this case, using $\hat{n}_l = n_l$ or $\hat{M} = M$ as an additional dynamical matching 
condition would be meaningless.

Finally, note that we may also introduce the bulk viscous pressure, 
but now with respect to the anisotropic state instead of the local equilibrium state, i.e., 
\begin{equation}
\Pi  \equiv P  - \hat{P} \left(\hat{\alpha}, \hat{\beta}_u, \hat{\beta}_l\right) 
=-\frac{1}{3}\left\langle \Delta ^{\mu \nu }k_{\mu
}k_{\nu }\right\rangle _{\hat{\delta}}=-\frac{m_{0}^{2}}{3}\,\hat{\rho}_{00}\;.  
\label{kinetic:bulk_aniso}
\end{equation}%
Using Eqs.\ (\ref{P_iso_relation}), (\ref{Pi_iso}), and (\ref{bulk_aniso}) we obtain 
\begin{equation}
\Pi\equiv \frac{1}{3}\left[ \left( P_{l}-\hat{P}_{l}\right) +2\left(
P_{\perp }-\hat{P}_{\bot }\right) \right] =\frac{1}{3}\left( P_{l}+2P_{\perp
}\right) -P_{0}-\hat{\Pi}\equiv \Pi _{iso}-\hat{\Pi}\;,
\label{kinetic:bulk_aniso_important}
\end{equation}
which can be used to express either $P_{l}$ or $P_{\perp }$  by $\Pi $ in the
conservation equations (\ref{conservation_eq_particle}) -- 
(\ref{conservation_eq_momenta_t}).
Also note that this equation is equivalent to Eq.\ (28) of Ref.\ \cite{Bazow:2013ifa} 
with the following identification of quantities: $\Pi \equiv \tilde{\Pi}$, 
$\hat{\Pi} \equiv \Pi_{\rm{AHYDRO}}$, and $\Pi_{iso} \equiv \Pi$.

With the matching conditions (\ref{Landau_matching_rho_10}) -- (\ref{matching_condition_rho_02}), 
Landau's choice (\ref{LR_Landau_rho}) for the LR frame, and the bulk viscous pressure
(\ref{kinetic:bulk_aniso_important}) replacing the transverse pressure $P_\perp$, the conservation
equations (\ref{conservation_eq_particle}) -- (\ref{conservation_eq_momenta_t}) simplify, 
\begin{align}
0& \equiv D\hat{n}+\hat{n} \left(l_{\mu }D_{l}u^{\mu }+\tilde{\theta}\right) -D_{l}n_{l}+n_{l}\left(\tilde{\theta}_{l}
- l_{\mu }Du^{\mu }\right) -V_{\perp}^{\mu }\left( Du_{\mu }
+D_{l}l_{\mu }\right)+\tilde{\nabla}_{\mu}V_{\perp }^{\mu }\;,  \label{conservation_eq_particle_b} \\
0& = D\hat{e}+\left( \hat{e}+\hat{P}_{l}\right) l_{\mu}D_{l}u^{\mu }+\left( \hat{e}+\hat{P}_{\perp}
+ \frac{3}{2}\, \Pi \right) \tilde{\theta}  +W_{\perp l}^{\mu }\left( D_{l}u_{\mu }
-l_{\nu }\tilde{\nabla}_{\mu}u^{\nu }\right)
-\pi _{\perp }^{\mu \nu } \tilde{\sigma}_{\mu \nu }\;,  \label{conservation_eq_energy_b} \\
0& = \left( \hat{e}+\hat{P}_{l}\right) l_{\mu }Du^{\mu }+ D_{l}\hat{P}_{l}+\left( \hat{P}_{\perp }-\hat{P}_{l}+
\frac{3}{2}\, \Pi \right) \tilde{\theta}_{l}
  +W_{\perp l}^{\mu }\left( Du_{\mu }+2\,D_{l}l_{\mu }+l_{\nu }\tilde{\nabla}_{\mu }u^{\nu } \right)-
\tilde{\nabla}_{\mu }W_{\perp l}^{\mu }-\pi _{\perp }^{\mu \nu }\tilde{\sigma}_{l,\mu \nu }\;, 
\label{conservation_eq_momenta_l_b} \\
0& = \left( \hat{e}+\hat{P}_{\perp } + \frac{3}{2}\, \Pi \right) \Xi^\alpha_\nu Du^{\nu } -\tilde{\nabla}^{\alpha } 
\left( \hat{P}_{\perp } + \frac{3}{2}\, \Pi\right)
+\left( \hat{P}_{\perp }-\hat{P}_{l}+ \frac{3}{2}\, \Pi\right) \Xi^\alpha_\nu D_{l}l^{\nu } 
-\Xi _{\nu }^{\alpha}D_{l}W_{\perp l}^{\nu }
+W_{\perp l}^{\alpha } \left( \frac{3}{2} \, \tilde{\theta}_{l}- l_{\mu }Du^{\mu } \right)  \notag \\
&+W_{\perp l,\nu }\left( \tilde{\sigma}_{l}^{\alpha \nu } -\tilde{\omega}_{l}^{\alpha \nu }\right)  
-\pi _{\perp }^{\mu \alpha }\left( Du_{\mu }+D_{l}l_{\mu }\right)
+\Xi _{\nu }^{\alpha }\tilde{\nabla}_{\mu }\pi _{\perp }^{\mu\nu }\;.  \label{conservation_eq_momenta_t_b}
\end{align}
These five equations contain 14 unknowns: the scalar quantities $\hat{\alpha},\, \hat{\beta}_u,\, \hat{\beta}_l$ 
(which determine the quantities $\hat{n}$, $\hat{n}_l$, $\hat{e}$, $\hat{P}_l$, $\hat{P}_\perp$), $n_l$, and $\Pi$, 
the three independent components of the fluid four-velocity $u^\mu$, the two independent components of
$V_\perp^\mu$, the two independent components of $W_{\perp l}^\mu$, and the two independent 
components of $\pi_\perp^{\mu \nu}$. Thus, we need to
specify nine additional equations of motion. This will be done in the next subsection, resorting again to the
system (\ref{eq_aniso_scalar}) -- (\ref{eq_aniso_tensor}) of equations of
motion for the irreducible moments.

\subsection{Relaxation equations in the 14-moment approximation}

In this subsection, we outline the derivation of the nine additional
equations of motion which close the system of conservation equations 
(\ref{conservation_eq_particle}) -- (\ref{conservation_eq_momenta_t}). 
As advertised above, to this end we again use the equations of motion 
(\ref{eq_aniso_scalar}) -- (\ref{eq_aniso_tensor}) for the
irreducible moments. 

We will also make use of the fact that
\begin{eqnarray}
d\hat{n} & = & \frac{\partial \hat{n}}{\partial \hat{\alpha}} \, d \hat{\alpha} + \frac{\partial \hat{n}}{\partial \hat{\beta}_u} 
\, d \hat{\beta}_u + \frac{\partial \hat{n}}{\partial \hat{\beta}_l} \, d \hat{\beta}_l \;, \label{dhatn} \\
d\hat{e} & = & \frac{\partial \hat{e}}{\partial \hat{\alpha}} \, d \hat{\alpha} + \frac{\partial \hat{e}}{\partial \hat{\beta}_u} 
\, d \hat{\beta}_u + \frac{\partial \hat{e}}{\partial \hat{\beta}_l} \, d \hat{\beta}_l \;. \label{dhate}
\end{eqnarray} 
Replacing the total derivatives with comoving derivatives, making use of the conservation equations
(\ref{conservation_eq_particle_b}) -- (\ref{conservation_eq_momenta_t_b}), and neglecting terms
of third order in dissipative quantities such as $\Pi,\, n_l,\, V_\perp^\mu,\, W_{\perp l}^\mu,
\, \pi_\perp^{\mu \nu}$, or in gradients of $\hat{n},\,\hat{n}_l,\, \hat{e},\, \hat{P}_l,\, \hat{P}_\perp,\, u^\mu,\, l^\mu$, 
we derive the following
equations for the comoving derivatives of $\hat{\alpha}$ and $\hat{\beta}_u$:
\begin{eqnarray}
D \hat{\alpha} & = & \left[ \frac{\partial (\hat{e}, \hat{n})}{\partial (\hat{\beta}_u,\hat{\alpha})} \right]^{-1}
\left\{  \left[ \frac{\partial \hat{n}}{\partial \hat{\beta}_u}\, (\hat{e} + \hat{P}_l) - \frac{\partial \hat{e}}{\partial \hat{\beta}_u}\,
\hat{n} \right] l_\mu D_l u^\mu 
+\left[ \frac{\partial \hat{n}}{\partial \hat{\beta}_u}\, (\hat{e} + \hat{P}_\perp) - \frac{\partial \hat{e}}{\partial \hat{\beta}_u}\,
\hat{n} \right] \tilde{\theta}  -\frac{\partial (\hat{e}, \hat{n})}{\partial (\hat{\beta}_u,\hat{\beta}_l)}\, D \hat{\beta}_l  
\right. \notag \\
&   &  \hspace*{2.3cm} + \, \frac{\partial \hat{e}}{\partial \hat{\beta}_u}
\left[ D_l n_l - \frac{\hat{e} + \hat{P}_\perp}{\hat{e} + \hat{P}_l}\, n_l \, \tilde{\theta}_l  
 -  \tilde{\nabla}_\mu V^\mu_\perp -  \frac{n_l}{\hat{e} + \hat{P}_l}
\left( \frac{\partial \hat{P}_l}{\partial \hat{\alpha}}\, D_l \hat{\alpha}
+ \frac{\partial \hat{P}_l}{\partial \hat{\beta}_u}\, D_l \hat{\beta}_u 
+ \frac{\partial \hat{P}_l}{\partial \hat{\beta}_l}\, D_l \hat{\beta}_l \right)  \right. \notag \\
&   &  \hspace*{3.5cm} +  \left. \frac{V^\mu_\perp}{\hat{e} + \hat{P}_\perp}
\left( \frac{\partial \hat{P}_\perp}{\partial \hat{\alpha}}\,  \tilde{\nabla}_\mu\hat{\alpha}
+ \frac{\partial \hat{P}_\perp}{\partial \hat{\beta}_u}\, \tilde{\nabla}_\mu \hat{\beta}_u 
+ \frac{\partial \hat{P}_\perp}{\partial \hat{\beta}_l}\, \tilde{\nabla}_\mu \hat{\beta}_l \right)
+ \frac{\hat{e} + \hat{P}_l}{\hat{e}+ \hat{P}_\perp}\,V_{\perp}^\mu D_l l_\mu\right] \notag \\
&    &  \hspace*{2.3cm} + \left. \frac{\partial \hat{n}}{\partial \hat{\beta}_u}\left[
 \frac{3}{2}\,\Pi\, \tilde{\theta} + W_{\perp l}^\mu \left(D_l u_\mu - l_\nu  \tilde{\nabla}_\mu u^\nu
\right)   -  \pi_\perp^{\mu \nu} \tilde{\sigma}_{\mu \nu} \right] \right\}\;, \label{Dhatalpha} \\
D \hat{\beta}_u & = & \left[ \frac{\partial (\hat{e}, \hat{n})}{\partial (\hat{\beta}_u,\hat{\alpha})} \right]^{-1}
\left\{  \left[ \frac{\partial \hat{e}}{\partial \hat{\alpha}}\,
\hat{n}- \frac{\partial \hat{n}}{\partial \hat{\alpha}}\, (\hat{e} + \hat{P}_l)  \right] l_\mu D_l u^\mu 
+\left[ \frac{\partial \hat{e}}{\partial \hat{\alpha}}\,
\hat{n}- \frac{\partial \hat{n}}{\partial \hat{\alpha}}\, (\hat{e} + \hat{P}_\perp)  \right] \tilde{\theta}  
-\frac{\partial (\hat{e}, \hat{n})}{\partial (\hat{\beta}_l,\hat{\alpha})}\, D \hat{\beta}_l  
\right. \notag \\
&   &  \hspace*{2.3cm} - \, \frac{\partial \hat{e}}{\partial \hat{\alpha}}
\left[ D_l n_l - \frac{\hat{e} + \hat{P}_\perp}{\hat{e} + \hat{P}_l}\, n_l \, \tilde{\theta}_l  
 -  \tilde{\nabla}_\mu V^\mu_\perp -  \frac{n_l}{\hat{e} + \hat{P}_l}
\left( \frac{\partial \hat{P}_l}{\partial \hat{\alpha}}\, D_l \hat{\alpha}
+ \frac{\partial \hat{P}_l}{\partial \hat{\beta}_u}\, D_l \hat{\beta}_u 
+ \frac{\partial \hat{P}_l}{\partial \hat{\beta}_l}\, D_l \hat{\beta}_l \right)  \right. \notag \\
&   &  \hspace*{3.5cm} +  \left. \frac{V^\mu_\perp}{\hat{e} + \hat{P}_\perp}
\left( \frac{\partial \hat{P}_\perp}{\partial \hat{\alpha}}\,  \tilde{\nabla}_\mu\hat{\alpha}
+ \frac{\partial \hat{P}_\perp}{\partial \hat{\beta}_u}\, \tilde{\nabla}_\mu \hat{\beta}_u 
+ \frac{\partial \hat{P}_\perp}{\partial \hat{\beta}_l}\, \tilde{\nabla}_\mu \hat{\beta}_l \right)
+ \frac{\hat{e} + \hat{P}_l}{\hat{e}+ \hat{P}_\perp}\,V_{\perp}^\mu D_l l_\mu\right] \notag \\
&    &  \hspace*{2.3cm} - \left. \frac{\partial \hat{n}}{\partial \hat{\alpha}}\left[
 \frac{3}{2}\,\Pi\, \tilde{\theta} + W_{\perp l}^\mu \left(D_l u_\mu - l_\nu  \tilde{\nabla}_\mu u^\nu
\right)   -  \pi_\perp^{\mu \nu} \tilde{\sigma}_{\mu \nu} \right] \right\}\;, \label{Dhatbetau}
\end{eqnarray}
where we have employed the notation
\begin{equation} \label{Jacobi}
\frac{\partial(X,Y)}{\partial(x,y)} \equiv \frac{\partial X}{\partial x}\, \frac{\partial Y}{\partial y} - 
\frac{\partial X}{\partial y}\, \frac{\partial Y}{\partial x}
\end{equation}
for the Jacobi determinant of partial derivatives.
Note that the comoving derivatives $D\hat{\alpha},\, D\hat{\beta}_u$ depend also
on the comoving derivative of the non-equilibrium parameter $D\hat{\beta}_l$, which is determined
by a relaxation equation, see below.

In this work, the form of $\hat{f}_{0{\bf k}}$ will not be explicitly specified. Thus, we cannot explicitly compute
the transport coefficients of anisotropic dissipative fluid dynamics. For this reason, we leave 
the establishment of a systematically improvable framework for this theory following the 
approach of Ref.\ \cite{Denicol:2012cn} to future work. Here, we restrict ourselves to the 
derivation of the equations in the simplest possible case: the 14-moment
approximation, i.e., $N_{0}=2,\,N_{1}=1,\,N_{2}=0$. This assumption simplifies the discussion tremendously,
since we no longer have to find the eigenmodes of the linearized Boltzmann equation as done
in Ref.\ \cite{Denicol:2012cn}. All we need to do is to close the system of equations of motion 
(\ref{eq_aniso_scalar}) -- (\ref{eq_aniso_tensor}) by employing
Eq.\ (\ref{kinetic:rho_ij_rho_nm}) in order to express the irreducible moments entering these equations in terms of the 
variables entering the conservation equations (\ref{conservation_eq_particle}) -- (\ref{conservation_eq_momenta_t}):
\begin{eqnarray}
\hat{\rho}_{ij} &\equiv &\hat{\rho}_{00}\,\gamma _{i0j0}^{\left( 0\right) }+
\hat{\rho}_{01}\,\gamma _{i0j1}^{\left( 0\right) }+\hat{\rho}_{02}\,\gamma_{i0j2}^{\left( 0\right) }
+\hat{\rho}_{10}\,\gamma _{i1j0}^{\left( 0\right)}+\hat{\rho}_{11}\,\gamma _{i1j1}^{\left( 0\right) }
+\hat{\rho}_{20}\,\gamma_{i2j0}^{\left( 0\right) }  \notag \\
&=&-\frac{3}{m_{0}^{2}}\,\Pi \,\gamma _{i0j0}^{\left( 0\right) }+(n_{l}-
\hat{n}_{l})\gamma _{i0j1}^{\left( 0\right) } - \hat{M} \,\gamma _{i1j1}^{\left(0\right) }\;,  \label{rho_ij_14} \\
\hat{\rho}_{ij}^{\mu } &\equiv &-\left( \hat{\rho}_{00}^{\mu }\,\gamma_{i0j0}^{\left( 1\right) }
+\hat{\rho}_{01}^{\mu }\,\gamma _{i0j1}^{\left(1\right) }+\hat{\rho}_{10}^{\mu }\,\gamma _{i1j0}^{\left( 1\right) }\right) 
=-\left( V_{\perp }^{\mu }\,\gamma_{i0j0}^{\left( 1\right) }+W_{\perp l}^{\mu }\,\gamma _{i0j1}^{\left( 1\right) }
\right) \;,  \label{rho_ij_mu_14} \\
\hat{\rho}_{ij}^{\mu \nu } &\equiv &2\hat{\rho}_{00}^{\mu \nu }\,\gamma_{i0j0}^{\left( 2\right) }
=2\pi _{\perp }^{\mu \nu }\,\gamma _{i0j0}^{\left(2\right) }\;,  \label{rho_ij_munu_14}
\end{eqnarray}
where we used Eqs.\ (\ref{hat_rho_n}) -- (\ref{hat_rho_pi_t}), the Landau
matching conditions (\ref{Landau_matching_rho_10}), (\ref{Landau_matching_rho_20}) 
together with the third matching condition (\ref{matching_condition_rho_02}), the choice 
(\ref{LR_Landau_rho}) of LR frame, and the definition (\ref{kinetic:bulk_aniso}) of the bulk viscosity.

In addition, since the conservation equations (\ref{conservation_eq_particle}) -- (\ref{conservation_eq_momenta_t})
only involve quantities up to tensor rank two, we neglect irreducible tensor moments of rank higher than two, 
i.e., $\hat{\rho}_{ij}^{\mu _{1}\cdots \mu _{\ell }}=0$ for $\ell \geq 3$ 
in Eqs.\ (\ref{eq_aniso_vector}), (\ref{eq_aniso_tensor}), cf.\ Ref.\ \cite{Denicol:2012es}.
Finally, since we want to derive relaxation equations for the dissipative quantities,
we need to choose pairs of indices $(i,j)$ in Eqs.\ (\ref{hat_rho_n}) -- (\ref{hat_rho_pi_t}),
for which the collision integrals $\mathcal{C}_{i-1,j}^{\left\{ \mu_1
\cdots \mu_\ell \right\} }$ do not vanish. This choice is not unique, as any values for the indices $(i,j)$ are possible
(except those that correspond to the conservation equations) and lead to a closed system of equations of motion. 
In the following, for the sake of simplicity the case where 
the indices $(i,j)$ take the lowest possible values will be considered.

Following this procedure, we obtain:
\begin{itemize}
\item[(i)] from Eq.\ (\ref{eq_aniso_scalar}) for $(i,j)=(0,0)$ the relaxation equation for the bulk viscous pressure $\Pi$, 
\begin{eqnarray}
D \Pi & = & - \frac{m_0^2}{3} \, \mathcal{C}_{-1,0} 
- \bar{\zeta}_l\, l_\mu D_l u^\mu - \bar{\zeta}_\perp\, \tilde{\theta} - \bar{\zeta}_{\perp l}\,  \tilde{\theta}_l
- \bar{\kappa}^{\Pi}_{\alpha} D_l \hat{\alpha}- \bar{\kappa}^{\Pi}_{u} D_l \hat{\beta}_u 
- \bar{\kappa}^{\Pi}_l D_l \hat{\beta}_l + \bar{\tau}^{\Pi}_l \, D \hat{\beta}_l \notag \\
&    & - \left( \bar{\beta}^{\Pi}_\Pi\, \Pi + \bar{\beta}^{\Pi}_{n}\, n_l \right) l_\mu D_l u^\mu
         - \left( \bar{\delta}^{\Pi}_{\Pi}\, \Pi   - \bar{\delta}^{\Pi}_n\, n_l \right) \tilde{\theta}
         - \left( \bar{\epsilon}^{\Pi}_\Pi\, \Pi -  \bar{\epsilon}^{\Pi}_{n}\, n_l \right) \tilde{\theta}_l \notag \\
&    & + \bar{\epsilon}^{\Pi}_{\Pi}\, D_l \Pi - \bar{\ell}^{\Pi}_n\, D_l n_l - \bar{\ell}^{\Pi}_V\, \tilde{\nabla}_\mu V_\perp^\mu
          - \bar{\ell}^{\Pi}_W\, \tilde{\nabla}_\mu W_{\perp l}^\mu \notag \\
&   & + \left( \bar{\tau}^\Pi_{\Pi \alpha}\, \Pi - \bar{\tau}^{\Pi}_{n \alpha}\, n_l\right) D_l \hat{\alpha}
         + \left( \bar{\tau}^{\Pi}_{\Pi u}\, \Pi - \bar{\tau}^{\Pi}_{n u}\, n_l \right) D_l \hat{\beta}_u
         + \left( \bar{\tau}^{\Pi}_{\Pi l}\, \Pi - \bar{\tau}^{\Pi}_{n l}\, n_l\right) D_l \hat{\beta}_l \notag \\
&   & -\left( \bar{\tau}^{\Pi}_{V \alpha}\, V_\perp^\mu +\bar{\tau}^{\Pi}_{W \alpha}\, W_{\perp l}^\mu \right)
            \tilde{\nabla}_\mu \hat{\alpha}
         - \left( \bar{\tau}^{\Pi}_{V u}\, V_\perp^\mu + \bar{\tau}^{\Pi}_{W u}\, W_{\perp l}^\mu \right)
             \tilde{\nabla}_\mu \hat{\beta}_u
         - \left( \bar{\tau}^{\Pi}_{V l}\, V_\perp^\mu  + \bar{\tau}^{\Pi}_{W l}\, W_{\perp l}^\mu \right)
             \tilde{\nabla}_\mu \hat{\beta}_l \notag \\
&   & + \left( \bar{\lambda}^{\Pi}_{V}\, V_\perp^\mu + \bar{\lambda}^{\Pi}_{W} \, W_{\perp l}^\mu \right) 
\left( D_l u_\mu - l_\nu \tilde{\nabla}_\mu u^\nu \right) 
+ \left( \bar{\lambda}^{\Pi}_{V l}\, V_\perp^\mu + \bar{\ell}^{\Pi}_{W } \, W_{\perp l}^\mu\right)  D_l l_\mu
+ \bar{\lambda}^{\Pi}_{\pi} \pi_\perp^{\mu \nu} \tilde{\sigma}_{\mu \nu} \;, \label{rel_Pi}
\end{eqnarray}
where we have made use of Eqs.\ (\ref{Dhatalpha}), (\ref{Dhatbetau}).
Due to the anisotropy, there are now two (instead of one) bulk viscosity coefficients  \cite{Huang:2009ue,Huang:2011dc}:
one coefficient multiplying $l_\mu D_l u^\mu$ for the fluid expansion in the direction of the anisotropy, and one 
multiplying $\tilde{\theta}$ for the expansion in the direction transverse to $l^\mu$. A third
bulk viscosity coefficient, $\bar{\zeta}_{\perp l}$, is related to the gradient of $l^\mu$ in the
direction transverse to both $u^\mu$ and $l^\mu$. 
Note that there are first-order terms 
in the equation for the bulk viscosity which are proportional to gradients in $l^\mu$-direction of the variables
$\hat{\alpha},\, \hat{\beta}_u$, and $\hat{\beta}_l$ and which do not appear in the isotropic case. There is also
a term $\sim D \hat{\beta}_l$ which couples the relaxation equation for the bulk viscous pressure to that
for the anisotropy parameter $\hat{\beta}_l$, see item (iii) below.
The transport coefficients appearing in Eq.\ (\ref{rel_Pi}) are listed in Appendix \ref{app_Pi}.

Finally, let us comment on the collision term $-m_0^2\, \mathcal{C}_{-1,0}/3$. Although it is complicated in
general, we expect that it is dominated by a term similar in structure to the RTA (\ref{RTA_1}),
\begin{equation}
-\frac{m_0^2}{3}\, \mathcal{C}_{-1,0} \sim  \frac{1}{\tau_\Pi} \left[  \frac{m_0^2}{3}\left(\hat{\mathcal{I}}_{00}
- \mathcal{I}_{00} \right)- \Pi \right]\;,
\end{equation}
where we have indicated the characteristic relaxation time for the bulk viscous pressure by $\tau_\Pi$.
The last term $\sim - \Pi/\tau_\Pi$ reflects the nature of Eq.\ (\ref{rel_Pi}) as relaxation equation for
the bulk viscous pressure: the bulk viscous pressure relaxes towards its Navier-Stokes value
(given by the first-order terms in Eq.\ (\ref{rel_Pi}) on a time scale $\sim \tau_\Pi$. This term already
appears in the isotropic case \cite{Denicol:2012cn}. However, the first
term $\sim \hat{\mathcal{I}}_{00}- \mathcal{I}_{00}$ is new: it represents an additional force that drives the
system towards the isotropic limit.

\item[(ii)] from Eq.\ (\ref{eq_aniso_scalar}) for $(i,j)=(0,1)$ the relaxation equation for the diffusion current  $n_l$ in the
direction of the anisotropy,
\begin{eqnarray}
D n_l & = &  \mathcal{C}_{-1,1} 
+ \bar{\kappa}^{n}_{\alpha} D_l \hat{\alpha}+ \bar{\kappa}^{n}_{u} D_l \hat{\beta}_u 
+ \bar{\kappa}^{n}_{l} D_l \hat{\beta}_l
+ \bar{\zeta}^{n}_{l}\, l_\mu D_l u^\mu - \bar{\zeta}^{n}_{\perp}\, \tilde{\theta} - \bar{\zeta}^{n}_{\perp l}\,  \tilde{\theta}_l
- V^\mu_\perp  D l_{\mu} \notag \\
&    & - \left( \bar{\beta}^{n}_{\Pi}\, \Pi   + \bar{\beta}^{n}_{n}\, n_l \right) l_\mu D_l u^\mu
         + \left( \bar{\delta}^{n}_{\Pi}\, \Pi   - \bar{\delta}^{n}_{n}\, n_l \right) \tilde{\theta}
         - \left( \bar{\epsilon}^{n}_{\Pi}\, \Pi  + \bar{\epsilon}^{n}_{n}\, n_l \right) \tilde{\theta}_l \notag \\
&    & - \bar{\ell}^{n}_{\Pi}\, D_l \Pi + \bar{\ell}^{n}_{n}\, D_l n_l + \bar{\ell}^{n}_{V}\, \tilde{\nabla}_\mu V_\perp^\mu
+ \bar{\ell}^{n}_{W}\, \tilde{\nabla}_\mu W_{\perp l}^\mu \notag \\
&   & - \left( \bar{\tau}^{n}_{\Pi \alpha}\, \Pi - \bar{\tau}^{n}_{n \alpha}\, n_l\right) D_l \hat{\alpha}
          - \left( \bar{\tau}^{n}_{\Pi u}\, \Pi - \bar{\tau}^{n}_{n u}\, n_l \right) D_l \hat{\beta}_u
         - \left( \bar{\tau}^{n}_{\Pi l}\, \Pi\ - \bar{\tau}^{n}_{n l}\, n_l\right)  D_l \hat{\beta}_l \notag \\
&   & +\left( \bar{\tau}^{n}_{V \alpha}\, V_\perp^\mu +\bar{\tau}^{n}_{W \alpha}\, W_{\perp l}^\mu \right) 
          \tilde{\nabla}_\mu \hat{\alpha}
         +\left( \bar{\tau}^{n}_{V u}\, V_\perp^\mu +\bar{\tau}^{n}_{W u}\, W_{\perp l}^\mu\right) 
           \tilde{\nabla}_\mu \hat{\beta}_u
         +\left( \bar{\tau}^{n}_{V l}\, V_\perp^\mu  +\bar{\tau}^{n}_{W l}\, W_{\perp l}^\mu \right)
          \tilde{\nabla}_\mu \hat{\beta}_l \notag \\
&   & - \left( \bar{\lambda}^{n}_{V u}\, V_\perp^\mu + \bar{\lambda}^{n}_{W u} \, W_{\perp l}^\mu \right) D_l u_\mu 
+\left( \bar{\lambda}^{n}_{V \perp}\, V_\perp^\mu + \bar{\lambda}^{n}_{W \perp} \, W_{\perp l}^\mu \right) 
 l_\nu \tilde{\nabla}_\mu u^\nu  
- \left( 2 \,\bar{\ell}^{n}_{V}\, V_\perp^\mu + \bar{\lambda}^{n}_{W l} \, W_{\perp l}^\mu\right)  D_l l_\mu \notag \\
&   & - \bar{\lambda}^{n}_{\pi} \pi_\perp^{\mu \nu} \tilde{\sigma}_{\mu \nu}
- \bar{\lambda}^{n}_{\pi l} \pi_\perp^{\mu \nu} \tilde{\sigma}_{l,\mu \nu} \;, \label{rel_n}
\end{eqnarray}
where we have employed Eqs.\ (\ref{conservation_eq_momenta_l_b}) and (\ref{conservation_eq_momenta_t_b}).
The transport coefficients in Eq.\ (\ref{rel_n}) are listed in App.\ \ref{app_n}. Note that the last term in the first line 
couples the relaxation equation for $n_l$ to the time evolution of $l^\mu$.

We also comment on the collision term $\mathcal{C}_{-1,1}$. Again, we make the assumption
that it is dominated by a term similar in structure to the RTA (\ref{RTA_1}),
\begin{equation}
 \mathcal{C}_{-1,1} \sim - \frac{1}{\tau_n} \left(\hat{\mathcal{I}}_{01} + \hat{\rho}_{01}
- \mathcal{I}_{01} \right) \equiv - \frac{1}{\tau_n} \left(n_l
- \mathcal{I}_{01} \right) \;,
\end{equation}
where we have indicated the characteristic relaxation time for the diffusion current $n_l$ by $\tau_n$.
For symmetry reasons $\mathcal{I}_{01} \equiv 0$, so that
we are left with a term $\sim - n_l/\tau_n$, which is needed such that Eq.\ (\ref{rel_n}) is a relaxation equation for $n_l$.

\item[(iii)] from Eq.\ (\ref{eq_aniso_scalar}) for $(i,j)=(0,2)$ an evolution equation for the anisotropic pressure in the
longitudinal direction $\hat{P}_l$, 
\begin{eqnarray}
D \hat{P}_l & = &  \mathcal{C}_{-1,2} 
+ \bar{\kappa}^{l}_{\alpha} D_l \hat{\alpha}+ \bar{\kappa}^{l}_{u} D_l \hat{\beta}_u + \bar{\kappa}^{l}_{l} D_l \hat{\beta}_l
+ \bar{\zeta}^{l}_{l}\, l_\mu D_l u^\mu - \bar{\zeta}^{l}_{\perp}\, \tilde{\theta} - \bar{\zeta}^{l}_{\perp l}\,  \tilde{\theta}_l
- 2 W^\mu_{\perp l}  D l_{\mu} \notag \\
&    & - \left( \bar{\beta}^{l}_{\Pi}\, \Pi - \bar{\beta}^{l}_{n}\, n_l \right) l_\mu D_l u^\mu
         + \left( \bar{\delta}^{l}_{\Pi}\, \Pi   - \bar{\delta}^{l}_{n}\, n_l \right) \tilde{\theta}
         + \left( \bar{\epsilon}^{l}_{\Pi}\, \Pi   - \bar{\epsilon}^{l}_{n}\, n_l \right) \tilde{\theta}_l \notag \\
&    & - \bar{\ell}^{l}_{\Pi}\, D_l \Pi + \bar{\ell}^{l}_{n}\, D_l n_l + \bar{\ell}^{l}_{V}\, \tilde{\nabla}_\mu V_\perp^\mu
+ \bar{\ell}^{l}_{W}\, \tilde{\nabla}_\mu W_{\perp l}^\mu \notag \\
&   & - \left( \bar{\tau}^{l}_{\Pi \alpha}\, \Pi - \bar{\tau}^{l}_{n \alpha}\, n_l\right) D_l \hat{\alpha} 
         - \left( \bar{\tau}^{l}_{\Pi u}\, \Pi - \bar{\tau}^{l}_{n u}\, n_l \right) D_l \hat{\beta}_u
         - \left( \bar{\tau}^{l}_{\Pi l}\, \Pi - \bar{\tau}^{l}_{n l}\, n_l\right) D_l \hat{\beta}_l \notag \\
&   & +\left( \bar{\tau}^{l}_{V \alpha}\, V_\perp^\mu +\bar{\tau}^{l}_{W \alpha}\, W_{\perp l}^\mu\right)
                \tilde{\nabla}_\mu \hat{\alpha}
         +\left( \bar{\tau}^{l}_{V u}\, V_\perp^\mu +\bar{\tau}^{l}_{W u}\, W_{\perp l}^\mu \right)
         \tilde{\nabla}_\mu \hat{\beta}_u
         +\left( \bar{\tau}^{l}_{V l}\, V_\perp^\mu  +\bar{\tau}^{l}_{W l}\, W_{\perp l}^\mu \right)
         \tilde{\nabla}_\mu \hat{\beta}_l \notag \\
&   & - \left( \bar{\lambda}^{l}_{V u}\, V_\perp^\mu + \bar{\lambda}^{l}_{W u} \, W_{\perp l}^\mu \right) D_l u_\mu 
+\left( \bar{\lambda}^{l}_{V u}\, V_\perp^\mu + \bar{\lambda}^{l}_{W \perp} \, W_{\perp l}^\mu \right) 
 l_\nu \tilde{\nabla}_\mu u^\nu  
- 3 \left( \bar{\ell}^{l}_{V}\, V_\perp^\mu + \bar{\ell}^{l}_{W } \, W_{\perp l}^\mu\right)  D_l l_\mu \notag \\
&   & - \bar{\lambda}^{l}_{\pi} \pi_\perp^{\mu \nu} \tilde{\sigma}_{\mu \nu}
- \bar{\lambda}^{l}_{\pi l} \pi_\perp^{\mu \nu} \tilde{\sigma}_{l,\mu \nu} \;. \label{rel_Pl}
\end{eqnarray}
The transport coefficients appearing in this equation are listed in App.\ \ref{app_Pl}. As in the previous case, 
there is a term (the last in the first line) which couples the relaxation equation for $\hat{P}_l$ to the time evolution
of $l^\mu$. Using
\begin{equation}
D\hat{P}_l = \frac{\partial \hat{P}_l}{\partial \hat{\alpha}} \, D\hat{\alpha} + \frac{\partial \hat{P}_l}{\partial \hat{\beta}_u} 
\, D\hat{\beta}_u + \frac{\partial \hat{P}_l}{\partial \hat{\beta}_l} \, D\hat{\beta}_l  \label{dhatPl}
\end{equation}
on the left-hand side as well as Eqs.\ (\ref{Dhatalpha}), (\ref{Dhatbetau}), it is straightforward to rewrite Eq.\ (\ref{rel_Pl}) 
into a relaxation equation for the parameter $\hat{\beta}_l$. 

Finally, let us comment on the collision integral $\mathcal{C}_{-1,2}$. Let us again assume that
this is dominated by an RTA-like term,
\begin{equation}
\mathcal{C}_{-1,2} \sim - \frac{1}{\tau_{l}} \left( \hat{P}_l - \mathcal{I}_{02} \right) \;.
\end{equation}
If we then neglect all other terms on the right-hand side of Eq.\ (\ref{rel_Pl}),
this equation tells us that the longitudinal pressure $\hat{P}_l$ relaxes towards its value 
in local thermodynamical equilibrium, $\mathcal{I}_{02}\equiv P_0/3$.

\item[(iv)] from Eq.\ (\ref{eq_aniso_vector}) for $(i,j)=(0,0)$ the relaxation equation for 
$V^{\mu}_{\perp} = \hat{\rho}_{00}^{\mu }$,
\begin{eqnarray}
D V_\perp^{\{\mu \}} & = &  \mathcal{C}_{-1,0}^{\{\mu \}}
+ \bar{\kappa}_{\alpha} \tilde{\nabla}^\mu \hat{\alpha}+ \bar{\kappa}_{u}  \tilde{\nabla}^\mu \hat{\beta}_u 
+ \bar{\kappa}_{l}  \tilde{\nabla}^\mu \hat{\beta}_l
+ \bar{\zeta}^V_{u} \left( \Xi^\mu_\nu D_l u^\nu - l_\nu  \tilde{\nabla}^\mu u^\nu \right)
+ \bar{\zeta}^V_{l} \, \Xi^\mu_\nu D_l l^\nu - n_l \Xi^\mu_\nu D l^\nu \notag \\
&    & - \bar{\beta}^V_{\Pi}\, \Pi  \left( \Xi^\mu_\nu D_l u^\nu - l_\nu  \tilde{\nabla}^\mu u^\nu \right)  
          - \left( \bar{\beta}^V_{\Pi l}\, \Pi - \bar{\beta}^V_{n l} \, n_l \right)  \Xi^\mu_\nu D_l l^\nu
         + \bar{\beta}^V_{n}\,  n_l \, \Xi^\mu_\nu D_l u^\nu - \bar{\beta}^V_{n \perp}\, n_l \, l_\nu  \tilde{\nabla}^\mu u^\nu
           \notag \\
&    & + \left( \bar{\delta}^V_{V}\, V_\perp^\mu + \bar{\delta}^V_{W}\, W_{\perp l}^\mu \right) \tilde{\theta}
          + \left( \bar{\epsilon}^V_{V}\, V_\perp^\mu + \bar{\epsilon}^V_{W} \,W_{\perp l}^\mu \right) \tilde{\theta}_l \notag \\
&    & + \bar{\ell}^V_{\Pi}\, \tilde{\nabla}^\mu \Pi - \bar{\ell}^V_{n}\, \tilde{\nabla}^\mu n_l 
          - \bar{\ell}^V_{V}\, \Xi^\mu_\nu D_l V_\perp^\nu -  \bar{\ell}^V_{W}\, \Xi^\mu_\nu D_l W_{\perp l}^\nu 
          - \bar{\ell}^V_{ \pi}\, \Xi^\mu_\nu \tilde{\nabla}_\alpha \pi_\perp^{\alpha \nu}
 \notag \\
&   & + \left( \bar{\tau}^V_{\Pi \alpha}\, \Pi - \bar{\tau}^V_{n \alpha}\, n_l\right) \tilde{\nabla}^\mu \hat{\alpha} 
         + \left( \bar{\tau}^V_{\Pi u}\, \Pi- \bar{\tau}^V_{n u}\, n_l \right) \tilde{\nabla}^\mu \hat{\beta}_u
         + \left( \bar{\tau}^V_{\Pi l}\, \Pi - \bar{\tau}^V_{n l}\, n_l\right)  \tilde{\nabla}^\mu \hat{\beta}_l \notag \\
&   & -\left( \bar{\tau}^V_{V \alpha}\, V_\perp^\mu + \bar{\tau}^V_{W \alpha}\, W_{\perp l}^\mu \right) D_l \hat{\alpha}
         -\left( \bar{\tau}^V_{V u}\, V_\perp^\mu + \bar{\tau}^V_{W u}\, W_{\perp l}^\mu \right) D_l \hat{\beta}_u
         -\left( \bar{\tau}^V_{V l}\, V_\perp^\mu + \bar{\tau}^V_{ W l}\, W_{\perp l}^\mu \right) D_l \hat{\beta}_l \notag \\
&   & - \left( \bar{\lambda}^V_{V u}\, V_{\perp}^{\mu} + \bar{\lambda}^V_{W u} \, W_{\perp l}^\mu \right) 
l_\nu D_l u^\nu
+ \left( \bar{\lambda}^V_{V \perp}\, V_{\perp, \nu} + \bar{\delta}^V_{W} \, W_{\perp l, \nu} \right) 
\tilde{\sigma}^{\mu \nu} \notag \\
&   & 
+ V_{\perp, \nu}\, \tilde{\omega}^{\mu \nu} + \left( \bar{\ell}^V_{V} \, V_{\perp, \nu}
+ \bar{\ell}^V_{W}\, W_{\perp l, \nu} \right) \left( \tilde{\sigma}^{\mu \nu}_l - \tilde{\omega}^{\mu \nu}_l \right) 
 \notag \\
&   & - \bar{\tau}^V_{\pi \alpha} \, \pi_\perp^{\mu \nu} \tilde{\nabla}_{\nu} \hat{\alpha} 
- \bar{\tau}^V_{\pi u}\, \pi_\perp^{\mu \nu} \tilde{\nabla}_{\nu} \hat{\beta}_u
- \bar{\tau}^V_{\pi l} \,\pi_\perp^{\mu \nu} \tilde{\nabla}_{\nu} \hat{\beta}_l 
+ \bar{\lambda}^V_{\pi u} \, \pi_\perp^{\mu \nu} \left( D_l u_\nu - l_\alpha \tilde{\nabla}_{\nu} u^\alpha \right)
+ \bar{\lambda}^V_{\pi l} \, \pi_\perp^{\mu \nu} D_l l_\nu \;, \label{rel_V}
\end{eqnarray}
where we have employed Eqs.\ (\ref{conservation_eq_momenta_l_b}) and (\ref{conservation_eq_momenta_t_b}).
The last term in the first line couples the relaxation equation for $V_\perp^\mu$ to the time evolution of $l^\mu$.
The transport coefficients appearing in Eq.\ (\ref{rel_V}) are listed in App.\ \ref{app_V}.
In RTA, the collision term $\mathcal{C}_{-1,0}^{\{ \mu \}} \sim - V_\perp^\mu/\tau_V$, such that Eq.\ (\ref{rel_V}) is
a relaxation-type equation for $V_\perp^\mu$.

\item[(v)] from Eq.\ (\ref{eq_aniso_vector}) for $(i,j)=(0,1)$ the relaxation equation for the energy-momentum flow in the 
direction of the anisotropy $W^{\mu}_{\perp l}$,
\begin{eqnarray}
\hspace*{-0.3cm} D W_{\perp l}^{\{\mu \}} & = &  \mathcal{C}_{-1,1}^{\{\mu \}}
- \bar{\kappa}^W_{\alpha} \tilde{\nabla}^\mu \hat{\alpha}- \bar{\kappa}^W_{u}  \tilde{\nabla}^\mu \hat{\beta}_u 
- \bar{\kappa}^W_{l}  \tilde{\nabla}^\mu \hat{\beta}_l
+2 \bar{\eta}^W_{u} \, \Xi^\mu_\nu D_l u^\nu -2 \bar{\eta}^W_\perp \,l_\nu  \tilde{\nabla}^\mu u^\nu 
+ 2 \bar{\eta}^W_{l} \, \Xi^\mu_\nu D_l l^\nu \notag \\
&    &  - \left[ \left( \bar{\tau}^W_{l} - \frac{3}{2}\, \Pi \right) \Xi^\mu_\nu 
+ \pi^{\mu \nu}_\perp \right] D l^\nu  \notag \\
&    & - \bar{\beta}^W_{\Pi}\, \Pi  \, \Xi^\mu_\nu D_l u^\nu 
          + \bar{\beta}^W_{\Pi \perp}\,\Pi\, l_\nu  \tilde{\nabla}^\mu u^\nu
          - \left( \bar{\beta}^W_{\Pi l}\, \Pi - \bar{\beta}^W_{n l} \, n_l \right)  \Xi^\mu_\nu D_l l^\nu
         + \bar{\beta}^W_{n}\,  n_l \left( \Xi^\mu_\nu D_l u^\nu -  l_\nu  \tilde{\nabla}^\mu u^\nu\right)
           \notag \\
&    & + \left( \bar{\delta}^W_{V}\, V_\perp^\mu + \bar{\delta}^W_{W}\, W_{\perp l}^\mu \right) \tilde{\theta}
          + \left( \bar{\epsilon}^W_{V}\, V_\perp^\mu + \bar{\epsilon}^W_{W} \,W_{\perp l}^\mu \right) \tilde{\theta}_l \notag \\
&    & + \bar{\ell}^W_{\Pi}\, \tilde{\nabla}^\mu \Pi - \bar{\ell}^W_{n}\, \tilde{\nabla}^\mu n_l 
          - \bar{\ell}^W_{V}\, \Xi^\mu_\nu D_l V_\perp^\nu -  \bar{\ell}^W_{W}\, \Xi^\mu_\nu D_l W_{\perp l}^\nu 
          - \bar{\ell}^W_{ \pi}\, \Xi^\mu_\nu \tilde{\nabla}_\alpha \pi_\perp^{\alpha \nu}
 \notag \\
&   & + \left( \bar{\tau}^W_{\Pi \alpha}\, \Pi - \bar{\tau}^W_{n \alpha}\, n_l\right) \tilde{\nabla}^\mu \hat{\alpha} 
         + \left( \bar{\tau}^W_{\Pi u}\, \Pi- \bar{\tau}^W_{n u}\, n_l \right) \tilde{\nabla}^\mu \hat{\beta}_u
         + \left( \bar{\tau}^W_{\Pi l}\, \Pi - \bar{\tau}^W_{n l}\, n_l\right)  \tilde{\nabla}^\mu \hat{\beta}_l \notag \\
&   & -\left( \bar{\tau}^W_{V \alpha}\, V_\perp^\mu + \bar{\tau}^W_{W \alpha}\, W_{\perp l}^\mu \right) D_l \hat{\alpha}
         -\left( \bar{\tau}^W_{V u}\, V_\perp^\mu + \bar{\tau}^W_{W u}\, W_{\perp l}^\mu \right) D_l \hat{\beta}_u
         -\left( \bar{\tau}^W_{V l}\, V_\perp^\mu + \bar{\tau}^W_{ W l}\, W_{\perp l}^\mu \right) D_l \hat{\beta}_l \notag \\
&   & - \left( \bar{\lambda}^W_{V u}\, V_{\perp}^{\mu} + \bar{\lambda}^W_{W u} \, W_{\perp l}^\mu \right) 
l_\nu D_l u^\nu
+ \left( \bar{\delta}^W_{V}\, V_{\perp, \nu} + \bar{\lambda}^W_{W \perp} \, W_{\perp l, \nu} \right) 
\tilde{\sigma}^{\mu \nu} \notag \\
&   & 
+ W_{\perp l, \nu}\, \tilde{\omega}^{\mu \nu} + \left( \bar{\lambda}^W_{V l} \, V_{\perp, \nu}
+ \bar{\lambda}^W_{W l}\, W_{\perp l, \nu} \right)  \tilde{\sigma}^{\mu \nu}_l - 
\left( \bar{\ell}^W_{V } \, V_{\perp, \nu}
+ \bar{\ell}^W_{W }\, W_{\perp l, \nu} \right) \tilde{\omega}^{\mu \nu}_l 
 \notag \\
&   & - \bar{\tau}^W_{\pi \alpha} \, \pi_\perp^{\mu \nu} \tilde{\nabla}_{\nu} \hat{\alpha} 
- \bar{\tau}^W_{\pi u}\, \pi_\perp^{\mu \nu} \tilde{\nabla}_{\nu} \hat{\beta}_u
- \bar{\tau}^W_{\pi l} \,\pi_\perp^{\mu \nu} \tilde{\nabla}_{\nu} \hat{\beta}_l 
+ \bar{\lambda}^W_{\pi u} \, \pi_\perp^{\mu \nu} \,  D_l u_\nu 
-  \bar{\lambda}^W_{\pi \perp}\, l_\alpha \tilde{\nabla}_{\nu} u^\alpha 
+ 2\bar{\ell}^W_{\pi} \, \pi_\perp^{\mu \nu} D_l l_\nu \;. \label{rel_W}
\end{eqnarray}
The transport coefficients in this equation are listed in App.\ \ref{app_W}. The term in the second line
couples the relaxation equation for $W_{\perp l}^\mu$ to the time evolution of $l^\mu$.
In RTA, the collision term $\mathcal{C}_{-1,1}^{\{ \mu \}} \sim - W_{\perp l}^\mu/\tau_W$, such that Eq.\ (\ref{rel_W}) is
a relaxation-type equation for $W_\perp^\mu$.

\item[(vi)] from Eq.\ (\ref{eq_aniso_tensor}) for $(i,j)=(0,0)$ the relaxation equation for the transverse shear-stress tensor 
$\hat{\pi}^{\mu \nu}_{\perp}$ ,
\begin{eqnarray}
D \pi_{\perp}^{\{\mu \nu \}} & = &  \mathcal{C}_{-1,0}^{\{\mu \nu \}}
+ 2\bar{\eta}\, \tilde{\sigma}^{\mu \nu} + 2\bar{\eta}_l\, \tilde{\sigma}_l^{\mu \nu} -
2\, W_{\perp l}^{\{ \mu} Dl^{\nu \}} \notag \\
&   & - \bar{\delta}^\pi_\pi\, \pi_\perp^{\mu \nu}\,\tilde{\theta} - 2 \bar{\ell}^\pi_\pi \, \pi_\perp^{\mu \nu}\,\tilde{\theta}_l
- \bar{\tau}^\pi_\pi \, \pi_\perp^{\lambda \{ \mu} \tilde{\sigma}_\lambda^{\nu\} }
- 2 \bar{\ell}^\pi_\pi \, \pi_\perp^{\lambda \{ \mu} \tilde{\sigma}_{l, \lambda}^{\nu\} } 
 + \left( \bar{\lambda}^\pi_\Pi\, \Pi - \bar{\lambda}^\pi_n\, n_l \right) \tilde{\sigma}^{\mu \nu}
+ \left( \bar{\lambda}^\pi_{\Pi l}\, \Pi - \bar{\lambda}^\pi_{n l}\, n_l \right) \tilde{\sigma}^{\mu \nu}_l \notag \\
&   & + 2 \pi_\perp^{\lambda \{\mu} \tilde{\omega}^{\nu\}}_{\hspace*{0.2cm} \lambda}
+ 2 \bar{\ell}^\pi_\pi\, \pi_\perp^{\lambda \{\mu} \tilde{\omega}^{\nu\}}_{l, \hspace*{0.1cm} \lambda}
+ \bar{\ell}^\pi_V\, \tilde{\nabla}^{\{ \mu} V_\perp^{\nu \}} + \bar{\ell}^\pi_W\, \tilde{\nabla}^{\{ \mu} W_{\perp l}^{\nu \}} 
+ \bar{\ell}^\pi_\pi \, \Xi^{\mu \nu}_{\alpha \beta} \, D_l \pi_\perp^{\alpha \beta} \notag \\
&   & +\left( \bar{\tau}^\pi_{V \alpha}\, V_\perp^{\{ \mu} + \bar{\tau}^\pi_{W \alpha}\, W_{\perp l}^{\{ \mu} \right) 
 \tilde{\nabla}^{\nu \} } \hat{\alpha}
 +\left( \bar{\tau}^\pi_{V u}\, V_\perp^{\{ \mu} + \bar{\tau}^\pi_{W u}\, W_{\perp l}^{\{ \mu} \right) 
 \tilde{\nabla}^{\nu \} } \hat{\beta}_u
 +\left( \bar{\tau}^\pi_{V l}\, V_\perp^{\{ \mu} + \bar{\tau}^\pi_{W l}\, W_{\perp l}^{\{ \mu} \right) 
 \tilde{\nabla}^{\nu \} } \hat{\beta}_l
 \notag \\
&   & + \bar{\tau}^\pi_{\pi \alpha}\, \pi_\perp^{\mu \nu}\, D_l \hat{\alpha}
+  \bar{\tau}^\pi_{\pi u}\, \pi_\perp^{\mu \nu}\, D_l \hat{\beta}_u
+  \bar{\tau}^\pi_{\pi l}\, \pi_\perp^{\mu \nu}\, D_l \hat{\beta}_l
+ \bar{\lambda}^\pi_\pi\, \pi_\perp^{\mu \nu}\, l_\alpha D_l u^\alpha \notag \\
&   & -  \bar{\lambda}^\pi_{V }\, V_{\perp}^{\{ \mu} \left( D_l u^{\nu \} }
- l_\alpha \tilde{\nabla}^{\nu \} } u^\alpha \right)  - \bar{\lambda}^\pi_{W u} \, W_{\perp l}^{\{ \mu}  
D_l u^{\nu \} } + \bar{\lambda}^\pi_{W \perp} \, W_{\perp l}^{\{ \mu} l_\alpha  \tilde{\nabla}^{\nu \} } u^\alpha  \notag \\
 &    & - \left( \bar{\lambda}^\pi_{V l}\, V_{\perp}^{\{ \mu}  + \bar{\lambda}^\pi_{W l} \, W_{\perp l}^{\{ \mu}  \right) 
D_l l^{\nu \} }\;. \label{rel_pi}
\end{eqnarray}
The transport coefficients in this equation are listed in App.\ \ref{app_pi}. The last term in the first line couples
the relaxation equation for $\pi^{\mu \nu}_\perp$ to the time evolution of $l^\mu$.
In RTA, the collision term $\mathcal{C}_{-1,0}^{\{ \mu \nu \}} \sim - \pi_{\perp}^{\mu \nu}/\tau_\pi$, 
such that Eq.\ (\ref{rel_pi}) is a relaxation-type equation for $\pi_\perp^{\mu \nu}$.
\end{itemize}

We conclude this section by a couple of remarks:
\begin{itemize}
\item[(a)] In the relaxation equations (\ref{rel_Pi}), (\ref{rel_n}), (\ref{rel_Pl}), (\ref{rel_V}), (\ref{rel_W}), and (\ref{rel_pi}),
at a given order (in gradients and dissipative quantities) all terms appear which are allowed by Lorentz symmetry.
In principle, the corresponding transport coefficients are all independent quantities. However, close inspection 
reveals that several of them are proportional (or identical) to other coefficients. At this stage, we perceive this to
be an artifact of the 14-moment approximation (but we cannot exclude that this will persist also at higher order
in the moment expansion).
\item[(b)] The power-counting scheme in the relaxation equations (\ref{rel_Pi}), (\ref{rel_n}), (\ref{rel_Pl}), (\ref{rel_V}),
(\ref{rel_W}), and (\ref{rel_pi}) is similar to that of Ref.\
\cite{Denicol:2012cn}, i.e., in terms of powers of Knudsen and inverse Reynolds numbers. Note, however, that
the latter are now proportional to the bulk viscous pressure (\ref{kinetic:bulk_aniso_important}) and the 
dissipative quantities defined in Eqs.\ (\ref{kinetic:n_l}) -- (\ref{kinetic:pit_munu}). At this point, let us remark 
that the collision integral (\ref{chat_coll_int}) drives the relaxation of 
$\hat{f}_{0\mathbf{k}}$ towards the local equilibrium distribution $f_{0\mathbf{k}}$. Thus, for a strong anisotropy it can
in principle become arbitrarily large and is thus not part of the above power-counting scheme.
\item[(c)] In the particular case of an anisotropic fluid specified by 
$\hat{f}_{0\mathbf{k}}$ alone, i.e., when the dissipative quantities 
with respect to the anisotropic reference state vanish $\hat{\rho}^{\mu_1 \cdots \mu_\ell}_{ij} = 0$, 
there is only one additional equation needed for closure. 
This is so, since in general $\hat{n}_l$, $\hat{P}_l$, 
and $\hat{\Pi}$ are not independent variables. Furthermore, in case the form of $\hat{f}_{0\mathbf{k}}$
leads to $\hat{n}_l=\hat{M}=\hat{\Pi}=0$ we are left with an equation for $\hat{P}_l$.
\item[(d)] Our approach is analogous to the method given in Ref.\ \cite{Denicol:2012cn}, 
with the exception that now the reference state is different from the local thermodynamic
equilibrium state.
Therefore, due to this specific functional difference, anisotropic fluid dynamics 
embodies an arbitrary local momentum-space anisotropy and hence extends 
the ideal fluid-dynamical approach of a local equilibrium distribution function.
This in turn also means that, when compared to ideal fluids, such anisotropic fluids correspond 
to a specific class of dissipative fluids with an additional independent variable.
More specifically, anisotropic fluids require the knowledge of the equation of state 
specified by $\hat{f}_{0\mathbf{k}}$ and only after that 
can one proceed with the solution of the more general equations of 
anisotropic dissipative fluid dynamics.
Furthermore, compared to the usual dissipative fluid-dynamical approach, in anisotropic dissipative 
fluid dynamics there are only eight additional dissipative quantities which "relax" to the anisotropic state 
characterized by six independent variables.
\end{itemize}

\section{Conclusions and outlook}
\label{conclusions}

Anisotropic fluid dynamics allows for a macroscopic description of systems when the
microscopic single-particle distribution exhibits a strong anisotropy in momentum space in
the local rest frame of the fluid.
In this paper, starting from the relativistic Boltzmann equation, we have
derived the equations of motion for the irreducible moments of the deviation of the single-particle 
distribution function from a given anisotropic reference state.
These equations of motion are given in
Eqs.\ (\ref{eq_aniso_scalar}) -- (\ref{eq_aniso_tensor}) up to tensor-rank two. 
For the derivation, we constructed a new orthogonal basis in terms of multivariate 
polynomials in {\em both\/} energy $E_{\mathbf{k}u}$ {\em and\/} in momentum $E_{\mathbf{k}l}$
in the direction of the anisotropy, specified by the space-like four-vector $l^\mu$,
as well as irreducible tensors in momentum space
orthogonal to both the fluid four-velocity $u^\mu$ and $l^\mu$.

We then derived the equations of anisotropic dissipative fluid dynamics in the Landau frame, 
Eqs.\ (\ref{conservation_eq_particle_b}) -- 
(\ref{conservation_eq_momenta_t_b}), (\ref{rel_Pi}), (\ref{rel_n}), (\ref{rel_Pl}),
(\ref{rel_V}), (\ref{rel_W}), and (\ref{rel_pi}),
from the equations of motion for the irreducible moments in the 14-moment approximation,
substituting the irreducible moments by the fluid-dynamical variables, Eqs.\ (\ref{rho_ij_munu_14}).

Our treatment is general in the sense that we did not specify the form of the anisotropic single-particle
distribution function. This, however, is mandatory before one can explicitly compute the transport coefficients 
and solve the equations of motion. We will leave this study to future work.

Besides applications to heavy-ion collisions, we envisage that the framework introduced here will also be useful
in formulating a theory of relativistic dissipative magneto-hydrodynamics \cite{Huang:2009ue,Huang:2011dc}. In this
case, the magnetic-field vector $B^\mu = \frac{1}{2} \epsilon^{\mu \nu \alpha \beta} F_{\alpha \beta} u_\nu$ assumes
the role of $l^\mu$. 

\begin{acknowledgments}
  The authors thank Gabriel Denicol and Pasi Huovinen for constructive comments.
  Furthermore we also thank Gyuri Wolf for the elegant proof of Eqs.\ (\ref{Wolf_1}), (\ref{Wolf_2}). 
  The work of E.M.\ was partially supported
  by the Alumni Program of the Alexander von Humboldt Foundation and by BMBF grant no.\
  05P15RFCA1.
  H.N.\ has received funding from the European Union's Horizon 2020 research and innovation
  programme under the Marie Sklodowska-Curie grant agreement no.\ 655285.
  E.M.\ and H.N.\ were partially supported by the Helmholtz International Center for
  FAIR within the framework of the LOEWE program launched by the State
  of Hesse. 
\end{acknowledgments}

\appendix

\section{Tensor decompositions with respect to fluid flow and the direction of anisotropy}
\label{Quantities}

The velocity of fluid-dynamical flow is specified in terms of the time-like four-vector 
\begin{equation}
u^{\mu }\left( t,\mathbf{x}\right)=\gamma \left( 1,v^{x},v^{y},v^{z}\right)\;,  \label{u_mu}
\end{equation}
which is taken to be normalized,
\begin{equation}
u^{\mu }u_{\mu }=1\; .  \label{u_mu_norm}
\end{equation}
Hence, the Lorentz-gamma factor is $\gamma =\left( 1-\mathbf{v}^{\,2}\right) ^{-1/2}$,
and the four-flow velocity contains only three independent components: the
three components of the fluid three-velocity ${\bf v}=(v^x,v^y,v^z)$.

In order to specify the direction of a possible anisotropy in a
given system, we define a space-like four-vector $l^{\mu }$,
\begin{equation}
l^{\mu }l_{\mu }=-l^2\; ,  \label{l_mu_norm}
\end{equation}
where $l^2>0$ characterizes the strength of the anisotropy. 
Note that $l^\mu \equiv z^{\mu }$ in the notation of Refs.\ \cite{Martinez:2012tu,Bazow:2013ifa}.
Furthermore, $l^\mu$ is taken to be orthogonal to the four-flow velocity,
\begin{equation}
u^{\mu }l_{\mu }=0\; .  \label{orthogonality}
\end{equation}
An anisotropy could be caused by an external magnetic field. In this case, $l^\mu$ would
be conveniently chosen to point into the direction of this field.
In the context of heavy-ion collisions, this four-vector would 
be chosen to point into the direction of the beam axis (usually the $z$--axis).

In general, a space-like four-vector can be written in the form 
\begin{equation}
l^{\mu }\left( t,\mathbf{x}\right)=l\, \gamma_{l} \left( 1,\ell^x,\ell^y,\ell^z\right)\; ,
\end{equation}%
where $\gamma_l \equiv (\Bell^{\,2}-1)^{-1/2}$ follows from the normalization
condition (\ref{l_mu_norm}). If one is only interested in the direction of the anisotropy,
and not its magnitude, $l^\mu$ can be normalized to one, i.e., 
$l\equiv 1$. The orthogonality of the normalized
$l^{\mu }$ to the flow velocity (\ref{orthogonality}) gives the constraint
\begin{equation}
u^{\mu }l_{\mu }\equiv \gamma \gamma _{l} (
1-\mathbf{v} \cdot \Bell\, ) =0\; , \label{orthogonality_2}
\end{equation}
which may serve to express one component of $l^\mu$ by the others,
provided the corresponding component of $u^\mu$ does not vanish. Thus,
in general a normalized $l^\mu$ has two independent components.
For instance, for purely longitudinal flow $u^\mu = \gamma_z (1, 0,0,v^z)$,
with $\gamma_{z}\equiv \left( 1-v_{z}^{2}\right) ^{-1/2}$, and one can
determine $\ell^z$ from Eq.\ (\ref{orthogonality_2}) as $\ell^z = 1/v^z$.
Without loss of generality it is possible to set $\ell^x=\ell^y = 0$, such that
$l^\mu$ is completely specified,
\begin{equation}
l^{\mu } =\gamma_{l}\left(1,0,0,1/v^z\right) \equiv\gamma_{z}\left(v^z,0,0,1\right) \; .
\label{l_mu}
\end{equation}
One may use this form even if the flow is three-dimensional, since
it still fulfills the requirements (\ref{l_mu_norm}) (with $l=1$) and (\ref{orthogonality})
\cite{Florkowski:2010cf,Ryblewski:2011aq,Ryblewski:2012rr}.
In the co-moving frame or LR frame of matter, $u_{LR}^{\mu }=\left( 1,0,0,0\right)$,
(independent of the physical meaning of the four-velocity), hence the direction of 
the anisotropy corresponds to the longitudinal
or $z$--direction of the coordinate system, $l_{LR}^{\mu}=\left( 0,0,0,1\right) $.

In fluid dynamics, one usually introduces a tensor
\begin{equation}
\Delta^{\mu \nu }\equiv g^{\mu \nu }-u^{\mu }u^{\nu }\; ,  \label{Delta_munu}
\end{equation}
which is symmetric $\Delta ^{\mu \nu }=\Delta ^{\nu \mu }$ and projects 
onto the three-dimensional space orthogonal to the four-flow velocity of matter, 
$\Delta^{\mu \nu }u_{\mu }=0$, $\Delta^{\mu}_{\mu} = 3$. 
Since $l^\mu$ is already orthogonal to $u^\mu$,
cf.\ Eq.\ (\ref{orthogonality}), we have $\Delta ^{\mu \nu }l_{\nu } \equiv l^{\mu }$.

An anisotropy in a system singles out another direction besides the direction
of fluid flow. In our case, this is the direction characterized by $l^\mu$. Thus,
it is natural to generalize the projection operator (\ref{Delta_munu}) to 
a new symmetric projection operator onto the two-dimensional subspace 
orthogonal to both $u^{\mu }$ and $l^{\mu }$ 
\cite{Gedalin_1991,Gedalin_1995,Huang:2009ue,Huang:2011dc}, 
\begin{equation}
\Xi ^{\mu \nu } \equiv g^{\mu \nu }-u^{\mu }u^{\nu }+l^{\mu }l^{\nu } 
= \Delta ^{\mu \nu }+l^{\mu }l^{\nu }\;,  \label{Xi_munu}
\end{equation}
where $\Xi ^{\mu \nu }=\Xi ^{\nu \mu }$, $\Xi ^{\mu \nu }u_{\nu }=\Xi ^{\mu
\nu }l_{\nu }=0$, and $\Xi _{\mu }^{\mu }=2$. Note also that
$\Delta ^{\mu \nu }\Xi _{\mu \nu }=2$.

Let us briefly remind the reader of our notational conventions (see Sec.\ \ref{Introduction}). 
The projection of an arbitrary four-vector $A^{\mu }$ orthogonal to $u^{\mu }$ 
will be denoted by 
\begin{equation}
A^{\left\langle \mu \right\rangle }=\Delta ^{\mu \nu }A_{\nu }\;,
\end{equation}
while the projection orthogonal to both $u^{\mu }$ and $l^{\mu }$ will be
denoted by
\begin{equation}
A^{\left\{ \mu \right\} }=\Xi ^{\mu \nu }A_{\nu }\; .
\end{equation}
The corresponding projections of arbitrary rank-two tensors are defined as 
\begin{eqnarray}
A^{\left\langle \mu \nu \right\rangle } &=&\Delta _{\alpha \beta }^{\mu \nu}A^{\alpha \beta }\; , \\
A^{\left\{ \mu \nu \right\} } &=&\Xi _{\alpha \beta }^{\mu \nu }A^{\alpha
\beta }\; ,
\end{eqnarray}
where the corresponding symmetric, orthogonal, and traceless projection
operators are
\begin{eqnarray}
\Delta _{\alpha \beta }^{\mu \nu } &=&\frac{1}{2}\left( \Delta _{\alpha}^{\mu }\Delta _{\beta }^{\nu }
+\Delta _{\beta }^{\mu }\Delta _{\alpha}^{\nu }\right) -\frac{1}{3}\Delta ^{\mu \nu }\Delta _{\alpha \beta }\;,
\label{Delta_traceless} \\
\Xi _{\alpha \beta }^{\mu \nu } &=&\frac{1}{2}\left( \Xi _{\alpha }^{\mu}\Xi _{\beta }^{\nu }
+\Xi _{\beta }^{\mu }\Xi _{\alpha }^{\nu }\right) -
\frac{1}{2}\Xi ^{\mu \nu }\Xi _{\alpha \beta }\;,  \label{Xi_traceless}
\end{eqnarray}
while the projection operators for higher-rank tensors following 
are specified in App.\ \ref{appendix_projections}.

We remark that one can trade the orthogonal projection (\ref{Delta_traceless})
for the projection (\ref{Xi_traceless}), 
\begin{equation}
\Delta _{\alpha \beta }^{\mu \nu }=\Xi _{\alpha \beta }^{\mu \nu
}-2l_{\left( \alpha \right. }\Xi _{\left. \beta \right) }^{\left( \mu
\right. }l^{\left. \nu \right) }+\frac{1}{6}\left( \Xi _{\alpha \beta
}+2l_{\alpha }l_{\beta }\right) \left( \Xi ^{\mu \nu }+2l^{\mu }l^{\nu}\right)\; ,  \label{Delta_Xi_relation}
\end{equation}
which can be obtained by inserting $\Delta ^{\mu \nu }=\Xi ^{\mu\nu }-l^{\mu }l^{\nu }$,
cf.\ Eq.\ (\ref{Xi_munu}), into Eq.\ (\ref{Delta_traceless}). 

The tensor decomposition of a four-vector with respect to $u^\mu$ and
$\Delta^{\mu \nu}$ reads
\begin{equation}
A^{\mu }= A^{\nu }u_{\nu }\, u^{\mu }+A^{\left\langle \mu
\right\rangle }\;,  \label{4_vector_decomposition_u}
\end{equation}
while that with respect to $u^\mu$, $l^\mu$, and $\Xi^{\mu \nu}$ reads
\begin{equation}
A^{\mu }= A^{\nu }u_{\nu }\, u^{\mu }- A^{\nu }l_{\nu}\, l^{\mu }+A^{\left\{ \mu \right\} }\; .
\label{4_vector_decomposition_u_l}
\end{equation}

Analogously, the tensor decomposition of the four-gradient with
respect to $u^\mu$ and $\Delta^{\mu \nu}$ reads
\begin{equation}
\partial _{\mu }=u_{\mu }D+\nabla _{\mu }\;,  \label{partial_mu_iso}
\end{equation}
where the comoving time-derivative and the four-gradient are defined as,%
\begin{eqnarray}
D &=&u^{\mu }\partial _{\mu }\; ,\  \\
\nabla _{\mu } &\equiv &\Delta _{\mu \nu }\partial ^{\nu }=\partial
_{\left\langle \mu \right\rangle }\; .
\end{eqnarray}
In the LR frame $D$ is the time derivative, $D_{LR}=\left( \partial /\partial t,0,0,0\right) $, 
while $\nabla _{LR}^{\mu }=\left( 0,\partial /\partial x,\partial /\partial y,\partial /\partial z\right) $ 
is the three-gradient.

Similarly, we decompose $\partial _{\mu }$ with respect to $u^{\mu }$, $l^{\mu }$,
and $\Xi^{\mu \nu}$ as
\begin{equation}
\partial _{\mu }=u_{\mu }D+l_{\mu }D_{l}+\tilde{\nabla}_{\mu }\; ,
\label{partial_mu_aniso}
\end{equation}
where 
\begin{eqnarray}
D_{l} &=&-l^{\mu }\partial _{\mu }\; ,\  \\
\tilde{\nabla}_{\mu } &\equiv &\Xi _{\mu \nu }\partial ^{\nu }=\partial_{\left\{ \mu \right\} }\; .
\end{eqnarray}
Comparing Eq.\ (\ref{partial_mu_iso}) to Eq.\ (\ref{partial_mu_aniso}) it is
immediately apparent that in the anisotropic case the usual gradient
operator is split into two parts, 
\begin{equation}
\nabla _{\mu }=l_{\mu }D_{l}+\tilde{\nabla}_{\mu }\; ,
\label{gradient_relation}
\end{equation}
where according to the specific choice of Eq.\ (\ref{l_mu}) in the LR frame 
$D_{l,LR}=\left( 0,0,0,\partial /\partial z\right) $ corresponds to the
derivative in the direction of the anisotropy, while $\tilde{\nabla}_{LR}^{\mu }
=\left( 0,\partial /\partial x,\partial /\partial y,0\right) $
is the spatial derivative in the remaining transverse directions orthogonal to
both $u^{\mu }$ and $l^{\mu }$.

Note that on account of the normalization and orthogonality conditions
(\ref{u_mu_norm}), (\ref{l_mu_norm}), and (\ref{orthogonality}), we have
the identities
\begin{align}
u^{\nu }\partial _{\mu }u_{\nu } & = u^{\nu }Du_{\nu } = u^{\nu }D_{l}u_{\nu }
=u^{\nu }\tilde{\nabla}_{\mu }u_{\nu}= 0\; ,  \label{u_identities} \\
l^{\nu }\partial _{\mu }l_{\nu }& =l^{\nu }Dl_{\nu }=l^{\nu }D_{l}l_{\nu }
=l^{\nu }\tilde{\nabla}_{\mu }l_{\nu }=0\;, 
\end{align}
as well as
\begin{equation}
u^\nu \partial_\mu l_\nu = - l^\nu \partial_\mu u_\nu\;, \;\;
u^\nu D l_\nu = - l^\nu D u_\nu\;, \;\;
u^\nu D_l l_\nu = - l^\nu D_l u_\nu\;, \;\;
u^\nu \tilde{\nabla}_\mu  l_\nu = - l^\nu  \tilde{\nabla}_\mu u_\nu\;.
\end{equation}

The decomposition of a rank-two tensor with respect to $u^{\mu }$ and 
$\Delta ^{\mu \nu }$ reads
\begin{equation}
A^{\mu \nu } =A^{\alpha \beta }u_{\alpha }u_{\beta }\, u^{\mu}u^{\nu }
+ \Delta^{\mu \alpha }A_{\alpha \beta }u^{\beta }\,u^{\nu }
+ \Delta^{\nu \beta }A_{\alpha \beta }u^{\alpha }\,u^{\mu }  
+\frac{1}{3}\, A^{\alpha \beta }\Delta _{\alpha \beta }\, \Delta^{\mu \nu }
+\Delta _{\alpha \beta }^{\mu \nu }A^{\alpha \beta }
+\Delta_{\alpha }^{\mu }\Delta _{\beta }^{\nu }A^{\left[ \alpha \beta \right] }\;.
\label{isotropic_decomposition}
\end{equation}
Applying this to the gradient of the four-flow velocity, we obtain with
Eq.\ (\ref{u_identities}) the well-known relativistic Cauchy-Stokes formula,
\begin{equation}
\partial_{\mu }u_{\nu }=u_{\mu }Du_{\nu }+\frac{1}{3}\, \theta\, \Delta_{\mu\nu}
+\sigma_{\mu \nu }+\omega_{\mu \nu }\;,  \label{Cauchy_iso}
\end{equation}
where
\begin{equation} \label{exp_iso}
\theta \equiv \nabla_{\mu }u^{\mu }
\end{equation}
is the expansion scalar, 
\begin{equation} \label{shear_iso}
\sigma^{\mu \nu }\equiv \partial^{\left\langle \mu \right. }u^{\left.\nu \right\rangle } 
=\Delta_{\alpha\beta }^{\mu \nu }\partial^{\alpha }u^{\beta }
=\frac{1}{2}\left( \nabla^{\mu }u^{\nu }+\nabla^{\nu}u^{\mu }\right) 
-\frac{1}{3}\,\theta\, \Delta^{\mu \nu}
\end{equation}
is the shear tensor and 
\begin{equation} \label{vort_iso}
\omega^{\mu \nu } \equiv \Delta_{\alpha }^{\mu }\Delta_{\beta }^{\nu}
\partial^{\left[ \alpha \right. }u^{\left. \beta \right] }
=\frac{1}{2} \left( \nabla^{\mu }u^{\nu }-\nabla ^{\nu }u^{\mu }\right)
\end{equation}
is the vorticity.

The tensor decomposition of a rank-two tensor with respect to $u^{\mu }$, $l^{\mu }$,
and $\Xi^{\mu \nu}$ reads 
\begin{align}
A^{\mu \nu }& = A^{\alpha \beta }u_{\alpha }u_{\beta }\, u^{\mu}u^{\nu }
+ A^{\alpha \beta }l_{\alpha }l_{\beta }\, l^{\mu}l^{\nu }
- A^{\alpha \beta }u_{\alpha }l_{\beta }\, u^{\mu}l^{\nu }
- A^{\alpha \beta }u_{\beta }l_{\alpha }\, u^{\nu}l^{\mu }  \notag \\
& + \Xi^{\mu \alpha }A_{\alpha \beta }u^{\beta }\, u^{\nu}
+ \Xi^{\nu \beta }A_{\alpha \beta }u^{\alpha }\, u^{\mu}
- \Xi^{\mu \alpha }A_{\alpha \beta }l^{\beta }\, l^{\nu}
- \Xi^{\nu \beta }A_{\alpha \beta }l^{\alpha }\, l^{\mu } 
\notag \\
& +\frac{1}{2}\, A^{\alpha \beta }\Xi_{\alpha \beta }\, \Xi ^{\mu\nu }
+\Xi_{\alpha \beta }^{\mu \nu }A^{\alpha \beta }
+\Xi_{\alpha }^{\mu}\Xi _{\beta }^{\nu }A^{\left[ \alpha \beta \right] }\;.
\label{anisotropic_decomposition}
\end{align}
With this decomposition we obtain the counterpart of the
relativistic Cauchy-Stokes formula (\ref{Cauchy_iso}),
\begin{equation}
\partial_{\mu }u_{\nu }=u_{\mu }Du_{\nu }+l_{\mu }D_{l}u_{\nu }
+\frac{1}{2}\,\tilde{\theta}\,\Xi _{\mu \nu }-l_{\beta }l_{\nu }\tilde{\nabla}_{\mu}u^{\beta }
+\tilde{\sigma}_{\mu \nu }+\tilde{\omega}_{\mu \nu }\;,
\label{Cauchy_aniso}
\end{equation}
where we have again made use of Eq.\ (\ref{u_identities}).
Note that $-l_{\beta }l_{\nu }\tilde{\nabla}_{\mu }u^{\beta }=-l_{\beta
}l_{\left( \mu \right. }\tilde{\nabla}_{\left. \nu \right) }u^{\beta
}+l_{\beta }l_{\left[ \mu \right. }\tilde{\nabla}_{\left. \nu \right]
}u^{\beta }$ can in principle also be further separated into a
symmetric and an antisymmetric part.
In Eq.\ (\ref{Cauchy_aniso}) we defined the following quantities in the subspace orthogonal 
to both $u^{\mu }$ and $l^{\mu }$: the transverse expansion scalar
\begin{equation} \label{exp_aniso}
\tilde{\theta} \equiv \tilde{\nabla}_{\mu }u^{\mu }\; ,
\end{equation}
the transverse shear tensor 
\begin{equation} \label{shear_aniso}
\tilde{\sigma}^{\mu \nu } \equiv \partial ^{\left\{ \mu \right. }u^{\left.\nu \right\} }
=\tilde{\nabla}^{\left( \mu \right. }u^{\left. \nu \right) }
-\frac{1}{2}\, \tilde{\theta}\, \Xi^{\mu \nu }
+l_{\beta }l^{\left( \mu \right. } \tilde{\nabla}^{\left. \nu \right) }u^{\beta }\; , 
\end{equation}
and the transverse vorticity
\begin{equation} \label{vort_aniso}
\tilde{\omega}^{\mu \nu } \equiv \Xi ^{\mu \alpha }\Xi ^{\nu \beta}
\partial _{\left[ \alpha \right. }u_{\left. \beta \right] }
=\tilde{\nabla}^{\left[ \mu \right. }u^{\left. \nu \right] }-l_{\beta }l^{\left[ \mu \right. }
\tilde{\nabla}^{\left. \nu \right] }u^{\beta }\;.
\end{equation}
Note that the shear tensor (\ref{shear_iso}) and
vorticity (\ref{vort_iso}) are orthogonal to the flow velocity, i.e., 
$\sigma ^{\mu \nu}u_{\nu }=\omega ^{\mu \nu }u_{\nu }=0$, but
$\sigma^{\mu \nu }l_{\nu }\neq 0$, $\omega ^{\mu \nu }l_{\nu }\neq 0$, 
while the transverse shear tensor (\ref{shear_aniso}) and the
transverse vorticity (\ref{vort_aniso}) are orthogonal to both four-vectors, 
$\tilde{\sigma}^{\mu \nu }u_{\nu }=\tilde{\sigma}^{\mu \nu }l_{\nu }=0$ 
and $\tilde{\omega}^{\mu \nu }u_{\nu }=\tilde{\omega}^{\mu \nu }l_{\nu }=0$.

The expansion scalars (\ref{exp_iso}) and (\ref{exp_aniso}) are related to each other
through Eq.\ (\ref{gradient_relation}),
\begin{equation}
\theta =l_{\mu }D_{l}u^{\mu }+\tilde{\theta}\;.  \label{expansion_relation}
\end{equation}
The shear tensors (\ref{shear_iso}) and (\ref{shear_aniso}) are symmetric 
by definition and related to each other through
Eq.\ (\ref{Delta_Xi_relation}),
\begin{equation}
\sigma ^{\mu \nu }=\tilde{\sigma}^{\mu \nu }
+l^{\left( \mu \right. }\Xi_{\beta }^{\left. \nu \right) }D_{l}u^{\beta }
-l_{\beta }l^{\left( \mu\right. }\tilde{\nabla}^{\left. \nu \right) }u^{\beta }
+\frac{1}{6}\left( \tilde{\theta}-2l_{\beta }D_{l}u^{\beta }\right) 
\left( \Xi ^{\mu \nu}+2l^{\mu }l^{\nu }\right)\;.  \label{sigma_l_mu_nu}
\end{equation}
Finally, the vorticities (\ref{vort_iso}) and (\ref{vort_aniso})
are antisymmetric and related via 
\begin{equation}
\omega^{\mu \nu }=\tilde{\omega}^{\mu \nu }+l^{\left[ \mu \right.}D_{l}u^{\left. \nu \right] }
+l_{\beta }l^{\left[ \mu \right. }\tilde{\nabla}^{\left. \nu \right] }u^{\beta }\;.
\end{equation}

Similarly to the Cauchy-Stokes formulae (\ref{Cauchy_iso}) and (\ref{Cauchy_aniso}) 
for $\partial _{\mu }u_{\nu }$ we also need
the decompositions of $\partial _{\mu }l_{\nu }$,
\begin{align}
\partial _{\mu }l_{\nu }& \equiv u_{\mu }Dl_{\nu }+\frac{1}{3}\theta
_{l}\Delta _{\mu \nu }+u_{\beta }u_{\nu }\nabla _{\mu }l^{\beta }+\sigma
_{l,\mu \nu }+\omega _{l,\mu \nu }  \notag \\
& =u_{\mu }Dl_{\nu }+l_{\mu }D_{l}l_{\nu }+\frac{1}{2}\,\tilde{\theta}_{l}\,\Xi_{\mu \nu }
+u_{\beta }u_{\nu }\tilde{\nabla}_{\mu }l^{\beta }+\tilde{\sigma}_{l,\mu \nu }
+\tilde{\omega}_{l,\mu \nu },
\end{align}
where $u_{\beta }u_{\nu }\nabla _{\mu }l^{\beta }=u_{\beta }u_{\left( \mu
\right. }\nabla _{\left. \nu \right) }l^{\beta }-u_{\beta }u_{\left[ \mu
\right. }\nabla _{\left. \nu \right] }l^{\beta }$ and we defined the
quantities
\begin{eqnarray}
\theta _{l} &\equiv&\nabla _{\mu }l^{\mu }\; , \\
\tilde{\theta}_{l} &\equiv &\tilde{\nabla}_{\mu }l^{\mu }\; , \\
\sigma _{l}^{\mu \nu } &\equiv &\partial ^{\left\langle \mu \right.}l^{\left. \nu \right\rangle }
=\nabla ^{\left( \mu \right. }l^{\left. \nu\right) }-\frac{1}{3}\,\theta _{l}\, \Delta ^{\mu \nu }
-u_{\beta }u^{\left( \mu\right. }\nabla ^{\left. \nu \right) }l^{\beta }\; , \\
\tilde{\sigma}_{l}^{\mu \nu } &\equiv &\partial ^{\left\{ \mu \right.}l^{\left. \nu \right\} }
=\tilde{\nabla}^{\left( \mu \right. }l^{\left. \nu\right) }-\frac{1}{2}\,\tilde{\theta}_{l}\,\Xi ^{\mu \nu }
-u_{\beta }u^{\left(\mu \right. }\tilde{\nabla}^{\left. \nu \right) }l^{\beta }\; , \\
\omega _{l}^{\mu \nu } &\equiv &\Delta ^{\mu \alpha }\Delta ^{\nu \beta}
\partial _{\left[ \alpha \right. }l_{\left. \beta \right] }=\nabla^{\left[\mu \right. }l^{\left. \nu \right] }
+u_{\beta }u^{\left[ \mu \right. }\nabla^{\left. \nu \right] }l^{\beta }\; , \\
\tilde{\omega}_{l}^{\mu \nu } &\equiv &\Xi ^{\mu \alpha }\Xi ^{\nu \beta}
\partial_{\left[ \alpha \right. }l_{\left. \beta \right] }=\tilde{\nabla}^{\left[ \mu \right. }l^{\left. \nu \right] }
+u_{\beta }u^{\left[ \mu \right. } \tilde{\nabla}^{\left. \nu \right] }l^{\beta }\; .
\end{eqnarray}
Note that $\sigma _{l}^{\mu \nu }u_{\nu }\neq 0$ and $\omega _{l}^{\mu \nu}u_{\nu }\neq 0$, 
but $\tilde{\sigma}_{l}^{\mu \nu }u_{\nu }=\tilde{\sigma}_{l}^{\mu \nu }l_{\nu }=0$ 
and $\tilde{\omega}_{l}^{\mu \nu }u_{\nu }=\tilde{\omega}_{l}^{\mu \nu }l_{\nu }=0$. 
The relationships between the various quantities are
\begin{eqnarray}
\theta _{l} &=&\tilde{\theta}_{l}\; , \\
\sigma _{l}^{\mu \nu } &=&\tilde{\sigma}_{l}^{\mu \nu }+l^{\left( \mu
\right. }\Xi _{\beta }^{\left. \nu \right) }D_{l}l^{\beta }+\frac{1}{6}
\tilde{\theta}_{l}\left( \Xi ^{\mu \nu }+2l^{\mu }l^{\nu }\right)\; , \\
\omega _{l}^{\mu \nu } &=&\tilde{\omega}_{l}^{\mu \nu }+l^{\left[ \mu\right. }
\Xi _{\beta }^{\left. \nu \right] }D_{l}l^{\beta }\; .
\end{eqnarray}

\section{Tensor decomposition of four-momentum}
\label{appB}

In this appendix, we apply the result of App.\ \ref{Quantities} to derive the
tensor decomposition of four-momentum 
$k^{\mu }=\left( k^{0},k^{x},k^{y},k^{z}\right)$. According to 
Eq.\ (\ref{4_vector_decomposition_u}), 
\begin{equation}
k^{\mu }=E_{\mathbf{k}u}u^{\mu }+k^{\left\langle \mu \right\rangle }\; ,
\label{k_mu_iso}
\end{equation}
where 
\begin{equation}
E_{\mathbf{k}u}=k^{\mu }u_{\mu }\;,\;\; k^{\left\langle \mu \right\rangle}=\Delta ^{\mu \nu }k_{\nu }\; .  \label{kinetic:E_u}
\end{equation}
The physical meaning of these quantities becomes apparent in the LR frame:
$E_{\mathbf{k}u,LR}=k^{0}$ is the energy while $k_{LR}^{\left\langle
\mu \right\rangle }=\left( 0,k^{x},k^{y},k^{z}\right) $ is the three-momentum.

Analogously, one can tensor-decompose $k^\mu$ using
Eq.\ (\ref{4_vector_decomposition_u_l}),
\begin{equation}
k^{\mu }=E_{\mathbf{k}u}u^{\mu }+E_{\mathbf{k}l}l^{\mu }+k^{\left\{ \mu\right\} }\; ,  \label{k_mu_aniso}
\end{equation}
where \begin{equation}
E_{\mathbf{k}l}=-k^{\mu }l_{\mu }\;,\;\; k^{\left\{ \mu \right\} }=\Xi ^{\mu \nu}k_{\nu }\;.  \label{kinetic:E_l}
\end{equation}
Comparing Eqs.\ (\ref{kinetic:E_u}) and (\ref{k_mu_aniso}), it is obvious that 
$k^{\left\langle \mu \right\rangle}=E_{\mathbf{k}l}l^{\mu }+k^{\left\{ \mu \right\} }$.
In the LR frame and with the choice (\ref{l_mu}) for $l^\mu$, 
the quantities defined in Eq.\ (\ref{kinetic:E_l}) are $E_{\mathbf{k}l,LR}=\left( 0,0,0,k^{z}\right) $,
i.e., the component of three-momentum in $l^\mu$--direction, 
and $k_{LR}^{\left\{ \mu \right\} }=\left( 0,k^{x},k^{y},0\right) $, i.e., the
components of three-momentum orthogonal to $l^\mu$.

For on-shell particles,
\begin{eqnarray}
k^{\mu }k_{\mu } &\equiv &E_{\mathbf{k}u}^{2}+k^{\left\langle \mu \right\rangle }
k_{\left\langle \mu \right\rangle }  \notag \\
&=&E_{\mathbf{k}u}^{2}-E_{\mathbf{k}l}^{2}+k^{\left\{ \mu \right\}}k_{\left\{ \mu \right\} }
=m_{0}^{2}\; ,  \label{kk_normalization}
\end{eqnarray}
where $m_{0}$ is the rest mass of the particle, while 
\begin{equation}
k^{\left\langle \mu \right\rangle }k_{\left\langle \mu \right\rangle}
=\Delta ^{\alpha \beta }k_{\alpha }k_{\beta }\; ,\;\; k^{\left\{\mu \right\} }k_{\left\{ \mu \right\} }
= \Xi ^{\alpha \beta }k_{\alpha
}k_{\beta }\; .  \label{kk_product}
\end{equation}

From Eq.\ (\ref{isotropic_decomposition})
\begin{equation} \label{B7}
k^{\mu }k^{\nu }=E_{\mathbf{k}u}^{2}u^{\mu }u^{\nu }+\frac{1}{3}\,
k^{\left\langle \alpha \right\rangle }k_{\left\langle \alpha \right\rangle}
\, \Delta ^{\mu \nu }+2E_{\mathbf{k}u}k^{\left( \left\langle \mu \right\rangle \right. }
u^{\left. \nu \right) }+k^{\left\langle \mu \right. }k^{\left. \nu \right\rangle }\;,
\end{equation}
while from Eq.\ (\ref{anisotropic_decomposition})
\begin{equation} \label{B8}
k^{\mu }k^{\nu }=E_{\mathbf{k}u}^{2}u^{\mu }u^{\nu }+E_{\mathbf{k}l}^{2}l^{\mu }l^{\nu }
+2E_{\mathbf{k}u}E_{\mathbf{k}l}u^{\left( \mu \right.}l^{\left. \nu \right) }
+2E_{\mathbf{k}u}k^{\left( \left\{ \mu\right\} \right. }u^{\left. \nu \right) }
+2E_{\mathbf{k}l}k^{\left( \left\{\mu \right\} \right. }l^{\left. \nu \right) }
+\frac{1}{2}\, k^{\left\{\alpha \right\}}k_{\left\{ \alpha\right\} }\, \Xi ^{\mu \nu }
+k^{\left\{ \mu \right.}k^{\left. \nu \right\} }\;,
\end{equation}
where we used $\Delta _{\alpha \beta }^{\mu \nu }k^{\alpha }k^{\beta}=
k^{\left\langle \mu \right. }k^{\left. \nu \right\rangle }$ and $\Xi_{\alpha \beta }^{\mu \nu }
k^{\alpha }k^{\beta }=k^{\left\{ \mu \right.}k^{\left. \nu \right\} }$. 
Higher-rank tensors formed from dyadic products of
$k^\mu$ can be decomposed in a similar manner.

\section{Thermodynamic integrals and properties}
\label{appendix_thermo_integrals}

In this appendix, we compute the thermodynamic integrals $I_{i+n,q}$ and $\hat{I}_{i+j+n,j+r,q}$ in
Eqs.\ (\ref{I_nq_moment}) and (\ref{I_nrq_moment}). They are obtained by suitable
projections of the tensors $\mathcal{I}_{i}^{\mu_1 \cdots \mu_n}$ and
$\hat{\mathcal{I}}_{ij}^{\mu_1 \cdots \mu_n}$.

Let us first focus on the integrals $I_{i+n,q}$.
The coefficient $b_{nq}$ in Eq.\ (\ref{I_nq_moment}) is defined as the
number of distinct terms in the symmetrized tensor product 
\begin{equation} \label{sym_tens_prod}
\Delta^{\left( \mu _{1}\mu _{2}\right. }\cdots \Delta ^{\mu _{2q-1}\mu_{2q}}
u^{\mu _{2q+1}}\cdots u^{\left. \mu _{n}\right) } \equiv \frac{1}{b_{nq}} \sum_{\mathcal{P}_\mu^n}
\Delta^{\mu _{1}\mu _{2}}\cdots \Delta ^{\mu _{2q-1}\mu_{2q}} u^{\mu _{2q+1}}\cdots u^{ \mu _{n} }\;,
\end{equation} 
where the sum runs over all distinct permutations of the $n$ indices $\mu_1, \ldots, \mu_n$.
The total number of permutations of $n$ indices
is $n!$. There are $q$ projection operators $\Delta^{\mu_i \mu_j}$ and $n-2q$ factors of $u^{\mu_k}$.
Permutations of the order of the $\Delta^{\mu_i \mu_j}$ and of the $u^{\mu_k}$ among themselves
do not lead to distinct terms, so we need to divide the total number $n!$ by
$q! (n-2q)!$. Finally, since $\Delta^{\mu_i \mu_j}$ is a symmetric projection operator, a permutation of its
indices does not lead to a distinct term. Since there are $q$ such projection
operators, there are $2^q$ permutations that also do not lead to distinct terms.
Hence, the total number of distinct terms in the symmetrized tensor product is
\begin{equation}
b_{nq}\equiv \frac{n!}{2^{q}q!\left( n-2q\right) !}=\frac{n!\left(2q-1\right) !!}{\left( 2q\right) !\left( n-2q\right) !}\; , 
 \label{a_nq}
\end{equation}%
which is identical to Eq.\ (A2) of Ref.\ \cite{Israel:1979wp}.

In order to obtain the thermodynamic integrals $I_{i+n,q}$  by projection of the
tensors $\mathcal{I}_{i}^{\mu_1 \cdots \mu_n}$, it is advantageous to use the 
orthogonality relation
\begin{equation} \label{contraction_Delta}
\Delta ^{\left( \mu_1 \mu_2\right.} \cdots \Delta^{\mu_{2q-1} \mu_{2q}}u^{\mu_{2q+1}} \cdots
u^{\left. \mu_n\right) }
\Delta_{\left( \mu_1 \mu_2\right.} \cdots \Delta_{\mu_{2q'-1} \mu_{2q'}}u_{\mu_{2q'+1}} \cdots
u_{\left. \mu_n\right) }
=\frac{\left( 2q+1\right) !!}{b_{nq}}\delta _{qq^{\prime }}\; ,
\end{equation}
cf.\ Eq.\ (A.3) of chapter VI.1 of Ref.\ \cite{deGroot}. Since later on we will generalize this result to the case
involving $l$'s and $\Xi$'s, we give the proof in some detail. First, it is clear that if $q \neq q'$ there are terms where
a $u^{\mu_i}$ gets contracted with a $\Delta_{\mu_i\mu_j}$, which gives zero. 
The existence of the Kronecker delta is thus easily explained and
we only need to prove Eq.\ (\ref{contraction_Delta}) for $q=q'$. Second, as the same set of indices
is symmetrized on both tensor products, it actually suffices to keep the set of indices fixed on one tensor,
say in the order $\mu_1, \ldots, \mu_{2q}, \mu_{2q+1}, \ldots, \mu_n$, and
symmetrize only the one on the other,
\begin{eqnarray}
\lefteqn{\Delta ^{\left( \mu_1 \mu_2\right.} \cdots \Delta^{\mu_{2q-1} \mu_{2q}}u^{\mu_{2q+1}} \cdots
u^{\left. \mu_n\right) }
\Delta_{\left( \mu_1 \mu_2\right.} \cdots \Delta_{\mu_{2q-1} \mu_{2q}}u_{\mu_{2q+1}} \cdots
u_{\left. \mu_n\right) } }  \nonumber \\
& = & \Delta^{ \mu_1 \mu_2} \cdots \Delta^{\mu_{2q-1} \mu_{2q}}u^{\mu_{2q+1}} \cdots
u^{ \mu_n } \, \frac{1}{b_{nq}} \sum_{\mathcal{P}_\mu^n}
\Delta_{ \mu_1 \mu_2} \cdots \Delta_{\mu_{2q-1} \mu_{2q}}u_{\mu_{2q+1}} \cdots
u_{ \mu_n } \;,
\end{eqnarray}
where we used Eq.\ (\ref{sym_tens_prod}).
Among the terms in the sum over all distinct permutations, only
those survive where the indices on the $u$'s are $\mu_{2q+1}, \ldots, \mu_n$, just as in the term
in front of the sum. (Otherwise, a $u_{\mu_i}$ will be contracted with a $\Delta^{\mu_i\mu_j}$, which gives zero.) 
Permutations among these indices do not lead to distinct terms. Using $u^{\mu} u_\mu =1$, we thus obtain
\begin{eqnarray}
\lefteqn{\Delta ^{\left( \mu_1 \mu_2\right.} \cdots \Delta^{\mu_{2q-1} \mu_{2q}}u^{\mu_{2q+1}} \cdots
u^{\left. \mu_n\right) }
\Delta_{\left( \mu_1 \mu_2\right.} \cdots \Delta_{\mu_{2q-1} \mu_{2q}}u_{\mu_{2q+1}} \cdots
u_{\left. \mu_n\right) } }  \nonumber \\
& = &\displaystyle \frac{1}{b_{nq}}\, \Delta^{ \mu_1 \mu_2} \cdots \Delta^{\mu_{2q-1} \mu_{2q}}
\sum_{\mathcal{P}_\mu^{2q}}
\Delta_{ \mu_1 \mu_2} \cdots \Delta_{\mu_{2q-1} \mu_{2q}}\;, \hspace*{4cm}
\end{eqnarray}
where the sum now runs only over the distinct permutations of $2q$ indices $\mu_1, \ldots, \mu_{2q}$ on
the $\Delta$ projectors. There are in total $(2q)!/(2^q q!) \equiv (2q-1)!!$ distinct terms, so that we obtain 
\begin{eqnarray}
\lefteqn{\Delta ^{\left( \mu_1 \mu_2\right.} \cdots \Delta^{\mu_{2q-1} \mu_{2q}}u^{\mu_{2q+1}} \cdots
u^{\left. \mu_n\right) }
\Delta_{\left( \mu_1 \mu_2\right.} \cdots \Delta_{\mu_{2q-1} \mu_{2q}}u_{\mu_{2q+1}} \cdots
u_{\left. \mu_n\right) } }  \nonumber \\
& = &\displaystyle \frac{(2q-1)!!}{b_{nq}} \, \Delta^{ \mu_1 \mu_2} \cdots \Delta^{\mu_{2q-1} \mu_{2q}}
\Delta_{\left( \mu_1 \mu_2\right.} \cdots \Delta_{\left.\mu_{2q-1} \mu_{2q}\right)} \;. \hspace*{3.2cm}
\end{eqnarray}
The proof of Eq.\ (\ref{contraction_Delta}) is completed by proving that
\begin{equation} \label{Delta_Delta}
\Delta^{ \mu_1 \mu_2} \cdots \Delta^{\mu_{2q-1} \mu_{2q}}
\Delta_{\left( \mu_1 \mu_2\right.} \cdots \Delta_{\left.\mu_{2q-1} \mu_{2q}\right)} = 2q+1\;,
\end{equation}
cf.\ Eq.\ (A.4) of chapter VI.1 of Ref.\ \cite{deGroot}. This is done by complete induction. Since $\Delta^{\mu_1\mu_2}
\Delta_{\mu_1 \mu_2} = \Delta^{\mu_1}_{\mu_1} = 3$, Eq.\ (\ref{Delta_Delta}) obviously holds for $q=1$. Now
suppose it holds for $q$. Then we have to show that it also holds for $q+1$. In this case, using the definition of
the symmetrized tensor,
\begin{eqnarray}
\lefteqn{\Delta^{ \mu_1 \mu_2} \cdots \Delta^{\mu_{2q+1} \mu_{2q+2}}
\Delta_{\left( \mu_1 \mu_2\right.} \cdots \Delta_{\left.\mu_{2q+1} \mu_{2q+2}\right)} }\nonumber \\
& = & \frac{2^{q+1}(q+1)!}{(2q+2)!}\, \Delta^{ \mu_1 \mu_2} \cdots \Delta^{\mu_{2q+1} \mu_{2q+2}}
\sum_{\mathcal{P}_\mu^{2q+2}} \Delta_{ \mu_1 \mu_2} \cdots \Delta_{\mu_{2q+1} \mu_{2q+2}}\;.
\end{eqnarray}
Consider the contraction of $\Delta^{\mu_{2q+1} \mu_{2q+2}}$ with the sum over distinct permutations
of $2q+2$ indices $\mu_1, \ldots, \mu_{2q+2}$.
There is one term in the sum where both indices are on the same $\Delta$ projector. This
term is $\sim \Delta^{\mu_{2q+1} \mu_{2q+2}} \Delta_{\mu_{2q+1} \mu_{2q+2}} \equiv 3$. Then, there are
$2q$ terms where the indices $\mu_{2q+1}$ and $\mu_{2q+2}$ are on different projectors, say
$\Delta_{\mu_{2q+1} \mu_j} \Delta_{\mu_i \mu_{2q+2}}$. Contracting with $\Delta^{\mu_{2q+1} \mu_{2q+2}}$
gives a term $\sim \Delta_{\mu_i \mu_j}$, where both indices are from the set $\mu_1, \ldots, \mu_{2q}$. Putting
this together and using Eq.\ (\ref{Delta_Delta}) gives
\begin{eqnarray}
\lefteqn{\Delta^{ \mu_1 \mu_2} \cdots \Delta^{\mu_{2q+1} \mu_{2q+2}}
\Delta_{\left( \mu_1 \mu_2\right.} \cdots \Delta_{\left.\mu_{2q+1} \mu_{2q+2}\right)} }\nonumber \\
& = & \frac{2(q+1)}{(2q+2)(2q+1)} \, \frac{2^q q!}{(2q)!}\,  \Delta^{ \mu_1 \mu_2} \cdots \Delta^{\mu_{2q-1} \mu_{2q}}
(2q+3) \sum_{\mathcal{P}_\mu^{2q}} \Delta_{ \mu_1 \mu_2} \cdots \Delta_{\mu_{2q-1} \mu_{2q}}\nonumber \\
& = & \frac{2q+3}{2q+1} \,  \Delta^{ \mu_1 \mu_2} \cdots \Delta^{\mu_{2q-1} \mu_{2q}}
\Delta_{\left( \mu_1 \mu_2\right.} \cdots \Delta_{\left. \mu_{2q-1} \mu_{2q}\right)}\equiv 2q+3\;, \;\;\; {\rm q.e.d.}\;.
\end{eqnarray}

With the orthogonality relation (\ref{contraction_Delta}), we now easily find by projecting Eq.\ (\ref{I_nq_moment}) that 
\begin{eqnarray}
I_{i+n,q} &\equiv &\frac{\left( -1\right) ^{q}}{\left( 2q+1\right) !!}\, \mathcal{I}_{i}^{\mu _{1}\cdots \mu _{n}}\,
\Delta _{\left( \mu _{1}\mu_{2}\right. }\cdots \Delta _{\mu _{2q-1}\mu _{2q}}u_{\mu _{2q+1}}\cdots
u_{\left. \mu _{n}\right) }  \notag \\
&=&\frac{\left( -1\right) ^{q}}{\left( 2q+1\right) !!}\int dK\, E_{\mathbf{k}u}^{i+n-2q}
\left( \Delta ^{\mu \nu }k_{\mu }k_{\nu }\right) ^{q}f_{0\mathbf{k}}\;,  \label{I_nq_appendix}
\end{eqnarray}
where we used the definition (\ref{I_n_tens}) of the tensor $\mathcal{I}_{i}^{\mu _{1}\cdots \mu _{n}}$.
With the definition of the thermodynamic average $\langle \ldots \rangle_0$, the second line yields Eq.\ (\ref{I_nq}).

Now we compute the thermodynamical integrals $\hat{I}_{i+j+n,j+r,q}$ in Eq.\ (\ref{I_nrq_moment}).
In that equation, we introduced the symmetrized tensor products 
\begin{equation}
\Xi^{\left( \mu _{1}\mu _{2}\right. }\cdots \Xi ^{\mu _{2q-1}\mu _{2q}}l^{\mu_{2q+1}}\cdots 
l^{\mu _{2q+r}}u^{\mu _{2q+r+1}}\cdots u^{\left. \mu_{n}\right) }
\equiv \frac{1}{b_{nrq}} \sum_{\mathcal{P}_\mu^n}
\Xi^{ \mu _{1}\mu _{2} }\cdots \Xi ^{\mu _{2q-1}\mu _{2q}}l^{\mu_{2q+1}}\cdots 
l^{\mu _{2q+r}}u^{\mu _{2q+r+1}}\cdots u^{ \mu_{n} }\;,
\end{equation}
where $b_{nrq}$ is the number of
terms in the sum over distinct permutations of the $n$ indices $\mu_1, \ldots, \mu_n$. 
Again there are in total $n!$ different permutations of the indices.
There are $q$ projection operators $\Xi ^{\mu _{i}\mu _{j}}$, $r$ factors of $l^{\mu_k}$, and
$n-r-2q$ factors of $u^{\mu_m}$. Permutations of the order of the $\Xi ^{\mu _{i}\mu _{j}}$, the $l^{\mu_k}$, and
of the $u^{\mu_m}$ among themselves to not lead to distinct terms. Likewise, permutations of the two indices
of the symmetric projection operator $\Xi ^{\mu _{i}\mu _{j}}$ do not lead to distinct terms. Thus the
total number of distinct terms in the symmetrized tensor product is
\begin{equation}
b_{nrq}\equiv \frac{n!}{2^{q}\, q!\, r!\, \left( n-r-2q\right) !}=
\frac{n!\left( 2q-1\right) !!}{\left( 2q\right) !\, r !\, \left(n-r-2q\right) !}\;.  \label{a_nrq}
\end{equation}

A suitable projection of the tensor $\hat{\mathcal{I}}_{ij}^{\mu_1 \cdots \mu_n}$ is now found by employing
the orthogonality relation
\begin{eqnarray}
\lefteqn{\Xi^{\left( \mu_1 \mu_2\right.} \cdots \Xi^{\mu_{2q-1} \mu_{2q}} l^{\mu_{2q+1}} \cdots
l^{ \mu_{2q+r}} u^{\mu_{2q+r+1}} \cdots u^{\left. \mu_n \right) }
\Xi_{\left( \mu_1 \mu_2\right.} \cdots \Xi_{\mu_{2q'-1} \mu_{2q'}} l_{\mu_{2q'+1}} \cdots
l_{ \mu_{2q'+r'}} u_{\mu_{2q'+r'+1}} \cdots u_{\left. \mu_n \right) }}\nonumber \\
& =& \left( -1\right) ^{r}\frac{(2q)!!}{b_{nrq}}\delta _{qq^{\prime}}\delta _{rr^{\prime }}\;. \hspace*{13cm}
\label{contraction_Xi}
\end{eqnarray}
In order to prove this relation, we first note that the Kronecker deltas are easily explained by the fact that
 if $q\neq q'$ or $r\neq r'$, there are terms where
a $u^{\mu_i}$ or an $l^{\mu_j}$ are either contracted with each other or with projection operators 
$\Xi_{\mu_i \mu_k}$, $\Xi_{\mu_j \mu_m}$, which gives zero. We thus need to prove Eq.\ (\ref{contraction_Xi})
only for $q=q'$, $r=r'$. Again, since both sets of indices are symmetrized, we may keep one set fixed, i.e., 
\begin{eqnarray}
\lefteqn{\Xi^{\left( \mu_1 \mu_2\right.} \cdots \Xi^{\mu_{2q-1} \mu_{2q}} l^{\mu_{2q+1}} \cdots
l^{ \mu_{2q+r}} u^{\mu_{2q+r+1}} \cdots u^{\left. \mu_n \right) }
\Xi_{\left( \mu_1 \mu_2\right.} \cdots \Xi_{\mu_{2q-1} \mu_{2q}} l_{\mu_{2q+1}} \cdots
l_{ \mu_{2q+r}} u_{\mu_{2q+r+1}} \cdots u_{\left. \mu_n \right) } } \nonumber \\
& =& \frac{1}{b_{nrq}}\, \Xi^{\mu_1 \mu_2} \cdots \Xi^{\mu_{2q-1} \mu_{2q}} l^{\mu_{2q+1}} \cdots
l^{ \mu_{2q+r}} u^{\mu_{2q+r+1}} \cdots u^{\mu_n } \sum_{\mathcal{P}_\mu^n} 
\Xi_{\mu_1 \mu_2} \cdots \Xi_{\mu_{2q-1} \mu_{2q}} l_{\mu_{2q+1}} \cdots
l_{ \mu_{2q+r}} u_{\mu_{2q+r+1}} \cdots u_{ \mu_n } \nonumber \\
& = &  \frac{1}{b_{nrq}}\, (-1)^r \Xi^{\mu_1 \mu_2} \cdots \Xi^{\mu_{2q-1} \mu_{2q}} \sum_{\mathcal{P}_\mu^{2q}} 
\Xi_{\mu_1 \mu_2} \cdots \Xi_{\mu_{2q-1} \mu_{2q}} \nonumber \\
& = &  \frac{1}{b_{nrq}}\, (-1)^r \frac{(2q)!}{2^q q!}\, 
\Xi^{\mu_1 \mu_2} \cdots \Xi^{\mu_{2q-1} \mu_{2q}}
\Xi_{\left( \mu_1 \mu_2\right.} \cdots \Xi_{\left. \mu_{2q-1} \mu_{2q}\right)}\;, \label{contraction_Xi_2}
\end{eqnarray}
where in the next-to-last step we used the fact that only those permutations in the sum are non-vanishing where
the indices $\mu_{2q+1}, \ldots, \mu_{2q+r}$ are on $l$'s and $\mu_{2q+r+1}, \ldots \mu_{2n}$ are on $u$'s.
Then, we exploited $u^\mu u_\mu=1$ and $l^\mu l_\mu =-1$. We now prove by complete induction that
\begin{equation} \label{Xi_Xi}
\Xi^{\mu_1 \mu_2} \cdots \Xi^{\mu_{2q-1} \mu_{2q}}
\Xi_{\left( \mu_1 \mu_2\right.} \cdots \Xi_{\left. \mu_{2q-1} \mu_{2q}\right)} = \frac{(2^q q!)^2}{(2q)!}\;.
\end{equation}
This holds obviously for $q=1$, since $\Xi^{\mu_1\mu_2} \Xi_{\mu_1\mu_2} = \Xi^{\mu_1}_{\mu_1} \equiv 2 \equiv 2^2/2$.
We now assume that Eq.\ (\ref{Xi_Xi}) holds for $q$ and prove it for $q+1$:
\begin{eqnarray}
\lefteqn{\Xi^{\mu_1 \mu_2} \cdots \Xi^{\mu_{2q+1} \mu_{2q+2}}
\Xi_{\left( \mu_1 \mu_2\right.} \cdots \Xi_{\left. \mu_{2q+1} \mu_{2q+2}\right)} } \nonumber \\
& = & \frac{2^{q+1}(q+1)!}{(2q+2)!}\,\Xi^{\mu_1 \mu_2} \cdots \Xi^{\mu_{2q+1} \mu_{2q+2}}
\sum_{\mathcal{P}_\mu^{2q+2}} \Xi_{ \mu_1 \mu_2} \cdots \Xi_{ \mu_{2q+1} \mu_{2q+2}} \;.
\end{eqnarray}
Consider the contraction of $\Xi^{\mu_{2q+1} \mu_{2q+2}}$ with the sum over distinct permutations
of $2q+2$ indices $\mu_1, \ldots, \mu_{2q+2}$.
There is one term in the sum where both indices are on the same $\Xi$ projector. This
term is $\sim \Xi^{\mu_{2q+1} \mu_{2q+2}} \Xi_{\mu_{2q+1} \mu_{2q+2}} \equiv 2$. Then, there are
$2q$ terms where the indices $\mu_{2q+1}$ and $\mu_{2q+2}$ are on different projectors, say
$\Xi_{\mu_{2q+1} \mu_j} \Xi_{\mu_i \mu_{2q+2}}$. Contracting with $\Xi^{\mu_{2q+1} \mu_{2q+2}}$
gives a term $\sim \Xi_{\mu_i \mu_j}$, where both indices are from the set $\mu_1, \ldots, \mu_{2q}$. Putting
this together and using Eq.\ (\ref{Xi_Xi}) gives
\begin{eqnarray}
\lefteqn{\Xi^{\mu_1 \mu_2} \cdots \Xi^{\mu_{2q+1} \mu_{2q+2}}
\Xi_{\left( \mu_1 \mu_2\right.} \cdots \Xi_{\left. \mu_{2q+1} \mu_{2q+2}\right)} } \nonumber \\
& = & \frac{2(q+1)}{(2q+2)(2q+1)}\, \frac{2^q q!}{(2q)!}\,\Xi^{\mu_1 \mu_2} \cdots \Xi^{\mu_{2q-1} \mu_{2q}}
(2q+2) \sum_{\mathcal{P}_\mu^{2q}} \Xi_{ \mu_1 \mu_2} \cdots \Xi_{ \mu_{2q-1} \mu_{2q}} \nonumber \\
& = & \frac{2q+2}{2q+1}\, \Xi^{\mu_1 \mu_2} \cdots \Xi^{\mu_{2q-1} \mu_{2q}} 
\Xi_{\left( \mu_1 \mu_2\right.} \cdots \Xi_{\left. \mu_{2q-1} \mu_{2q}\right)} \nonumber \\
& = & \frac{2(q+1)}{2q+1}\, \frac{(2^q q!)^2}{(2q)!} = 
\frac{[2(q+1)]^2}{(2q+2)(2q+1)} \, \frac{(2^q q!)^2}{(2q)!} = \frac{[2^{q+1} (q+1)!]^2}{(2q+2)!}\;,
\end{eqnarray}
which is Eq.\ (\ref{Xi_Xi}) for $q+1$. Now we insert Eq.\ (\ref{Xi_Xi}) into Eq.\ (\ref{contraction_Xi_2}) and
obtain
\begin{eqnarray}
\lefteqn{\Xi^{\left( \mu_1 \mu_2\right.} \cdots \Xi^{\mu_{2q-1} \mu_{2q}} l^{\mu_{2q+1}} \cdots
l^{ \mu_{2q+r}} u^{\mu_{2q+r+1}} \cdots u^{\left. \mu_n \right) }
\Xi_{\left( \mu_1 \mu_2\right.} \cdots \Xi_{\mu_{2q-1} \mu_{2q}} l_{\mu_{2q+1}} \cdots
l_{ \mu_{2q+r}} u_{\mu_{2q+r+1}} \cdots u_{\left. \mu_n \right) } } \nonumber \\
& =&  \frac{1}{b_{nrq}}\, (-1)^r \frac{(2q)!}{2^q q!}\,  \frac{(2^q q!)^2}{(2q)!}
= \frac{1}{b_{nrq}}\, (-1)^r 2^q q!\;. \hspace*{8cm}
\end{eqnarray}
With $(2q)!! = 2^q q!$, we obtain Eq.\ (\ref{contraction_Xi}), q.e.d.

The thermodynamic integrals $\hat{I}_{i+j+n,j+r,q}$ are now obtained by a projection
of Eq.\ (\ref{I_nrq_moment}). The orthogonality relation (\ref{contraction_Xi}) leads to the result
\begin{eqnarray}
\hat{I}_{i+j+n,j+r,q} &\equiv &\frac{\left( -1\right) ^{q+r}}{\left(
2q\right) !!}\hat{\mathcal{I}}_{ij}^{\mu _{1}\cdots \mu _{n}}\Xi _{\left(
\mu _{1}\mu _{2}\right. }\cdots \Xi _{\mu _{2q-1}\mu _{2q}}l_{\mu
_{2q+1}}\cdots l_{\mu _{2q+r}}u_{\mu _{2q+r+1}}\cdots u_{\left. \mu
_{n}\right) }  \notag \\
&=&\frac{\left( -1\right) ^{q}}{\left( 2q\right) !!}\int dKE_{\mathbf{k}u}^{i+n-r-2q}
E_{\mathbf{k}l}^{j+r}\left( \Xi ^{\mu \nu }k_{\mu }k_{\nu
}\right) ^{q}\hat{f}_{0\mathbf{k}}\;,  \label{I_nrq_appendix}
\end{eqnarray}
where we used the definition (\ref{I_ij_tens}) of the tensor $\hat{\mathcal{I}}_{ij}^{\mu _{1}\cdots \mu _{n}}$.
With the definition of the average $\langle \ldots \rangle_{\hat{0}}$, the second line yields Eq.\ (\ref{I_nrq}).

Note that since $E_{\mathbf{k}u}^{i}=\left( k^{\mu}u_{\mu }\right)^{i}$ 
and $E_{\mathbf{k}l}^{j}=\left( -k^{\mu }l_{\mu}\right)^{j}$ the tensors (\ref{I_n_tens}), (\ref{I_ij_tens}) 
immediately follow from the projection of higher-rank tensors
on tensors built from $u$'s and $l$'s,
\begin{eqnarray}
\mathcal{I}_{i}^{\mu _{1}\cdots \mu _{n}} &=&u_{\left( \mu _{1}\right.}\cdots u_{\left. \mu _{i}\right) }
\mathcal{I}_{0}^{\left( \mu _{1}\cdots\mu _{n+i}\right) }\; ,  \label{I_i_tensor_appendix_1} \\
\hat{\mathcal{I}}_{ij}^{\mu _{1}\cdots \mu _{n}} &=&\left( -1\right)^{j} u_{\left( \alpha _{1}\right. }\cdots u_{\alpha _{i}}
l_{\beta _{1}}\cdots l_{\left. \beta _{j}\right) }\hat{\mathcal{I}}_{00}^{\left( \alpha_{1}\cdots \alpha _{i}\beta _{1}\cdots 
\beta _{j}\mu _{1}\cdots \mu_{n}\right) }\;.  \label{I_ij_tensor_appendix_1}
\end{eqnarray}

Other useful relations are obtained from contracting two indices of the tensors (\ref{I_n_tens}), (\ref{I_ij_tens})
with a $\Delta$ resp.\ a $\Xi$ projector,
\begin{eqnarray}
\mathcal{I}_{i}^{\mu _{1}\cdots \mu _{n}}\Delta _{\mu _{n-1}\mu _{n}}
&=&m_{0}^{2}\, \mathcal{I}_{i}^{\mu _{1}\cdots \mu _{n-2}}-\mathcal{I}_{i+2}^{\mu _{1}\cdots \mu _{n-2}}\; ,  
\label{I_i_tensor_appendix_2} \\
\hat{\mathcal{I}}_{ij}^{\mu _{1}\cdots \mu _{n}}\Xi _{\mu _{n-1}\mu _{n}}
&=&m_{0}^{2}\, \hat{\mathcal{I}}_{ij}^{\mu _{1}\cdots \mu _{n-2}}-\hat{\mathcal{I}}_{i+2,j}^{\mu _{1}\cdots \mu _{n-2}}
+\hat{\mathcal{I}}_{i,j+2}^{\mu_{1}\cdots \mu _{n-2}}\;.  \label{I_ij_tensor_appendix_2}
\end{eqnarray}

The conventional and the generalized thermodynamic integrals (\ref{I_nq}), (\ref{J_nq}), (\ref{I_nrq}), and (\ref{J_nrq})
obey useful recursion relations, which are given here. 
Replacing $\left( \Delta ^{\alpha \beta }k_{\alpha}k_{\beta }\right)^{q+1}
=\left( \Delta ^{\alpha \beta }k_{\alpha }k_{\beta}\right) ^{q}\left( m_{0}^{2}-E_{\mathbf{k}u}^{2}\right)$ 
in Eq.\ (\ref{I_nq}) we obtain for $0\leq q\leq n/2$,
\begin{eqnarray}
I_{n+2,q} &=&m_{0}^{2}\,I_{nq}+\left( 2q+3\right) I_{n+2,q+1}\;, \label{E22}\\
J_{n+2,q} &=&m_{0}^{2}\,J_{nq}+\left( 2q+3\right) J_{n+2,q+1}\; .
\end{eqnarray}%
Correspondingly, using $ \Xi ^{\mu \nu }k_{\mu }k_{\nu }
=m_{0}^{2}-E_{\mathbf{k}u}^{2}+E_{\mathbf{k}l}^{2}$ in Eq.\ (\ref{I_nrq}) we get
\begin{eqnarray}
\hat{I}_{n+2,r,q}-\hat{I}_{n+2,r+2,q} &=&m_{0}^{2}\,\hat{I}_{nrq}+\left(2q+2\right) \hat{I}_{n+2,r,q+1}\;, \label{E24}\\
\hat{J}_{n+2,r,q}-\hat{J}_{n+2,r+2,q} &=&m_{0}^{2}\,\hat{J}_{nrq}+\left(2q+2\right) \hat{J}_{n+2,r,q+1}\;.
\end{eqnarray}%
For $n=r=q=0$ Eqs.\ (\ref{E22}), (\ref{E24}) read
\begin{eqnarray}
I_{20} &=&m_{0}^{2}\,I_{00}+3I_{21}\;, \\
\hat{I}_{200} &=&m_{0}^{2}\,\hat{I}_{000}+\hat{I}_{220}+2\hat{I}_{201}\;.
\end{eqnarray}%
In the massless limit this leads to the familiar
relations $e_{0}=3P_{0}$ and $\hat{e}=\hat{P}_{l}+2\hat{P}_{\perp }$.

\section{Thermodynamic integrals in the equilibrium limit}
\label{appendix_thermo_integrals_equilibrium}

In this appendix we derive some properties of the
generalized moments (\ref{I_ij_tens}) and the corresponding
generalized thermodynamic integrals (\ref{I_nrq}) in the limit of local
thermodynamic equilibrium, see Eq.\ (\ref{I_ij_moment_equilibrium}). 

The generalized moments (\ref{I_ij_moment_equilibrium}) can also be expanded in terms of
the four-vectors $u^{\mu }$, $l^{\mu }$, and $\Xi ^{\mu \nu }$ , just as in
Eq.\ (\ref{I_nrq_moment}), where the corresponding thermodynamic integrals are defined
similarly to Eq.\ (\ref{I_nrq}),
\begin{equation}
I_{nrq}=\frac{\left( -1\right) ^{q}}{\left( 2q\right) !!}\int dKE_{\mathbf{k}%
u}^{n-r-2q}E_{\mathbf{k}l}^{r}\left( \Xi ^{\mu \nu }k_{\mu }k_{\nu }\right)
^{q}f_{0\mathbf{k}}\;.  \label{I_nrq_equilibirum}
\end{equation}
Making use of Eq.\ (\ref{Xi_munu}), of the binomial theorem, and of the definition of the
double factorial for even arguments it is
straightforward to obtain a relation between the thermodynamical integrals 
$I_{nq}$ and $I_{nrq}$,
\begin{eqnarray}
I_{nq} & = &\frac{\left( -1\right) ^{q}}{\left( 2q+1\right) !!}\int dK\, 
E_{\mathbf{k}u}^{n-2q}\left( \Xi ^{\mu \nu }k_{\mu }k_{\nu }
-E_{\mathbf{k}l}^{2}\right)^{q}f_{0\mathbf{k}}  \notag \\
&=&\frac{1}{\left( 2q+1\right) !!} \sum_{r=0}^q \frac{2^{q-r} q!}{r!}
\, I_{n,2r,q-r} \;.  \label{IMPORTANT}
\end{eqnarray}%
E.g.\ for $q=0,1,2$ we have
\begin{eqnarray}
I_{n0} &=&I_{n00}\;, \\
I_{n1} &=&\frac{1}{3}\left( 2I_{n01}+I_{n20}\right) \;, \\
I_{n2} &=&\frac{1}{15}\left( 8I_{n02}+4I_{n21}+I_{n40}\right) \;.
\end{eqnarray}%
Note that the corresponding auxiliary thermodynamical integrals $J_{nrq}$ may
be defined similarly to Eq.\ (\ref{J_nrq}), and obviously will lead to
analogous relations.

Furthermore, using 
$d f_{0\mathbf{k}}/ d E_{\mathbf{k}u} = - \beta_0 f_{0\mathbf{k}}(1-af_{0\mathbf{k}})$,
after an integration by parts Eq.\ (\ref{J_nq}) can be rewritten as a relation
between the conventional thermodynamic and auxiliary integrals,
\begin{equation}
\beta _{0}J_{nq}=I_{n-1,q-1}+\left( n-2q\right) I_{n-1,q}\;.
\label{thermo_int_rel_iso}
\end{equation}
Similarly, for the auxiliary thermodynamical integrals $J_{nrq}$ we obtain (as long as $r \geq 2$, $q\geq1$)
\begin{eqnarray}
\beta _{0}J_{nrq} &\equiv &I_{n-1,r,q-1}+\left( n-r-2q\right) I_{n-1,r,q}
\label{thermo_int_rel_aniso} \\
&=&\left( r-1\right) I_{n-1,r-2,q}+\left( n-r-2q\right) I_{n-1,r,q}\; .  \notag
\end{eqnarray}%
Comparing the right-hand sides, we obtain the identity
\begin{equation}
I_{n-1,r,q-1}=\left( r-1\right) I_{n-1,r-2,q}\;.  \label{VERY_IMPORTANT}
\end{equation}%
E.g., for $n=3$, $r=2$, and $q=1$ we obtain the equivalence of the longitudinal and
transverse pressures in thermodynamical equilibrium, 
\begin{equation}
I_{220}=I_{201}\; .
\end{equation}

The main thermodynamic relations are also obtained from integration by parts, namely
\begin{align}
dI_{nq}\left( \alpha _{0},\beta _{0}\right) & \equiv \left( \frac{\partial
I_{nq}}{\partial \alpha _{0}}\right)_{\beta_0} d\alpha _{0}+\left( \frac{\partial
I_{nq}}{\partial \beta _{0}}\right)_{\alpha _{0}} d\beta _{0}  \notag \\
& =J_{nq}d\alpha _{0}-J_{n+1,q}d\beta _{0}\;,  \label{d_Inq}
\end{align}%
and similarly
\begin{align}
dI_{nrq}\left( \alpha _{0},\beta _{0}\right) & \equiv \left( \frac{\partial
I_{nrq}}{\partial \alpha _{0}}\right)_{\beta_0} d\alpha _{0}+\left( \frac{\partial
I_{nrq}}{\partial \beta _{0}}\right)_{\alpha_0} d\beta _{0}  \notag \\
& =J_{nrq}d\alpha _{0}-J_{n+1,r,q}d\beta _{0}\;.  \label{d_Inrq}
\end{align}

\section{Irreducible projection operators}
\label{appendix_projections}

In this appendix, we present the irreducible projection operators necessary to
derive the irreducible moments of $\delta f_{\bf k}$ or $\delta \hat{f}_{\bf k}$.
We start by recalling the definition of the irreducible projection operators 
in the first case \cite{deGroot,Denicol:2012cn,Molnar:2013lta},
\begin{equation}
\Delta^{\mu _{1}\cdots \mu _{n}}_{\nu _{1}\cdots \nu _{n}}
=\sum_{q=0}^{[n/2]}  C(n,q)  \, \frac{1}{\mathcal{N}_{nq}} 
\sum_{\mathcal{P}_\mu^n \mathcal{P}_\nu^n} 
\Delta^{\mu _{1}\mu _{2}}\cdots \Delta^{\mu_{2q-1}\mu _{2q}}  
\Delta_{\nu_{1}\nu _{2}}\cdots \Delta_{\nu _{2q-1}\nu_{2q}}
\Delta^{\mu_{2q+1}}_{\nu _{2q+1}} \cdots \Delta^{\mu_{n}}_{\nu _{n}}\;.
\label{Deltaproj}
\end{equation}
Here, $[n/2]$ denotes the largest integer less than or equal to $n/2$, the coefficients
$C(n,q)$ are defined as
\begin{equation} \label{Ccoeff}
C(n,q) = (-1)^q \frac{(n!)^2}{(2n)!} \, \frac{(2n-2q)!}{q! (n-q)! (n-2q)!}\;,
\end{equation}
and the second sum in Eq.\ (\ref{Deltaproj}) runs over all {\em distinct\/} permutations 
$\mathcal{P}_\mu^n \mathcal{P}_\nu^n$ of $\mu$- and $\nu$-type indices.
The coefficient in front of this sum is just the inverse of the total number of these distinct permutations,
\begin{equation}
\mathcal{N}_{nq} \equiv \frac{1}{(n-2q)!}\, \left( \frac{n!}{2^q q!} \right)^2\;.
\end{equation} 
This number can be explained as follows: 
$(n!)^2$ is the number of {\em all\/} permutations of $\mu$- and $\nu$-type indices. In order to obtain
the number of {\em distinct\/} permutations, one has to divide this by the number $(2^q)^2$ of permutations of 
$\mu$- and $\nu$-type indices on the same $\Delta$ projectors (where only projectors with only $\mu$- and
only $\nu$-type indices are considered), and by the number $(q!)^2$ of 
trivial reorderings of the sequence of these projectors. Finally, one also has to divide by the number $(n-2q)!$
of trivial reorderings of the sequence of projectors with mixed indices.

The projectors (\ref{Deltaproj}) are symmetric under exchange of $\mu$- and $\nu$-type indices,
\begin{equation}
\Delta^{\mu _{1}\cdots \mu_{n}}_{\nu _{1}\cdots \nu _{n}}
=\Delta^{\left( \mu _{1}\cdots \mu _{n}\right)}_{\left( \nu _{1}\cdots \nu _{n}\right) }\;,
\end{equation}
and traceless with respect to contraction of either $\mu$- or $\nu$-type indices,
\begin{equation}
\Delta^{\mu _{1}\cdots \mu _{n}}_{\nu _{1}\cdots \nu _{n}}g_{\mu _{i}\mu _{j}}
=\Delta^{\mu _{1}\cdots \mu _{n}}_{\nu _{1}\cdots \nu _{n}}g^{\nu _{i}\nu_{j}}=0\;\;\; \mbox{for any}\; i,j \;. 
\label{Delta_complete_traceless} 
\end{equation}
Moreover, upon complete contraction,
\begin{equation}
\Delta_{\mu _{1}\cdots \mu _{n}}^{\mu _{1}\cdots \mu _{n}} \equiv 
\Delta^{\mu _{1}\cdots \mu _{n}\nu _{1}\cdots \nu _{n}}g_{\mu _{1}\nu _{1}}\cdots
g_{\mu _{n}\nu _{n}}=2n+1\;,  \label{Delta_complete_contraction} 
\end{equation}
cf.\ Eq.\ (23) in chapter VI.2 of Ref.\ \cite{deGroot}.

Analogously, in the case where we decompose
tensors with respect to both $u^\mu$ and $l^\mu$ the irreducible projection operators read
\begin{equation}
\Xi^{\mu _{1}\cdots \mu _{n}}_{\nu _{1}\cdots \nu _{n}}
=\sum_{q=0}^{[n/2]} \hat{C}(n,q)  \, \frac{1}{\mathcal{N}_{nq}} 
\sum_{\mathcal{P}_\mu^n \mathcal{P}_\nu^n} 
\Xi^{\mu _{1}\mu _{2}}\cdots \Xi^{\mu _{2q-1}\mu_{2q}}  \Xi_{\nu _{1}\nu _{2}}\cdots \Xi_{\nu _{2q-1}\nu _{2q}}
\Xi^{\mu _{2q+1}}_{\nu _{2q+1}}\cdots \Xi^{\mu _{n}}_{\nu _{n}}\;.  \label{Xiproj}
\end{equation}
The coefficient in front of the sum over distinct permutations is the same as in Eq.\ (\ref{Deltaproj}). The
coefficient $\hat{C}(n,q)$ will be determined below.
Just as the projectors (\ref{Deltaproj}), the projectors (\ref{Xiproj}) are also symmetric under the interchange of
$\mu $- and $\nu $-type indices, 
\begin{equation}
\Xi^{\mu _{1}\cdots \mu_{n}}_{\nu _{1}\cdots \nu _{n}}
=\Xi^{\left( \mu _{1}\cdots \mu _{n}\right)}_{\left( \nu _{1}\cdots \nu _{n}\right) }\;,
\end{equation}
and traceless with respect to contraction of either $\mu$- or $\nu$-type indices,
\begin{equation} \label{Xi_complete_traceless}
\Xi^{\mu _{1}\cdots \mu _{n}}_{\nu _{1}\cdots \nu _{n}}g_{\mu _{i}\mu _{j}}
=\Xi^{\mu _{1}\cdots \mu _{n}}_{\nu _{1}\cdots \nu _{n}}g^{\nu _{i}\nu_{j}}=0\;\;\; \mbox{for any}\; i,j \;.
\end{equation}
This relation forms the basis for the determination of the coefficients $\hat{C}(n,q)$, as we shall show now. 
[The calculation closely follows that given in Ref.\ \cite{deGroot} for the coefficients $C(n,q)$, cf.\ Eq.\ (\ref{Ccoeff}).]

Without loss of generality,
let us consider the contraction (\ref{Xi_complete_traceless})
of the projector (\ref{Xiproj}) with respect to the two indices $\mu_1,\, \mu_2$.
For the following arguments, it is advantageous to replace the sum over distinct permutations 
$\mathcal{P}_\mu^n \mathcal{P}_\nu^n$ of the $n$ $\mu$-- and $\nu$--type indices by the sum over
{\em all\/} permutations $\bar{\mathcal{P}}_\mu^n \bar{\mathcal{P}}_\nu^n$ (with in total
$(n!)^2$ different terms). In the various terms of this sum, the
indices $\mu_1,\,\mu_2$ can appear in four different ways (for the sake of notational simplicity, we omit the other 
$\Xi$ projectors in these terms):
\begin{align*}
{\rm (i)} \hspace*{2cm} & \Xi^{\mu_1 \mu_i} \Xi^{\mu_j \mu_2}  & 
\stackrel{\displaystyle \times g_{\mu_1\mu_2}}{\xrightarrow{\hspace*{1.5cm}}}
& \hspace*{1cm} \Xi^{\mu_i \mu_j} & 2q(2q-2)\; \mbox{terms}\;, \\
{\rm (ii)} \hspace*{2cm} & \Xi^{\mu_1 \mu_2} & 
{\xrightarrow{\hspace*{1.5cm}}}
& \hspace*{1cm} 2 & 2q\; \mbox{terms}\; , \\
{\rm (iii)} \hspace*{2cm} & \Xi^{\mu_1 \mu_i} \Xi^{\mu_2}_{\nu_j} & 
{\xrightarrow{\hspace*{1.5cm}}}
& \hspace*{1cm} \Xi^{\mu_i}_{\nu_j} & 2 \cdot 2q(n-2q)\; \mbox{terms}\;, \\
{\rm (iv)} \hspace*{2cm} & \Xi^{\mu_1}_{\nu_i} \Xi^{\mu_2}_{\nu_j} & 
{\xrightarrow{\hspace*{1.5cm}}}
& \hspace*{1cm} \Xi_{\nu_i\nu_j} & (n-2q)(n-2q-1)\; \mbox{terms}\; .
\end{align*}
The arrow symbolizes contraction with $g_{\mu_1 \mu_2}$, with the corresponding result shown
to the right of the arrow.
On the far right we also denoted the number of times that such terms occur in the sum over
all permutations. Computing this number is a simple combinatorial exercise:
for case (i), the index $\mu_1$ can appear in $2q$ different positions and the index
$\mu_2$ in the remaining $2q-2$ positions. For case (ii), there are $2q$ different positions for the index $\mu_1$, but
then the position of $\mu_2$ is fixed. For case (iii), there are $2q$ positions for $\mu_1$ and $n-2q$ positions
for $\mu_2$. Interchanging $\mu_1 \leftrightarrow \mu_2$ gives another factor of 2. Finally, for case (iv) there
are $n-2q$ positions for $\mu_1$ and $n-2q-1$ remaining positions for $\mu_2$.

The cases (i) -- (iii) generate terms of the form
\begin{equation}
\Xi^{\mu _{3}\mu _{4}}\cdots \Xi^{\mu _{2q-1}\mu_{2q}}  \Xi_{\nu _{1}\nu _{2}}\cdots \Xi_{\nu _{2q-1}\nu _{2q}}
\Xi^{\mu _{2q+1}}_{\nu _{2q+1}}\cdots \Xi^{\mu _{n}}_{\nu _{n}}\;,
\end{equation}
while case (iv) generates terms of the form
\begin{equation}
\Xi^{\mu _{3}\mu _{4}}\cdots \Xi^{\mu _{2q+1}\mu_{2q+2}}  \Xi_{\nu _{1}\nu _{2}}\cdots \Xi_{\nu _{2q+1}\nu _{2q+2}}
\Xi^{\mu _{2q+3}}_{\nu _{2q+3}}\cdots \Xi^{\mu _{n}}_{\nu _{n}}\;.
\end{equation}
Note that, because of the contraction of $\mu_1, \mu_2$, only $n-2$ indices of the $\mu$--type are to be
permutated.
Collecting all terms (i) -- (iv) with the correct prefactors, Eq.\ (\ref{Xi_complete_traceless}) reads
\begin{eqnarray}
0 & = & \sum_{q=0}^{[n/2]} \hat{C}(n,q)\, \frac{1}{(n!)^2} \sum_{\bar{\mathcal{P}}_\mu^{n-2} \bar{\mathcal{P}}_\nu^n}
\left\{ 2q \left[ 2q-2 +2 + 2(n-2q) \right] 
\Xi^{\mu _{1}\mu _{2}}\cdots \Xi^{\mu _{2q-3}\mu_{2q-2}}  \Xi_{\nu _{1}\nu _{2}}\cdots \Xi_{\nu _{2q-1}\nu _{2q}}
\Xi^{\mu _{2q-1}}_{\nu _{2q+1}}\cdots \Xi^{\mu _{n-2}}_{\nu _{n}} \right. \nonumber \\
&    & \hspace*{3.7cm}+ \left. (n-2q) (n-2q-1) 
\Xi^{\mu _{1}\mu _{2}}\cdots \Xi^{\mu _{2q-1}\mu_{2q}}  \Xi_{\nu _{1}\nu _{2}}\cdots \Xi_{\nu _{2q+1}\nu _{2q+2}}
\Xi^{\mu _{2q+1}}_{\nu _{2q+3}} \cdots \Xi^{\mu _{n-2}}_{\nu _{n}} \right\}\;,
\end{eqnarray}
where we relabelled the $\mu$--type indices $\mu_i \rightarrow \mu_{i-2}$. We now observe that the $q=0$--term
does not contribute to the first term in curly brackets, while the $q=[n/2]$--term does not contribute to the last term 
(no matter whether $n$ is even or odd). Thus, we may write
\begin{eqnarray}
0 & = &  \frac{1}{(n!)^2} \sum_{\bar{\mathcal{P}}_\mu^{n-2} \bar{\mathcal{P}}_\nu^n}
\left\{ \sum_{q=1}^{[n/2]} \hat{C}(n,q)\, 2q \left(2n- 2q \right) 
\Xi^{\mu _{1}\mu _{2}}\cdots \Xi^{\mu _{2q-3}\mu_{2q-2}}  \Xi_{\nu _{1}\nu _{2}}\cdots \Xi_{\nu _{2q-1}\nu _{2q}}
\Xi^{\mu _{2q-1}}_{\nu _{2q+1}}\cdots \Xi^{\mu _{n-2}}_{\nu _{n}} \right. \nonumber \\
&    & \hspace*{1cm}+ \left. \sum_{q=0}^{[n/2]-1} \hat{C}(n,q)\,(n-2q) (n-2q-1) 
\Xi^{\mu _{1}\mu _{2}}\cdots \Xi^{\mu _{2q-1}\mu_{2q}}  \Xi_{\nu _{1}\nu _{2}}\cdots \Xi_{\nu _{2q+1}\nu _{2q+2}}
\Xi^{\mu _{2q+1}}_{\nu _{2q+3}} \cdots \Xi^{\mu _{n-2}}_{\nu _{n}} \right\}\;.
\end{eqnarray}
Substituting $q \rightarrow q-1$ in the second sum, we observe that the product of the $\Xi$--projectors is actually
identical in both sums, so that we obtain
\begin{eqnarray}
0 & = &  \sum_{q=1}^{[n/2]} \left[ \hat{C}(n,q)\, 2q \left(2n- 2q \right) + \hat{C}(n,q-1)\,(n-2q+2) (n-2q+1) \right]
\nonumber \\
&   & \times 
\frac{1}{(n!)^2} \sum_{\bar{\mathcal{P}}_\mu^{n-2} \bar{\mathcal{P}}_\nu^n}\Xi^{\mu _{1}\mu _{2}}\cdots 
\Xi^{\mu _{2q-3}\mu_{2q-2}}  \Xi_{\nu _{1}\nu _{2}}\cdots \Xi_{\nu _{2q-1}\nu _{2q}}
\Xi^{\mu _{2q-1}}_{\nu _{2q+1}}\cdots \Xi^{\mu _{n-2}}_{\nu _{n}}\;.
\end{eqnarray}
In order to fulfill this relation, we have to demand that the term in brackets vanishes, which leads to the
recursion relation
\begin{equation}
\hat{C}(n,q) = - \frac{(n-2q+2)(n-2q+1)}{2q (2n-2q)}\, \hat{C}(n,q-1)\;.
\end{equation}
If we set 
\begin{equation}
\hat{C}(n,0) \equiv 1\;\;\; {\rm  for all}\;\, n\;,
\end{equation} 
the solution is
\begin{equation} \label{Chatcoeff}
\hat{C}(n,q) = (-1)^q \, \frac{1}{2^q q!}\, \frac{n!}{(n-2q)!} \frac{(2n-2q-2)!!}{(2n-2)!!}
= (-1)^q \frac{1}{4^q}\, \frac{(n-q)!}{q!(n-2q)!}\, \frac{n}{n-q}\;,
\end{equation}
where we used the definition of the double factorial for even numbers.

Upon complete contraction,
\begin{equation}\label{Xi_complete_contraction}
\Xi_{\mu_{1}\cdots \mu _{n}}^{\mu _{1}\cdots \mu _{n}} \equiv \Xi ^{\mu_{1}\cdots \mu _{n}}_{\nu _{1}\cdots \nu _{n}}
g_{\mu_1}^{\nu_{1}}\cdots g_{\mu_n}^{\nu _{n}}=2\;.
\end{equation}
In order to prove this relation, it is advantageous to first prove the recursion relation
\begin{equation} \label{recursionXi}
\Xi_{\nu_{1}\cdots \nu_{n}}^{\mu _{1}\cdots \mu_{n}} g_{\mu_i}^{\nu_j} =
\Xi_{\nu_{1}\cdots \nu_{n-1}}^{\mu _{1}\cdots \mu_{n-1}}\;,
\end{equation}
valid for any $i,j$ and $n>1$. Equation (\ref{Xi_complete_contraction}) then immediately follows
on account of $\Xi^{\mu_1 \cdots \mu_n}_{\mu_1 \cdots \mu_n} = \Xi^{\mu_1}_{\mu_1} = 2$.

Equation (\ref{recursionXi}) is proved as follows. First note that, since 
$\Xi_{\nu_{1}\cdots \nu_{n}}^{\mu _{1}\cdots \mu_{n}}$ is symmetric in all indices, we may without
loss of generality choose $i=j=n$ in Eq.\ (\ref{recursionXi}). 
Then, as already done in the derivation of Eq.\ (\ref{Chatcoeff}), it is advantageous to 
replace in Eq.\ (\ref{Xiproj}) the sum over distinct permutations $\mathcal{P}_\mu^n \mathcal{P}_\nu^n$  
by the sum over all permutations $\bar{\mathcal{P}}_\mu^n \bar{\mathcal{P}}_\nu^n$ 
(with in total $(n!)^2$ different terms). Inserting this into the left-hand side of Eq.\ (\ref{recursionXi}), 
we see that contraction of the indices $\mu_n,\nu_n$ generates five different types of terms:
\begin{align*}
{\rm (i)} \hspace*{2cm} & \Xi^{\mu_i \mu_n} \Xi_{\nu_j \nu_n} & 
\stackrel{\displaystyle \times g_{\mu_n}^{\nu_n}}{\xrightarrow{\hspace*{1.5cm}}}
& \hspace*{1cm} \Xi^{\mu_i }_{\nu_j} & 2q \cdot 2q\; \mbox{terms}\;, \\
{\rm (ii)} \hspace*{2cm} & \Xi^{\mu_i \mu_n}  \Xi^{\mu_j}_{\nu_n}& 
{\xrightarrow{\hspace*{1.5cm}}}
& \hspace*{1cm} \Xi^{\mu_i \mu_j} & 2q (n-2q)\; \mbox{terms}\; , \\
{\rm (iii)} \hspace*{2cm} & \Xi_{\nu_i \nu_n} \Xi^{\mu_n}_{\nu_j} & 
{\xrightarrow{\hspace*{1.5cm}}}
& \hspace*{1cm} \Xi_{\nu_i \nu_j} &  2q(n-2q)\; \mbox{terms}\;, \\
{\rm (iv)} \hspace*{2cm} & \Xi^{\mu_n}_{\nu_j} \Xi^{\mu_i}_{\nu_n} & 
{\xrightarrow{\hspace*{1.5cm}}}
& \hspace*{1cm} \Xi^{\mu_i}_{\nu_j} & (n-2q)(n-2q-1)\; \mbox{terms}\;, \\
{\rm (v)} \hspace*{2cm} & \Xi^{\mu_n}_{\nu_n} & 
{\xrightarrow{\hspace*{1.5cm}}}
& \hspace*{1cm} 2 & (n-2q)\; \mbox{terms}\; .
\end{align*}
The number of terms is easily explained as follows: in case (i), there are $2q$ possible positions for
the index $\mu_n$ and $2q$ possible positions for the index $\nu_n$. In cases (ii) and (iii), there are $2q$ resp.\
$n-2q$ positions for the index $\mu_n$ and $n-2q$ resp.\ $2q$ positions for the index $\nu_n$. In case (iv), there
are $n-2q$ positions for the index $\mu_n$ and a remaining $n-2q-1$ positions for the index $\nu_n$. Finally, in
case (v) there are $n-2q$ possibilities (equal to the number of projectors with mixed indices) to have
the indices $\mu_n$ and $\nu_n$ occurring at the same projector.
One observes that upon contraction, in case (i) (and after suitably relabelling indices) one generates terms of the form
\begin{displaymath}
\Xi^{\mu _{1}\mu _{2}}\cdots \Xi^{\mu _{2q-3}\mu_{2q-2}}  \Xi_{\nu _{1}\nu _{2}}\cdots \Xi_{\nu _{2q-3}\nu _{2q-2}}
\Xi^{\mu _{2q-1}}_{\nu _{2q-1}}\cdots \Xi^{\mu _{n-1}}_{\nu _{n-1}}\;,
\end{displaymath}
i.e., terms with $q-1$ projectors $\Xi^{\mu_i \mu_j}$, $q-1$ projectors $\Xi_{\nu_i \nu_j}$, and
$n-2q+1$ projectors $\Xi^{\mu_i}_{\nu_j}$. On the other hand,
in all other cases (ii) -- (v) one generates terms of the form
\begin{displaymath}
\Xi^{\mu _{1}\mu _{2}}\cdots \Xi^{\mu _{2q-1}\mu_{2q}}  \Xi_{\nu _{1}\nu _{2}}\cdots \Xi_{\nu _{2q-1}\nu _{2q}}
\Xi^{\mu _{2q+1}}_{\nu _{2q+1}}\cdots \Xi^{\mu _{n-1}}_{\nu _{n-1}}\;,
\end{displaymath}
i.e., terms with $q$ projectors $\Xi^{\mu_i \mu_j}$, $q$ projectors $\Xi_{\nu_i \nu_j}$, and
$n-2q-1$ projectors $\Xi^{\mu_i}_{\nu_j}$.

To proceed, it is advantageous to consider
the case of $n$ even and $n$ odd separately. Let us first focus on the (somewhat simpler) case of $n$ even,
where $[n/2] \equiv n/2$.
Collecting the results obtained so far, the left-hand-side of Eq.\ (\ref{recursionXi}) can be written as
\begin{eqnarray}
\Xi_{\nu_{1}\cdots \nu_{n}}^{\mu _{1}\cdots \mu_{n}} g_{\mu_n}^{\nu_n} & = & 
\sum_{q=0}^{n/2} \hat{C}(n,q)\, \frac{1}{(n!)^2}\!\!\! \sum_{\bar{\mathcal{P}}_\mu^{n-1} \bar{\mathcal{P}}_\nu^{n-1}}
\!\!\! \left[ (n-2q)(n+2q+1) \,
\Xi^{\mu _{1}\mu _{2}}\cdots \Xi^{\mu _{2q-1}\mu_{2q}}  \Xi_{\nu _{1}\nu _{2}}\cdots \Xi_{\nu _{2q-1}\nu _{2q}}
\Xi^{\mu _{2q+1}}_{\nu _{2q+1}}\cdots \Xi^{\mu _{n-1}}_{\nu _{n-1}} \right. \nonumber \\
&   & \hspace*{3.7cm} + \left. 
4q^2\, \Xi^{\mu _{1}\mu _{2}}\cdots \Xi^{\mu _{2q-3}\mu_{2q-2}}  \Xi_{\nu _{1}\nu _{2}}\cdots \Xi_{\nu _{2q-3}\nu _{2q-2}}
\Xi^{\mu _{2q-1}}_{\nu _{2q-1}}\cdots \Xi^{\mu _{n-1}}_{\nu _{n-1}} \right]\;.
\end{eqnarray}
We now note that the first term in brackets does not contribute for $q=n/2$, while the second term does not
contribute for $q=0$,
\begin{eqnarray}
\Xi_{\nu_{1}\cdots \nu_{n}}^{\mu _{1}\cdots \mu_{n}} g_{\mu_n}^{\nu_n} & = & 
\sum_{q=0}^{n/2-1} \hat{C}(n,q)\, (n-2q)(n+2q+1)  
\frac{1}{(n!)^2}\!\!\! \sum_{\bar{\mathcal{P}}_\mu^{n-1} \bar{\mathcal{P}}_\nu^{n-1}}\!\!\!
\Xi^{\mu _{1}\mu _{2}}\cdots \Xi^{\mu _{2q-1}\mu_{2q}}  \Xi_{\nu _{1}\nu _{2}}\cdots \Xi_{\nu _{2q-1}\nu _{2q}}
\Xi^{\mu _{2q+1}}_{\nu _{2q+1}}\cdots \Xi^{\mu _{n-1}}_{\nu _{n-1}} \nonumber \\
&  + &  \sum_{q=1}^{n/2} \hat{C}(n,q)\,  4q^2 
\frac{1}{(n!)^2}\!\!\! \sum_{\bar{\mathcal{P}}_\mu^{n-1} \bar{\mathcal{P}}_\nu^{n-1}}\!\!\!
\Xi^{\mu _{1}\mu _{2}}\cdots \Xi^{\mu _{2q-3}\mu_{2q-2}}  \Xi_{\nu _{1}\nu _{2}}\cdots \Xi_{\nu _{2q-3}\nu _{2q-2}}
\Xi^{\mu _{2q-1}}_{\nu _{2q-1}}\cdots \Xi^{\mu _{n-1}}_{\nu _{n-1}} \;.
\end{eqnarray}
Substituting the summation index $q \rightarrow q+1$ in the second sum, we observe that the product of
projectors becomes identical to the one in the first sum, so that we can write
\begin{eqnarray}
\Xi_{\nu_{1}\cdots \nu_{n}}^{\mu _{1}\cdots \mu_{n}} g_{\mu_n}^{\nu_n} & = & 
\sum_{q=0}^{n/2-1} \left[ \hat{C}(n,q)\, (n-2q)(n+2q+1) + \hat{C}(n,q+1)\,  4(q+1)^2 \right] \nonumber \\
&    & \times
\frac{1}{(n!)^2}\!\!\! \sum_{\bar{\mathcal{P}}_\mu^{n-1} \bar{\mathcal{P}}_\nu^{n-1}}\!\!\!
\Xi^{\mu _{1}\mu _{2}}\cdots \Xi^{\mu _{2q-1}\mu_{2q}}  \Xi_{\nu _{1}\nu _{2}}\cdots \Xi_{\nu _{2q-1}\nu _{2q}}
\Xi^{\mu _{2q+1}}_{\nu _{2q+1}}\cdots \Xi^{\mu _{n-1}}_{\nu _{n-1}}\;. \label{recursion_2}
\end{eqnarray}
With the definition (\ref{Chatcoeff}) one now convinces oneself that
\begin{equation} \label{recursion_3}
\frac{1}{n^2} \left[ \hat{C}(n,q)\, (n-2q)(n+2q+1) + \hat{C}(n,q+1)\,  4(q+1)^2 \right] = \hat{C}(n-1,q)\;.
\end{equation}
Reverting the sum over all permutations of $n-1$ indices of $\mu$-- and of $\nu$--type
to the one over distinct permutations and using the definition (\ref{Xiproj}) and the fact that for even $n$ one
has $n/2-1 = [(n-1)/2]$, one then arrives at Eq.\ (\ref{recursionXi}).

Let us now consider the case of $n$ odd, where $[n/2] = [(n-1)/2] = (n-1)/2$.
For $q < (n-1)/2$, the arguments of the previous case of $n$ even can be taken over unchanged. However,
when $q=(n-1)/2$ things become more subtle: in the cases (ii), (iii), and (iv), one observes that after contraction of
the indices, no further projectors of the type $\Xi^{\mu_i}_{\nu_j}$ occur. We thus treat the case 
$q=(n-1)/2$ separately. 
Collecting the results obtained so far, the left-hand-side of Eq.\ (\ref{recursionXi}) can be written as
\begin{eqnarray}
\Xi_{\nu_{1}\cdots \nu_{n}}^{\mu _{1}\cdots \mu_{n}} g_{\mu_n}^{\nu_n} & = & \!\!\!
\sum_{q=0}^{(n-1)/2-1} \!\!\! 
\hat{C}(n,q)\, \frac{1}{(n!)^2}\!\!\! \sum_{\bar{\mathcal{P}}_\mu^{n-1} \bar{\mathcal{P}}_\nu^{n-1}}
\!\!\! \left[ (n-2q)(n+2q+1) \,
\Xi^{\mu _{1}\mu _{2}}\cdots \Xi^{\mu _{2q-1}\mu_{2q}}  \Xi_{\nu _{1}\nu _{2}}\cdots \Xi_{\nu _{2q-1}\nu _{2q}}
\Xi^{\mu _{2q+1}}_{\nu _{2q+1}}\cdots \Xi^{\mu _{n-1}}_{\nu _{n-1}} \right. \nonumber \\
&   & \hspace*{4.3cm} + \left. 
4q^2\, \Xi^{\mu _{1}\mu _{2}}\cdots \Xi^{\mu _{2q-3}\mu_{2q-2}}  \Xi_{\nu _{1}\nu _{2}}\cdots \Xi_{\nu _{2q-3}\nu _{2q-2}}
\Xi^{\mu _{2q-1}}_{\nu _{2q-1}}\cdots \Xi^{\mu _{n-1}}_{\nu _{n-1}} \right] \nonumber \\
& + & \hat{C}\left(n, \frac{n-1}{2}\right)\frac{1}{(n!)^2}\!\!\! \sum_{\bar{\mathcal{P}}_\mu^{n-1} \bar{\mathcal{P}}_\nu^{n-1}}
\!\!\! \left[ 2n \,\Xi^{\mu _{1}\mu _{2}}\cdots \Xi^{\mu _{n-2}\mu_{n-1}}  
\Xi_{\nu _{1}\nu _{2}}\cdots \Xi_{\nu _{n-2}\nu _{n-1}} \right. \nonumber \\
&   & \hspace*{4.3cm} + \left. 
(n-1)^2\, \Xi^{\mu _{1}\mu _{2}}\cdots \Xi^{\mu _{n-4}\mu_{n-3}}  \Xi_{\nu _{1}\nu _{2}}\cdots \Xi_{\nu _{n-4}\nu _{n-3}}
\Xi^{\mu _{n-2}}_{\nu _{n-2}}\cdots \Xi^{\mu _{n-1}}_{\nu _{n-1}} \right] \;.
\end{eqnarray}
Noting that the term in the second line does not contribute for $q=0$, we may combine the terms in the second
and fourth line to a sum that runs from $q=1$ to $(n-1)/2$. Substituting $q \rightarrow q+1$ in that sum, we observe
that this sum can be combined with the one in the first line, similar to Eq.\ (\ref{recursion_2}). With Eq.\ (\ref{recursion_3})
this yields
\begin{eqnarray}
\Xi_{\nu_{1}\cdots \nu_{n}}^{\mu _{1}\cdots \mu_{n}} g_{\mu_n}^{\nu_n} & = & \!\!\!
\sum_{q=0}^{(n-1)/2-1} \!\!\! 
\hat{C}(n-1,q)\, \frac{1}{[(n-1)!]^2}\!\!\! \sum_{\bar{\mathcal{P}}_\mu^{n-1} \bar{\mathcal{P}}_\nu^{n-1}}
\!\!\! \Xi^{\mu _{1}\mu _{2}}\cdots \Xi^{\mu _{2q-1}\mu_{2q}}  \Xi_{\nu _{1}\nu _{2}}\cdots \Xi_{\nu _{2q-1}\nu _{2q}}
\Xi^{\mu _{2q+1}}_{\nu _{2q+1}}\cdots \Xi^{\mu _{n-1}}_{\nu _{n-1}}  \nonumber \\
& + & \hat{C}\left(n, \frac{n-1}{2}\right)\frac{2}{n}\,
\frac{1}{[(n-1)!]^2} \!\!\! \sum_{\bar{\mathcal{P}}_\mu^{n-1} \bar{\mathcal{P}}_\nu^{n-1}}
\!\!\!\Xi^{\mu _{1}\mu _{2}}\cdots \Xi^{\mu _{n-2}\mu_{n-1}}  
\Xi_{\nu _{1}\nu _{2}}\cdots \Xi_{\nu _{n-2}\nu _{n-1}}\;. \label{recursion_4}
\end{eqnarray}
With Eq.\ (\ref{Chatcoeff}) one proves that
\begin{equation}
\frac{2}{n}\, \hat{C}\left(n, \frac{n-1}{2}\right)  = \hat{C} \left(n-1, \frac{n-1}{2}\right)\;.
\end{equation}
Then, the last term in Eq.\ (\ref{recursion_4}) just represents the missing $q = (n-1)/2$--term of the sum and we again
obtain Eq.\ (\ref{recursionXi}), q.e.d.

Note that relations (\ref{Delta_complete_contraction}) and (\ref{Xi_complete_contraction})
mean that the projection of an arbitrary tensor of rank $n$ with respect to either
(\ref{Deltaproj}) or (\ref{Xiproj}), i.e.,
$A^{\nu _{1}\cdots \nu _{n}}\Delta _{\nu_{1}\cdots \nu _{n}}^{\mu _{1}\cdots \mu _{n}}$ 
or $A^{\nu _{1}\cdots \nu_{n}}\Xi _{\nu _{1}\cdots \nu _{n}}^{\mu _{1}\cdots \mu _{n}}$,
has $2n+1 $ or $2$ independent tensor components, respectively. 

In order to prove this, we note that an arbitrary tensor  
$A_{d}^{\mu _{1}\cdots \mu _{n}}$ of rank $n$ in $d$-dimensional space-time
has $d^n$ independent components, because each of the $n$ indices
can assume $d$ distinct values. Now consider a rank-$n$ tensor which is
completely symmetric with respect to the interchange of indices. This tensor
can be constructed from the arbitrary tensor $A_{d}^{\mu _{1}\cdots \mu _{n}}$
via symmetrization,
\begin{equation}
A_{d}^{\left( \mu _{1}\cdots \mu _{n}\right) }=\frac{1}{n!}\sum_{%
\mathcal{P}_{\mu}}A_{d}^{\mu _{1}\cdots \mu _{n}}\;,
\end{equation} 
where the sum over $\mathcal{P}_{\mu}$ runs over all 
$n!$ permutations of the $\mu$-type indices.
The number of independent tensor components of such a symmetric tensor is 
given by the number of combinations with repetition to draw $n$ elements from a 
set of $d$ elements, 
\begin{equation}
N_{dn}\left( A_{d}^{\left( \mu _{1}\cdots \mu _{n}\right) }\right) =\frac{%
\left( n+d-1\right) !}{n!\left( d-1\right) !}\;.  \label{Wolf_1}
\end{equation}
Let us now demand in addition that this tensor is traceless,
\begin{equation}
0 = A_{d,tr}^{\left( \mu _{1}\cdots \mu _{n}\right)}g_{\mu _{n-1}\mu _{n}}
\equiv A_d^{\left( \mu _{1}\cdots \mu _{n-2}\right)}\;,
\end{equation}
where the right-hand side defines a new symmetric tensor of rank $n-2 $. 
According to Eq.\ (\ref{Wolf_1}), this tensor has
\begin{equation} \label{Wolf_1a}
N_{dn}\left( A_{d}^{\left( \mu _{1}\cdots \mu _{n-2}\right) }\right) =\frac{%
\left( n+d-3\right) !}{(n-2)!\left( d-1\right) !}
\end{equation}
independent components. This is also the number
of constraints by which the number of independent components of 
the original symmetric tensor $A_{d}^{\left( \mu _{1}\cdots \mu _{n}\right) }$ is reduced, if
we demand that it is traceless in addition to being symmetric.
Thus, the number of independent components of a symmetric traceless tensor is
\begin{eqnarray}
N_{dn}\left( A_{d,tr}^{\left( \mu _{1}\cdots \mu _{n}\right) }\right)
& = & N_{dn}\left( A_{d}^{\left( \mu _{1}\cdots \mu _{n}\right) }\right)
-N_{dn}\left( A_{d}^{\left( \mu _{1}\cdots \mu _{n-2}\right) }\right) \nonumber \\
& = & \frac{\left( n+d-1\right) !}{n!\left( d-1\right) !} 
- \frac{\left( n+d-3\right) !}{(n-2)!\left( d-1\right) !} 
 =  \frac{\left( n+d-3\right) !}{n!\left( d-2\right) !} (2n+d-2)\;.\label{Wolf_2}
\end{eqnarray}
Let us now require in addition that such a symmetric traceless
tensor is orthogonal to a given four-vector $u^\mu$,
\begin{equation} \label{tensor_ortho}
0= A_{d,tr,ortho}^{\left( \mu _{1}\cdots \mu _{n}\right) }u_{\mu _{n}}
\equiv A_{d,tr}^{\left( \mu _{1}\cdots \mu _{n-1}\right) }\;.
\end{equation}
The right-hand side defines a new symmetric traceless tensor of rank $n-1$ which,
according to Eq.\ (\ref{Wolf_2}), has
\begin{equation}
N_{dn}\left( A_{d,tr}^{\left( \mu _{1}\cdots \mu _{n-1}\right) }\right)
= \frac{\left( n+d-4\right) !}{(n-1)!\left( d-2\right) !} (2n+d-4)
\end{equation}
independent components. This number reduces the number of independent
components of the original symmetric traceless tensor, if we demand in addition
that it is orthogonal to $u^\mu$; thus the latter has
\begin{eqnarray}
N_{dn}\left( A_{d,tr,ortho}^{\left( \mu _{1}\cdots \mu _{n}\right) }\right)
& = & N_{dn}\left( A_{d,tr}^{\left( \mu _{1}\cdots \mu _{n}\right) }\right)
-N_{dn}\left( A_{d,tr}^{\left( \mu _{1}\cdots \mu _{n-1}\right) }\right) \nonumber \\
& = & \frac{\left( n+d-3\right) !}{n!\left( d-2\right) !} (2n+d-2)
- \frac{\left( n+d-4\right) !}{(n-1)!\left( d-2\right) !} (2n+d-4) \nonumber \\
& =  &\frac{\left( n+d-4\right) !}{n!\left( d-3\right) !} (2n+d-3)\label{Wolf_3}
\end{eqnarray}
independent components. Comparing this equation to Eq.\ (\ref{Wolf_2}) we
realize that the orthogonality constraint (\ref{tensor_ortho}) has effectively reduced
the number of dimensions by one unit, $d \rightarrow d-1$. Subsequently
demanding orthogonality to another four-vector $l^\mu$ would reduce the number
of dimensions by another unit, etc.

Now taking $d=4$, Eq.\ (\ref{Wolf_3}) tells us that any symmetric traceless tensor of rank $n$, 
which is orthogonal to $u^{\mu }$, has 
$N_{4n}(A_{4,tr,ortho}^{\left( \mu _{1}\cdots\mu_{n}\right) })=2n+1$ 
independent components. If this tensor is in addition orthogonal to another four-vector
$l^\mu$, then Eq.\ (\ref{Wolf_3}) applies replacing $d=4$ by $d=3$, and we
obtain $N_{3n}(A_{3,tr,ortho}^{\left(\mu _{1}\cdots\mu _{n}\right) })=2$
independent components. This result is independent of the tensor rank
$n$. 

In the following, we list the irreducible
projection operators which are necessary for the derivation of 
Eqs.\ (\ref{eq_aniso_scalar}) -- (\ref{eq_aniso_tensor}). 
In the case of rank-one tensors, the irreducible projection operator (\ref{Deltaproj}) is trivially given by
\begin{equation}
\Delta^{\mu _{1}\nu _{1}} = C(1,0) \, \frac{1}{{\cal N}_{10}}\, \Delta^{\mu_1 \nu_1}\;,
\end{equation}
with $C(1,0) = {\cal N}_{10} = 1$.
Analogously, the irreducible projection operator (\ref{Xiproj}) is
\begin{equation}
\Xi^{\mu _{1}\nu _{1}} = \hat{C}(1,0)\, \frac{1}{{\cal N}_{10}}\, \Xi^{\mu_1 \nu_1}\;,
\end{equation}
with $\hat{C}(1,0) = {\cal N}_{10} = 1$.

For rank-two tensors, the irreducible projection operator (\ref{Deltaproj}) is
\begin{eqnarray}
\Delta^{\mu _{1}\mu _{2}\nu _{1}\nu _{2}}& =& C(2,0)\, \frac{1}{\mathcal{N}_{20}}\left(\Delta ^{\mu_{1} \nu _{1} }
\Delta ^{ \nu _{2} \mu_{2}}+ \Delta^{\mu_1 \nu_2} \Delta^{\nu_1 \mu_2} \right)
+C(2,1)\,\frac{1}{\mathcal{N}_{21}}\, \Delta ^{\mu _{1}\mu _{2}}\Delta ^{\nu _{1}\nu _{2}} \notag \\
& = & 
\Delta ^{\mu _{1}\left( \nu
_{1}\right. }\Delta ^{\left. \nu _{2}\right) \mu _{2}}-\frac{1}{3}\Delta
^{\mu _{1}\mu _{2}}\Delta ^{\nu _{1}\nu _{2}}\;.\label{Delta_1234_1}
\end{eqnarray}
Analogously, the irreducible projection operator (\ref{Xiproj}) reads
\begin{eqnarray}
\Xi ^{\mu _{1}\mu _{2}\nu _{1}\nu _{2}}& =&\hat{C}(2,0)\, \frac{1}{\mathcal{N}_{20}}\left(\Xi^{\mu_{1} \nu _{1} }
\Xi^{ \nu _{2} \mu_{2}}+ \Xi^{\mu_1 \nu_2} \Xi^{\nu_1 \mu_2} \right)
+\hat{C}(2,1)\,\frac{1}{\mathcal{N}_{21}}\, \Xi^{\mu _{1}\mu _{2}}\Xi^{\nu _{1}\nu _{2}}  \nonumber \\
& = & \Xi ^{\mu _{1}\left( \nu _{1}\right.}
\Xi ^{\left. \nu _{2}\right) \mu _{2}}-\frac{1}{2}\Xi ^{\mu _{1}\mu_{2}}\Xi ^{\nu _{1}\nu _{2}}\;.
\end{eqnarray}

We also give the irreducible projection operators for rank-three tensors:
\begin{align}
\Delta^{\mu _{1}\mu _{2}\mu _{3}\nu _{1}\nu _{2}\nu _{3}}& =\frac{1}{3}
\left( \Delta^{\mu _{1}\nu _{1}}\Delta^{\mu _{2}\left( \nu _{2}\right.}
\Delta^{\left. \nu _{3}\right) \mu _{3}}+\Delta^{\mu _{1}\nu _{2}}\Delta^{\mu _{2}
\left( \nu _{1}\right. }\Delta^{\left. \nu _{3}\right) \mu_{3}}
+\Delta^{\mu _{1}\nu _{3}}\Delta^{\mu _{2}\left( \nu _{2}\right.}
\Delta^{\left. \nu _{1}\right) \mu _{3}}\right)  \notag \\
& -\frac{3}{5}\Delta^{\left( \mu _{1}\mu _{2}\right. }\Delta^{\left. \mu_{3}\right) 
\left( \nu _{3}\right. }\Delta^{\left. \nu _{1}\nu _{2}\right)}\; ,
\end{align}
and
\begin{align}
\Xi^{\mu _{1}\mu _{2}\mu _{3}\nu _{1}\nu _{2}\nu _{3}}& =\frac{1}{3}\left(
\Xi^{\mu _{1}\nu _{1}}\Xi^{\mu _{2}\left( \nu _{2}\right. }\Xi^{\left.\nu _{3}\right) \mu _{3}}
+\Xi^{\mu _{1}\nu _{2}}\Xi^{\mu _{2}\left( \nu_{1}\right. }\Xi^{\left. \nu _{3}\right) \mu _{3}}
+\Xi^{\mu _{1}\nu_{3}}\Xi^{\mu _{2}\left( \nu _{2}\right. }\Xi^{\left. \nu _{1}\right) \mu_{3}}\right)  \notag \\
& -\frac{3}{4}\Xi^{\left( \mu _{1}\mu _{2}\right. }\Xi^{\left. \mu_{3}\right) 
\left( \nu _{3}\right. }\Xi^{\left. \nu _{1}\nu _{2}\right) }\;.
\end{align}

Finally, for rank-four tensors, we obtain for the irreducible projection operators:
\begin{align}
\Delta^{\mu _{1}\mu _{2}\mu _{3}\mu _{4}\nu _{1}\nu _{2}\nu _{3}\nu _{4}}& =
\frac{1}{4!}\sum_{\mathcal{P}_\mu^4 \mathcal{P}_\nu^4}\Delta^{\mu _{1}\nu _{1}}\Delta^{\mu _{2}\nu_{2}}
\Delta^{\mu _{3}\nu _{3}}\Delta ^{\mu _{4}\nu _{4}}
-\frac{3}{14}\Delta^{\left( \mu _{1}\mu _{2}\right. }\Delta^{\left. \mu _{3}\right) \left( \nu_{3}\right. }
\Delta^{\nu _{1}\nu _{2}}\Delta ^{\left. \nu _{4}\right) \mu_{4}}  \notag \\
& -\frac{3}{14}\Delta^{\left( \mu _{1}\mu _{2}\right. }\Delta ^{\left. \mu_{4}\right) 
\left( \nu _{3}\right. }\Delta^{\nu _{1}\nu _{2}}\Delta^{\left. \nu _{4}\right) \mu _{3}}
-\frac{3}{14}\Delta^{\left( \mu _{1}\mu_{3}\right. }\Delta^{\left. \mu _{4}\right) \left( \nu _{3}\right. }
\Delta^{\nu _{1}\nu _{2}}\Delta^{\left. \nu _{4}\right) \mu _{2}}  \notag \\
& -\frac{3}{14}\Delta^{\left( \mu _{2}\mu _{3}\right. }\Delta^{\left. \mu_{4}\right) 
\left( \nu _{3}\right. }\Delta^{\nu _{1}\nu _{2}}\Delta^{\left. \nu _{4}\right) \mu _{1}}
+\frac{3}{35}\Delta^{\mu _{1}\left( \mu_{2}\right. }\Delta^{\left. \mu _{3}\mu _{4}\right) }
\Delta^{\nu_{1}\left(\nu _{2}\right. }\Delta^{\left. \nu _{3}\nu _{4}\right) }\;,
\end{align}
and 
\begin{align}
\Xi^{\mu _{1}\mu _{2}\mu _{3}\mu _{4}\nu _{1}\nu _{2}\nu _{3}\nu _{4}}& =
\frac{1}{4!}\sum_{\mathcal{P}_\mu^4 \mathcal{P}_\nu^4}
\Xi^{\mu _{1}\nu _{1}}\Xi ^{\mu _{2}\nu _{2}}\Xi^{\mu _{3}\nu _{3}}\Xi ^{\mu _{4}\nu _{4}}
-\frac{1}{4}\Xi^{\left( \mu_{1}\mu _{2}\right. }\Xi^{\left. \mu _{3}\right) \left( \nu _{3}\right.}
\Xi^{\nu _{1}\nu _{2}}\Xi^{\left. \nu _{4}\right) \mu _{4}}  \notag \\
& -\frac{1}{4}\Xi^{\left( \mu _{1}\mu _{2}\right. }\Xi^{\left. \mu_{4}\right) 
\left( \nu _{3}\right. }\Xi^{\nu _{1}\nu _{2}}\Xi ^{\left. \nu_{4}\right) \mu _{3}}
-\frac{1}{4}\Xi^{\left( \mu _{1}\mu _{3}\right. }\Xi^{\left. \mu _{4}\right) 
\left( \nu _{3}\right. }\Xi^{\nu _{1}\nu _{2}}\Xi^{\left. \nu _{4}\right) \mu _{2}}  \notag \\
& -\frac{1}{4}\Xi^{\left( \mu _{2}\mu _{3}\right. }\Xi^{\left. \mu_{4}\right) 
\left( \nu _{3}\right. }\Xi^{\nu _{1}\nu _{2}}\Xi^{\left. \nu_{4}\right) \mu _{1}}
+\frac{1}{8}\Xi^{ \mu _{1}\left( \mu _{2}\right. }\Xi^{\left. \mu _{3}\mu _{4}\right) }
\Xi^{ \nu _{1}\left(\nu _{2}\right. }\Xi^{\left. \nu _{3}\nu _{4}\right) }\;.
\end{align}
In both expressions, the first sum runs over all distinct
permutations of the four $\mu$- and the four $\nu$-type indices. 

The irreducible projection operators (\ref{Deltaproj}) acting on tensors formed by
the $n-$adic product of a four-vector, i.e., $A^{\mu _{1}\cdots \mu
_{n}}=A^{\mu _{1}}\cdots A^{\mu _{n}}$, lead to the following expressions for $n=1, \ldots, 4$
\begin{align}
A^{\left\langle \mu _{1}\right\rangle }& =\Delta ^{\mu _{1}\nu _{1}}A_{\nu_{1}}\;, \\
A^{\left\langle \mu _{1}\right. }A^{\left. \mu _{2}\right\rangle }& =
\Delta ^{\mu _{1}\mu _{2}\nu _{1}\nu _{2}}A_{\nu _{1}}A_{\nu_{2}}
=A^{\left\langle \mu _{1}\right\rangle }A^{\left\langle \mu_{2}\right\rangle }
-\frac{1}{3}\Delta ^{\mu _{1}\mu _{2}}\left( \Delta
^{\alpha \beta }A_{\alpha }A_{\beta }\right)\;, \\
A^{\left\langle \mu _{1}\right. }A^{\mu _{2}}A^{\left. \mu _{3}\right\rangle} & =
\Delta ^{\mu _{1}\mu _{2} \mu_3\nu _{1}\nu _{2}\nu_3}A_{\nu _{1}}A_{\nu_{2}} A_{\nu_3} =
A^{\left\langle \mu _{1}\right\rangle }A^{\left\langle \mu_{2}\right\rangle }
A^{\left\langle \mu _{3}\right\rangle }-\frac{3}{5}\Delta^{\left( \mu _{1}\mu _{2}\right. }
A^{\left. \left\langle \mu_{3}\right\rangle \right) }\left( \Delta ^{\alpha \beta }A_{\alpha }A_{\beta}\right) \;, \\
A^{\left\langle \mu _{1}\right. }A^{\mu _{2}}A^{\mu _{3}}A^{\left. \mu_{4}\right\rangle }& =
\Delta ^{\mu _{1}\mu _{2} \mu_3 \mu_4\nu _{1}\nu _{2}\nu_3\nu_4}A_{\nu _{1}}A_{\nu_{2}} A_{\nu_3}A_{\nu_4} 
\notag \\
& = A^{\left\langle \mu _{1}\right\rangle }A^{\left\langle\mu _{2}\right\rangle }
A^{\left\langle \mu _{3}\right\rangle}A^{\left\langle \mu _{4}\right\rangle }-\frac{3}{14}
\Delta ^{\left( \mu_{1}\mu _{2}\right. }A^{\left. \left\langle \mu _{3}\right\rangle \right)}
A^{\left\langle \mu _{4}\right\rangle }\left( \Delta ^{\alpha \beta}A_{\alpha }A_{\beta }\right)  \notag \\
& -\frac{3}{14}\Delta ^{\left( \mu _{1}\mu _{2}\right. }A^{\left.\left\langle \mu _{4}\right\rangle \right) }
A^{\left\langle \mu_{3}\right\rangle }\left( \Delta ^{\alpha \beta }A_{\alpha }A_{\beta}\right) 
-\frac{3}{14}\Delta ^{\left( \mu _{1}\mu _{4}\right. }A^{\left.\left\langle \mu _{3}\right\rangle \right) }
A^{\left\langle \mu_{2}\right\rangle }\left( \Delta ^{\alpha \beta }A_{\alpha }A_{\beta }\right)
\notag \\
& -\frac{3}{14}\Delta ^{\left( \mu _{4}\mu _{2}\right. }A^{\left.\left\langle \mu _{3}\right\rangle \right) }
A^{\left\langle \mu_{1}\right\rangle }\left( \Delta ^{\alpha \beta }A_{\alpha }A_{\beta}\right) 
+\frac{3}{35}\Delta^{ \mu _{1}\left(\mu _{2}\right. }\Delta^{\left. \mu _{3}\mu _{4}\right) }\left( \Delta ^{\alpha \beta }
A_{\alpha}A_{\beta }\right)^{2}\;.
\end{align}
Similarly, for the irreducible projection operators (\ref{Xiproj}) we obtain
\begin{align}
A^{\left\{ \mu _{1}\right\} } & =\Xi ^{\mu _{1}\nu _{1}}A_{\nu _{1}} \;, \\
A^{\left\{ \mu _{1}\right. }A^{\left. \mu _{2}\right\} }& = \Xi ^{\mu_{1}\mu _{2}\nu _{1}\nu _{2}}
A_{\nu _{1}}A_{\nu _{2}}=A^{\left\{ \mu_{1}\right\} }A^{\left\{ \mu _{2}\right\} }
-\frac{1}{2}\Xi ^{\mu _{1}\mu_{2}}\left( \Xi ^{\alpha \beta }A_{\alpha }A_{\beta }\right)\; , \\
A^{\left\{ \mu \right. _{1}}A^{\mu _{2}}A^{\left. \mu _{3}\right\}} & =
\Xi^{\mu _{1}\mu _{2} \mu_3\nu _{1}\nu _{2}\nu_3}A_{\nu _{1}}A_{\nu_{2}} A_{\nu_3}=
A^{\left\{ \mu _{1}\right\} }A^{\left\{ \mu _{2}\right\} }A^{\left\{ \mu_{3}\right\} }
-\frac{3}{4}\Xi ^{\left( \mu _{1}\mu _{2}\right. }A^{\left.\left\{ \mu _{3}\right\} \right) }
\left( \Xi ^{\alpha \beta }A_{\alpha}A_{\beta }\right) \;, \\
A^{\left\{ \mu _{1}\right. }A^{\mu _{2}}A^{\mu _{3}}A^{\left. \mu_{4}\right\} }& =
\Xi^{\mu _{1}\mu _{2} \mu_3 \mu_4\nu _{1}\nu _{2}\nu_3\nu_4}A_{\nu _{1}}A_{\nu_{2}} A_{\nu_3}A_{\nu_4} 
\notag \\
& 
=A^{\left\{ \mu _{1}\right\} }A^{\left\{ \mu _{2}\right\}}A^{\left\{ \mu _{3}\right\} }A^{\left\{ \mu _{4}\right\} }
-\frac{1}{4}\Xi^{\left( \mu _{1}\mu _{2}\right. }A^{\left. \left\{ \mu _{3}\right\} \right)}
A^{\left\{ \mu _{4}\right\} }\left( \Xi ^{\alpha \beta }A_{\alpha }A_{\beta}\right)  \notag \\
& -\frac{1}{4}\Xi ^{\left( \mu _{1}\mu _{2}\right. }A^{\left. \left\{ \mu_{4}\right\} \right) }
A^{\left\{ \mu _{3}\right\} }\left( \Xi ^{\alpha \beta}A_{\alpha }A_{\beta }\right) 
-\frac{1}{4}\Xi ^{\left( \mu _{1}\mu_{4}\right. }A^{\left. \left\{ \mu _{3}\right\} \right) }
A^{\left\{ \mu_{2}\right\} }\left( \Xi ^{\alpha \beta }A_{\alpha }A_{\beta }\right)  \notag\\
& -\frac{1}{4}\Xi ^{\left( \mu _{4}\mu _{2}\right. }A^{\left. \left\{ \mu_{3}\right\} \right) }
A^{\left\{ \mu _{1}\right\} }\left( \Xi ^{\alpha \beta}A_{\alpha }A_{\beta }\right) 
+\frac{1}{8}\Xi ^{ \mu _{1}\left(\mu_{2}\right. }\Xi ^{\left. \mu _{3}\mu _{4}\right) }
\left( \Xi ^{\alpha \beta}A_{\alpha }A_{\beta }\right)^{2}\;.
\end{align}

\section{Orthogonality properties and conditions}
\label{appendix_orhogonality} 

In this appendix, we derive the orthogonality conditions
(\ref{normalization_isotropic}) and (\ref{normalization_anisotropic}). 
The derivation utilizes the relations
\begin{equation} \label{completecontraction_k_Delta}
k^{\left\langle \mu _{1}\right. }\cdots k^{\left. \mu _{\ell }\right\rangle}
k_{\left\langle \mu _{1}\right. }\cdots k_{\left. \mu _{\ell }\right\rangle}
=\frac{\ell !}{\left( 2\ell -1\right) !!}\left( \Delta ^{\alpha \beta}k_{\alpha }k_{\beta }\right) ^{\ell }\; ,
\end{equation}
and
\begin{equation}\label{completecontraction_k_Xi}
k^{\left\{ \mu _{1}\right. }\cdots k^{\left. \mu _{\ell }\right\}}
k_{\left\{ \mu _{1}\right. }\cdots k_{\left. \mu _{\ell }\right\} }
=\frac{1}{2^{\ell -1}}\left( \Xi ^{\alpha \beta }k_{\alpha }k_{\beta }\right) ^{\ell}\;.
\end{equation}
The first relation is proved as follows. We first note that
\begin{equation}
k^{\left\langle \mu _{1}\right. }\cdots k^{\left. \mu _{\ell }\right\rangle}
k_{\left\langle \mu _{1}\right. }\cdots k_{\left. \mu _{\ell }\right\rangle} =
\Delta^{\mu_1 \cdots \mu_\ell}_{\beta_1 \cdots \beta_\ell} \Delta_{\mu_1 \cdots \mu_\ell}^{\alpha_1 \cdots \alpha_\ell}
k_{\alpha_1} \cdots k_{\alpha_\ell} k^{\beta_1} \cdots k^{\beta_\ell}
= \Delta^{\alpha_1 \cdots \alpha_\ell}_{\beta_1 \cdots \beta_\ell} k_{\alpha_1} \cdots k_{\alpha_\ell} 
k^{\beta_1} \cdots k^{\beta_\ell}\;.
\end{equation}
Now we insert the explicit form (\ref{Deltaproj}) of the projection operator and note that the contraction
of all indices with the momenta reduces the second sum (including the prefactor $1/\mathcal{N}_{nq}$)
to just a factor of $(\Delta^{\alpha \beta}k_\alpha k_\beta)^\ell$,
\begin{equation} \label{sum_C}
k^{\left\langle \mu _{1}\right. }\cdots k^{\left. \mu _{\ell }\right\rangle}
k_{\left\langle \mu _{1}\right. }\cdots k_{\left. \mu _{\ell }\right\rangle} =
\sum_{q=0}^{[\ell/2]} C(\ell,q) \left(\Delta^{\alpha \beta}k_\alpha k_\beta \right)^\ell\;.
\end{equation}
The Legendre polynomials $P_\ell(z)$ have the representation \cite{GR8.911.1}
\begin{equation}
P_\ell(z) = \frac{1}{2^\ell } \sum_{q=0}^{[\ell/2]} (-1)^q \frac{(2\ell - 2q)!}{q! (\ell - q)! (\ell - 2q)! }\, z^{\ell - 2q}
\equiv \frac{1}{2^\ell } \frac{(2 \ell)!}{(\ell !)^2} \sum_{q=0}^{[\ell/2]} C(\ell,q) \, z^{\ell-2q}\;.
\end{equation}
Since for all $\ell$
\begin{equation}
1 \equiv P_\ell(1) =  \frac{1}{2^\ell} \frac{(2 \ell)!}{(\ell !)^2} \sum_{q=0}^{[\ell/2]} C(\ell,q)\;,
\end{equation}
we derive the identity
\begin{equation}
\sum_{q=0}^{[\ell/2]} C(\ell,q) = \frac{2^\ell (\ell !)^2}{(2 \ell)!} = \frac{\ell ! \, 2^{\ell-1} (\ell - 1)!}{(2\ell-1)!} 
\equiv \frac{\ell !}{(2 \ell -1)!!}\;,
\end{equation}
where we have used the definition of the double factorial for odd numbers. Inserting this
into Eq.\ (\ref{sum_C}) proves Eq.\ (\ref{completecontraction_k_Delta}).

We now prove Eq.\ (\ref{completecontraction_k_Xi}). Analogously to Eq.\ (\ref{sum_C}) we obtain
\begin{equation}
k^{\left\{ \mu _{1}\right. }\cdots k^{\left. \mu _{\ell }\right\}}
k_{\left\{ \mu _{1}\right. }\cdots k_{\left. \mu _{\ell }\right\} }
=\sum_{q=0}^{[\ell/2]} \hat{C}(\ell,q) \left(\Xi^{\alpha \beta}k_\alpha k_\beta \right)^\ell\;.
\end{equation}
The Chebyshev polynomial of the first kind $T_\ell(z)$ has the representation \cite{AS22.3.6}
\begin{equation}
T_\ell(z) = \frac{\ell}{2} \sum_{q=0}^{[\ell/2]} (-1)^q \frac{(\ell - q- 1)!}{q! (\ell - 2q)!} (2z)^{\ell-2q}
= 2^{\ell-1}  \sum_{q=0}^{[\ell/2]} (-1)^q \frac{1}{4^q}\, \frac{(\ell - q)!}{q! (\ell - 2q)!}\, \frac{\ell}{\ell - q}\,
z^{\ell-2q}\;.
\end{equation}
Since $T_\ell(1)=1$  for all $\ell$ \cite{AS22.3.6}, we obtain with Eq.\ (\ref{Chatcoeff})
\begin{equation}
\sum_{q=0}^{[\ell/2]} \hat{C}(\ell,q) = \frac{1}{2^{\ell-1}}\;, \;\;\; {\rm q.e.d.}\;.
\end{equation}

The orthogonality condition (\ref{normalization_isotropic}) is
obtained from an integral of the type 
\begin{equation}
M^{\left\langle \mu _{1}\cdots \mu _{\ell }\right\rangle}_{\left\langle\nu _{1}\cdots \nu_{n}\right\rangle }
=\int dK\text{ }\mathrm{F}(E_{\mathbf{k}u})\
k^{\left\langle \mu _{1}\right. }\cdots k^{\left. \mu _{\ell }\right\rangle
}k_{\left\langle \nu _{1}\right. }\cdots k_{\left. \nu _{n}\right\rangle }\; ,
\end{equation}
which is a tensor of rank ($\ell +n$) that is (separately) symmetric
under the permutation of $\mu $-type and $\nu $-type indices.
In Appendix A of Ref.\  \cite{Denicol:2012cn} it is proven that tensors of this type must obey the relation
\begin{equation} \label{MtensorMscalar}
M^{\left\langle \mu _{1}\cdots \mu _{\ell }\right\rangle}_{\left\langle\nu _{1}\cdots \nu_{n}\right\rangle }
=\delta _{\ell n}\, \mathcal{M}\, \Delta^{\mu_{1}\cdots \mu _{\ell }}_{\nu _{1}\cdots \nu _{n}}\;,
\end{equation}%
where $\mathcal{M}$ is an invariant scalar that can be computed by completely contracting the indices of
$M^{\left\langle \mu _{1}\cdots \mu _{\ell }\right\rangle}_{\left\langle\nu _{1}\cdots \nu_{n}\right\rangle }$,
\begin{eqnarray}
\mathcal{M} &\mathcal{\equiv }&\frac{1}{\Delta _{\mu _{1}\cdots \mu _{\ell}}^{\mu _{1}\cdots \mu _{\ell }}}
\int dK\text{ }\mathrm{F}(E_{\mathbf{k}u}) \, k^{\left\langle \mu _{1}\right. }\cdots k^{\left. \mu _{\ell }\right\rangle}
k_{\left\langle \mu _{1}\right. }\cdots k_{\left. \mu _{\ell }\right\rangle}  \notag \\
&=&\frac{\ell !}{\left( 2\ell +1\right) !!}\int dK\text{ }\mathrm{F}(E_{\mathbf{k}u})
\left( \Delta ^{\alpha \beta }k_{\alpha }k_{\beta }\right)^{\ell }\; ,
\end{eqnarray}
where we have used Eqs.\ (\ref{Delta_complete_contraction}) and (\ref{completecontraction_k_Delta}). 
This proves Eq.\ (\ref{normalization_isotropic}).

Similarly one derives Eq.\ (\ref{normalization_anisotropic}). We define a rank-$(\ell +n)$ tensor
\begin{equation}
\hat{M}^{\left\{ \mu _{1}\cdots \mu _{\ell }\right\}}_{\left\{\nu _{1}\cdots \nu _{n}\right\}}
=\int dK\text{ }\mathrm{\hat{F}}(E_{\mathbf{k}u},E_{\mathbf{k}l})\, k^{\left\{ \mu _{1}\right. }\cdots 
k^{\left. \mu _{\ell }\right\} } k_{\left\{ \nu _{1}\right. }\cdots k_{\left. \nu _{n}\right\} }\; ,
\end{equation}
which is (separately) symmetric under permutations of the $\mu $- and $\nu $-type indices 
and depends solely on the fluid four-velocity $u^{\mu }$ and the four-vector $l^{\mu }$. 
Therefore, $\hat{M}^{\left\{ \mu _{1}\cdots \mu _{\ell } \right\}}_{\left\{ \nu _{1}\cdots \nu _{n}\right\} }$ 
must be constructed from tensor structures made of $u^{\mu }$, $l^{\mu }$,
and $\Xi ^{\mu \nu }$. Furthermore, 
$\hat{M}^{\left\{ \mu _{1}\cdots \mu_{\ell }\right\}}_{\left\{\nu _{1}\cdots \nu _{n}\right\}} $ 
must be orthogonal to $u^{\mu }$ as well as to $l^{\mu }$, which implies that it can only be constructed from
combinations of the projection operators $\Xi ^{\mu \nu }$, and
henceforth the rank of the tensor, $\ell +n$, must be an even number. Now,
following the arguments presented in Appendix A of Ref.\ \cite{Denicol:2012cn} one can
prove that
\begin{equation}
\hat{M}^{\left\{ \mu _{1}\cdots \mu _{\ell }\right\} }_{\left\{\nu _{1}\cdots \nu _{n}\right\}}
=\delta _{\ell n}\, \mathcal{\hat{M}}\, \Xi ^{\mu _{1}\cdots \mu _{\ell}}_{ \nu _{1}\cdots \nu _{n} }\;,
\label{M_hat_property_appendix}
\end{equation}%
where $\mathcal{\hat{M}}$ is a scalar. This is the analogue to Eq.\ (\ref{MtensorMscalar}). 
Using Eqs.\ (\ref{Xi_complete_contraction}) and (\ref{completecontraction_k_Xi}), we finally obtain
\begin{eqnarray}
\mathcal{\hat{M}} &\mathcal{\equiv }&\frac{1}{\Xi _{\mu _{1}\cdots \mu
_{\ell }}^{\mu _{1}\cdots \mu _{\ell }}}\int dK\text{ }\mathrm{\hat{F}}(E_{\mathbf{k}u},E_{\mathbf{k}l})\ 
k^{\left\{ \mu _{1}\right. }\cdots k^{\left.\mu _{\ell }\right\} }k_{\left\{ \mu _{1}\right. }\cdots 
k_{\left. \mu_{\ell }\right\} }  \notag \\
&=&\frac{1}{2^{\ell }}\int dK\text{ }\mathrm{\hat{F}}(E_{\mathbf{k}u},E_{\mathbf{k}l})\ 
\left( \Xi ^{\alpha \beta }k_{\alpha }k_{\beta }\right)^{\ell }\;.
\end{eqnarray}

Finally, we also prove Eq.\ (\ref{L_ij_simplified}). We first rewrite the collision integral
(\ref{L_ij_full}) as
\begin{equation}
\hat{\mathcal{L}}_{ij}^{\left\{ \mu _{1}\cdots \mu _{\ell }\right\} }\equiv
\sum_{r=0}^{\infty }\sum_{n=0}^{N_{r}}\sum_{m=0}^{N_{r}-n}\hat{\rho}_{nm}^{\nu _{1}\cdots \nu _{r}}
\left( \mathcal{A}_{injm}\right) _{\nu_{1}\cdots \nu _{r}}^{\mu _{1}\cdots \mu _{\ell }}\;,
\label{L_ij_simplified_appendix}
\end{equation}%
where, following similar arguments as above, the tensor 
$\left( \mathcal{A}_{injm}\right) _{\nu _{1}\cdots \nu _{r}}^{\mu_{1}\cdots \mu _{\ell }}$ can be shown 
to possess the property%
\begin{equation}
\left( \mathcal{A}_{injm}\right) _{\nu _{1}\cdots \nu _{r}}^{\mu _{1}\cdots\mu _{\ell }}
=\delta _{\ell r}\, \mathcal{A}_{injm}^{\left( \ell \right) }\, \Xi_{\nu _{1}\cdots \nu _{r}}^{\mu _{1}\cdots \mu _{\ell }}\;.
\end{equation}%
Substituting this into Eq.\ (\ref{L_ij_simplified_appendix}) leads to Eq.\ (\ref{L_ij_simplified}).

\section{The polynomial coefficients in the 14-moment approximation}
\label{polynomial_coefficients}

Here we construct the complete set of orthonormal polynomials in both $E_{\mathbf{k}u}$ and $E_{\mathbf{k}l}$ 
using the Gram-Schmidt orthogonalization
procedure in the 14-moment approximation, i.e., where $N_{0}=2$, $N_{1}=1$, and $N_{2}=0$, cf.\
Refs.\ \cite{Denicol:2012cn,Denicol:2012es}. With Eq.\ (\ref{kinetic:phi}), one observes that we only
need to determine the polynomials
$P_{\mathbf{k}00}^{\left( 0\right) },\, P_{\mathbf{k}01}^{\left( 0\right) },\, P_{\mathbf{k}02}^{\left( 0\right) },\,
P_{\mathbf{k}10}^{\left( 0\right) },\, P_{\mathbf{k}11}^{\left( 0\right) },\, P_{\mathbf{k}20}^{\left( 0\right) },
P_{\mathbf{k}00}^{\left( 1\right) },\, P_{\mathbf{k}01}^{\left( 1\right) },$
$ P_{\mathbf{k}10}^{\left( 1\right) }$, and $P_{\mathbf{k}00}^{\left( 2\right) }$.

Using the orthonormality condition (\ref{kinetic:orthonormality}) for $n=m=n'=m'=0$ and Eq.\ (\ref{J_nrq})
we first obtain the value of the normalization constant in Eq.\ (\ref{kinetic:weight}):
\begin{equation}
\hat{W}^{\left( \ell \right) }=\frac{\left( -1\right) ^{\ell }}{\hat{J}_{2\ell ,0,\ell }}\; .
\end{equation}
Then, the orthonormality condition (\ref{kinetic:orthonormality}) can be written with the help of 
Eqs.\ (\ref{J_nrq}) and (\ref{kinetic:P_nm}) as
\begin{equation} \label{orthohatJ}
\hat{J}_{2\ell,0,\ell}\, \delta_{nn'} \delta_{mm'} = \sum_{i=0}^n\sum_{j=0}^m\sum_{r=0}^{n'}\sum_{s=0}^{m'}
a_{nimj}^{(\ell)}\, a_{n'rm's}^{(\ell)}\, \hat{J}_{i+r+j+s+2\ell, j+s, \ell}\;.
\end{equation}
From this equation, we will successively construct the polynomials.
\begin{itemize}
\item[(i)] $P_{\mathbf{k}00}^{\left( 0\right) },\,P_{\mathbf{k}00}^{\left( 1\right) }$, $P_{\mathbf{k}00}^{\left( 2\right) }$:
For any $\ell$ and $n=n'=m=m'=0$, we obtain from Eq.\ (\ref{orthohatJ})
\begin{equation}
1 = a_{0000}^{(\ell )} \equiv P_{\mathbf{k}00}^{\left( \ell \right) } \;.
\end{equation}
This determines the polynomials $P_{\mathbf{k}00}^{\left( 0\right) },\,P_{\mathbf{k}00}^{\left( 1\right) }$, 
and $P_{\mathbf{k}00}^{\left( 2\right) }$.
\item[(ii)] $P_{\mathbf{k}10}^{\left( 0\right) }$:
Consider Eq.\ (\ref{orthohatJ}) for $\ell =0$, $n=n'=1$, $m=m'=0$, as well as for $\ell = 0$, $n=1$, $n'=0$, $m=m'=0$.
Solving these two equations for the coefficients $a_{1000}^{(0)}$ and $a_{1100}^{(0)}$ leads to the result
\begin{equation}
\frac{a_{1000}^{(0)}}{a_{1100}^{(0)}} =-\frac{\hat{J}_{100}}{\hat{J}_{000}}\;,\;\;
\left( a_{1100}^{(0)}\right) ^{2}=\frac{\hat{J}_{000}^{2}}{\hat{D}_{10}}\;,
\end{equation}
where
\begin{equation}
\hat{D}_{nq} =\hat{J}_{n-1,0,q}\hat{J}_{n+1,0,q}-\hat{J}_{n0q}^{2}\;.
\label{D_nq} 
\end{equation}
This uniquely determines the polynomial
$P_{\mathbf{k}10}^{\left( 0\right) } =a_{1000}^{(0)}+a_{1100}^{(0)}E_{\mathbf{k}u}$.
\item[(iii)] $P_{\mathbf{k}20}^{\left( 0\right) }$:
Consider Eq.\ (\ref{orthohatJ}) for $\ell =0$, $n=n'=2$, $m=m'=0$,  for $\ell = 0$, $n=2$, $n'=1$, $m=m'=0$,
and for $\ell = 0$, $n=2$, $n'=0$, $m=m'=0$. From these three equations one obtains
\begin{equation}
\frac{a_{2000}^{(0)}}{a_{2200}^{(0)}}=\frac{\hat{D}_{20}}{\hat{D}_{10}}\;,\;\;
\frac{a_{2100}^{(0)}}{a_{2200}^{(0)}}=\frac{\hat{G}_{12}}{\hat{D}_{10}}\;, \;\;
\left( a_{2200}^{(0)}\right)^{2} =\frac{\hat{J}_{000}\hat{D}_{10}}{\hat{J}_{200}\hat{D}_{20}+\hat{J}_{300}
\hat{G}_{12}+\hat{J}_{400}\hat{D}_{10}}\;,
\end{equation}
where
\begin{equation}
\hat{G}_{nm} =\hat{J}_{n00}\hat{J}_{m00}-\hat{J}_{n-1,0,0}\hat{J}_{m+1,0,0}\; .  \label{G_nm}
\end{equation}
This uniquely determines the polynomial $P_{\mathbf{k}20}^{\left( 0\right) }=a_{2000}^{(0)}+a_{2100}^{(0)}
E_{\mathbf{k}u}+a_{2200}^{(0)}E_{\mathbf{k}u}^{2}$.
\item[(iv)] $P_{\mathbf{k}10}^{\left( 1\right) }$: Consider Eq.\ (\ref{orthohatJ}) for 
$\ell =1$, $n=n'=1$, $m=m'=0$, as well as for $\ell = 1$, $n=1$, $n'=0$, $m=m'=0$. From these two equations one 
obtains
\begin{equation}
\frac{a_{1000}^{(1)}}{a_{1100}^{(1)}}=-\frac{\hat{J}_{301}}{\hat{J}_{201}}\;, \;\;
\left( a_{1100}^{(1)}\right) ^{2}=\frac{\hat{J}_{201}^{2}}{\hat{D}_{31}}\; ,
\end{equation}
which uniquely determines the polynomial 
$P_{\mathbf{k}10}^{\left( 1\right) }=a_{1000}^{(1)}+a_{1100}^{(1)}E_{\mathbf{k}u}$.
\item[(v)] $P_{\mathbf{k}01}^{\left( 0\right) }$: Consider Eq.\ (\ref{orthohatJ}) for 
$\ell =0$, $n=n'=0$, $m=m'=1$, as well as for $\ell = 0$, $n=n'=0$, $m=1$, $m'=0$.
From these two equations one obtains
\begin{equation}
\frac{a_{0010}^{(0)}}{a_{0011}^{(0)}} =-\frac{\hat{J}_{110}}{\hat{J}_{000}}\; ,\;\;
\left( a_{0011}^{(0)}\right) ^{2}=\frac{\hat{J}_{000}^{2}}{\hat{D}_{110}}\;,
\end{equation}
where we defined
\begin{equation}
\hat{D}_{nrq} =\hat{J}_{n-1,r-1,q}\hat{J}_{n+1,r+1,q}-\hat{J}_{nrq}^{2}\;.
\label{D_nrq}
\end{equation}
This uniquely determines the polynomial 
$P_{\mathbf{k}01}^{\left( 0\right) } =a_{0010}^{(0)}+a_{0011}^{(0)}E_{\mathbf{k}l}$.
\item[(vi)] $P_{\mathbf{k}02}^{\left( 0\right) }$: Consider Eq.\ (\ref{orthohatJ}) for 
$\ell =0$, $n=n'=0$, $m=m'=2$, for $\ell = 0$, $n=n'=0$, $m=2$, $m'=1$, and for $\ell = 0$, $n=n'=0$, $m=2$, $m'=0$.
From these three equations one obtains
\begin{equation}
\frac{a_{0020}^{(0)}}{a_{0022}^{(0)}} =\frac{\hat{D}_{220}}{\hat{D}_{110}}\;,\;\;
\frac{a_{0021}^{(0)}}{a_{0022}^{(0)}}=\frac{\hat{G}_{1122}}{\hat{D}_{110}}\;, \;\;
\left( a_{0022}^{(0)}\right) ^{2} =\frac{\hat{J}_{000}\hat{D}_{110}}{\hat{J}_{220}\hat{D}_{220}
+\hat{J}_{330}\hat{G}_{1122}+\hat{J}_{440}\hat{D}_{110}}\;,
\end{equation}
where we defined
\begin{equation}
\hat{G}_{nrmp} =\hat{J}_{nr0}\hat{J}_{mp0}-\hat{J}_{n-1,r-1,0}\hat{J}_{m+1,p+1,0}\;.  \label{G_nrmp}
\end{equation}
This uniquely determines the polynomial 
$P_{\mathbf{k}02}^{\left( 0\right) } =a_{0020}^{(0)}+a_{0021}^{(0)}E_{\mathbf{k}l}+a_{0022}^{(0)}E_{\mathbf{k}l}^{2}$.
\item[(vii)] $P_{\mathbf{k}01}^{\left( 1\right) }$: Consider Eq.\ (\ref{orthohatJ}) for 
$\ell =1$, $n=n'=0$, $m=m'=1$ and for $\ell = 1$, $n=n'=0$, $m=1$, $m'=0$. From these two equations
one obtains
\begin{equation}
\frac{a_{0010}^{(1)}}{a_{0011}^{(1)}}=-\frac{\hat{J}_{311}}{\hat{J}_{201}}\;,\;\;
\left( a_{0011}^{(1)}\right) ^{2}=\frac{\hat{J}_{201}^{2}}{\hat{D}_{311}}\;,
\end{equation}
which uniquely determines the polynomial 
$P_{\mathbf{k}01}^{\left( 1\right) } = a_{0010}^{(1)}+a_{0011}^{(1)}E_{\mathbf{k}l}$.
\item[(viii)] $P_{\mathbf{k}11}^{\left( 0\right) }$: Consider Eq.\ (\ref{orthohatJ}) for 
$\ell =0$, $n=n'=1$, $m=m'=1$, for $\ell = 0$, $n=n'=1$, $m=1$, $m'=0$, 
for $\ell =0$, $n=1$, $n'=0$, $m=m'=1$, and for $\ell = 0$, $n=1$, $n'=0$, $m=1$, $m'=0$. From these
four equations one obtains
\begin{eqnarray}
\frac{a_{1010}^{(0)}}{a_{1111}^{(0)}} &=& -\frac{\hat{J}_{310}\hat{G}_{2210}
-\hat{J}_{210}\hat{D}_{210}-\hat{J}_{200}\hat{G}_{2221}}{\hat{J}_{210}\hat{G}_{2100}
-\hat{J}_{200}\hat{D}_{110}-\hat{J}_{100}\hat{G}_{2111}}\;, \label{G14} \\
\frac{a_{1011}^{(0)}}{a_{1111}^{(0)}} &=&- \frac{\hat{J}_{310}\hat{G}_{2100}
+\hat{J}_{200}\hat{G}_{1121}+\hat{J}_{100}\hat{D}_{210}}{\hat{J}_{210}\hat{G}_{2100}
-\hat{J}_{200}\hat{D}_{110}-\hat{J}_{100}\hat{G}_{2111}}\; , \\
\frac{a_{1110}^{(0)}}{a_{1111}^{(0)}} &=&-\frac{\hat{J}_{320}\hat{G}_{2100}
-\hat{J}_{310}\hat{D}_{110}-\hat{J}_{210}\hat{G}_{2111}}{\hat{J}_{210}\hat{G}_{2100}
-\hat{J}_{200}\hat{D}_{110}-\hat{J}_{100}\hat{G}_{2111}}\; , \label{G16}
\end{eqnarray}
and 
\begin{align}
\frac{1}{\left( a_{1111}^{(0)}\right) ^{2}}& =\frac{\hat{J}_{420}}{\hat{J}_{000}}
+\left( \frac{a_{1010}^{(0)}}{a_{1111}^{(0)}}\right) ^{2}+\frac{\hat{J}_{220}}{\hat{J}_{000}}
\left( \frac{a_{1011}^{(0)}}{a_{1111}^{(0)}}\right)^{2}+\frac{\hat{J}_{200}}{\hat{J}_{000}}
\left( \frac{a_{1110}^{(0)}}{a_{1111}^{(0)}}\right)^{2}  \notag \\
& +2\frac{\hat{J}_{210}}{\hat{J}_{000}}\, \frac{a_{1010}^{(0)}}{a_{1111}^{(0)}}
+2\frac{\hat{J}_{320}}{\hat{J}_{000}}\, \frac{a_{1011}^{(0)}}{a_{1111}^{(0)}}
+2\frac{\hat{J}_{310}}{\hat{J}_{000}}\, \frac{a_{1110}^{(0)}}{a_{1111}^{(0)}}  \notag \\
& +2\frac{\hat{J}_{100}}{\hat{J}_{000}}\, \frac{a_{1010}^{(0)}a_{1110}^{(0)}}{\left(a_{1111}^{(0)}\right)^2}
 +2\frac{\hat{J}_{110}}{\hat{J}_{000}}
\, \frac{a_{1010}^{(0)}a_{1011}^{(0)}}{\left(a_{1111}^{(0)}\right)^2}
+2\frac{\hat{J}_{210}}{\hat{J}_{000}}
\, \frac{a_{1011}^{(0)}a_{1110}^{(0)}}{\left(a_{1111}^{(0)}\right)^2}\; . \label{G17}
\end{align}
In this equation, we refrained from explicitly inserting the coefficients (\ref{G14}) -- (\ref{G16}), because the resulting
expression becomes too unwieldy. With Eqs.\ (\ref{G14}) -- (\ref{G17}), 
the polynomial
$P_{\mathbf{k}11}^{\left( 0\right) } =a_{1010}^{(0)}+a_{1011}^{(0)}E_{\mathbf{k}l}
+a_{1110}^{(0)}E_{\mathbf{k}u}+a_{1111}^{(0)}E_{\mathbf{k}u}E_{\mathbf{k}l}$ is uniquely determined.
\end{itemize}
We remark that the results for $m=0$ are formally similar to the polyomials 
$P_{\mathbf{k}n}^{\left( \ell \right) }$ with coefficients $a_{ni}^{(\ell)}$
from Eq.\ (\ref{P_n_polynomials}), see for example Eqs.\ (91) -- (99) of Ref.\ 
\cite{Denicol:2012es}. The reason is that the polynomials $P_{\mathbf{k}n}^{\left( \ell \right) }$ 
are the $m=0$ case of the more general multivariate polynomials $P_{\mathbf{k}n0}^{\left( \ell \right) }$.

Finally, with these results we can explicitly compute the $\hat{\mathcal{H}}_{\mathbf{k}nm}^{(\ell )}$
coefficients from Eq. (\ref{kinetic:H_nm}) for the cases relevant for the
14-moment approximation.

\section{Transport coefficients}
\label{transportcoefficients}

In this appendix, we list the expressions for the transport coefficients in Eqs.\ (\ref{rel_Pi}), (\ref{rel_n}), (\ref{rel_Pl}),
(\ref{rel_V}), (\ref{rel_W}), and (\ref{rel_pi}), as obtained in the 14-moment approximation.

\subsection{Bulk viscous pressure, Eq.\ (\ref{rel_Pi})}
\label{app_Pi}

Using Eq.\ (\ref{Jacobi}) the three bulk viscosity coefficients in Eq.\ (\ref{rel_Pi}) are
\begin{eqnarray}
\bar{\zeta}_l &=& \frac{m_0^2}{3} \left\{ \hat{\mathcal{I}}_{-2,2} - \hat{\mathcal{I}}_{00}
- \hat{n}_l \gamma_{-2,0,2,1}^{(0)} - \hat{M}\, \gamma_{-2,1,2,1}^{(0)}
+ \! \left[ \frac{\partial (\hat{e},\hat{n})}{\partial (\hat{\beta}_u,\hat{\alpha})} \right]^{-1}\!\! \left[
 \frac{\partial (\hat{\mathcal{I}}_{00}, \hat{n})}{\partial (\hat{\beta}_u,\hat{\alpha})} (\hat{e}+\hat{P}_l)
- \frac{\partial (\hat{\mathcal{I}}_{00}, \hat{e})}{\partial (\hat{\beta}_u,\hat{\alpha})}\, \hat{n} \right]\!
\right\}, \\
\bar{\zeta}_\perp &=& \frac{m_0^2}{6} \left\{ - m_0^2\, \hat{\mathcal{I}}_{-2,0} - \hat{\mathcal{I}}_{00} - 
\hat{\mathcal{I}}_{-2,2} + \hat{n}_l\, \left( m_0^2\, \gamma_{-2,0,0,1}^{(0)}+ \gamma_{-2,0,2,1}^{(0)}\right) 
+ \hat{M}\, \left( m_0^2\, \gamma_{-2,1,0,1}^{(0)} + \gamma_{-2,1,2,1}^{(0)} \right) \right. \notag \\
&    & \hspace*{1cm} +\left.  2\, \left[ \frac{\partial (\hat{e},\hat{n})}{\partial (\hat{\beta}_u,\hat{\alpha})} \right]^{-1}\!\! 
\left[ \frac{\partial (\hat{\mathcal{I}}_{00}, \hat{n})}{\partial (\hat{\beta}_u,\hat{\alpha})} (\hat{e}+\hat{P}_\perp)
- \frac{\partial (\hat{\mathcal{I}}_{00}, \hat{e})}{\partial (\hat{\beta}_u,\hat{\alpha})}\, \hat{n} \right]
\right\}\; ,  \\
\bar{\zeta}_{\perp l} & = & \frac{m_0^2}{3} \left(- \hat{\mathcal{I}}_{-1,1} + \hat{n}_l\, \gamma_{-1,0,1,1}^{(0)} 
+ \hat{M}\, \gamma_{-1,1,1,1}^{(0)} \right)\;.
\end{eqnarray}
The diffusion coefficients are
\begin{equation}
\bar{\kappa}^{\Pi}_{\alpha}  =   \frac{\partial \bar{\zeta}_{\perp l}}{\partial \hat{\alpha}} \;, \;\;\;
\bar{\kappa}^{\Pi}_{u}  =   \frac{\partial \bar{\zeta}_{\perp l}}{\partial \hat{\beta}_u} \;, \;\;\;
\bar{\kappa}^{\Pi}_{l}  =   \frac{\partial \bar{\zeta}_{\perp l}}{\partial \hat{\beta}_l} \;.
\end{equation}
The coefficient which couples the relaxation equation for the bulk viscous pressure to the
one for the variable $\hat{\beta}_l$ is
\begin{equation}
\bar{\tau}^{\Pi}_{l} 
\equiv \frac{m_0^2}{3} \left[ \frac{\partial (\hat{e},\hat{n})}{\partial (\hat{\beta}_u,\hat{\alpha})} \right]^{-1}
\frac{\partial (\mathcal{\hat{I}}_{00},\hat{e},\hat{n})}{\partial(\hat{\beta}_l,\hat{\beta}_u,\hat{\alpha})}\;,
\end{equation}
where the last factor is a generalization of Eq.\ (\ref{Jacobi}) to $(3 \times 3)$ matrices,
\begin{equation}
\frac{\partial (\mathcal{\hat{I}}_{00},\hat{e},\hat{n})}{\partial(\hat{\beta}_l,\hat{\beta}_u,\hat{\alpha})}
= \frac{\partial \mathcal{\hat{I}}_{00}}{\partial \hat{\alpha}}  \frac{\partial (\hat{e},\hat{n})}{\partial (\hat{\beta}_l,\hat{\beta}_u)}
- \frac{\partial \mathcal{\hat{I}}_{00}}{\partial \hat{\beta}_u}  \frac{\partial (\hat{e},\hat{n})}{\partial (\hat{\beta}_l,\hat{\alpha})}
+  \frac{\partial \mathcal{\hat{I}}_{00}}{\partial \hat{\beta}_l} \frac{\partial (\hat{e},\hat{n})}{\partial (\hat{\beta}_u,\hat{\alpha})}\; .
\end{equation}
Finally, the coefficients in front of the second-order terms are
\begin{eqnarray}
\bar{\beta}^{\Pi}_{\Pi}  & = & 1 - \gamma_{-2,0,2,0}^{(0)}\;, \;\; 
 \bar{\beta}^{\Pi}_{n}   = \frac{m_0^2}{3}\, \gamma_{-2,0,2,1}^{(0)}\;,   \\
 \bar{\delta}^{\Pi}_{\Pi}  & =& \frac{1}{2} \left\{ 1 + m_0^2 \, \gamma_{-2,0,0,0}^{(0)} + \gamma_{-2,0,2,0}^{(0)}
 + m_0^2 \left[ \frac{\partial (\hat{e},\hat{n})}{\partial (\hat{\beta}_u,\hat{\alpha})} \right]^{-1}
\frac{\partial (\hat{\mathcal{I}}_{00}, \hat{n})}{\partial (\hat{\beta}_u, \hat{\alpha})} \right\} \;, \\
 \bar{\delta}^{\Pi}_{n}  & = &  \frac{m_0^2}{6} \left( m_0^2 \, \gamma_{-2,0,0,1}^{(0)} +
\gamma_{-2,0,2,1}^{(0)} \right) \;,  \\  
\bar{\epsilon}^{\Pi}_{\Pi} & = & \gamma_{-1,0,1,0}^{(0)}\;, \;\;
\bar{\epsilon}^{\Pi}_{n} = \frac{m_0^2}{3} \left\{ \gamma_{-1,0,1,1}^{(0)} + \frac{\hat{e}+\hat{P}_\perp}{\hat{e}+
\hat{P}_l}  \left[ \frac{\partial (\hat{e},\hat{n})}{\partial (\hat{\beta}_u,\hat{\alpha})} \right]^{-1} 
 \frac{\partial (\hat{\mathcal{I}}_{00}, \hat{e})}{\partial (\hat{\beta}_u, \hat{\alpha})} \right\} \;, \\
\bar{\ell}^{\Pi}_{n} & = & \frac{m_0^2}{3} \left\{ \gamma_{-1,0,1,1}^{(0)} 
+  \left[ \frac{\partial (\hat{e},\hat{n})}{\partial (\hat{\beta}_u,\hat{\alpha})} \right]^{-1} 
 \frac{\partial (\hat{\mathcal{I}}_{00}, \hat{e})}{\partial (\hat{\beta}_u, \hat{\alpha})} \right\} \;, \\
\bar{\ell}^{\Pi}_{V} & = & \frac{m_0^2}{3} \left\{ \gamma_{-1,0,0,0}^{(1)} 
-  \left[ \frac{\partial (\hat{e},\hat{n})}{\partial (\hat{\beta}_u,\hat{\alpha})} \right]^{-1} 
 \frac{\partial (\hat{\mathcal{I}}_{00}, \hat{e})}{\partial (\hat{\beta}_u, \hat{\alpha})} \right\} \;, \;\;
 \bar{\ell}^{\Pi}_{W} = \frac{m_0^2}{3}  \gamma_{-1,0,0,1}^{(1)}  \;, \\
\bar{\tau}^{\Pi}_{\Pi \alpha} & = & \frac{\partial \bar{\epsilon}^{\Pi}_{\Pi}}{\partial \hat{\alpha}}\;, \;\;
 \bar{\tau}^{\Pi}_{\Pi u}   =  \frac{\partial \bar{\epsilon}^{\Pi}_{\Pi}}{\partial \hat{\beta}_u}\;, \;\;
 \bar{\tau}^{\Pi}_{\Pi l}   =  \frac{\partial \bar{\epsilon}^{\Pi}_{\Pi}}{\partial \hat{\beta}_l}\;, \\
\bar{\tau}^{\Pi}_{n \alpha}  & =  & \frac{m_0^2}{3} \left\{ \frac{\partial \gamma_{-1,0,1,1}^{(0)}}{\partial \hat{\alpha}}
- \frac{\partial \hat{P}_l/\partial \hat{\alpha}}{\hat{e} + \hat{P}_l}\,
\left[ \frac{\partial (\hat{e},\hat{n})}{\partial (\hat{\beta}_u,\hat{\alpha})} \right]^{-1} 
 \frac{\partial (\hat{\mathcal{I}}_{00}, \hat{e})}{\partial (\hat{\beta}_u, \hat{\alpha})} \right\} 
\;, \\
\bar{\tau}^{\Pi}_{n u}  & =  & \frac{m_0^2}{3} \left\{ \frac{\partial \gamma_{-1,0,1,1}^{(0)}}{\partial \hat{\beta}_u}
- \frac{\partial \hat{P}_l/\partial \hat{\beta}_u}{\hat{e} + \hat{P}_l}\,
\left[ \frac{\partial (\hat{e},\hat{n})}{\partial (\hat{\beta}_u,\hat{\alpha})} \right]^{-1} 
 \frac{\partial (\hat{\mathcal{I}}_{00}, \hat{e})}{\partial (\hat{\beta}_u, \hat{\alpha})} \right\} 
\;, \\
\bar{\tau}^{\Pi}_{n l}  & =  & \frac{m_0^2}{3} \left\{ \frac{\partial \gamma_{-1,0,1,1}^{(0)}}{\partial \hat{\beta}_l}
- \frac{\partial \hat{P}_l/\partial \hat{\beta}_l}{\hat{e} + \hat{P}_l}\,
\left[ \frac{\partial (\hat{e},\hat{n})}{\partial (\hat{\beta}_u,\hat{\alpha})} \right]^{-1} 
 \frac{\partial (\hat{\mathcal{I}}_{00}, \hat{e})}{\partial (\hat{\beta}_u, \hat{\alpha})} \right\} 
\;, \\
\bar{\tau}^{\Pi}_{V \alpha}  & =  & \frac{m_0^2}{3} \left\{ \frac{\partial \gamma_{-1,0,0,0}^{(1)}}{\partial \hat{\alpha}}
+ \frac{\partial \hat{P}_\perp/\partial \hat{\alpha}}{\hat{e} + \hat{P}_\perp}\,
\left[ \frac{\partial (\hat{e},\hat{n})}{\partial (\hat{\beta}_u,\hat{\alpha})} \right]^{-1} 
 \frac{\partial (\hat{\mathcal{I}}_{00}, \hat{e})}{\partial (\hat{\beta}_u, \hat{\alpha})} \right\} 
\;, \\
\bar{\tau}^{\Pi}_{V u}  & =  & \frac{m_0^2}{3} \left\{ \frac{\partial \gamma_{-1,0,0,0}^{(1)}}{\partial \hat{\beta}_u}
+ \frac{\partial \hat{P}_\perp/\partial \hat{\beta}_u}{\hat{e} + \hat{P}_\perp}\,
\left[ \frac{\partial (\hat{e},\hat{n})}{\partial (\hat{\beta}_u,\hat{\alpha})} \right]^{-1} 
 \frac{\partial (\hat{\mathcal{I}}_{00}, \hat{e})}{\partial (\hat{\beta}_u, \hat{\alpha})} \right\} 
\;, \\
\bar{\tau}^{\Pi}_{V l}  & =  & \frac{m_0^2}{3} \left\{ \frac{\partial \gamma_{-1,0,0,0}^{(1)}}{\partial \hat{\beta}_l}
+ \frac{\partial \hat{P}_\perp/\partial \hat{\beta}_l}{\hat{e} + \hat{P}_\perp}\,
\left[ \frac{\partial (\hat{e},\hat{n})}{\partial (\hat{\beta}_u,\hat{\alpha})} \right]^{-1} 
 \frac{\partial (\hat{\mathcal{I}}_{00}, \hat{e})}{\partial (\hat{\beta}_u, \hat{\alpha})} \right\} 
\;, \\
\bar{\tau}^{\Pi}_{W \alpha} & = & \frac{\partial \bar{\ell}^{\Pi}_{W}}{\partial \hat{\alpha}}\;, \;\;
\bar{\tau}^{\Pi}_{W u}  =  \frac{\partial \bar{\ell}^{\Pi}_{W}}{\partial \hat{\beta}_u}\;, \;\;
\bar{\tau}^{\Pi}_{W l}  =  \frac{\partial \bar{\ell}^{\Pi}_{W}}{\partial \hat{\beta}_l}\;, \\
\bar{\lambda}^{\Pi}_{V} &= &  \frac{m_0^2}{3}\, \gamma_{-2,0,1,0}^{(1)}\;, \;\; 
\bar{\lambda}^{\Pi}_{ W} =  \frac{m_0^2}{3}\left\{ \gamma_{-2,0,1,1}^{(1)}
- \left[ \frac{\partial (\hat{e},\hat{n})}{\partial (\hat{\beta}_u,\hat{\alpha})} \right]^{-1} 
 \frac{\partial (\hat{\mathcal{I}}_{00}, \hat{n})}{\partial (\hat{\beta}_u, \hat{\alpha})} \right\}\;,  \\
\bar{\lambda}^{\Pi}_{V l} & = & \frac{m_0^2}{3}\left\{ \gamma_{-1,0,0,0}^{(1)}
- \frac{\hat{e}+ \hat{P}_l}{\hat{e}+ \hat{P}_\perp}
\left[ \frac{\partial (\hat{e},\hat{n})}{\partial (\hat{\beta}_u,\hat{\alpha})} \right]^{-1} 
 \frac{\partial (\hat{\mathcal{I}}_{00}, \hat{e})}{\partial (\hat{\beta}_u, \hat{\alpha})} \right\}\;, \\
\bar{\lambda}^{\Pi}_{\pi} & = & \frac{m_0^2}{3} \left\{ 2\, \gamma_{-2,0,0,0}^{(2)} + 
\left[ \frac{\partial (\hat{e},\hat{n})}{\partial (\hat{\beta}_u,\hat{\alpha})} \right]^{-1} 
 \frac{\partial (\hat{\mathcal{I}}_{00}, \hat{n})}{\partial (\hat{\beta}_u, \hat{\alpha})} \right\}\;.
\end{eqnarray}

\subsection{Diffusion current in $l^\mu$-direction, Eq.\ (\ref{rel_n})}
\label{app_n}

The diffusion coefficients in Eq.\ (\ref{rel_n}) are
\begin{eqnarray}
\bar{\kappa}^{n}_{\alpha} &=&  \frac{\partial \hat{\mathcal{I}}_{-1,2}}{\partial \hat{\alpha}}  -
\frac{\hat{n}}{\hat{e} + \hat{P}_l}\, \frac{\partial \hat{P}_l}{\partial \hat{\alpha}} -
\frac{\partial}{\partial \hat{\alpha}} \left( \hat{n}_l \, \gamma_{-1,0,2,1}^{(0)} + \hat{M}\, \gamma_{-1,1,2,1}^{(0)} \right)
\;, \\
\bar{\kappa}^{n}_{u} &=& \frac{\partial \hat{\mathcal{I}}_{-1,2}}{\partial \hat{\beta}_u}  -
\frac{\hat{n}}{\hat{e} + \hat{P}_l}\, \frac{\partial \hat{P}_l}{\partial \hat{\beta}_u} -
\frac{\partial}{\partial \hat{\beta}_u} \left( \hat{n}_l \, \gamma_{-1,0,2,1}^{(0)} + \hat{M}\, \gamma_{-1,1,2,1}^{(0)} \right)
\;, \\
\bar{\kappa}^{n}_{l} &=& \frac{\partial \hat{\mathcal{I}}_{-1,2}}{\partial \hat{\beta}_l}  -
\frac{\hat{n}}{\hat{e} + \hat{P}_l}\, \frac{\partial \hat{P}_l}{\partial \hat{\beta}_l} -
\frac{\partial}{\partial \hat{\beta}_l} \left( \hat{n}_l \, \gamma_{-1,0,2,1}^{(0)} + \hat{M}\, \gamma_{-1,1,2,1}^{(0)} \right)
\;. 
\end{eqnarray}
The transport coefficients that couple the evolution of $n_l$ to the expansion scalars are
\begin{eqnarray}
\bar{\zeta}^{n}_{l}  & = &  \hat{\mathcal{I}}_{-2,3} - \hat{n}_l \, \gamma_{-2,0,3,1}^{(0)} - \hat{M}\, \gamma_{-2,1,3,1}^{(0)} 
\;, \\
\bar{\zeta}^{n}_{\perp}  & = & \frac{1}{2} \left[ m_0^2\, \hat{\mathcal{I}}_{-2,1} + \hat{\mathcal{I}}_{-2,3}
-   \hat{n}_l \left( m_0^2\, \gamma_{-2,0,1,1}^{(0)} + \gamma_{-2,0,3,1}^{(0)} \right)
-   \hat{M}\left( m_0^2\, \gamma_{-2,1,1,1}^{(0)} + \gamma_{-2,1,3,1}^{(0)} \right) \right] \;, \\
\bar{\zeta}^{n}_{\perp l}  & = &  \frac{1}{2} \left[  
m_0^2\, \hat{\mathcal{I}}_{-1,0} - \hat{n}  + 3\, \hat{\mathcal{I}}_{-1,2}
-   \hat{n}_l \left( m_0^2\, \gamma_{-1,0,0,1}^{(0)} + 3\, \gamma_{-1,0,2,1}^{(0)} \right)
-   \hat{M}\left( m_0^2\, \gamma_{-1,1,0,1}^{(0)} + 3\, \gamma_{-1,1,2,1}^{(0)} \right) \right] \notag \\
&   & + \frac{\hat{n} (\hat{P}_\perp - \hat{P}_l)}{\hat{e} + \hat{P}_l} \;.
\end{eqnarray}
The second-order transport coefficients are
\begin{eqnarray}
\bar{\beta}^{n}_{\Pi}  & = & \frac{3}{m_0^2}\, \gamma_{-2,0,3,0}^{(0)}\;, \;\; 
 \bar{\beta}^{n}_{n}   = 2 -  \gamma_{-2,0,3,1}^{(0)}\;,   \\
 \bar{\delta}^{n}_{\Pi}  & =& \frac{3}{2 m_0^2} \left( m_0^2 \, \gamma_{-2,0,1,0}^{(0)} + \gamma_{-2,0,3,0}^{(0)} \right)
\;, \;\;  \bar{\delta}^{n}_{n}   =   \frac{1}{2} \left( 1+ m_0^2 \, \gamma_{-2,0,1,1}^{(0)} +
\gamma_{-2,0,3,1}^{(0)} \right) \;,  \\  
\bar{\epsilon}^{n}_{\Pi} & = & \frac{3}{2}\left[ \frac{\hat{n}}{\hat{e}+ \hat{P}_l}
- \frac{1}{m_0^2} \left( m_0^2\, \gamma_{-1,0,0,0}^{(0)} + 3\, \gamma_{-1,0,2,0}^{(0)} \right) \right] \;, \;\;
\bar{\epsilon}^{n}_{n} = \frac{1}{2} \left( m_0^2\, \gamma_{-1,0,0,1}^{(0)} + 3\, \gamma_{-1,0,2,1}^{(0)} \right) \;, \\
\bar{\ell}^{n}_{\Pi} & = & \frac{3}{m_0^2} \, \gamma_{-1,0,2,0}^{(0)}  \;, \;\;
\bar{\ell}^{n}_{n} = \gamma_{-1,0,2,1}^{(0)}\;, \;\; 
\bar{\ell}^{n}_{V}  =   \gamma_{-1,0,1,0}^{(1)} \; , \;\;
\bar{\ell}^{n}_{W} =  \gamma_{-1,0,1,1}^{(1)} +  \frac{\hat{n}}{\hat{e}+ \hat{P}_l} \;, \\
\bar{\tau}^{n}_{\Pi \alpha} & = & \frac{\partial \bar{\ell}^{n}_{\Pi}}{\partial \hat{\alpha}}\;, \;\;
 \bar{\tau}^{n}_{\Pi u}   =  \frac{\partial \bar{\ell}^{n}_{\Pi}}{\partial \hat{\beta}_u}\;, \;\;
 \bar{\tau}^{n}_{\Pi l}   =  \frac{\partial \bar{\ell}^{n}_{\Pi}}{\partial \hat{\beta}_l}\;, \;\;
 \bar{\tau}^{n}_{n \alpha}   =    \frac{\partial \bar{\ell}^{n}_{n}}{\partial \hat{\alpha}}\;, \;\;
 \bar{\tau}^{n}_{n u}   =  \frac{\partial \bar{\ell}^{n}_{n}}{\partial \hat{\beta}_u}\;, \;\;
 \bar{\tau}^{n}_{n l}   =  \frac{\partial \bar{\ell}^{n}_{n}}{\partial \hat{\beta}_l}\;, \\
\bar{\tau}^{n}_{V \alpha}  & =  &  \frac{\partial \bar{\ell}^{n}_{V}}{\partial \hat{\alpha}}\;, \;\;
 \bar{\tau}^{n}_{V u}   =  \frac{\partial \bar{\ell}^{n}_{V}}{\partial \hat{\beta}_u}\;, \;\;
 \bar{\tau}^{n}_{V l}   =  \frac{\partial \bar{\ell}^{n}_{V}}{\partial \hat{\beta}_l}\;, \;\;
\bar{\tau}^{n}_{W \alpha}  =  \frac{\partial \gamma_{-1,0,1,1}^{(1)}}{\partial \hat{\alpha}}
- \frac{\hat{n}}{(\hat{e} + \hat{P}_l)(\hat{e}+\hat{P}_\perp)}\, \frac{\partial \hat{P}_\perp}{\partial \hat{\alpha}}\;, \\
\bar{\tau}^{n}_{W u}  & = & \frac{\partial \gamma_{-1,0,1,1}^{(1)}}{\partial \hat{\beta}_u}
- \frac{\hat{n}}{(\hat{e} + \hat{P}_l)(\hat{e}+\hat{P}_\perp)}\, \frac{\partial \hat{P}_\perp}{\partial \hat{\beta}_u}\;, \;\;
\bar{\tau}^{n}_{W l}  =  \frac{\partial \gamma_{-1,0,1,1}^{(1)}}{\partial \hat{\beta}_l}
- \frac{\hat{n}}{(\hat{e} + \hat{P}_l)(\hat{e}+\hat{P}_\perp)}\, \frac{\partial \hat{P}_\perp}{\partial \hat{\beta}_l}\;, \\
\bar{\lambda}^{n}_{V u} &= &  \gamma_{-2,0,2,0}^{(1)}\;, \;\; 
\bar{\lambda}^{n}_{W u} =  \gamma_{-2,0,2,1}^{(1)}\;, \;\;  
\bar{\lambda}^{n}_{V \perp} = 1+ \bar{\lambda}^{n}_{Vu}\;, \;\;
\bar{\lambda}^{n}_{W \perp} = \bar{\lambda}^{n}_{Wu} - \frac{\hat{n}}{\hat{e}+ \hat{P}_l}\;, \\
\bar{\lambda}^{n}_{W l} & = & 2\, \bar{\ell}^{n}_{W} - \frac{\hat{n}}{\hat{e}+ \hat{P}_l}\, \frac{\hat{P}_\perp- \hat{P}_l}{
\hat{e}+ \hat{P}_\perp}\;, \;\;
\bar{\lambda}^{n}_{\pi}  =   2\, \gamma_{-2,0,1,0}^{(2)} \;, \;\;
\bar{\lambda}^{n}_{\pi l} = 2\, \gamma_{-1,0,0,0}^{(2)} - \frac{\hat{n}}{\hat{e}+ \hat{P}_l}\;.
\end{eqnarray}

\subsection{Longitudinal pressure, Eq.\ (\ref{rel_Pl})}
\label{app_Pl}

The first-order diffusion coefficients in Eq.\ (\ref{rel_Pl}) are
\begin{eqnarray}
\bar{\kappa}^{l}_{\alpha} &=&  \frac{\partial \hat{\mathcal{I}}_{-1,3}}{\partial \hat{\alpha}}  -
\frac{\partial}{\partial \hat{\alpha}} \left( \hat{n}_l \, \gamma_{-1,0,3,1}^{(0)} + \hat{M}\, \gamma_{-1,1,3,1}^{(0)} \right)
\;, \\
\bar{\kappa}^{l}_{u} &=& \frac{\partial \hat{\mathcal{I}}_{-1,3}}{\partial \hat{\beta}_u}  -
\frac{\partial}{\partial \hat{\beta}_u} \left( \hat{n}_l \, \gamma_{-1,0,3,1}^{(0)} + \hat{M}\, \gamma_{-1,1,3,1}^{(0)} \right)
\;, \\
\bar{\kappa}^{l}_{l} &=& \frac{\partial \hat{\mathcal{I}}_{-1,3}}{\partial \hat{\beta}_l}  -
\frac{\partial}{\partial \hat{\beta}_l} \left( \hat{n}_l \, \gamma_{-1,0,3,1}^{(0)} + \hat{M}\, \gamma_{-1,1,3,1}^{(0)} \right)
\;, 
\end{eqnarray}
while the viscosity coefficients are
\begin{eqnarray}
\bar{\zeta}^{l}_{l}  & = &  \hat{\mathcal{I}}_{-2,4} - 3\,\hat{P}_l- \hat{n}_l \, \gamma_{-2,0,4,1}^{(0)} 
- \hat{M}\, \gamma_{-2,1,4,1}^{(0)} 
\;, \\
\bar{\zeta}^{l}_{\perp}  & = & \frac{1}{2} \left[ m_0^2\, \hat{\mathcal{I}}_{-2,2} + \hat{P}_l + \hat{\mathcal{I}}_{-2,4}
-   \hat{n}_l \left( m_0^2\, \gamma_{-2,0,2,1}^{(0)} + \gamma_{-2,0,4,1}^{(0)} \right)
-   \hat{M}\left( m_0^2\, \gamma_{-2,1,2,1}^{(0)} + \gamma_{-2,1,4,1}^{(0)} \right) \right] \;, \\
\bar{\zeta}^{l}_{\perp l}  & = &  m_0^2\, \hat{\mathcal{I}}_{-1,1}  + 2\, \hat{\mathcal{I}}_{-1,3}
-   \hat{n}_l \left( m_0^2\, \gamma_{-1,0,1,1}^{(0)} + 2\, \gamma_{-1,0,3,1}^{(0)} \right)
-   \hat{M}\left( m_0^2\, \gamma_{-1,1,1,1}^{(0)} + 2\, \gamma_{-1,1,3,1}^{(0)} \right) \;.
\end{eqnarray}
The second-order transport coefficients are
\begin{eqnarray}
\bar{\beta}^{l}_{\Pi}  & = & \frac{3}{m_0^2}\, \gamma_{-2,0,4,0}^{(0)}\;, \;\; 
 \bar{\beta}^{l}_{n}   =  \gamma_{-2,0,4,1}^{(0)}\;,   \\
 \bar{\delta}^{l}_{\Pi}  & =& \frac{3}{2 m_0^2} \left( m_0^2 \, \gamma_{-2,0,2,0}^{(0)} + \gamma_{-2,0,4,0}^{(0)} \right)
\;, \;\;  \bar{\delta}^{l}_{n}   =   \frac{1}{2} \left( m_0^2 \, \gamma_{-2,0,2,1}^{(0)} +
\gamma_{-2,0,4,1}^{(0)} \right) \;,  \\  
\bar{\epsilon}^{l}_{\Pi} & = & \frac{3}{m_0^2}
\left( m_0^2\, \gamma_{-1,0,1,0}^{(0)} + 2\, \gamma_{-1,0,3,0}^{(0)} \right) \;, \;\;
\bar{\epsilon}^{l}_{n} =  m_0^2\, \gamma_{-1,0,1,1}^{(0)} + 2\, \gamma_{-1,0,3,1}^{(0)}  \;, \\
\bar{\ell}^{l}_{\Pi} & = & \frac{3}{m_0^2} \, \gamma_{-1,0,3,0}^{(0)}  \;, \;\;
\bar{\ell}^{l}_{n} = \gamma_{-1,0,3,1}^{(0)}\;, \;\; 
\bar{\ell}^{l}_{V}  =   \gamma_{-1,0,2,0}^{(1)} \; , \;\;
\bar{\ell}^{l}_{W} =  \gamma_{-1,0,2,1}^{(1)} \;, \\
\bar{\tau}^{l}_{\Pi \alpha} & = & \frac{\partial \bar{\ell}^{l}_{\Pi}}{\partial \hat{\alpha}}\;, \;\;
 \bar{\tau}^{l}_{\Pi u}   =  \frac{\partial \bar{\ell}^{l}_{\Pi}}{\partial \hat{\beta}_u}\;, \;\;
 \bar{\tau}^{l}_{\Pi l}   =  \frac{\partial \bar{\ell}^{l}_{\Pi}}{\partial \hat{\beta}_l}\;, \;\;
 \bar{\tau}^{l}_{n \alpha}   =    \frac{\partial \bar{\ell}^{l}_{n}}{\partial \hat{\alpha}}\;, \;\;
 \bar{\tau}^{l}_{n u}   =  \frac{\partial \bar{\ell}^{l}_{n}}{\partial \hat{\beta}_u}\;, \;\;
 \bar{\tau}^{l}_{n l}   =  \frac{\partial \bar{\ell}^{l}_{n}}{\partial \hat{\beta}_l}\;, \\
\bar{\tau}^{l}_{V \alpha}  & =  &  \frac{\partial \bar{\ell}^{l}_{V}}{\partial \hat{\alpha}}\;, \;\;
 \bar{\tau}^{l}_{V u}   =  \frac{\partial \bar{\ell}^{l}_{V}}{\partial \hat{\beta}_u}\;, \;\;
 \bar{\tau}^{l}_{V l}   =  \frac{\partial \bar{\ell}^{l}_{V}}{\partial \hat{\beta}_l}\;, \;\;
\bar{\tau}^{l}_{W \alpha}  =  \frac{\partial \bar{\ell}^{l}_{W}}{\partial \hat{\alpha}}\;, \;\;
\bar{\tau}^{l}_{W u}  =  \frac{\partial \bar{\ell}^{l}_{W}}{\partial \hat{\beta}_u}\;, \;\;
\bar{\tau}^{l}_{W l}  =  \frac{\partial \bar{\ell}^{l}_{W}}{\partial \hat{\beta}_l}\;,  \\
\bar{\lambda}^{l}_{V u} &= &  \gamma_{-2,0,3,0}^{(1)}\;, \;\; 
\bar{\lambda}^{l}_{W u} =  \gamma_{-2,0,3,1}^{(1)}\;, \;\;  
\bar{\lambda}^{l}_{W \perp} = 2+\bar{\lambda}^{l}_{Wu}\;, \;\;
\bar{\lambda}^{l}_{\pi}   =   2\, \gamma_{-2,0,2,0}^{(2)} \;, \;\;
\bar{\lambda}^{l}_{\pi l} = 4\, \gamma_{-1,0,1,0}^{(2)}\;.
\end{eqnarray}

\subsection{Diffusion current in the direction perpendicular to $u^\mu$ and $l^\mu$, Eq.\ (\ref{rel_V})}
\label{app_V}

The diffusion coefficients in Eq.\ (\ref{rel_V}) are
\begin{eqnarray}
\bar{\kappa}_{\alpha} &=&  \frac{1}{2} \frac{\partial}{\partial \hat{\alpha}}  
\left[- m_0^2\, \hat{\mathcal{I}}_{-1,0} + \hat{n} - \hat{\mathcal{I}}_{-1,2} 
+ \hat{n}_l \left( m_0^2\, \gamma_{-1,0,0,1}^{(0)} + \gamma_{-1,0,2,1}^{(0)} \right)
+ \hat{M} \left( m_0^2\, \gamma_{-1,1,0,1}^{(0)} + \gamma_{-1,1,2,1}^{(0)} \right)\right] \notag \\
&    & - \frac{\hat{n}}{\hat{e}+\hat{P}_\perp}\, \frac{\partial \hat{P}_\perp }{\partial \hat{\alpha}}
\;, \\
\bar{\kappa}_{u} &=&  \frac{1}{2} \frac{\partial}{\partial \hat{\beta}_u}  
\left[- m_0^2\, \hat{\mathcal{I}}_{-1,0} + \hat{n} - \hat{\mathcal{I}}_{-1,2} 
+ \hat{n}_l \left( m_0^2\, \gamma_{-1,0,0,1}^{(0)} + \gamma_{-1,0,2,1}^{(0)} \right)
+ \hat{M} \left( m_0^2\, \gamma_{-1,1,0,1}^{(0)} + \gamma_{-1,1,2,1}^{(0)} \right)\right] \notag \\
&    & - \frac{\hat{n}}{\hat{e}+\hat{P}_\perp}\, \frac{\partial \hat{P}_\perp }{\partial \hat{\beta}_u}
\;, \\
\bar{\kappa}_{l} &=&  \frac{1}{2} \frac{\partial}{\partial \hat{\beta}_l}  
\left[-m_0^2\, \hat{\mathcal{I}}_{-1,0} + \hat{n} - \hat{\mathcal{I}}_{-1,2} 
+ \hat{n}_l \left( m_0^2\, \gamma_{-1,0,0,1}^{(0)} + \gamma_{-1,0,2,1}^{(0)} \right)
+ \hat{M} \left( m_0^2\, \gamma_{-1,1,0,1}^{(0)} + \gamma_{-1,1,2,1}^{(0)} \right)\right] \notag \\
&    & - \frac{\hat{n}}{\hat{e}+\hat{P}_\perp}\, \frac{\partial \hat{P}_\perp }{\partial \hat{\beta}_l}
\;, 
\end{eqnarray}
while the viscosity coefficients are
\begin{eqnarray}
\bar{\zeta}^{V}_{u}  & = &   \frac{1}{2} 
\left[m_0^2\, \hat{\mathcal{I}}_{-2,1}  + \hat{\mathcal{I}}_{-2,3} 
- \hat{n}_l \left( m_0^2\, \gamma_{-2,0,1,1}^{(0)} + \gamma_{-2,0,3,1}^{(0)} \right)
- \hat{M} \left( m_0^2\, \gamma_{-2,1,1,1}^{(0)} + \gamma_{-2,1,3,1}^{(0)} \right)\right] 
\;, \\
\bar{\zeta}^{V}_{l}  & = & \frac{1}{2} \left[ m_0^2\, \hat{\mathcal{I}}_{-1,0} - \hat{n} + 3 \hat{\mathcal{I}}_{-1,2}
-   \hat{n}_l \left( m_0^2\, \gamma_{-1,0,0,1}^{(0)} + 3 \gamma_{-1,0,2,1}^{(0)} \right)
-   \hat{M}\left( m_0^2\, \gamma_{-1,1,0,1}^{(0)} + 3\gamma_{-1,1,2,1}^{(0)} \right) \right] \notag \\
&    & + \frac{\hat{n}}{\hat{e}+\hat{P}_\perp} (\hat{P}_\perp - \hat{P}_l) \;.
\end{eqnarray}
The second-order transport coefficients are
\begin{eqnarray}
\bar{\beta}^{V}_{\Pi}  & = & \frac{3}{2m_0^2}\left( m_0^2 \gamma_{-2,0,1,0}^{(0)} + \gamma_{-2,0,3,0}^{(0)} \right)\;, \;\; 
 \bar{\beta}^{V}_{\Pi l}   =  \frac{3}{2m_0^2}\left( m_0^2 \gamma_{-1,0,0,0}^{(0)} + 3 \gamma_{-1,0,2,0}^{(0)} \right) 
 - \frac{3}{2}\,\frac{\hat{n}}{\hat{e}+\hat{P}_\perp}\,
\frac{\hat{e}+\hat{P}_l}{\hat{e}+\hat{P}_\perp} \;,   \\
\bar{\beta}^{V}_{n}  & = & \frac{1}{2}\left( 1+ m_0^2 \gamma_{-2,0,1,1}^{(0)} + \gamma_{-2,0,3,1}^{(0)} \right)\;, \;\;
\bar{\beta}^V_{n \perp} = 1 -  \bar{\beta}^{V}_{n}\;, \;\; 
 \bar{\beta}^{V}_{n l}   =  \frac{1}{2}\left( m_0^2 \gamma_{-1,0,0,1}^{(0)} + 3 \gamma_{-1,0,2,1}^{(0)} \right)  \;,   \\
 \bar{\delta}^{V}_{V}  & =& \frac{1}{2} \left( m_0^2 \, \gamma_{-2,0,0,0}^{(1)} + \gamma_{-2,0,2,0}^{(1)} \right) -1
\;, \;\;  \bar{\delta}^{V}_{W}   =   \frac{1}{2} \left( m_0^2 \, \gamma_{-2,0,0,1}^{(1)} +\gamma_{-2,0,2,1}^{(1)} \right) \;,  \\  
\bar{\epsilon}^{V}_{V} & = & \frac{3}{2}\,  \gamma_{-1,0,1,0}^{(1)}  \;, \;\;
\bar{\epsilon}^{V}_{W} =  \frac{3}{2}\, \gamma_{-1,0,1,1}^{(1)} + \frac{1}{2}\,\frac{\hat{n}}{\hat{e}+\hat{P}_\perp}
+\frac{\hat{n}}{\hat{e}+\hat{P}_l}  \;, \\
\bar{\ell}^{V}_{\Pi} & = & \frac{3}{2 m_0^2} \left( m_0^2\, \gamma_{-1,0,0,0}^{(0)} + \gamma_{-1,0,2,0}^{(0)}
\right) - \frac{3}{2}\,\frac{\hat{n}}{\hat{e}+\hat{P}_\perp}   \;, \;\;
\bar{\ell}^{V}_{n} = \frac{1}{2} \left( m_0^2\, \gamma_{-1,0,0,1}^{(0)} + \gamma_{-1,0,2,1}^{(0)}\right)\;, \\
\bar{\ell}^{V}_{V}  & = & \frac{2}{3}\, \bar{\epsilon}^V_V \; , \;\;
\bar{\ell}^{V}_{W} =  \gamma_{-1,0,1,1}^{(1)} + \frac{\hat{n}}{\hat{e}+\hat{P}_\perp}   \;, \;\;
\bar{\ell}^{V}_{\pi} =  2 \gamma_{-1,0,0,0}^{(2)} - \frac{\hat{n}}{\hat{e}+\hat{P}_\perp} \;, \\
\bar{\tau}^{V}_{\Pi \alpha} & = & \frac{3}{2 m_0^2}\, \frac{\partial}{\partial \hat{\alpha}}
\left( m_0^2\, \gamma_{-1,0,0,0}^{(0)} + \gamma_{-1,0,2,0}^{(0)} \right) + \frac{3}{2}\, \frac{\hat{n}}{(\hat{e}+
\hat{P}_\perp)^2}\, \frac{\partial \hat{P}_\perp}{\partial \hat{\alpha}}\;, \\
\bar{\tau}^{V}_{\Pi u} & = & \frac{3}{2 m_0^2}\, \frac{\partial}{\partial \hat{\beta}_u}
\left( m_0^2\, \gamma_{-1,0,0,0}^{(0)} + \gamma_{-1,0,2,0}^{(0)} \right) + \frac{3}{2}\, \frac{\hat{n}}{(\hat{e}+
\hat{P}_\perp)^2}\, \frac{\partial \hat{P}_\perp}{\partial \hat{\beta}_u}\;, \\
\bar{\tau}^{V}_{\Pi l} & = & \frac{3}{2 m_0^2}\, \frac{\partial}{\partial \hat{\beta}_l}
\left( m_0^2\, \gamma_{-1,0,0,0}^{(0)} + \gamma_{-1,0,2,0}^{(0)} \right) + \frac{3}{2}\, \frac{\hat{n}}{(\hat{e}+
\hat{P}_\perp)^2}\, \frac{\partial \hat{P}_\perp}{\partial \hat{\beta}_l}\;, \\
\bar{\tau}^{V}_{n \alpha}  &  =  &  \frac{\partial \bar{\ell}^{V}_{n}}{\partial \hat{\alpha}}\;, \;\;
 \bar{\tau}^{V}_{n u}   =  \frac{\partial \bar{\ell}^{V}_{n}}{\partial \hat{\beta}_u}\;, \;\;
 \bar{\tau}^{V}_{n l}   =  \frac{\partial \bar{\ell}^{V}_{n}}{\partial \hat{\beta}_l}\;, \;\;
\bar{\tau}^{V}_{V \alpha}   =    \frac{\partial \bar{\ell}^{V}_{V}}{\partial \hat{\alpha}}\;, \;\;
 \bar{\tau}^{V}_{V u}   =  \frac{\partial \bar{\ell}^{V}_{V}}{\partial \hat{\beta}_u}\;, \;\;
 \bar{\tau}^{V}_{V l}   =  \frac{\partial \bar{\ell}^{V}_{V}}{\partial \hat{\beta}_l}\;, \\
\bar{\tau}^{V}_{W \alpha} &  =  & \frac{\partial \gamma_{-1,0,1,1}^{(1)}}{\partial \hat{\alpha}}
-\frac{\hat{n}}{\hat{e}+\hat{P}_\perp}\, \frac{1}{\hat{e}+\hat{P}_l}\, \frac{\partial \hat{P}_l}{\partial \hat{\alpha}}\; \;\;
\bar{\tau}^{V}_{W u}   =   \frac{\partial \gamma_{-1,0,1,1}^{(1)}}{\partial \hat{\beta}_u}
-\frac{\hat{n}}{\hat{e}+\hat{P}_\perp}\, \frac{1}{\hat{e}+\hat{P}_l}\, \frac{\partial \hat{P}_l}{\partial \hat{\beta}_u}\;, \\
\bar{\tau}^{V}_{W l} &  =  & \frac{\partial \gamma_{-1,0,1,1}^{(1)}}{\partial \hat{\beta}_l}
-\frac{\hat{n}}{\hat{e}+\hat{P}_\perp}\, \frac{1}{\hat{e}+\hat{P}_l}\, \frac{\partial \hat{P}_l}{\partial \hat{\beta}_l}\;,\;\; 
\bar{\lambda}^{V}_{V u}   = 1 + \gamma_{-2,0,2,0}^{(1)}\;, \;\;
\bar{\lambda}^{V}_{W u}  =  \gamma_{-2,0,2,1}^{(1)}\;, \;\; 
\bar{\lambda}^{V}_{V \perp} =  \frac{1}{2} + \bar{\delta}^V_V \;, 
\end{eqnarray}
\begin{eqnarray} 
\bar{\tau}^{V}_{\pi \alpha} &  =  & 2\, \frac{\partial \gamma_{-1,0,0,0}^{(2)}}{\partial \hat{\alpha}}
+\frac{\hat{n}}{(\hat{e}+\hat{P}_\perp)^2}\,  \frac{\partial \hat{P}_\perp}{\partial \hat{\alpha}}\; \;\;
\bar{\tau}^{V}_{\pi u}   =   2\, \frac{\partial \gamma_{-1,0,0,0}^{(2)}}{\partial \hat{\beta}_u}
+\frac{\hat{n}}{(\hat{e}+\hat{P}_\perp)^2}\,  \frac{\partial \hat{P}_\perp}{\partial \hat{\beta}_u}\;, \\
\bar{\tau}^{V}_{\pi l} &  =  & 2\, \frac{\partial \gamma_{-1,0,0,0}^{(2)}}{\partial \hat{\beta}_l}
+\frac{\hat{n}}{(\hat{e}+\hat{P}_\perp)^2}\,  \frac{\partial \hat{P}_\perp}{\partial \hat{\beta}_l}\;,\;\; 
\bar{\lambda}^{V}_{\pi u}   = 2\, \gamma_{-2,0,1,0}^{(2)}\;, \;\;
\bar{\lambda}^{V}_{\pi l}  = 2\, \gamma_{-1,0,0,0}^{(2)} - \frac{\hat{n}}{\hat{e}+\hat{P}_\perp}\,
\frac{\hat{e}+\hat{P}_l}{\hat{e}+\hat{P}_\perp}\;.
\end{eqnarray}

\subsection{Shear-tensor component in $l^\mu$-direction, Eq.\ (\ref{rel_W})}
\label{app_W}

The diffusion coefficients appearing in Eq.\ (\ref{rel_W}) are
\begin{eqnarray}
\bar{\kappa}^W_{\alpha} &=&  \frac{1}{2} \frac{\partial}{\partial \hat{\alpha}}  
\left[m_0^2\, \hat{\mathcal{I}}_{-1,1}  + \hat{\mathcal{I}}_{-1,3} 
- \hat{n}_l \left( m_0^2\, \gamma_{-1,0,1,1}^{(0)} + \gamma_{-1,0,3,1}^{(0)} \right)
- \hat{M} \left( m_0^2\, \gamma_{-1,1,1,1}^{(0)} + \gamma_{-1,1,3,1}^{(0)} \right)\right] 
\;, \\
\bar{\kappa}^W_{u} &=&  \frac{1}{2} \frac{\partial}{\partial \hat{\beta}_u}  
\left[m_0^2\, \hat{\mathcal{I}}_{-1,1}  + \hat{\mathcal{I}}_{-1,3} 
- \hat{n}_l \left( m_0^2\, \gamma_{-1,0,1,1}^{(0)} + \gamma_{-1,0,3,1}^{(0)} \right)
- \hat{M} \left( m_0^2\, \gamma_{-1,1,1,1}^{(0)} + \gamma_{-1,1,3,1}^{(0)} \right)\right] 
\;, \\
\bar{\kappa}^W_{l} &=&  \frac{1}{2} \frac{\partial}{\partial \hat{\beta}_l}  
\left[m_0^2\, \hat{\mathcal{I}}_{-1,1}  + \hat{\mathcal{I}}_{-1,3} 
- \hat{n}_l \left( m_0^2\, \gamma_{-1,0,1,1}^{(0)} + \gamma_{-1,0,3,1}^{(0)} \right)
- \hat{M} \left( m_0^2\, \gamma_{-1,1,1,1}^{(0)} + \gamma_{-1,1,3,1}^{(0)} \right)\right] 
\;, 
\end{eqnarray}
while the viscosity coefficients are
\begin{eqnarray}
\bar{\eta}^{W}_{u}  & = &   \frac{1}{4} 
\left[ m_0^2\, \hat{\mathcal{I}}_{-2,2}  + \hat{P}_l +\hat{\mathcal{I}}_{-2,4} 
- \hat{n}_l \left( m_0^2\, \gamma_{-2,0,2,1}^{(0)} + \gamma_{-2,0,4,1}^{(0)} \right)
- \hat{M} \left( m_0^2\, \gamma_{-2,1,2,1}^{(0)} + \gamma_{-2,1,4,1}^{(0)} \right)\right] 
\;, \\
\bar{\eta}^{W}_{\perp}  & = &   \bar{\eta}^W_u - \frac{1}{4} \left( m_0^2\,  \hat{\mathcal{I}}_{00} - \hat{e}
+3 \hat{P}_l \right)  \;, \\
\bar{\eta}^{W}_{l}  & = &  \frac{1}{2} 
\left[ m_0^2\, \hat{\mathcal{I}}_{-1,1}  + 2 \,\hat{\mathcal{I}}_{-1,3}
-   \hat{n}_l \left( m_0^2\, \gamma_{-1,0,1,1}^{(0)} + 2 \gamma_{-1,0,3,1}^{(0)} \right)
-   \hat{M}\left( m_0^2\, \gamma_{-1,1,1,1}^{(0)} + 2\gamma_{-1,1,3,1}^{(0)} \right) \right] \;.
\end{eqnarray}
The coefficient coupling Eq.\ (\ref{rel_W}) to the time evolution of $l^\mu$ is
\begin{equation}
\bar{\tau}^W_l = \frac{1}{2} \left( m_0^2\, \hat{\mathcal{I}}_{00} - \hat{e} + 3 \hat{P}_l  \right)\;.
\end{equation}
The second-order transport coefficients are
\begin{eqnarray}
\bar{\beta}^{W}_{\Pi}  & = & \frac{3}{2m_0^2}\left( m_0^2 \gamma_{-2,0,2,0}^{(0)} + \gamma_{-2,0,4,0}^{(0)} \right)\;, \;\; 
\bar{\beta}^W_{\Pi \perp} =\bar{\beta}^{W}_{\Pi}- \frac{3}{2} \;, \;\; 
 \bar{\beta}^{W}_{\Pi l}   =  \frac{3}{2m_0^2}\left( m_0^2 \gamma_{-1,0,1,0}^{(0)} + 2 \gamma_{-1,0,3,0}^{(0)} \right) \;,   \\
\bar{\beta}^{W}_{n}  & = & \frac{1}{2}\left( m_0^2 \gamma_{-2,0,2,1}^{(0)} + \gamma_{-2,0,4,1}^{(0)} \right)\;, \;\;
 \bar{\beta}^{W}_{n l}   =   m_0^2 \gamma_{-1,0,1,1}^{(0)} + 2 \gamma_{-1,0,3,1}^{(0)}  \;,   \\
 \bar{\delta}^{W}_{V}  & =& \frac{1}{2} \left( m_0^2 \, \gamma_{-2,0,1,0}^{(1)} + \gamma_{-2,0,3,0}^{(1)} \right) 
\;, \;\;  \bar{\delta}^{W}_{W}   =   \frac{1}{2} \left( m_0^2 \, \gamma_{-2,0,1,1}^{(1)} +\gamma_{-2,0,3,1}^{(1)} \right) -1\;,  \\  
\bar{\epsilon}^{W}_{V} & = & \frac{1}{2} \left(m_0^2\, \gamma_{-1,0,0,0}^{(1)} + 4 \gamma_{-1,0,2,0}^{(1)} \right) \;, \;\;
\bar{\epsilon}^{W}_{W} =  \frac{1}{2}\left(m_0^2\, \gamma_{-1,0,0,1}^{(1)} + 4 \gamma_{-1,0,2,1}^{(1)} \right) \;, \\
\bar{\ell}^{W}_{\Pi} & = & \frac{3}{2 m_0^2} \left( m_0^2\, \gamma_{-1,0,1,0}^{(0)} + \gamma_{-1,0,3,0}^{(0)}
\right) \;, \;\;
\bar{\ell}^{W}_{n} = \frac{1}{2} \left( m_0^2\, \gamma_{-1,0,1,1}^{(0)} + \gamma_{-1,0,3,1}^{(0)}\right)\;, \\
\bar{\ell}^{W}_{V}  & = & \gamma_{-1,0,2,0}^{(1)} \; , \;\;
\bar{\ell}^{W}_{W} =  \gamma_{-1,0,2,1}^{(1)}  \;, \;\;
\bar{\ell}^{W}_{\pi} =  2 \gamma_{-1,0,1,0}^{(2)}  \;, \\
\bar{\tau}^{W}_{\Pi \alpha} & = & \frac{\partial \bar{\ell}^{W}_{\Pi}}{\partial \hat{\alpha}}\;, \;\;
\bar{\tau}^{W}_{\Pi u} = \frac{\partial \bar{\ell}^{W}_{\Pi}}{\partial \hat{\beta}_u}\;, \;\;
\bar{\tau}^{W}_{\Pi l}  = \frac{\partial \bar{\ell}^{W}_{\Pi}}{\partial \hat{\beta}_l} \;, \;\;
\bar{\tau}^{W}_{n \alpha}    =    \frac{\partial \bar{\ell}^{W}_{n}}{\partial \hat{\alpha}}\;, \;\;
 \bar{\tau}^{W}_{n u}   =  \frac{\partial \bar{\ell}^{W}_{n}}{\partial \hat{\beta}_u}\;, \;\;
 \bar{\tau}^{W}_{n l}   =  \frac{\partial \bar{\ell}^{W}_{n}}{\partial \hat{\beta}_l}\;, \\
\bar{\tau}^{W}_{V \alpha}  & = &   \frac{\partial \bar{\ell}^{W}_{V}}{\partial \hat{\alpha}}\;, \;\;
 \bar{\tau}^{W}_{V u}   =  \frac{\partial \bar{\ell}^{W}_{V}}{\partial \hat{\beta}_u}\;, \;\;
 \bar{\tau}^{W}_{V l}   =  \frac{\partial \bar{\ell}^{W}_{V}}{\partial \hat{\beta}_l}\;, \;\;
\bar{\tau}^{W}_{W \alpha}   =   \frac{\partial \bar{\ell}^{W}_{W} }{\partial \hat{\alpha}}\;, \; \;
\bar{\tau}^{W}_{W u}   =   \frac{\partial \bar{\ell}^{W}_{W} }{\partial \hat{\beta}_u}\;, \;\;
\bar{\tau}^{W}_{W l}   =   \frac{\partial \bar{\ell}^{W}_{W}}{\partial \hat{\beta}_l}\;,\\ 
\bar{\lambda}^{W}_{V u}  &= & \gamma_{-2,0,3,0}^{(1)}\;, \;\;
\bar{\lambda}^{W}_{W u}  =  2+\gamma_{-2,0,3,1}^{(1)}\;, \;\; 
\bar{\lambda}^{W}_{W \perp} =  \frac{1}{2} + \bar{\delta}^W_W  \;, \\
\bar{\lambda}^{W}_{V l}  & = &  \frac{1}{2}\left(m_0^2\, \gamma_{-1,0,0,0}^{(1)} + 3 \gamma_{-1,0,2,0}^{(1)} \right)\;, \;\;
\bar{\lambda}^{W}_{W l}  =   \frac{1}{2}\left(m_0^2\, \gamma_{-1,0,0,1}^{(1)} + 3 \gamma_{-1,0,2,1}^{(1)} \right)\;, \;\; \\
\bar{\tau}^{W}_{\pi \alpha} &  =  &  \frac{\partial \bar{\ell}^W_\pi}{\partial \hat{\alpha}} \; ,\;\;
\bar{\tau}^{W}_{\pi u}   =    \frac{\partial  \bar{\ell}^W_\pi}{\partial \hat{\beta}_u} \;, \;\;
\bar{\tau}^{W}_{\pi l}   =    \frac{\partial  \bar{\ell}^W_\pi}{\partial \hat{\beta}_l} \;,\;\; 
\bar{\lambda}^{W}_{\pi u}   = 2\, \gamma_{-2,0,2,0}^{(2)} \;, \;\;
\bar{\lambda}^{W}_{\pi \perp}  = \bar{\lambda}^{W}_{\pi u}-1\;.
\end{eqnarray}

\subsection{Shear tensor in the direction perpendicular to $u^\mu$ and $l^\mu$, Eq.\ (\ref{rel_pi})}
\label{app_pi}

The shear viscosity coefficients in Eq.\ (\ref{rel_pi}) are
\begin{eqnarray}
\bar{\eta} &=&  \frac{1}{8} \left[ 3 \hat{e} - 2 \hat{P}_l - m_0^4\, \hat{\mathcal{I}}_{-2,0}
- 2m_0^2 \left( \hat{\mathcal{I}}_{00} + \hat{\mathcal{I}}_{-2,2} \right) - \hat{\mathcal{I}}_{-2,4}
+ \hat{n}_l \left( m_0^4\, \gamma_{-2,0,0,1}^{(0)} + 2m_0^2 \, \gamma_{-2,0,2,1}^{(0)}
+ \gamma_{-2,0,4,1}^{(0)} \right)\right. \notag \\
&    & \hspace*{0.4cm} + \left. \hat{M} \left( m_0^4\, \gamma_{-2,1,0,1}^{(0)} + 2m_0^2 \, \gamma_{-2,1,2,1}^{(0)}
+ \gamma_{-2,1,4,1}^{(0)} \right) \right] 
\;, \\
\bar{\eta}_l &=&  \frac{1}{2} \left[ - m_0^2 \,\hat{\mathcal{I}}_{-1,1} - \hat{\mathcal{I}}_{-1,3} 
+ \hat{n}_l \left( m_0^2 \, \gamma_{-1,0,1,1}^{(0)}+ \gamma_{-1,0,3,1}^{(0)} \right)
+ \hat{M} \left( m_0^2 \, \gamma_{-1,1,1,1}^{(0)}+ \gamma_{-1,1,3,1}^{(0)} \right)\right] 
\;.
\end{eqnarray}
The second-order transport coefficients are
\begin{eqnarray}
\bar{\delta}^{\pi}_{\pi}  & =& \frac{3}{2} +  m_0^2 \, \gamma_{-2,0,0,0}^{(2)} + \gamma_{-2,0,2,0}^{(2)} \; , \; \;
\bar{\ell}^\pi_\pi = 2 \gamma_{-1,0,1,0}^{(2)} \;, \;\;  
\bar{\tau}^\pi_\pi = \frac{4}{3} \left( \bar{\delta}^\pi_\pi - \frac{1}{2} \right) \;, \\
\bar{\lambda}^{\pi}_{\Pi} & = & \frac{3}{4 m_0^2} \left[m_0^4\, \gamma_{-2,0,0,0}^{(0)} + 2m_0^2 \left(
\gamma_{-2,0,2,0}^{(0)}+1 \right) + \gamma_{-2,0,4,0}^{(0)}  \right] \;, \\
\bar{\lambda}^{\pi}_{n} & = & \frac{1}{4} \left(m_0^4\, \gamma_{-2,0,0,1}^{(0)} + 2m_0^2 \,
\gamma_{-2,0,2,1}^{(0)} + \gamma_{-2,0,4,1}^{(0)}  \right) \;, \\
\bar{\lambda}^{\pi}_{\Pi l} & = & \frac{3}{m_0^2} \left(m_0^2 \, \gamma_{-1,0,1,0}^{(0)} + \gamma_{-1,0,3,0}^{(0)}  
\right) \;, \;\;
\bar{\lambda}^{\pi}_{n l}  =  m_0^2 \,\gamma_{-1,0,1,1}^{(0)} + \gamma_{-1,0,3,1}^{(0)} \;, \\
\bar{\ell}^{\pi}_{V}  & = & \frac{1}{2} \left( m_0^2\, \gamma_{-1,0,0,0}^{(1)} + \gamma_{-1,0,2,0}^{(1)} \right) \; , \;\;
\bar{\ell}^{\pi}_{W} = \frac{1}{2} \left( m_0^2\, \gamma_{-1,0,0,1}^{(1)} + \gamma_{-1,0,2,1}^{(1)} \right)    \;, \\
\bar{\tau}^{\pi}_{V \alpha}  & = &   \frac{\partial \bar{\ell}^{\pi}_{V}}{\partial \hat{\alpha}}\;, \;\;
 \bar{\tau}^{\pi}_{V u}   =  \frac{\partial \bar{\ell}^{\pi}_{V}}{\partial \hat{\beta}_u}\;, \;\;
 \bar{\tau}^{\pi}_{V l}   =  \frac{\partial \bar{\ell}^{\pi}_{V}}{\partial \hat{\beta}_l}\;, \;\;
\bar{\tau}^{\pi}_{W \alpha}   =   \frac{\partial \bar{\ell}^{\pi}_{W} }{\partial \hat{\alpha}}\;, \; \;
\bar{\tau}^{\pi}_{W u}   =   \frac{\partial \bar{\ell}^{\pi}_{W} }{\partial \hat{\beta}_u}\;, \;\;
\bar{\tau}^{\pi}_{W l}   =   \frac{\partial \bar{\ell}^{\pi}_{W}}{\partial \hat{\beta}_l}\;,\\ 
\bar{\tau}^{\pi}_{\pi \alpha} &  =  &  \frac{\partial \bar{\ell}^\pi_\pi}{\partial \hat{\alpha}} \; ,\;\;
\bar{\tau}^{\pi}_{\pi u}   =    \frac{\partial  \bar{\ell}^\pi_\pi}{\partial \hat{\beta}_u} \;, \;\;
\bar{\tau}^{\pi}_{\pi l}   =    \frac{\partial  \bar{\ell}^\pi_\pi}{\partial \hat{\beta}_l} \;,\;\;
\bar{\lambda}^\pi_\pi = 2 \gamma_{-2,0,2,0}^{(2)} -1\;, \\
\bar{\lambda}^{\pi}_{V}  &= & \frac{1}{2} \left( m_0^2\, \gamma_{-2,0,1,0}^{(1)} + \gamma_{-2,0,3,0}^{(1)} \right) \;, \;\;
\bar{\lambda}^{\pi}_{W u}  = \frac{1}{2} \left( m_0^2\, \gamma_{-2,0,1,1}^{(1)} + \gamma_{-2,0,3,1}^{(1)} \right) 
- \frac{3}{2} \;, \;\; 
\bar{\lambda}^{\pi}_{W \perp} = \bar{\lambda}^{\pi}_{W u} + 2   \;, \\
\bar{\lambda}^{\pi}_{V l}  & = &  \frac{1}{2}\left(m_0^2\, \gamma_{-1,0,0,0}^{(1)} + 5 \gamma_{-1,0,2,0}^{(1)} \right)\;, \;\;
\bar{\lambda}^{\pi}_{W l}  =   \frac{1}{2}\left(m_0^2\, \gamma_{-1,0,0,1}^{(1)} + 5 \gamma_{-1,0,2,1}^{(1)} \right)\;.
\end{eqnarray}


\end{document}